\documentclass[
  acmsmall,
  screen,
  nonacm,
  authorversion
]{acmart}

\usepackage{
  bbm,
  stmaryrd
}
\usepackage{
  multirow,
  longtable,
  wrapfig
}
\usepackage{cleveref}
\usepackage{adjustbox}
\usepackage[inline]{enumitem}
\usepackage{microtype}
\usepackage{xspace}

\DisableLigatures[q]{encoding = T1, shape = sc}

\usepackage{listings}

\definecolor{clrRosewater}{HTML}{DC8A78}
\definecolor{clrFlamingo}{HTML}{DD7878}
\definecolor{clrPink}{HTML}{EA76CB}
\definecolor{clrMauve}{HTML}{8839EF}
\definecolor{clrRed}{HTML}{D20F39}
\definecolor{clrMaroon}{HTML}{E64553}
\definecolor{clrPeach}{HTML}{FE640B}
\definecolor{clrYellow}{HTML}{DF8E1D}
\definecolor{clrGreen}{HTML}{40A02B}
\definecolor{clrTeal}{HTML}{179299}
\definecolor{clrSky}{HTML}{04A5E5}
\definecolor{clrSapphire}{HTML}{209FB5}
\definecolor{clrBlue}{HTML}{1E66F5}
\definecolor{clrLavender}{HTML}{7287FD}
\definecolor{clrText}{HTML}{4C4F69}
\definecolor{clrSubtext}{HTML}{6C6F85}
\definecolor{clrOverlay}{HTML}{9CA0B0}
\definecolor{clrSurface}{HTML}{CCD0DA}
\definecolor{clrBase}{HTML}{EFF1F5}

\lstdefinestyle{haskell-catppuccin}{
  language=Haskell,
  basicstyle=\ttfamily\small\color{clrText},
  keywordstyle=\color{clrMauve},
  keywords={
    let,in,case,of,where,data,type,newtype,if,then,else,
    class,instance,do,import,module
  },
  emphstyle=\color{clrYellow},
  emph={
    Int,Integer,Float,Double,Bool,Char,String,
    Maybe,Either,IO,Bad,Void,Typ,Val,Ix,Pos,Nat,
    Rou,Tree
  },
  emph={[2]
    MkBad,Just,Nothing,Left,Right,TyIota,TyArr,
    VIota,VArr,MkPos,Next,Over,Leaf,Node,TArr,TIota
  },
  emphstyle={[2]\color{clrPeach}},
  emph={[3]
    reflect,stratified,measure,lazy
  },
  emphstyle={[3]\color{clrMaroon}},
  emph={[4]
    prop
  },
  emphstyle={[4]\color{clrPink}},
  commentstyle=\color{clrOverlay},
  stringstyle=\color{clrGreen},
  numbers=none,
  showstringspaces=false,
  breaklines=true,
  columns=fullflexible,
  keepspaces=true,
  literate=
    {\{-@}{{{\color{clrOverlay}\{-@}}}3
    {@-\}}{{{\color{clrOverlay}@-\}}}}3
    {->}{{{\color{clrSky}$\rightarrow$}}}2
    {::}{{{\color{clrSky}::}}}2,
  moredelim=**[s][\color{clrText}]{\{-@}{@-\}},
  xleftmargin=4em
}

\newcommand\semtypem{\ensuremath{{\color{\semcolor}s_\tyalt}}}
\newcommand\semdeclm{\ensuremath{{\color{\sembotcolor}s_D}}}
\newcommand\semidxm{\ensuremath{{\color{\sembotcolor}s_\ty}}}

\newcommand\Fstar{\ensuremath{F^\star}\xspace}
\newcommand\coreFstar{\ensuremath{\mu \Fstar}\xspace}
\newcommand\coreLH{\ensuremath{\lambda_{RF}}\xspace}

\newcommand\sparagraph[1]{\paragraph{#1\textcolor{white}{.}\hspace{-0.3em}}}

\NewDocumentCommand{\Crefsub}{m m}{
  \Cref{#1}~\ref{#1:#2}
}

\NewDocumentCommand{\NewMeta}{m >{\SplitList{,}}m}{
  \edef\suffix{}
  \edef\meta{#1}
  \DeclareDocumentCommand{\declaremeta}{m}{
    \expandafter\NewDocumentCommand\csname
      \meta\suffix
    \endcsname{}{
      \ensuremath{##1}
    }
    \edef\suffix{\suffix alt}
  }
  \ProcessList{#2}{\declaremeta}
}

\NewDocumentCommand{\MathSym}{m m}{
  \expandafter\NewDocumentCommand\csname #1\endcsname{}{
    \ensuremath{#2}
  }
}

\NewDocumentCommand{\MathCmd}{m m m}{
  \expandafter\NewDocumentCommand\csname #1\endcsname{#2}{
    \ensuremath{#3}
  }
}

\NewDocumentCommand{\samewidth}{m m}{
  \makebox[0pt][l]{\ensuremath{#1}}\phantom{\ensuremath{#2}}
}

\MathSym{Bigcap}{
  \mathop{\scalebox{2}{\ensuremath{\bigcap}}}
}

\MathSym{Bigcup}{
  \mathop{\scalebox{2}{\ensuremath{\bigcup}}}
}

\MathSym{bots}{
  _{\negspace{1}\scalebox{0.6}{$\bot$}}
}

\MathSym{defeq}{\triangleq}
\MathSym{cons}{,\,}
\MathCmd{subst}{m m}{
  \left[#1 / #2 \right]
}

\MathCmd{ioverline}{s m m}{
  \IfBooleanTF{#1}{
    \smash{\overline{#3}^{\, #2}}
  }{
    \overline{#3}^{\, #2}
  }
}

\MathCmd{iunderline}{s m m}{
  \IfBooleanTF{#1}{
    \smash{\underline{#3}_{\, #2}}
  }{
    \underline{#3}_{\, #2}
  }
}

\MathCmd{iforall}{m}{
  \forall #1.\,
}

\MathCmd{fix}{m}{
  \mathsf{fix}\,#1\,
}
\MathCmd{lfp}{m}{
  \mathsf{lfp}\,#1\,
}
\MathCmd{gfp}{m}{
  \mathsf{gfp}\,#1\,
}

\NewDocumentCommand{\semfull}{m m m m m}{
  \vcenter{\hbox to \textwidth{%
    $\setbox0=\hbox{$\displaystyle #5$}%
      {\color{#1}%
        \vrule height\ht0 depth\dp0 width0.4pt
        \kern0.6pt
        \color{#2}%
        \vrule height\ht0 depth\dp0 width0.4pt}$\hfil
    $\displaystyle #5$\hfil
    $\setbox0=\hbox{$\displaystyle #5$}%
      {\color{#3}%
        \vrule height\ht0 depth\dp0 width0.4pt
        \kern0.6pt
        \color{#4}%
        \vrule height\ht0 depth\dp0 width0.4pt}$%
  }}%
}

\NewDocumentCommand{\propDefinition}{m}{
  \semfull{\semprpcolor}{\semprpcolor}{\semprpcolor}{\semprpcolor}{#1}
}

\NewDocumentCommand{\botDefinition}{m}{
  \semfull{\sembotcolor}{\sembotcolor}{\sembotcolor}{\sembotcolor}{#1}
}

\NewDocumentCommand{\runtimeDefinition}{m}{
  \semfull{\semtmcolor}{\semtmcolor}{\semtmcolor}{\semtmcolor}{#1}
}

\NewDocumentCommand{\refinedDefinition}{m}{
  \semfull{\semcolor}{\semcolor}{\semcolor}{\semcolor}{#1}
}

\NewDocumentCommand{\redBotDefinition}{m}{
  \semfull{\sembotcolor}{\semcolor}{\semcolor}{\sembotcolor}{#1}
}

\MathCmd{pow}{m}{
  \wp{\left(#1\right)}
}

\MathCmd{pair}{s m m}{
  \IfBooleanF{#1}{
    \left
  }
  \langle #2 \cons #3
  \IfBooleanF{#1}{
    \right
  }
  \rangle
}

\MathCmd{vpair}{m m}{
  \left\langle
  \begin{array}{c}
    #1, \\
    #2
  \end{array}
  \right\rangle
}

\MathSym{lub}{\sqcup}
\MathSym{Lub}{\bigsqcup}
\MathSym{glb}{\sqcap}
\MathSym{Glb}{\bigsqcap}

\NewDocumentCommand{\ie}{}{\textit{i.e.}\xspace}
\NewDocumentCommand{\eg}{}{\textit{e.g.}\xspace}
\NewDocumentCommand{\ih}{}{I. H.}
\NewDocumentCommand{\ihs}{}{I. Hs.}
\NewDocumentCommand{\corelan}{}{CIRT\xspace}

\MathCmd{dpair}{s m m}{
  \IfBooleanF{#1}{
    \left
  }
  \langle #2 \cons #3
  \IfBooleanF{#1}{
    \right
  }
  \rangle
}

\MathSym{iso}{\cong}
\MathSym{suniv}{\mathcal{U}}
\MathCmd{diag}{m}{\Delta\left(#1\right)}
\MathSym{app}{\cdot}

\NewDocumentEnvironment{setbuild}{m O{c}}{
  \setlength\arraycolsep{0pt}
  \left\{\,#1\,\middle|\,\begin{array}{#2}
}{
  \end{array}\,\right\}
}

\MathSym{snat}{\mathbbm{N}}
\MathSym{snatbot}{\mathbbm{N}\bots}
\MathSym{sint}{\mathbbm{Z}}
\MathSym{sintbot}{\mathbbm{Z}\bots}
\MathSym{ssing}{\mathbbm{1}}
\MathSym{ssingbot}{\mathbbm{1}\bots}
\MathSym{sunit}{\mathbbm{\star}}
\MathSym{sreal}{\mathbbm{R}}
\MathSym{sbool}{\mathbbm{2}}
\MathSym{sboolbot}{\mathbbm{2}\bots}
\MathSym{strue}{\mathit{t\!t}}
\MathSym{sfalse}{\mathit{f\!\!f}}
\MathSym{slabel}{\mathcal{L}}
\MathSym{slabelbot}{\mathcal{L}\bots}

\MathCmd{per}{m}{\mathcal{PER}\left(#1\right)}
\MathCmd{pper}{m}{\mathcal{CUPER}\left(#1\right)}
\MathCmd{dom}{s m}{
  \mathsf{dom}
  \IfBooleanF{#1}{\left}
  (#2
  \IfBooleanF{#1}{\right}
  )
}

\MathSym{dinf}{\mathcal{D}}
\MathCmd{dfun}{m m}{\left[ #1 \to #2 \right]}
\MathSym{dsum}{\oplus}
\MathCmd{demb}{m m}{i_{#1}\left(#2\right)}
\MathCmd{dret}{m m}{r_{#1}\left(#2\right)}

\MathCmd{inference}{m m o}{
  \frac{\begin{array}{c}
    #1
  \end{array}}{\begin{array}{c}
    #2
  \end{array}}
  \IfValueT{#3}{
    \;\text{\scshape{#3}}
  }
}

\MathCmd{ann}{m m o}{
  #1%
  \text{:}%
  \IfValueT{#3}{^#3}%
  #2
}
\def\annsz{0.45}
\MathCmd{annty}{m m}{#1 \smash{:_{\scalebox{\annsz}{T}}} #2}
\MathCmd{annkd}{m m}{#1 \smash{:_{\scalebox{\annsz}{K}}} #2}

\MathSym{ctxemp}{\cdot}
\NewMeta{ctx}{\Gamma, \Delta, \Theta}

\MathCmd{tynamedec}{m}{
  \mathsf{#1}
}
\MathSym{tynat}{\tynamedec{Nat}}
\MathSym{tyint}{\tynamedec{Int}}
\MathSym{tyunit}{\tynamedec{Unit}}
\MathSym{tybool}{\tynamedec{Bool}}
\MathSym{tybase}{\mathcal{B}}
\MathCmd{tyref}{s m m m}{
  \IfBooleanTF{#1}{
    \begin{setbuild}{\ann{#2}{#3}}%
      #3%
    \end{setbuild}
  }{
    \{\ann{#2}{#3}\,|\,#4\}
  }
}
\MathCmd{tyfun}{s o m m}{
  \IfValueTF{#2}{
    \IfBooleanT{#1}{\left(}
    \ann{#2}{#3}
    \IfBooleanT{#1}{\right)}
  }{
    #3
  }
  \to #4
}
\MathCmd{tyall}{s m m m}{
  \forall
  \IfBooleanT{#1}{\left(}
  \ann{#2}{#3}
  \IfBooleanT{#1}{\right)}
  .\,#4
}

\MathSym{kdtype}{\tynamedec{Ty}}
\MathCmd{kdfun}{s o m m}{
  \IfValueTF{#2}{
    \IfBooleanT{#1}{\left(}
    \ann{#2}{#3}
    \IfBooleanT{#1}{\right)}
  }{
    #3
  }
  \to #4
}
\MathSym{kdapp}{\cdot}

\MathSym{kind}{\mathsf{Kd}}

\MathSym{tmtrue}{\mathit{tt}}
\MathSym{tmfalse}{\mathit{ff}}
\MathSym{tmunit}{\mathit{unit}}
\MathCmd{tmlam}{m m}{\lambda #1.\,#2}
\MathCmd{tmLam}{m m}{\Lambda #1.\,#2}
\MathSym{tmapp}{\,\,}
\MathSym{tmApp}{\cdot}
\MathCmd{tmif}{m m m}{
  \mathsf{if}\:#1\:\mathsf{then}\:#2\:\mathsf{else}\:#3
}
\MathSym{tmle}{<}
\MathCmd{tmelim}{m m}{
  \mathit{elim}_{#1}
  \left( #2 \right)
}

\MathCmd{tyel}{m m m m}{
  \mathsf{El}_{#1;\,#2}\left(#3 \cons #4\right)
}

\MathCmd{tyelp}{m m m m}{
  \mathsf{El}^+_{#1;\,#2}\left(#3 \cons #4\right)
}

\MathCmd{tyelm}{m m m m}{
  \mathsf{El}^-_{#1;\,#2}\left(#3 \cons #4\right)
}

\MathCmd{tmel}{m m m m m}{
  \mathsf{tmEl}_{#1;\,#5}\left(#2 \cons #3 \cons #4\right)
}

\MathCmd{tmelp}{m m m m m}{
  \mathsf{tmEl}^+_{#1}\left(#2 \cons #3 \cons #4 \cons #5\right)
}

\MathCmd{tmeln}{m m m m m}{
  \mathsf{tmEl}^-_{#1}\left(#2 \cons #3 \cons #4 \cons #5\right)
}

\MathSym{propeq}{\equiv}
\MathSym{propand}{\wedge}
\MathSym{propnot}{\lnot}
\MathSym{propimpl}{\implies}

\MathCmd{arity}{m}{
  \mathsf{arity}\left(#1\right)
}

\MathCmd{defind}{m m m m}{
  \mu\, #1^{
    \left\langle #3 \right\rangle
  } \left\{\, #4 \,\right\}
}
\MathSym{varpos}{+}
\MathSym{varstrat}{\stratsmall}
\newcommand{\varstratR}{m}

\MathCmd{positive}{m m}{
  \mathsf{pol}_{#1}\left(#2\right)
}
\MathSym{pospos}{+}
\MathSym{posneg}{-}
\MathSym{posboth}{\pm}
\MathSym{posnone}{\circ}
\MathSym{posleq}{\leq}
\MathSym{poslub}{\sqcup}
\MathCmd{posconv}{m}{
  \overline{#1}
}

\MathCmd{stratifiedLong}{m m m m m m m}{
  \mathsf{stratifiable}_{#5;\, #1;\, #6}\left(#2, #3, #4, #7\right)
}

\MathCmd{stratified}{m m m m m m m}{
  \mathsf{strat}_{#2;\, #4;\, #5;\, #1;\, #6}\left(#3, #7\right)
}
\MathSym{stratsmall}{\sqsubset}

\MathCmd{rank}{m}{
  rank\left(#1\right)
}
\MathSym{osup}{\mathsf{sup}\;}
\MathSym{osuc}{\mathsf{suc}\;}

\NewMeta{tm}{t, u, v, m}
\NewMeta{tmrec}{r}
\NewMeta{var}{x, y, z, e}
\NewMeta{prop}{\phi, \psi, \chi}
\NewMeta{ty}{\tau, \sigma, \rho, \pi}
\NewMeta{kd}{\kappa, \mu, \nu}
\NewMeta{lvl}{i, j, k}
\NewMeta{tyctor}{K, L, M}
\NewMeta{ctor}{C, D, E	}
\NewMeta{variance}{\nu, \xi, \zeta}

\def\semcolor{clrMauve}
\def\sembotcolor{clrMaroon}
\def\semtmcolor{clrGreen}
\def\semprpcolor{clrBlue}
\def\semstupidcolor{clrYellow}

\MathCmd{sem}{s m o}{
  \mathcolor{\semcolor}{%
    \IfBooleanF{#1}{\left}
    \lbrack%
    \mkern-5.5mu
    \IfBooleanF{#1}{\left}
    \lbrack%
  }%
      #2%
  \mathcolor{\semcolor}{%
    \IfBooleanF{#1}{\right}
    \rbrack%
    \mkern-4.9mu
    \IfBooleanF{#1}{\right}
    \rbrack%
  }%
  \IfValueT{#3}{%
    \IfBooleanF{#1}{\left}
    (#3
    \IfBooleanF{#1}{\right}
    )
  }
}

\MathCmd{sembot}{s m o}{
  \mathcolor{\sembotcolor}{%
    \IfBooleanF{#1}{\left}
    \lfloor
    \mkern-6.4mu
    \IfBooleanF{#1}{\left}
    \lfloor
  }%
      #2%
  \mathcolor{\sembotcolor}{
    \IfBooleanF{#1}{\right}
    \rfloor
    \mkern-6mu
    \IfBooleanF{#1}{\right}
    \rfloor
  }%
  \IfValueT{#3}{%
    \IfBooleanF{#1}{\left}
    (#3
    \IfBooleanF{#1}{\right}
    )
  }
}

\MathCmd{semtm}{s m o}{
  \mathcolor{\semtmcolor}{
    \IfBooleanF{#1}{\left}
    (
    \mkern-5.7mu
    \IfBooleanF{#1}{\left}
    (
  }%
      #2%
  \mathcolor{\semtmcolor}{
    \IfBooleanF{#1}{\right}
    )
    \mkern-5mu
    \IfBooleanF{#1}{\right}
    )
  }%
  \IfValueT{#3}{%
    \IfBooleanF{#1}{\left}
    (
    #3
    \IfBooleanF{#1}{\right}
    )
  }
}

\MathCmd{semprp}{s m o}{
  \mathcolor{\semprpcolor}{
    \IfBooleanF{#1}{\left}
    \{
    \!\!
    \IfBooleanF{#1}{\left}
    \{
  }%
      #2%
  \mathcolor{\semprpcolor}{
    \IfBooleanF{#1}{\right}
    \}
    \!\!
    \IfBooleanF{#1}{\right}
    \}
  }%
  \IfValueT{#3}{%
    \IfBooleanF{#1}{\left}
    (#3
    \IfBooleanF{#1}{\right}
    )
  }
}

\MathCmd{unrefass}{m m}{
  \mathcolor{\semstupidcolor}{
    \left|
    \!
    \left|
  }%
      #1%
  \mathcolor{\semstupidcolor}{
    \right|
    \!
    \right|
  }%
  \left(#2\right)
}

\MathCmd{projctx}{m m}{
  \iota_{#1}\left(#2\right)
}
\MathCmd{projty}{m m m m}{
  \pi_{#1 ;\; #3}\left(#2 \cons #4 \right)
}
\MathCmd{projrun}{m m}{
  \rho_{#1}\left(#2\right)
}
\MathCmd{runctx}{m}{
  \mathcolor{\semtmcolor}{%
    \left(%
    \mkern-5.7mu
    \left(%
  }%
      #1%
  \mathcolor{\semtmcolor}{%
    \right)%
    \mkern-5mu
    \right)%
  }%
}

\NewMeta{sctx}{\gamma, \delta, \epsilon, \theta}
\NewMeta{sfun}{f, g, h}
\NewMeta{sval}{v, w, q, n}
\NewMeta{srec}{r}
\NewMeta{sty}{P, Q, R}
\NewMeta{skd}{\mathit{P}, \mathit{Q}, \mathit{R}}
\NewMeta{tyop}{\mathcal{F}, \mathcal{G}, \mathcal{H}}

\MathCmd{wfctx}{m}{\vdash #1 \; \mathsf{ctx}
}
\MathCmd{wftm}{m m m}{#1 \vdash \ann{#2}{#3}}
\MathCmd{wfty}{m m}{\wftm{#1}{#2}{\kdtype}}
\MathCmd{wfkd}{m m}{\wftm{#1}{#2}{\kind}}
\MathCmd{wfprop}{m m}{#1 \vdash #2 \; \mathsf{prp}}
\MathCmd{wfsub}{m m m}{#1 \vdash #2 \preceq #3}
\MathCmd{wfsmt}{m m}{#1 \models #2}

\NewDocumentCommand{\DeclareRule}{m m m m}{%
  \expandafter\NewDocumentCommand\csname #1\endcsname{m}{%
    \IfStrEq{##1}{full}{%
      \ensuremath{\inference{#3}{#4}[#2]}%
    }{%
      \IfStrEq{##1}{no label}{%
        \ensuremath{\inference{#3}{#4}}%
      }{%
        \IfStrEq{##1}{label}{%
          #2%
        }{%
          \ensuremath{#4}%
        }%
      }%
    }%
  }%
}

\MathCmd{decl}{O{\tyctor}}{
  \defind{#1}{
    \kdfun{\ty}{\kdtype}
  }{\variance}{
    \iunderline{i}{\ann{\ctor}{
      \tyfun{\left(
        \ioverline*{j}{\ann{\varalt}{\tyalt}}
      \right)}{
        #1 \tmApp \ioverline*{k}{\tm}
      }
    }}
  }
}

\MathCmd{declpos}{O{\tyctor}}{
  \defind{#1}{
    \kdfun{\ty}{\kdtype}
  }{\varpos}{
    \iunderline{i}{\ann{\ctor}{
      \tyfun{\left(
        \ioverline*{j}{\ann{\varalt}{\tyalt}}
      \right)}{
        #1 \tmApp \ioverline*{k}{\tm}
      }
    }}
  }
}

\MathCmd{declstrat}{O{\tyctor}}{
  \defind{#1}{
    \kdfun{\ty}{\kdtype}
  }{m}{
    \iunderline{i}{\ann{\ctor}{
      \tyfun{\left(
        \ioverline*{j}{\ann{\varalt}{\tyalt}}
      \right)}{
        #1 \tmApp \ioverline*{k}{\tm}
      }
    }}
  }
}

\DeclareRule{RuleCtxEmpty}{Ctx-Empty}{
  \
}{
  \wfctx{\ctxemp}
}
\DeclareRule{RuleCtxTy}{Ctx-Ty}{
  \wfctx{\ctx} \quad
  \wfty{\ctx}{\ty}
}{
  \wfctx{\ctx \cons \annty{\var}{\ty}}
}
\DeclareRule{RuleCtxKd}{Ctx-Kd}{
  \wfctx{\ctx} \quad
  \wfkd{\ctx}{\kd}
}{
  \wfctx{\ctx \cons \annkd{\var}{\kd}}
}

\DeclareRule{RuleKdType}{Kd-Type}{
  \wfctx{\ctx}
}{
  \wfkd{\ctx}{\kdtype}
}
\DeclareRule{RuleKdFun}{Kd-Fun}{
  \wfty{\ctx}{\ty} \quad
  \wfkd{\ctx \cons \annty{\var}{\ty}}{\kd}
}{
  \wfkd{\ctx}{\kdfun*[{\var}]{\ty}{\kd}}
}

\DeclareRule{RuleTyNat}{Ty-Nat}{
  \wfctx{\ctx}
}{
  \wfty{\ctx}{\tynat}
}
\DeclareRule{RuleTyBool}{Ty-Bool}{
  \wfctx{\ctx}
}{
  \wfty{\ctx}{\tybool}
}
\DeclareRule{RuleTyFun}{Ty-Fun}{
  \wfty{\ctx}{\ty} \quad
  \wfty{\ctx \cons \annty{\var}{\ty}}{\tyalt}
}{
  \wfty{\ctx}{\tyfun*[{\var}]{\ty}{\tyalt}}
}
\DeclareRule{RuleTyAll}{Ty-All}{
  \wfkd{\ctx}{\kd} \quad
  \wfty{\ctx \cons \annkd{\var}{\kd}}{\ty}
}{
  \wfty{\ctx}{\tyall{\var}{\kd}{\ty}}
}
\DeclareRule{RuleTyVar}{Ty-Var}{
  \wfctx{\ctx} \quad
  \ctx \ni \annkd{\var}{\kd}
}{
  \wftm{\ctx}{\var}{\kd}
}
\DeclareRule{RuleTyRef}{Ty-Ref}{
  \wfty{\ctx}{\ty} \quad
  \wfprop{\ctx \cons \annty{\var}{\ty}}{\prop}
}{
  \wfty{\ctx}{\tyref{\var}{\ty}{\prop}}
}
\DeclareRule{RuleTyApp}{Ty-App}{
  \wftm{\ctx}{\tm}{\ty} \quad
  \wftm{\ctx}{\tyalt}{\kdfun*[{\var}]{\ty}{\kd}}
}{
  \wftm{\ctx}{\tyalt \tmapp \tm}{\kd\subst{\tm}{\var}}
}

\newcommand\propEqName{\textsc{Prop-Eq}}
\DeclareRule{RulePropEq}{\propEqName}{
  \wftm{\ctx}{\tm}{\ty} \quad
  \wftm{\ctx}{\tmalt}{\ty}
}{
  \wfprop{\ctx}{\tm \propeq_\ty \tmalt}
}
\DeclareRule{RulePropAnd}{Prop-And}{
  \wfprop{\ctx}{\prop} \quad
  \wfprop{\ctx}{\propalt}
}{
  \wfprop{\ctx}{\prop \propand \propalt}
}
\DeclareRule{RulePropNot}{Prop-Not}{
  \wfprop{\ctx}{\prop}
}{
  \wfprop{\ctx}{\propnot \prop}
}

\DeclareRule{RuleTmNat}{Tm-Nat}{
  \wfctx{\ctx} \quad n \in \snat
}{
  \wftm{\ctx}{n}{\tynat}
}
\DeclareRule{RuleTmBool}{Tm-Bool}{
  \wfctx{\ctx} \quad b \in \sbool
}{
  \wftm{\ctx}{b}{\tybool}
}
\DeclareRule{RuleTmVar}{Tm-Var}{
  \wfctx{\ctx} \quad
  \ctx \ni \annty{\var}{\ty}
}{
  \wftm{\ctx}{\var}{\ty}
}
\DeclareRule{RuleTmLam}{Tm-Lam}{
  \wftm{\ctx \cons \annty{\var}{\ty}}{\tm}{\tyalt}
}{
  \wftm{\ctx}{\tmlam{\var}{\tm}}{\tyfun*[{\var}]{\ty}{\tyalt}}
}
\DeclareRule{RuleTmApp}{Tm-App}{
  \wftm{\ctx}{\tm}{\tyfun*[{\var}]{\ty}{\tyalt}} \quad
  \wftm{\ctx}{\tmalt}{\ty}
}{
  \wftm{\ctx}{\tm \app \tmalt}{\tyalt\subst{\tmalt}{\var}}
}
\DeclareRule{RuleTmTyLam}{Tm-TyLam}{
  \wftm{\ctx \cons \annkd{\var}{\kd}}{\tm}{\ty}
}{
  \wftm{\ctx}{\tmLam{\var}{\tm}}{\tyall{\var}{\kd}{\ty}}
}
\DeclareRule{RuleTmTyApp}{Tm-TyApp}{
  \wftm{\ctx}{\tm}{\tyall{\var}{\kd}{\ty}} \quad
  \wftm{\ctx}{\tyalt}{\kd}
}{
  \wftm{\ctx}{\tm \tmapp \tyalt}{\ty\subst{\tyalt}{\var}}
}
\DeclareRule{RuleTmRefl}{Tm-Refl}{
  \wftm{\ctx}{\tm}{\ty}
}{
  \wftm{\ctx}{\tm}{\tyref{\var}{\ty}{\var \propeq_\ty \tm}}
}
\newcommand\tmSub{Tm-Sub}
\DeclareRule{RuleTmSub}{\tmSub}{
  \wftm{\ctx}{\tm}{\ty} \quad
  \wfsub{\ctx}{\ty}{\tyalt} \quad
  \wfty{\ctx}{\tyalt}
}{
  \wftm{\ctx}{\tm}{\tyalt}
}

\DeclareRule{RuleSubRefl}{Sub-Refl}{
  \
}{
  \wfsub{\ctx}{\ty}{\ty}
}
\DeclareRule{RuleSubTop}{Sub-Top}{
  \
}{
  \wfsub{\ctx}{\ty}{\tyref{\var}{\ty}{\tmtrue \propeq_\tybool \tmtrue}}
}
\DeclareRule{RuleSubRefine}{Sub-Refine}{
  \wfsub{\ctx}{\ty}{\tyalt}
}{
  \wfsub{\ctx}{
    \tyref{\var}{\ty}{\prop}
  }{
    \tyref{\var}{\tyalt}{\prop}
  }
}
\newcommand\subFun{Sub-Fun}
\DeclareRule{RuleSubFun}{\subFun}{
  \wfsub{\ctx}{\tyalt}{\ty} \quad
  \wfsub{\ctx \cons \annty{\var}{\tyalt}}{\tyaltalt}{\tyaltaltalt}
}{
  \wfsub{\ctx}{
    \tyfun*[{\var}]{\ty}{\tyaltalt}
  }{
    \tyfun*[{\var}]{\tyalt}{\tyaltaltalt}
  }
}
\DeclareRule{RuleSubFlatten}{Sub-Flat}{
  \
}{
  \wfsub{\ctx}{
    \tyref{\var}{\tyref{\varalt}{\ty}{\propalt}}{\prop}
  }{
    \tyref{\varalt}{\ty}{\propalt \propand \prop\subst{\varalt}{\var}}
  }
}
\DeclareRule{RuleSubForget}{Sub-Forget}{
  \
}{
  \wfsub{\ctx}{\tyref{\var}{\ty}{\prop}}{\ty}
}
\DeclareRule{RuleSubTrans}{Sub-Trans}{
  \wfsub{\ctx}{\ty}{\tyalt} \quad
  \wfty{\ctx}{\tyalt} \quad
  \wfsub{\ctx}{\tyalt}{\tyaltalt}
}{
  \wfsub{\ctx}{\ty}{\tyaltalt}
}
\DeclareRule{RuleSubAll}{Sub-All}{
  \wfsub{\ctx \cons \annkd{\var}{\kd}}{\ty}{\tyalt}
}{
  \wfsub{\ctx}{\tyall{\var}{\kd}{\ty}}{\tyall{\var}{\kd}{\tyalt}}
}
\DeclareRule{RuleSubImpl}{Sub-Impl}{
  \wfsmt{\ctx \cons \annty{\var}{\ty}}{\prop \propimpl \propalt}
}{
  \wfsub{\ctx}{
    \tyref{\var}{\ty}{\prop}
  }{
    \tyref{\var}{\ty}{\propalt}
  }
}

\DeclareRule{RuleKdIndPos}{Ty-IndPos}{
  \wfkd{\ctx}{\kdfun{\ioverline{k}{\ty}}{\kdtype}} \quad
  \iforall{l, j} \positive{\tyctor}{\ty_{l, j}} \posleq \pospos \\
  \iforall{l} \wfty{
    \ctx \cons \annkd{\tyctor}{\kdfun{\ioverline{k}{\ty}}{\kdtype}}
  }{
    \tyfun{\ioverline{j}{\ty_l}}{\tyctor \tmapp \ioverline{k}{\tm_l}}
  }
}{
  \wftm{\ctx}{
    \defind{\tyctor}{
      \kdfun{\ioverline{k}{\ty}}{\kdtype}
    }{\varpos}{
      \iunderline{l}{\ann{\ctor}{
        \tyfun{\ioverline{j}{\ty}}{\tyctor \tmapp \ioverline{k}{\tm}}
      }}
    }
  }{
    \kdfun{\ioverline{k}{\ty}}{\kdtype}
  }
}

\DeclareRule{RuleSubIx}{Sub-Ix}{
  \iforall{k} \wfsmt{\ctx}{\tm_k \propeq_{\ty_k} \tmalt_k}
}{
  \wfsub{\ctx}{
    \ty \tmapp \ioverline{k}{\tm}
  }{
    \ty \tmapp \ioverline{k}{\tmalt}
  }
}

\DeclareRule{RuleKdIndStrat}{Ty-IndStrat}{
  \wfkd{\ctx}{\kdfun{\ioverline{k}{\ann{\var}{\ty}}}{\kdtype}} \quad
  \wftm{\ctx}{\varstratR}{\tyfun{\ioverline{k}{\ann{\var}{\ty}}}{\tynat}} \\
  \iforall{l, j}
  \stratifiedLong{\varstratR}{
    (\ctx \cons \annkd{\tyctor}{\kdfun{\ioverline*{k}{\ann{\var}{\ty}}}{\kdtype}} \cons
    \ioverline*{j-1}{\annty{\varalt}{\tyalt}})
  }{\ctxemp}{
    (\ioverline*{\geq j}{\annty{\varalt}{\tyalt}})
  }
  {\tyctor}{\ioverline*{k}{\tm_l}}{\ty_{l, j}} \\
  \iforall{l} \wfty{
    \ctx \cons \annkd{\tyctor}{\kdfun{\ioverline{k}{\ty}}{\kdtype}}
  }{
    \tyfun{\ioverline{j}{\ty_l}}{\tyctor \tmapp \ioverline{k}{\tm_l}}
  }
}{
  \wftm{\ctx}{
    \defind{\tyctor}{
      \kdfun{\ioverline{k}{\ty}}{\kdtype}
    }{\varstratR}{
      \iunderline{l}{\ann{\ctor}{
        \tyfun{\ioverline*{j}{(\ann{\varalt}{\tyalt})}}{\tyctor \tmapp \ioverline{k}{\tm}}
      }}
    }
  }{
    \kdfun{\ioverline{k}{\ty}}{\kdtype}
  }
}

\DeclareRule{RuleTmIntro}{Tm-Intro}{
  D \defeq \decl \quad
  \wftm{\ctx}{D}{\kdfun{\ioverline{k}{\ty}}{\kdtype}}
}{
  \wftm{\ctx}{\ctor(i)_D}{
    \tyfun{
      (\ioverline{j}{\ann{\varalt}{\tyalt\subst{D}{\tyctor}}})
    }{
      D \tmApp \ioverline*{k}{\tm}
    }
  }
}

\DeclareRule{RuleTyElim}{Tm-Elim}{
  D \defeq \decl \quad
  \wfty{
    \ctx \cons
    \ioverline{k}{\annty{\var}{\ty}} \cons
    \annty{\varalt}{D \tmApp \ioverline{k}{\var}}
  }{\tyaltalt} \\
  \iforall{i}
  \wftm{
    \ctx \cons
    \ioverline{j}{\annty{\varaltalt}{\tyalt}\subst{D}{\tyctor}} \cons
    \ioverline{j}{\tyel{\tyctor}{\tyaltalt}{\varaltalt}{\tyalt}%
      \subst{D}{\tyctor}
    }
  }{c_i}{
    \tyaltalt%
    \subst{\ioverline*{k}{\tm}}{\ioverline*{k}{\var}}%
    \subst{\ctor(i)_D \tmapp \ioverline*{j}{\varaltalt}}{\varalt}
  }
}{
  \wftm{\ctx}{\tmelim{D}{\ioverline*{i}{c}}}{
    \tyfun{\left(\ioverline*{k}{\annty{\var}{\ty}}\right)}{
      \tyfun*[{\varalt}]{D \tmApp {\ioverline*{k}{\var}}}{\tyaltalt}
    }
  }
}

\usepackage{todonotes}
\newcommand{\af}[1]{}
\newcommand{\ak}[1]{}
\newcommand{\nv}[1]{}
\newcommand{\nvold}[1]{}

\NewDocumentCommand{\side}{O{;} m >{\SplitList{#1}}m}{%
  \DeclareDocumentCommand{\sidecontent}{m}{%
    \begin{minipage}{1\linewidth / #2}\begin{center}%
      ##1%
    \end{center}\end{minipage}%
  }%
  \begin{center}%
    \ProcessList{#3}{\sidecontent}%
  \end{center}%
}

\NewDocumentCommand{\sidesized}{O{;} >{\SplitList{#1}}m}{%
  \DeclareDocumentCommand{\sideop}{m m}{%
    \begin{minipage}{##1\linewidth}\begin{center}%
      ##2%
    \end{center}\end{minipage}%
  }%
  \DeclareDocumentCommand{\sidecontent}{>{\SplitArgument{1}{|}}m}{%
    \sideop##1%
  }%
  \begin{center}%
    \ProcessList{#2}{\sidecontent}%
  \end{center}%
}

\newcount \negscount

\NewDocumentCommand{\negspace}{m}{
  \negscount 0
  \loop \ifnum \negscount < #1
    \!
  \advance \negscount by 1
  \repeat
}

\newlength{\ListTopSep}
\newlength{\ListItemSep}
\newlength{\ListPartopSep}

\newtheorem{definition}{Definition}
\newtheorem{assumption}{Assumption}
\crefname{assumption}{Assumption}{Assumptions}
\newtheorem{corollary}{Corollary}
\newtheorem{example}{Example}

\title{CIRT: Calculus of Inductive Refinement Types}
\subtitle{Realizability Semantics of Refinement Type Theory}

\author{Alessio Ferrarini}
\orcid{0009-0006-8919-3022}
\affiliation{%
  \institution{IMDEA Software Institute}
  \city{Madrid}
  \country{Spain}
}
\affiliation{%
  \institution{Universidad Politécnica de Madrid}
  \city{Madrid}
  \country{Spain}
}
\email{alessio.ferrarini@imdea.org}

\author{G. A. Kavvos}
\orcid{0000-0001-7953-7975}
\affiliation{%
  \institution{University of Bristol}
  \city{Bristol}
  \country{United Kingdom}
}
\email{alex.kavvos@bristol.ac.uk}

\author{Niki Vazou}
\orcid{0000-0003-0732-5476}
\affiliation{%
  \institution{IMDEA Software Institute}
  \city{Madrid}
  \country{Spain}
}
\email{niki.vazou@imdea.org}

\begin{document}

\begin{abstract}

Refinement types are a powerful mechanism for statically verifying properties of
programs. They have been successfully integrated into practical languages such
as Haskell, Scala, and Rust. However, their metatheory is not yet fully
understood. In this paper we present the first semantic model for a refinement
type system that supports the full feature combination of subtyping, SMT-style
equality, impredicative polymorphism, and (stratified) inductive types.
Interactions between these features have led to inconsistencies in the past, and
existing formalizations of refinement type systems omit at least one of them. We
present a realizability model based on partial equivalence relations (PERs) over
a Scott domain, which overcomes these obstacles. We use our model to prove the
consistency of a core refinement type system and to give a semantic account of
the annotation-independence property of refinement types. Our semantics uses two
interpretations: a refined one, which interprets types and their refinements as
PERs; and an unrefined one, consisting of complete uniform PERs with a bottom
element (CUPERs), which models computational behaviour independently of
refinements.  We extend the model to positive and stratified inductive types,
showing that both are consistent in the presence of impredicativity.

\end{abstract}

\maketitle

\section{Introduction}

Refinement types~\cite{jhalaRefinementTypesTutorial2020} extend a host language
with logical predicates that constrain the values inhabiting types. For
instance, the type $\tyref{\var}{\tyint}{\var \neq 0}$ \emph{refines} the type
of integers, allowing us to statically forbid division by zero. Such systems
have been integrated into practical languages, including Haskell (Liquid
Haskell)~\cite{vazouRefinementTypesHaskell2014}, Scala
(Stainless)~\cite{hamzaSystemFRFormalized2019}, and Rust
(Flux)~\cite{lehmannFluxLiquidTypes2023}. Unlike most foundational type
theories, refinement type systems were developed primarily as engineering
artifacts: their design was driven by the goal of verifying existing programs
with minimal disruption to the underlying language, often accepting a degree of
``soundiness''~\cite{livshitsDefenseSoundinessManifesto2015} in return for
practicability.

Despite their practical success, the metatheory of refinement types remains only
partially understood. Over the last decade practical refinement type systems
have accumulated several features that interact in non-trivial ways at the
semantic level, and potentially lead to inconsistencies if not handled with
care. For example, \emph{1) subtyping} combined with \emph{2) unrefined equality}
(\ie an SMT-style equality that we describe in \Cref{sec:untyped-equality})
and function extensionality has led to inconsistencies in both Liquid
Haskell and F*~\cite{%
IncompatibilitiesFunctionalExtensionality}. Additionally, \emph{3) impredicative
polymorphism} combined with some classes of \emph{4) non-strictly positive
inductive} definitions as supported in Liquid Haskell can be
shown to be inconsistent in standard type-theoretic
settings~\cite{coquandInductivelyDefinedTypes1990}.

The desire to prevent inconsistency has led to several formalizations of
refinement type systems. For example,
\citet{swamyDependentTypesMultimonadic2016} prove normalization for a model of
F*, but their calculus has no impredicative polymorphism. \citet{hamzaSystemFRFormalized2019}
prove normalization for a model of Stainless which lacks subtyping.
\citet{ghalayiniExplicitRefinementTypes2023} construct a model for a
system with explicit refinement types, but do not consider inductive types.
Finally, \citet{borkowskiMechanizingRefinementTypes2024} prove type soundness
for a model of Liquid Haskell without higher-order unrefined equalities. As a
result, \emph{there is currently no single semantic account that combines the
four key features of refinement types}. If we truly want to understand
refinement types and prove their soundness,
we need a semantic model that supports
all of these features simultaneously.

Nevertheless, discovering such a semantic model is challenging. The
set-theoretic semantics given
by~\citet{ghalayiniExplicitRefinementTypes2023}, which interprets each
refined type as a subset of the interpretation of its erased type,
does not extend to impredicative polymorphism, which has no
set-theoretic model in classical set
theory~\cite{reynoldsPolymorphismNotSettheoretic1984}; subtyping and
unrefined equality further constrain how a model can interpret refined
types, as we discuss in~\Cref{sec:set-models}. Second, modeling the
non-strictly-positive inductive types of Liquid Haskell is not
straightforward, as they do not have a direct analogue in type theory.

In this paper we present a model based on \emph{partial equivalence relations
(PERs)} over a Scott domain. In broad strokes, our model follows much of the
standard structure of PER models: types are interpreted as PERs over the domain
and terms as well-defined elements of these PERs
\cite{macqueenIdealModelRecursive1986}. Thus, the semantic counterpart of
typing is membership in the domain of the PER, and equality is interpreted as
the PER itself. The subtyping relation is interpreted as inclusions between
PERs, so no cast operations are needed. We will show that the notion of
\emph{unrefined equality} used in refinements may also be modelled by the
unrefined PER interpreting a type. Finally, the PERs offer enough structure to
model impredicative polymorphism in the usual manner, viz. as set-theoretic
intersection over the set of all PERs.

The realizability model that we designed
interprets the key features of refinement type systems
and is
deliberately \emph{elementary}  so that
it can serve  as a reference model against which
candidate features are checked \emph{before} they are
implemented. Concretely, our contributions are:
\begin{itemize}[leftmargin=*, topsep=0.2pt, itemsep=0pt]
\item \emph{A model of refinement types as they are, not as they are
  convenient to model.} We isolate four features of practical
  refinement type systems (implicit subtyping, SMT-style unrefined
  equality, impredicative polymorphism, and non-standard inductive
  types) whose interaction is precisely what has caused soundness
  bugs in real systems, and we model their \emph{combination} without
  simplifying any of them away. No existing formalization supports all
  four (\Cref{sec:existing-formalizations}).

\item \emph{Modeling unrefined equality.} Every type receives
  a refined interpretation as a PER, included in an unrefined
  interpretation as a CUPER over a Scott domain
  (\Cref{sec:model-overview,sec:model-basic,sec:model-refinement-types}).
  This model explains how the SMT-style equality
  interacts with higher-order refined terms.
  So, our model does not merely avoid the extensionality
  inconsistencies of Liquid Haskell and \Fstar{}, it \emph{explains}
  them: they validate refined equalities that cannot coexist with
  unrefined equality (\Cref{sec:untyped-equality}).

\item \emph{Consistency of positive and stratified inductive types
  under impredicativity} (\Cref{sec:model-inductive,thm:consistency}).
  The model shows that positive and stratified types are
  consistent in a refined setting even with impredicative polymorphism. Our
  techniques are not specific to refinement types: they can be
  adapted to any type theory that admits a realizability model, for
  example MLTT.
  Ours is moreover the first \emph{denotational} semantics of
  stratified indexed types, previously accounted for only by
  normalization or step-indexing
  arguments~\cite{jacob-raoIndexStratifiedTypes2018,
  hamzaSystemFRFormalized2019}.
  The interpretation of an inductive
  type is an initial algebra: folding and unfolding form an isomorphism,
  and the eliminator may expose the recursive arguments, in contrast to
  Mendler-style \emph{weakly} initial algebras. So, both positive and
  stratified types share a single, uniform introduction and elimination
  discipline (\Cref{sec:model-inductive}). The message is constructive:
  positive and stratified indexed types are a straightforward addition
  to refinement systems.
\end{itemize}

The rest of the paper is structured as follows.
In \Cref{sec:overview} we present the key features of the system,
compare them with existing formalizations,
and explain why they require a
semantic treatment beyond set-theoretic models.
\Cref{sec:model-basic,sec:model-refinement-types,sec:model-inductive}
develop the formal construction of the model \corelan{} and prove
consistency.
Finally, we present related works and conclude.

Every theorem in this paper is accompanied by a detailed proof in the
supplementary material.

\section{Refinement types: Syntax \& Semantics}%
\label{sec:overview}

We start (\Cref{sec:features}) by presenting the key ingredients of practical refinement type
systems that are relevant for the semantic development of this
paper.
Next (\Cref{sec:existing-formalizations}), we describe how these features are
formalized in the literature and summarize their respective
limitations.
Finally (\Cref{sec:set-models}), we derive from these features the
requirements that guide the design of our model.

\subsection{Core Features of refinement type systems}%
\label{sec:features}

A refinement type $\tyref{\var}{\ty}{\prop}$ consists of a base type
$\ty$, a proposition $\prop$ that constrains the values inhabiting
$\ty$ as the ones satisfying $\prop$.
The language of propositions $\prop$ includes boolean-valued
expressions of the host language, equality predicates over terms
(including equality that may not be decidable), and closure under
standard logical connectives.
Refinements are not restricted to base types, but may also appear throughout the type structure, for example in (dependent) function types, polymorphic types, and inductive types.

A key aspect of refinement type systems is that refinements act as an
\emph{annotation layer} over an existing language: they enrich static checking
while leaving the operational semantics unchanged. This design enables
deployment without runtime overhead, and further ensures that the behaviour of a
program does not depend on refinement annotations. We call the independence of
program behavior from verification annotations the
\emph{annotation-independence} property.

Refinement type systems are extended with several orthogonal mechanisms that
interact in non-trivial ways at the semantic level. Next, we present the
four main ingredients that appear in practical systems and that are
relevant for the semantic issues studied in this paper.

\subsubsection{Subtyping}%
\label{sec:semantic-subtyping}

A refinement that arbitrarily constrains the elements of a type may be
naturally regarded as a \emph{subtype} of another type; for example, positive
integers are integers. This allows a term to be used at a less-refined or
unrefined type without requiring explicit casts. Semantically, subtyping
expresses inclusion between the meanings of types: every value inhabiting a
subtype must also inhabit its supertype.
Importantly, subtyping is \emph{implicit}, meaning that, as presented in the
rule \textsc{\tmSub} below, no type casts are required to use a term at a
supertype. Function types, in turn, satisfy the standard
contravariant/covariant subtyping rule
\textsc{\subFun}~\cite{cardelli_1984,cardelli_1985}.
\begin{align*}
\inference{
  \wftm{\ctx}{\tm}{\ty}
  \quad
  \wfsub{\ctx}{\ty}{\tyalt}
}{
  \wftm{\ctx}{\tm}{\tyalt}
}[\textsc{\tmSub}]
\qquad
\inference{
  \wfsub{\ctx}{\tyalt}{\ty}
  \quad
  \wfsub{\ctx}{\tyaltalt}{\tyaltaltalt}
}{
  \wfsub{\ctx}{\ty \to \tyaltalt}{\tyalt \to \tyaltaltalt}
}[\textsc{\subFun}]
\end{align*}

\subsubsection{Unrefined equality}%
\label{sec:untyped-equality}
Refinements often use equality as an unrefined relation on terms, including
terms of function type. This reflects
the view of the SMT solver, which operates on a representation of these unrefined
terms, and derives equalities without refined type
information~\cite{vazouRefinementReflectionComplete2017}.

This notion of \emph{unrefined equality} is the cause of many subtle soundness
questions. For example, function extensionality was introduced in both Liquid
Haskell~\cite{vazouHowSafelyUse2022} and
F*~\cite{IncompatibilitiesFunctionalExtensionality} and later shown to be
inconsistent. The inconsistency arises from the interaction between unrefined
equality and subtyping: in a typed setting, equality is indexed by type and
extensionality can be formulated so as to be harmless; with unrefined equality,
the same axiom has greater force and \citet{vazouHowSafelyUse2022} showed that admitting it leads to a proof of false.

\subsubsection{Parametric Polymorphism}%
  \label{sec:polymorphism}

Most refinement type systems are built on top of languages that have parametric
polymorphism, such as Haskell or Scala. We model such languages through
impredicative type quantification \`{a} la System~F
\citep{girard_1972,reynolds_1984}. Thus, a quantified type has the same kind as
the types over which it quantifies. For example, the type
$\tyall{\var}{\kdtype}{\tyint}$ itself has kind $\kdtype$, and may therefore be
instantiated with arbitrary types of kind $\kdtype$, including itself. This
contrasts with predicative polymorphism, where the quantified type lives at a
strictly higher level than the types it quantifies over, in a stratified
hierarchy \citep{leivant_1991}.

Refinement types should interact smoothly with polymorphism: it should be
possible to refine type variables, instantiate them with arbitrary types, and
quantify over indexed type families.

\subsubsection{Non-standard Inductive Types}%
\label{sec:inductive-types}

To preserve consistency, general recursion must be excluded from a system of
refinement types. However, simply replacing it with some flavour of structural
recursion does not suffice, as it is well-known that recursive type definitions
can also be used to encode general
recursion~\citep{pierce_2002, harper_2016}. The canonical counterexample is:
\begin{lstlisting}
data Bad where MkBad :: (Bad -> Void) -> Bad
\end{lstlisting}
where \lstinline{Void} is the empty type. This type enables encoding
self-application and yields a contradiction:
\begin{lstlisting}
bad :: Bad -> Void             worst :: Void
bad (MkBad f) = f (MkBad f)    worst = bad (MkBad bad)
\end{lstlisting}
This term is a typed instance of Curry's paradox; that admitting such negative
datatypes breaks normalisation---and hence consistency---in the second-order
$\lambda$-calculus is due to~\citet{mendler_1987}.
While this code is totally valid Haskell, it is rightfully rejected by
Liquid Haskell, which aims to construct a consistent refinement type
system. However, strict positivity is too strong a restriction:
positive types and stratified negative types could be admitted without
compromising consistency.

\sparagraph{Positive types}% Positive types
allow the inductive type to
occur negatively, provided that each
negative occurrence is itself under a negative function arrow; a canonical example is
Hofmann's routine type~\cite{
  HofmannNonStrictlyPositive,%
  bergerMartinHofmannsCase2019%
}:
\begin{lstlisting}
data Rou where Over :: Rou; Next :: ((Rou -> [Nat]) -> [Nat]) -> Rou
\end{lstlisting}
While it is not immediately obvious, the class of positive types can also lead
to inconsistencies in dependently typed systems built on
CIC~\cite{coquandInductivelyDefinedTypes1990}. In contrast, our denotational
model proves that they are harmless in a refinement type system.

\paragraph{Stratified types,}%
\label{sec:stratified-types}
%Stratified types,
introduced by~\citet{jacob-raoIndexStratifiedTypes2018} and
also used by~\citet{hamzaSystemFRFormalized2019}, are indexed inductive types
whose recursive occurrences appear at `strictly smaller indices' both in
positive and negative positions, ensuring well-foundedness. As an example,
consider the shallow embedding of the simply-typed lambda calculus (STLC) in
Liquid Haskell (from~\cite{ferrariniLargeEliminationIndexed,ferrariniPLEXNormalizationRefinement2026})
where the data types \lstinline{Typ} and \lstinline{Val}, respectively, embed types and values of STLC:
\begin{lstlisting}
data Typ = TIota | TArr Typ Typ
data Val where {-@ stratified Val @-}
{-@ VIota :: Int -> Ix Val TIota @-}
    VIota :: Int ->    Val
{-@ VArr :: t:Typ -> u:Typ -> (Ix Val t -> Ix Val u) -> Ix Val (TArr t u) @-}
    VArr ::   Typ ->   Typ -> (   Val   ->    Val  ) ->    Val
\end{lstlisting}
The type \lstinline{Typ} contains the base type (\lstinline{TIota}) and the function type (\lstinline{TArr}). The family \lstinline{Val} is indexed by such types;
intuitively, \lstinline{Val t} contains exactly the values of
object-language type \lstinline{t}. The \lstinline{Ix} annotation is
Liquid Haskell's notation for attaching such an index to a datatype.

Ignoring the refinement annotations, \lstinline{Val} can be viewed as
the solution of the recursive domain equation of the untyped
$\lambda$-calculus, which is sufficiently expressive to encode general
recursion. The refinement annotations, however, equip values with a
type index. A value indexed by \lstinline{TIota} must be constructed
using \lstinline{VIota}, whereas a value indexed by
\lstinline{TArr t u} must be constructed using \lstinline{VArr}. The
recursive occurrences of \lstinline{Val} appearing in the argument of
\lstinline{VArr} are indexed by \lstinline{t} and \lstinline{u}, which
are strict subcomponents of the result index
\lstinline{TArr t u}. Consequently, recursive occurrences only appear
at smaller indices, allowing the family \lstinline{Val} to be
interpreted by induction on the structure of the index rather than as
an arbitrary recursive domain equation.
The \lstinline{stratified Val} annotation enables the stratification
checker to verify the well-formedness of the type: as explained
in~\cref{sec:compilation}, it checks that the indices of the recursive
occurrences are indeed strictly smaller than the index of the result
type, and accepts the definition of \lstinline{Val}.

\subsection{Existing formalizations}%
\label{sec:existing-formalizations}
We give a high-level overview of existing formalizations of refinement type
systems, and the extent to which they model the features described in
\Cref{sec:features}.  \Cref{fig:existing-formalizations}
summarizes the four formalizations that we
discuss:
\coreFstar~\cite{swamyDependentTypesMultimonadic2016}
that models \Fstar,
System FR~\cite{hamzaSystemFRFormalized2019} that models Stainless,
\coreLH~\cite{borkowskiMechanizingRefinementTypes2024} that models Liquid Haskell, and
explicit refinement types~\cite{ghalayiniExplicitRefinementTypes2023} that models a system with explicit refinement types.
For each system we indicate the extent to which it supports the four features of refinement types, and the semantics used to model them.

\paragraph{Subtyping}

Many formal models of refinement types, such
as explicit refinement types and System FR,
do not include explicit subtyping. Instead they rely on a \emph{reflection}
rule:
\[
\inference{
  \wftm{\ctx}{\tm}{\ty} \quad \ctx \vdash \prop(\tm)\; ok
}{
  \wftm{\ctx}{\tm}{\tyref{\var}{\ty}{\prop}}
}[\textsc{Tm-Refl}]
\]
Subtyping is then \emph{implicitly} possible, as whenever $\phi$ implies $\psi$,
any term $\wftm{\ctx}{\tm}{\tyref{\var}{\ty}{\prop}}$ is also a term of type
$\wftm{\ctx}{\tm}{\tyref{\var}{\ty}{\psi}}$. However, this is \emph{not}
equivalent to the explicit formulation. To see why, suppose we have a free
variable $f : \tyfun*{\tyint}{\tyref{\var}{\tyint}{\var \geq 0}}$ which we
wish to use as a function of type $\tyfun{\tyref{\var}{\tyint}{\var \geq
0}}{\tyint}$. This
cannot be derived from reflection alone, as refinement information
cannot flow through type formers. The eta-expansion
$\tmlam{\var}{f \app \var}$ does have this type, a technique that goes back
to~\citet{barendregt_1983}, but it is a different term from $f$. Explicit subtyping, as
presented in~\cref{sec:semantic-subtyping}, is instead designed with specific
rules for each type former, \eg contravariance on the domain of the
function.

\paragraph{Equality.}
Most systems, including those of \coreLH and \coreFstar,
restrict equality to base types, such as $\tyint$. This is sound as the refined
equality coincides with the unrefined one on base types. However, it is
too restrictive, as it rules out equalities between higher-order terms
entirely.

In System FR  equality is
\emph{reduction-based}, so that $\tm \propeq \tmalt$ if and only if $\tm$ and
$\tmalt$ reduce to a common value. This is untyped and by consequence
also unrefined, but too fine-grained. For example, reduction almost always
uses weak head normal form, and natural equalities at higher-order
types do not hold, \eg making $\tyref{\var}{\tyunit}{\tmlam{\varalt}{1
+ 2} \propeq \tmlam{\varalt}{3}} \to \bot$ inhabited. Unfortunately,
such equalities arise naturally in refinement types
when encoding the equivalence of terminating functional programs.\looseness=-1

In explicit refinement types

equality is the typed $\beta\eta$-equality of the underlying simply-typed
terms; \citet{ghalayiniExplicitRefinementTypes2023} already suggest moving to a
semantics based on partial equivalence relations, which our model also uses.

\paragraph{Polymorphism}

\coreLH allows the refinement of type
variables, but  they can only be instantiated with base types.
Similarly, the calculus of System FR includes indexed
types, but quantification ranges only over fully applied types. As the system
of \coreFstar is dependent, they only have
\emph{predicative} polymorphism: $\tyall{\var}{\kdtype}{\tyint}$ is only a type
at the strictly higher level than $\kdtype$, and hence it cannot be instantiated
with itself. Finally, in explicit refinement types there is no
polymorphism.

\paragraph{Inductive types.}
In System FR they have step-indexed inductive
types, which are similar to stratified types in the sense
of~\citet{jacob-raoIndexStratifiedTypes2018}, but they do not have a
direct analogue of positive types. In the rest of the systems there
are no inductive types at all, so no direct analogue of either
positive or stratified types.

\begin{table}
\caption{High level comparison of the formalizations of
  refinement type systems of:
  \coreFstar~\cite{swamyDependentTypesMultimonadic2016},
  System FR~\cite{hamzaSystemFRFormalized2019},
  \coreLH~\cite{borkowskiMechanizingRefinementTypes2024}, and
  explicit refinement types~\cite{ghalayiniExplicitRefinementTypes2023}.
}%
\label{fig:existing-formalizations}
\resizebox{\linewidth}{!}{
\begin{tabular}{llccccc}
\toprule
System & Models
  & \begin{tabular}{c}Subtyping\\\ref{sec:semantic-subtyping}\end{tabular}
  & \begin{tabular}{c}Equality\\\ref{sec:untyped-equality}\end{tabular}
  & \begin{tabular}{c}Polymorphism\\\ref{sec:polymorphism}\end{tabular}
  & \begin{tabular}{c}Inductive types\\\ref{sec:inductive-types}\end{tabular}
  & \begin{tabular}{c}Construction\end{tabular}
  \\
\midrule
  $\mu \Fstar$ & (\Fstar)
  % \cite{swamyDependentTypesMultimonadic2016}
  & Yes
  & Base types
  & Predicative
  & No
  & Normalization
  \\
  System FR & (Stainless)
  % \cite{hamzaSystemFRFormalized2019}
  & No
  & Reduction
  & Yes
  & Step-Indexed
  & Normalization
  \\
  \coreLH &(Liquid Haskell)
%  \cite{borkowskiMechanizingRefinementTypes2024}
  & Yes
  & Base types
  & Yes, Limited
  & No
  & Safety
  \\
  \multicolumn{2}{l}{Explicit Refinement Types}
%  \cite{ghalayiniExplicitRefinementTypes2023}
  & No
  & Typed $\beta\eta$
  & No
  & No
  & Set model
  \\
\bottomrule
\end{tabular}
}
\end{table}

\subsection{Design requirements for a realizability model}%
\label{sec:set-models}

The simplest models of type theories are set-theoretic: types are interpreted as
sets and terms as elements of these sets \citep{henkin_1950}. The core
features described above impose three requirements on the design of
our model.

The \textbf{first} requirement is support for impredicative polymorphism.
Set-theoretic models of impredicative polymorphism cannot exist because of size
issues~\cite{reynoldsPolymorphismNotSettheoretic1984}, at least not within
classical logic~\cite{pitts_1987}.
So our model interprets types as PERs over a fixed domain: these form a set,
so type quantification can be interpreted as an intersection over them.

The \textbf{second} requirement is that subtyping be interpreted as inclusion.
It fails in models that interpret each function type
as the set of all functions between the interpretations of its domain and
codomain: in such a model, subtyping between function types requires \emph{cast
operations}. For example, type $\tyfun{\tyint}{\tyint}$
is interpreted as the set of total functional relations $R \subseteq \sint \times \sint$.
Computationally, this type is a subtype of $\tyfun{\tyref{\var}{\tyint}{x \geq
0}}{\tyint}$. However, the latter is interpreted as the set of total functional relations
$R \subseteq \snat \times \sint$. The first set is \emph{not} a subset of the second, as
it contains pairs $\pair{n}{m}$ with $n < 0$. To pass from one to the other
we must \emph{cast}, discarding the offending pairs. This example exhibits the
general principle of \emph{coercive subtyping}, which interprets each subtyping
judgment as such a cast inserted at every use of subsumption
\citep{cardelli_1985}. This is unfortunate on two counts. First, the denotation
of a term comes to depend not on the term alone but on the shape of its typing
derivation; this is the classical \emph{coherence problem}, and well-definedness
can be recovered only by a theorem showing that all derivations of a judgment
agree~\cite{reynolds_1980,breazutannen_1991}. Second, and more seriously for us,
the cast genuinely \emph{removes} elements, so that the meaning of a term comes
to depend on the refinements appearing in its type, and annotation-independence %(explained in~\Cref{sec:refinements})
is lost. Interpreting subtyping as inclusion requires no
casts~\cite{cardelli_1984,macqueenIdealModelRecursive1986,frisch_2008}. This is the case in
the subset model of refinement types~\citep{melliesFunctorsAreType2015}, where
a refined type is a subset of its underlying type; our model has the same
structure, with PERs in place of raw sets.

The \textbf{third} requirement is that the meaning of a term not depend on the
refinements in its type, so that unrefined equality can observe its behavior
outside them. Unrefined equality cannot be modeled in the presence of the
cast operations described above: a cast removes elements from the
interpretation of a functional term, and there is no way to recover them to
check whether two terms are equal on their unrefined domain.

\subsection{Our approach: PERs over a Scott domain}%
\label{sec:approach}
To meet these requirements, we use a model based on \emph{partial
equivalence relations} (PERs) over a Scott domain $\dinf$. Our terms
are interpreted as elements of $\dinf$ and types are interpreted as
PERs over $\dinf$, so that the semantic counterpart of typing is
membership in a PER rather than membership in a set, and the equality
is interpreted by the relation described by the PER itself. The
subtyping relation is interpreted as set inclusion between PERs, so no
cast operations are needed and annotation independence holds
by construction. The unrefined equality can be modeled by carefully
designing the unrefined PERs so that two terms are related by
unrefined equality exactly when they are identified by the unrefined
interpretation of their type. Finally, the PERs offer enough structure
to model impredicative polymorphism as set-theoretic intersection over
the set of all PERs.

\section{Background}%
\label{sec:background}

This section collects the domain-theoretic preliminaries on which our
model rests. We follow the  standard accounts of \citet{davey_2002},
\citet{abramsky_1994}, \citet{gunter_1992}, \citet{amadio_1998}, and
\citet{streicher_2006}, to which we refer the reader for proofs.

\begin{definition}[PER]\label{def:per}
A partial equivalence relation (PER) over a set $A$ is a symmetric and
transitive relation over $A$, in other words it is a set $S \subseteq A \times
A$ such that
\begin{enumerate*}[label={(\roman*)}, itemjoin={{; }}, itemjoin*={{, and }}]
\item\label{def:per:sym}
  if $\pair{a}{b} \in{} S$ then $\pair{b}{a} \in{} S$
\item\label{def:per:tra}
  if $\pair{a}{b} \in S$ and $\pair{b}{c} \in{} S$ then $\pair{a}{c} \in{} S$.
\end{enumerate*}
\end{definition}

\begin{definition}[Domain of a PER]
  Given a PER $S$ over $A$, its \emph{domain} is the set of elements of $A$
  which are related to themselves, \ie $\dom{S} = \{ a \in A \mid \pair{a}{a}
  \in{} S \}$.
\end{definition}

\begin{definition}[Diagonal PER]
  The \emph{diagonal PER} of any set $A$ is given by $\diag{A} =
  \begin{setbuild}{\pair{a}{a}} a \in A \end{setbuild}$.
\end{definition}

We will use the notation $\per{A}$ to denote the set of PERs over
$A$. The PER structure will be used to model types, following a long
line of work that interprets types as partial equivalence relations
over a domain or a combinatory algebra
\citep{abadiModelPolymorphismRecursive1990,
amadioRecursionRealizabilityStructures1991, cardone_1989}: the domain
of a PER gives us the membership information, while the relation itself
gives us the equality relation. While the definition of a PER does not
give us reflexivity, it satisfies a weaker property that we call
partial reflexivity: if $\pair{a}{b} \in{} S$ then $\pair{a}{a} \in{}
S$.

\begin{definition}[Pointed PER]\label{def:pointedPER}
Let $A$ be a set with a distinguished element $\bot \in A$. A \emph{pointed PER
over $A$} is a PER $S \in \per{A}$ such that $\pair{\bot}{\bot} \in{} S$
\end{definition}
The point will be used to model a nonterminating element, in the manner of
\citet{amadioRecursionRealizabilityStructures1991}.

\begin{theorem}[Knaster-Tarski fixed point theorem]\label{thm:knaster-tarski}
Let $(L, \leq)$ be a complete lattice and $f: L \to L$ be a monotone
function. Then the set of fixed points of $f$ forms a complete lattice,
and in particular there is a least fixed point $\lfp f = \Glb \{ x \in L
\mid f(x) \leq x \}$ and a greatest fixed point $\gfp f = \Lub \{ x \in L
\mid x \leq f(x) \}$ \citep{tarski_1955}.
\end{theorem}

\begin{definition}[Scott domain]\label{def:domain}
A \emph{Scott domain} is a complete partial order that is moreover
\emph{$\omega$-algebraic} and \emph{bounded-complete}. It is
$\omega$-algebraic when every element is the directed least upper bound
of the countably many \emph{compact} elements below it---an element
$c$ is compact just if $c \leq \Lub X$ for a directed $X$ implies $c
\leq x$ for some $x \in X$. It is bounded-complete when every subset
that has an upper bound has a least upper bound \citep{scott_1976,
abramsky_1994, gunter_1992}.
\end{definition}

\begin{definition}[Chain of projections]\label{def:projections}
  A \emph{chain of projections} on a Scott domain $D$ is an increasing family of
  continuous functions ${(\pi_n)}_{n \in \snat}$, where each $\pi_n: D \to D$ is
  idempotent and below the identity, with $\pi_0$ the constant function $\bot$,
  and which moreover satisfies $\Lub_n \pi_n = id$.

  Given an element $d \in D$ we define the \emph{$n$-th approximation of $d$} to
  be $d_n \defeq \pi_n(d)$.
\end{definition}
In the rest of this section $D$ is a Scott domain equipped with a chain of
projections.

To model types we restrict attention to PERs that interact well with
this order- and approximation structure, in the manner of the ideal
and realizability models of \citet{macqueenIdealModelRecursive1986},
\citet{abadiModelPolymorphismRecursive1990}, and
\citet{amadioRecursionRealizabilityStructures1991}. The
\emph{uniformity} condition below is the analogue for PERs of their
uniform ideals.

\begin{definition}[Complete, uniform, and pointed PERs] \hfill
  \begin{enumerate}[topsep=\ListTopSep,itemsep=\ListItemSep,partopsep=\ListPartopSep]

    \item

    A PER $S \subseteq D \times D$ over a domain $D$ is \emph{complete} whenever
    every directed subset ${\{\pair{a_i}{b_i}\}}_{i \in I} \subseteq S$ (with
    respect to the product order on $D \times D$) has a least upper bound
    $\pair{\Lub_{i \in I} a_i}{\Lub_{i \in I} b_i}$ in $S$.

    \item

    A PER $S \subseteq D \times D$ over a domain $D$ is \emph{uniform} whenever
    for every $\pair{a}{b} \in S$ and every $n \in \snat$ we have
    $\pair{\pi_n(a)}{\pi_n(b)} \in S$.

    \item

    In this paper, a CUPER is a pointed, complete, and uniform PER. We write
    $\pper{D}$ for the collection of such PERs over a domain $D$.
  \end{enumerate}
\end{definition}

Completeness asks that a type be closed under the directed least upper bounds
of $D \times D$, so that the limits along which recursive computations
converge remain within it and least fixed points can be interpreted there
\citep{amadioRecursionRealizabilityStructures1991}. Uniformity asks instead
that membership be stable under approximation: a related pair remains related
at every stage $\pi_n$. In the presence of completeness, an element
then belongs to a type exactly when all of its approximations do.

Recursive and inductive types will be interpreted as fixed points
(\cref{sec:model-inductive}). We adapt the contractive construction
of~\citet{amadioRecursionRealizabilityStructures1991}: because
$\Lub_n \pi_n = id$, a CUPER is determined by its
approximations, and a contractive operator---one that sharpens agreement
by one approximation per unfolding---has a unique fixed point.

\begin{definition}[Approximation of a relation]\label{def:rel-approx}
The \emph{$n$-th approximation} of a relation $A \subseteq D \times D$ is its
pairwise image under the $n$-th projection,
$A_n \defeq
  \begin{setbuild}{\pair{\pi_n(\sval)}{\pi_n(\svalalt)}}
    \pair{\sval}{\svalalt} \in A
  \end{setbuild}$.
\end{definition}

\begin{lemma}[Restriction]\label[lemma]{lem:cuper-restriction}
Let $S$ be a CUPER over $D$ and write $D_n \defeq \pi_n(D)$. Then $S_n = S \cap
(D_n \times D_n)$. Consequently each $S_n$ is a pointed PER, and $S_j \subseteq
S_n$ whenever $j \leq n$.
\end{lemma}

\begin{lemma}[A CUPER is determined by its approximations]\label[lemma]{lem:cuper-det}
If $A$ and $B$ are CUPERs over $D$ with $A_n = B_n$ for all $n$, then $A = B$.
\end{lemma}

\begin{definition}[Non-expansive and contractive operators]\label{def:contractive-op}
An operator $\Phi : \pper{D} \to \pper{D}$ is \emph{non-expansive} when $A_n =
B_n$ implies $\Phi(A)_n = \Phi(B)_n$ for all $n$, and \emph{contractive} when
$A_n = B_n$ implies $\Phi(A)_{n+1} = \Phi(B)_{n+1}$ for all $n$.
\end{definition}

\begin{lemma}[Contractive fixed point]\label[lemma]{lem:guarded-fixpoint}
Every contractive operator $\Phi : \pper{D} \to \pper{D}$ has a unique fixed
point.
\end{lemma}

\section{Overview of the model}%
\label{sec:model-overview}

\NewDocumentCommand{\scottd}{}{\mathcal{D}}

The basis of our model is a Scott domain $\scottd$, over which we will define PERs.

\begin{definition}\label{def:dinf}
We take $\scottd$ to be the solution to the recursive domain equation
\[
\scottd
  \iso
           \dfun{\scottd}{\scottd}
    \dsum (\scottd \times{} \scottd)
    \dsum \snatbot{}
    \dsum \ssingbot{}
    \dsum \sboolbot{}
    \dsum \slabelbot{}
\]
where $\dfun{\scottd}{\scottd}$ is the domain of continuous functions, $\dsum$
is the coalesced sum, $\times$ is the cartesian product, and $\snatbot{}$,
$\ssingbot{}$, $\sboolbot{}$, $\slabelbot{}$ are the flat domains of
natural numbers, the unit, booleans, and labels.
\end{definition}

As usual, this domain is obtained as the bilimit of the $\omega$-chain
$\mathcal{D}_i$ given by
\begin{align*}
  \mathcal{D}_0      = \{\bot\} \qquad
  \mathcal{D}_{n + 1} = \dfun{\mathcal{D}_n}{\mathcal{D}_n}
                \dsum (\mathcal{D}_n \times{} \mathcal{D}_n)
                \dsum \snat{}
                \dsum \ssing{}
                \dsum \sbool{}
                \dsum \slabel{}
                \dsum \{\bot\}
\end{align*}
along the embedding-projection pairs $i_{n, m}:
\dfun{\mathcal{D}_n}{\mathcal{D}_m}$ and $j_{n, m}:
\dfun{\mathcal{D}_m}{\mathcal{D}_n}$ satisfying $j_{a} \circ i_{a} = id$ and
$i_{a} \circ j_{a} \leq id$ \citep{scott_1976, smyth_1982, streicher_2006}.
Each constructor thus carries an embedding-retraction pair, e.g.
$i_{\to}: \dfun{\dfun{\scottd}{\scottd}}{\scottd}$ and
$r_{\to}: \dfun{\scottd}{\dfun{\scottd}{\scottd}}$, satisfying $r_{\to} \circ
i_{\to} = id$ and $i_{\to} \circ r_{\to} \leq id$, and similarly for products,
$\snatbot$, $\ssingbot$, $\sboolbot$, and $\slabelbot$. The projections of \cref{def:projections}
are $\pi_n = i_n \circ j_n$, so that the $n$-th approximation of $d \in \scottd$
is $d_n = i_n(j_n(d))$.
They form a chain of projections on $\scottd$: in particular
$\Lub_n \pi_n = id$, which follows from the \emph{minimal invariance} of
$\scottd$~\citep{pitts_1996}, namely that $id$ is the least fixed point of
$e \mapsto F(e, e)$, for $F$ the functor of the domain equation.

A central feature of our model is that every type is equipped with two
semantics:
\begin{itemize}
  \item The \emph{refined semantics}, which interprets types as PERs over
    $\scottd$ and respects refinements.

  \item The \emph{unrefined semantics}, which interprets types as CUPERs over
    $\scottd$, ignoring refinements.
\end{itemize}

Following~\citet{amadioRecursionRealizabilityStructures1991}, the distinguished
bottom element will account for non-termination and general recursive types that
are present in the unrefined semantics.

\emph{The two semantics will be related by inclusion}: the refined interpretation of
each type will be included in its unrefined interpretation. This reflects the
intuition that refinements only add logical information without changing the
underlying computation. Several constructions in the model, notably function
types and equality predicates, rely on the interaction between these two layers.

Since types may be indexed by terms, kinds cannot be interpreted
merely as syntactic classifications. Instead, kinds will be interpreted
set-theoretically as collections of PER-valued functions respecting
the PER structure of their arguments. Context interpretations
therefore store both term-level information and type-level information
together with projections connecting the refined and unrefined worlds.

\section{Basic system}%
\label{sec:model-basic}

We start by presenting the syntax of (\Cref{sec:model-basic-syntax}) and semantics
(\Cref{sec:model-basic-semantics}) of the basis
of the \corelan calculus, which is
a dependent version of System~F~\citep{longo_1991,barendregt_1991}.
We prove the soundness of our semantics
(\Cref{thm:per-respecting-terms})
and consistency of the system
(\Cref{thm:consistency}).
Later we extend this core calculus with refinements
(\Cref{sec:model-refinement-types}) and indexed types
(\Cref{sec:model-inductive}).
All theorems are stated once and proved for the complete calculus.

\subsection{Syntax}%
\label{sec:model-basic-syntax}
\begin{figure}
  {\small
  \textbf{Contexts} \hfill \boxed{$\wfctx{\ctx}$}
  \begin{align*}
  \RuleCtxEmpty{full}
  \qquad
  \RuleCtxTy{full}
  \qquad
  \RuleCtxKd{full}
  \end{align*}
  \textbf{Kinds} \hfill \boxed{$\wfkd{\ctx}{\kd}$}
  \begin{align*}
  \RuleKdType{full}
  \end{align*}
  \textbf{Types} \hfill \boxed{$\wftm{\ctx}{\ty}{\kd}$}
  \begin{gather*}
  \RuleTyNat{full}
  \qquad
  \RuleTyBool{full}
  \qquad
  \RuleTyVar{full}
  \\
  \RuleTyFun{full}
  \qquad
  \RuleTyAll{full}
  \end{gather*}
  \textbf{Terms} \hfill \boxed{$\wftm{\ctx}{\tm}{\ty}$}
  \begin{gather*}
  \RuleTmNat{full}
  \qquad
  \RuleTmBool{full}
  \qquad
  \RuleTmVar{full}
  \\
  \RuleTmLam{full}
  \qquad
  \RuleTmApp{full}
  \\
  \RuleTmTyLam{full}
  \qquad
  \RuleTmTyApp{full}
  \end{gather*}
  }
  \caption{Typing rules of basic \corelan.}%
  \label{fig:syntax-basic}
\end{figure}

\Cref{fig:syntax-basic} presents the typing rules for the kinds, contexts, types, and terms of the basic system.

\sparagraph{Kinds} comprise just one base kind $\kdtype$, which classifies
types. Their presence will become useful in~\cref{sec:model-inductive}, where they are extended to include
indexed kinds for types indexed by terms.

\sparagraph{Contexts} are built by extension with \emph{term variables}
$\annty{\var}{\ty}$ and \emph{type variables} $\annkd{\var}{\kd}$.

\sparagraph{Types} are constructed from type variables, base types
(naturals and booleans), dependent functions, and System-F style impredicative
quantification. The dependent functions are only allowed to abstract over
\emph{term} variables. In this section none of our base types truly exercise
dependence, but we will add refinements and indexed types in~\cref{%
  sec:model-refinement-types,%
  sec:model-inductive%
}.

\sparagraph{Terms} include variables, $\lambda$-abstraction and
application, type abstraction and application, and base terms for naturals and
booleans. We omit the operations on base types (arithmetic,
comparisons, conditionals): they can be added as constants interpreted by
continuous functions that map related arguments to related results, in both
the refined and the unrefined interpretation of their type
(as will be required by \Cref{thm:per-respecting-terms}).
In the application rules, $\tmalt\subst{\tm}{\var}$ denotes the
capture avoiding substitution of $\tm$ for $\var$ in $\tmalt$, as is usual in
dependently-typed systems. Note that the terms are Curry-style terms, and do
\emph{not} contain type annotations.

\subsection{Semantics}
\label{sec:model-basic-semantics}

As mentioned in \cref{sec:model-overview}, every type has two interpretations:
the \emph{unrefined semantics} $\sembot{-}$ interprets types as CUPERs over
$\scottd$, forgetting refinements. In contrast, the \emph{refined semantics}
$\sem{-}$ interprets types as PERs over $\scottd$, retaining the logical
information needed by refinements and indexed types. Both are defined by mutual
induction on typing derivations. In summary:

\begin{itemize}[leftmargin=*,labelsep=.25em,topsep=\ListTopSep,itemsep=\ListItemSep,partopsep=\ListPartopSep]
  \item Contexts are interpreted as PERs over refined or unrefined environments.
  \item Kinds are interpreted as dependent PERs families in $\sem{-}$ and
    as the constant CUPER family in $\sembot{-}$.\looseness=-1
  \item Types are interpreted as context-indexed PERs in $\sem{-}$ and
    context-indexed CUPERs in $\sembot{-}$.
\end{itemize}

Terms have only one interpretation, viz.\ as elements of $\scottd$. It is
defined (in~\cref{sec:term-semantics}) by induction on the raw syntax of Curry-style terms and is denoted by
$\semtm{-}$. Concretely, for each context
$\ctx$ such that $\tm$ is typed,
the interpretation of $\tm$ has the following signature:
\[
  \semtm{\tm}[-] : \runctx{\wfctx{\ctx}} \to \scottd
\]
where $\runctx{\wfctx{\ctx}}$ is the set of \emph{runtime environments}
determined by $\ctx$, which only contain values:
\[\runtimeDefinition{
\begin{aligned}
  \runctx{\RuleCtxEmpty{concl}}
  &\defeq \ssing,
  &
  \runctx{\RuleCtxTy{concl}}
  &\defeq \runctx{\wfctx{\ctx}} \times \scottd,
  &
  \runctx{\RuleCtxKd{concl}}
  &\defeq \runctx{\wfctx{\ctx}}
\end{aligned}
}
\]
\emph{Notation:} In the rest of the paper, we
use the $\vrule\! \vrule$ delimiter to mark the
definitions of interpretations.

We keep this carrier separate because terms are Curry-style: their raw
denotation is independent of typing derivations. Later we will equip runtime
environments with a PER $\unrefass{\ctx}{-}$, which we use to state type
soundness: for any $\srec, \srec' \in \runctx{\wfctx{\ctx}}$,
\[
  \wftm{\ctx}{\tm}{\ty},\
  \sctx \in \dom{\sembot{\wfctx{\ctx}}},\
  \pair{\srec}{\srec'} \in \unrefass{\ctx}{\sctx}
  \quad\Longrightarrow\quad
  \pair{\semtm{\tm}[\srec]}{\semtm{\tm}[\srec']}
  \in \sembot{\wfty{\ctx}{\ty}}[\sctx].
\]
Thus the two type semantics constrain the same `runtime' computations.

This distinction also appears in contexts. The unrefined context semantics keeps
type-variable assignments, but does not store term-variable values. This
reflects that unrefined type interpretation erases term dependency. The refined
context semantics keeps term assignments, since refinements and indexed types
may depend on them.

\subsubsection{Contexts}
In the (un)refined semantics contexts are interpreted as PERs over
\emph{environments}. The empty context is interpreted as the diagonal PER on the
empty environment, i.e.\ the singleton:
\[\redBotDefinition{
\begin{aligned}
  \sembot{\RuleCtxEmpty{concl}} &\defeq \diag{\ssing}
  &
  \sem{\RuleCtxEmpty{concl}} &\defeq \diag{\ssing}
\end{aligned}
}\]
Context extension by a term variable is interpreted as the usual dependent sum
in the refined interpretation, whereas it is silently dropped in the unrefined one:
\[\redBotDefinition{
\begin{array}{@{}r@{\;}c@{\;}l@{}}
  \sembot{\RuleCtxTy{concl}} &\defeq&
    \sembot{\wfctx{\ctx}}
  \\
  \sem{\RuleCtxTy{concl}} &\defeq&
    \begin{setbuild}{\pair{\pair{\sctx}{\sval}}{\pair{\sctxalt}{\svalalt}}}
      \pair{\sctx}{\sctxalt} \in \sem{\wfctx{\ctx}},
      \pair{\sval}{\svalalt} \in \sem{\wfty{\ctx}{\ty}}[\sctx]
    \end{setbuild}
\end{array}
}
\]
This distinction captures the fact that the computational behaviour of this
calculus is actually not dependent at all and that the refinement types
exercise dependence only on the logical level.

Finally, extending a context with a type variable in the unrefined
semantics simply stores an environment compatible with the
interpretation of the kind.
\[\botDefinition{
\begin{aligned}
  \sembot{\RuleCtxKd{concl}}
    &\defeq
  \begin{setbuild}{\pair{\pair{\sctx}{\skd}}{\pair{\sctxalt}{\skd}}}
    \pair{\sctx}{\sctxalt} \in \sembot{\wfctx{\ctx}},
    \skd \in \sembot{\wfkd{\ctx}{\kd}}[\sctx]
  \end{setbuild}
\end{aligned}
}
\]

The refined semantics must also determine the compatible unrefined
environment. Thus, a type-variable extension stores both a refined kind element
and an unrefined over-approximation of the same type variable; the former
belongs to the refined interpretation of the kind, while the latter belongs to
the unrefined one.
\[\refinedDefinition{
\begin{aligned}
  \sem{\RuleCtxKd{concl}} &\defeq
  \begin{setbuild}{
    \pair{
      \pair{\sctx}{\pair{\skd}{\skd\bots}}
    }{
      \pair{\sctxalt}{\pair{\skd}{\skd\bots}}
    }
  }[l]
    \pair{\sctx}{\sctxalt} \in \sem{\wfctx{\ctx}},
    \skd \in \sem{\wfkd{\ctx}{\kd}}[\sctx],
    \\
    {}
    \skd\bots \in \sembot{\wfkd{\ctx}{\kd}}[\projctx{\ctx}{\sctx}],
    \projty{\ctx}{\sctx}{\kd}{\skd} \subseteq \skd\bots
  \end{setbuild}
\end{aligned}
}\]

The function $\projctx{\ctx}{-}$ is a \emph{context projection} operation, which
maps each refined environment to its unrefined version. Given a context
$\wfctx{\ctx}$ it is defined by induction on $\ctx$ as follows:
\[\begin{array}{l @{\;} c @{\;} l@{\quad} l @{\;} c @{\;} l @{\quad} l @{\;} c @{\;} l}
  \multicolumn{9}{c}{\projctx{\ctx}{-} : \dom{\sem{\wfctx{\ctx}}} \to
  \dom{\sembot{\wfctx{\ctx}}}} \\[0.4em]
  \projctx{\ctxemp}{\sunit}
  &\defeq& \sunit &
  \projctx{\ctx \cons \annty{\var}{\ty}}{\pair{\sctx}{\sval}}
  &\defeq& \projctx{\ctx}{\sctx} &
  \projctx{\ctx \cons \annkd{\var}{\kd}}{\pair{\sctx}{\pair{\skd}{\skd\bots}}}
  &\defeq& \pair{\projctx{\ctx}{\sctx}}{\skd\bots}
\end{array}
\]
The empty context is projected to the empty context; values are removed; and a
(refined, unrefined) pair of PERs is mapped to the unrefined PER. It satisfies
the expected compatibility condition:
\[
  \pair{\sctx}{\sctxalt} \in \sem{\wfctx{\ctx}}
  \quad\Longrightarrow\quad
  \pair{\projctx{\ctx}{\sctx}}{\projctx{\ctx}{\sctxalt}}
  \in \sembot{\wfctx{\ctx}}.
\]

\begin{definition}[Type projection]%
\label{def:type-projection}
The companion type projection $\projty{\ctx}{-}{\kd}{-}$ extracts the
underlying relation on $\scottd$ from a refined type interpretation, so that it can be
compared with an unrefined interpretation. In the basic system there
is only the kind $\kdtype$, so the projection is just the identity:
\[
\begin{array}{@{}lcl@{}}
  \projty{\ctx}{-}{\kd}{-}
  &:&
    (\sctx : \dom{\sem{\wfctx{\ctx}}})
    \to \sem{\wfkd{\ctx}{\kd}}[\sctx]
    \to \mathcal{P}(\scottd \times \scottd)
  \\% [0.4em]
  \projty{\ctx}{\sctx}{\kdtype}{\skd}
  &\defeq& \skd
\end{array}
\]
For the base kind $\kdtype$ the projection is the identity, so its value is in
fact a PER.
\end{definition}

\subsubsection{Kinds} Kinds are interpreted as families of PERs in the refined
semantics and as constant families of CUPERs in the unrefined semantics. Since
$\scottd$ is a set, the PERs over it form a set $\per{\scottd}$, which is a
complete lattice under inclusion, and the CUPERs form a subset $\pper{\scottd}$ of it.

The semantics of kinds takes as argument an environment and returns the set of
all possible interpretations of types of that kind in that context:
\begin{align*}
  \sem{\wfkd{\ctx}{\kd}}[-]    : \dom{\sem{\wfctx{\ctx}}}   \to \mathbf{Set}
  \qquad
  \sembot{\wfkd{\ctx}{\kd}}[-] : \dom{\sembot{\wfctx{\ctx}}} \to \mathbf{Set}
\end{align*}
For the base kind this is just all PERs and CUPERs respectively:
\[\redBotDefinition{
\begin{aligned}
  \sem{\RuleKdType{concl}}[\sctx]    \defeq \per{\scottd}
  \qquad
  \sembot{\RuleKdType{concl}}[\sctx] \defeq \pper{\scottd}
\end{aligned}
}\]

The PERs and CUPERs are meant to be a notion of equality between realizers. We
can show that this semantics respects this notion of equality, mapping related
environments to related kinds:

\begin{theorem}[Kinds respect PERs]%
\label{thm:semantics-per-respecting-kinds}
Let $\wfctx{\ctx}$ and $\wfkd{\ctx}{\kd}$. Then:
\begin{enumerate}[label={(\roman*)},topsep=\ListTopSep,itemsep=\ListItemSep,partopsep=\ListPartopSep]
\item{Refined:\quad}\label{thm:semantics-per-respecting-kinds:refined}
  If $\pair{\sctx}{\sctxalt} \in \sem{\wfctx{\ctx}}$, then
  $\sem{\wfkd{\ctx}{\kd}}[\sctx] = \sem{\wfkd{\ctx}{\kd}}[\sctxalt]$.
\item{Unrefined:}\label{thm:semantics-per-respecting-kinds:unrefined}
  If $\pair{\sctx}{\sctxalt} \in \sembot{\wfctx{\ctx}}$, then
  $\sembot{\wfkd{\ctx}{\kd}}[\sctx] = \sembot{\wfkd{\ctx}{\kd}}[\sctxalt]$.
\end{enumerate}
\end{theorem}

One of our goals is to prove that the refined interpretation of types
is included in the unrefined one, but we first define how we relate
them. The type projection of \cref{def:type-projection} maps a
refined interpretation at a kind to the relation over $\scottd$ that it
determines. Since $\pper{\scottd}$ are relations over $\scottd$, this
lets us ask whether a refined interpretation is included in an
unrefined one.

\subsubsection{Types}

Types are interpreted as dependent functions or sections over those families:
they map environments to an element of the appropriate kind:
\begin{align*}
  \sem{\wftm{\ctx}{\ty}{\kd}}[-]    &:
    (\sctx : \dom{\sem{\wfctx{\ctx}}})
    \to \sem{\wfkd{\ctx}{\kd}}[\sctx]
  \\
  \sembot{\wftm{\ctx}{\ty}{\kd}}[-] &:
    (\sctx : \dom{\sembot{\wfctx{\ctx}}})
    \to \sembot{\wfkd{\ctx}{\kd}}[\sctx]
\end{align*}

\sparagraph{Base types} are interpreted as the PERs over $\scottd$ that relate exactly the
elements of $\scottd$ that correspond to the values of the type.
\[\refinedDefinition{
\begin{aligned}
  \sem{\RuleTyNat{concl}}[\sctx]  \defeq
  \begin{setbuild}{\pair{\demb{\snat}{n}}{\demb{\snat}{n}}}
    n \in \snat
  \end{setbuild}\;
  \sem{\RuleTyBool{concl}}[\sctx] \defeq
  \begin{setbuild}{\pair{\demb{\sbool}{b}}{\demb{\sbool}{b}}}
    b \in \sbool
  \end{setbuild}
\end{aligned}
}\]
where $\tybase$ ranges over the base sets $\snat$, $\ssing$, $\sbool$, and
$\slabel$, and $\demb{\tybase}{\sval}$ denotes the continuous embedding of $\sval \in
\tybase$ into the domain $\scottd$ (per \cref{def:dinf}).

The unrefined interpretation extends the basic PERs by adding the
pair $\pair{\bot}{\bot}$:
\[\botDefinition{
\begin{aligned}
  \sembot{\RuleTyNat{concl}}[\sctx] &\defeq
  \begin{setbuild}{\pair{\demb{\snat}{n}}{\demb{\snat}{n}}}
    n \in \snat
  \end{setbuild} \cup \{ \pair{\bot}{\bot} \}
  \\
  \sembot{\RuleTyBool{concl}}[\sctx] &\defeq
  \begin{setbuild}{\pair{\samewidth{\demb{\sbool}{b}}{\demb{\snat}{n}}}{\samewidth{\demb{\sbool}{b}}{\demb{\snat}{n}}}}
    b \in \sbool
  \end{setbuild} \cup \{ \pair{\bot}{\bot} \}
\end{aligned}
}\]
Since basic types are flat domains in the domain equation that
describes $\scottd$, adding the pair $\pair{\bot}{\bot}$ is enough to
obtain pointedness, completeness and uniformity.

\sparagraph{Function types} have a more interesting
interpretation, as they relate functions that are `extensionally' equal, \ie
map related arguments to related outputs. In the unrefined case:
\[\botDefinition{
\vcenter{\hbox{\resizebox{0.97\textwidth}{!}{$\displaystyle
\begin{aligned}
  \sembot{\RuleTyFun{concl}}[\sctx]
    \defeq
\begin{setbuild}{\pair{\sfun}{\sfunalt}}[l]
  \forall \pair{\sval}{\svalalt} \in
    \sembot{\wfty{\ctx}{\ty}}[\sctx].\;
  \pair{\sfun \app \sval}{\sfunalt \app \svalalt} \in
    \sembot{\wfty{\ctx \cons \annty{\var}{\ty}}{\tyalt}}[\sctx]
\end{setbuild}
\end{aligned}
$}}}
}\]
One would na\"{i}vely expect the refined interpretation to mimic the unrefined
one, replacing $\sembot{-}$ with $\sem{-}$. However, this would equate functions
whose domains have been restricted by refinements. If the refined
function PER ignored the unrefined behavior, it would be impossible to
prove that the refined interpretation is included in the unrefined one
(per \cref{thm:refined-sub-unrefined}), as it would equate
functions that are not related in the unrefined interpretation. Hence,
we impose in
the refined interpretation the extra condition
$(ii)$ below that ensures compatibility with the unrefined
interpretation: %  of equality:
% to the simple relatedness.
\[\refinedDefinition{
\begin{aligned}
  & \sem{\RuleTyFun{concl}}[\sctx]
  \\
  & \quad \defeq
  \begin{setbuild}{\pair{\sfun}{\sfunalt}}[lc@{\;}llc@{\;}ll@{\;\;}lc]
      \forall&\pair{\sval}{\svalalt}
      &\in{}& \sem{\wfty{\ctx}{\ty}}[\sctx].\;
      &\pair{\sfun \app \sval}{\sfunalt \app \svalalt} &\in{}&
      \sem{\wfty{
        \ctx \cons \annty{\var}{\ty}
      }{\tyalt}}[\pair{\sctx}{\sval}]
      & \textcolor{gray}{\text{(i)}}
    \\
      \forall&\pair{\sval\bots}{\svalalt\bots}
      &\in{}& \sembot{\wfty{\ctx}{\ty}}[\projctx{\ctx}{\sctx}].\;
      &\pair{\sfun \app \sval\bots}{\sfunalt \app \svalalt\bots} &\in{}&
      \sembot{\wfty{
        \ctx \cons \annty{\var}{\ty}
      }{\tyalt}}[\projctx{\ctx}{\sctx}]
      & \textcolor{gray}{\text{(ii)}}
  \end{setbuild}
\end{aligned}
}
\]
The unrefined relatedness condition also excludes semantic functions
that do not correspond to any term that is well typed in the
underlying type system. For example consider a function that
returns values of  different types depending on the input.
Concretely $f(0) = 0$ and $f(n) = \strue$ for $n \geq 1$.
$f$ is not in the refined interpretation of the type
$\tyfun{\tyref{\var}{\tynat}{x \geq 1}}{\tybool}$ even though it
satisfies the refined condition because it violates the unrefined
relatedness condition.

\paragraph{Type Quantification}

Impredicative type quantification is interpreted by intersection over all
candidates in the interpretation of the quantified kind. In the unrefined case,
when $\kd = \kdtype$, it is defined to be the intersection over all CUPERs.
\[\botDefinition{
\begin{aligned}
  \sembot{\RuleTyAll{concl}}[\sctx] \defeq
  \begin{setbuild}{\pair{\sval}{\svalalt}}[l]
    \forall \skd \in \sembot{\wfkd{\ctx}{\kd}}[\sctx].\;
    \pair{\sval}{\svalalt} \in
    \sembot{\wfty{
      \ctx \cons \annkd{\var}{\kd}
    }{\ty}}[\pair{\sctx}{\skd}]
  \end{setbuild}
\end{aligned}
}\]
For example, in the unrefined interpretation, an element belongs to the
polymorphic identity type
$\tyall{\var}{\kdtype}{\tyfun{\var}{\var}}$ exactly when it behaves as an
identity function (or one of its approximations $\pi_n(id)$) for every
CUPER chosen as the interpretation of $\var$.

The refined interpretation follows the all-PER reading of
System~F~\citep{reynolds_1984}, but adds a compatibility constraint
between the refined
candidates and their unrefined over-approximations.

\[\refinedDefinition{
\begin{aligned}
  \sem{\RuleTyAll{concl}}[\sctx]
   \defeq
  \begin{setbuild}{\pair{\sval}{\svalalt}}[c]
    \forall \skd \in \sem{\wfkd{\ctx}{\kd}}[\sctx],\;
    \forall \skd\bots \in
    \sembot{\wfkd{\ctx}{\kd}}[\projctx{\ctx}{\sctx}] \\
    \projty{\ctx}{\sctx}{\kd}{\skd} \subseteq \skd\bots
    \Rightarrow
    \pair{\sval}{\svalalt} \in
    \sem{\wfty{
      \ctx \cons \annkd{\var}{\kd}
    }{\ty}}[\pair{\sctx}{\pair{\skd}{\skd\bots}}]
  \end{setbuild}
\end{aligned}
}\]

Type variables are interpreted by lookup in the context
interpretation; while in the refined semantics the refined component is projected:
\[\redBotDefinition{
\begin{aligned}
  \sem{\RuleTyVar{concl}}[\sctx] \defeq
  \skd
  \quad\text{where } \sctx(\var) = \pair{\skd}{\skd\bots}
  \qquad
  \sembot{\RuleTyVar{concl}}[\sctx] \defeq
  \skd
  \quad\text{where } \sctx(\var) = \skd
\end{aligned}
}\]

In both semantics, the interpretation of types
belongs to the interpretation of their kind:
\begin{theorem}[Type interpretation is well-formed]%
\label{thm:type-interpretation-well-formed}
Let $\wfctx{\ctx}$, $\wfkd{\ctx}{\kd}$, $\wftm{\ctx}{\ty}{\kd}$ be
well-formed context, kind, and type judgments then:
\begin{enumerate}[label={(\roman*)},topsep=\ListTopSep,itemsep=\ListItemSep,partopsep=\ListPartopSep]
\item{Refined:\quad}\label{thm:type-interpretation-well-formed:refined}
  If $\sctx \in \dom{\sem{\wfctx{\ctx}}}$, then
  $\sem{\wftm{\ctx}{\ty}{\kd}}[\sctx] \in \sem{\wfkd{\ctx}{\kd}}[\sctx]$.
\item{Unrefined:}\label{thm:type-interpretation-well-formed:unrefined}
  If $\sctx \in \dom{\sembot{\wfctx{\ctx}}}$, then
  $\sembot{\wftm{\ctx}{\ty}{\kd}}[\sctx] \in \sembot{\wfkd{\ctx}{\kd}}[\sctx]$.
\end{enumerate}
\end{theorem}

Following the same reasoning behind
\Cref{thm:semantics-per-respecting-kinds},
the semantics respects the notion of equality,
since equal contexts map to equal interpretations of types.
\begin{theorem}[Types respect PERs]%
\label{thm:per-respecting-types}
Let $\wfctx{\ctx}$, $\wfkd{\ctx}{\kd}$, $\wftm{\ctx}{\ty}{\kd}$ be
well-formed context, kind, and type judgments, then:
\begin{enumerate}[label={(\roman*)},topsep=\ListTopSep,itemsep=\ListItemSep,partopsep=\ListPartopSep]
\item{Refined:\quad}\label{thm:per-respecting-types:refined}
  If $\pair{\sctx}{\sctxalt} \in \sem{\wfctx{\ctx}}$, then
  $\sem{\wftm{\ctx}{\ty}{\kd}}[\sctx] = \sem{\wftm{\ctx}{\ty}{\kd}}[\sctxalt]$.
\item{Unrefined:}\label{thm:per-respecting-types:unrefined}
  If $\pair{\sctx}{\sctxalt} \in \sembot{\wfctx{\ctx}}$, then
  $\sembot{\wftm{\ctx}{\ty}{\kd}}[\sctx] = \sembot{\wftm{\ctx}{\ty}{\kd}}[\sctxalt]$.
\end{enumerate}
\end{theorem}

Last, we prove that indeed the refined interpretation of types is
included in the unrefined one, since in general kinds are interpreted
as families of PERs, where the type projection operation is used to
relate the two interpretations.
\begin{theorem}[Refined interpretation included in unrefined
  interpretation]%
\label{thm:refined-sub-unrefined}
Let $\wfctx{\ctx}$, $\wftm{\ctx}{\ty}{\kd}$ be well-formed context and
type judgments, and $\sctx \in \dom{\sem{\wfctx{\ctx}}}$, then
\begin{align*}
  \projty{\ctx}{\sctx}{\kd}{\sem{\wftm{\ctx}{\ty}{\kd}}[\sctx]}
  \subseteq
  \sembot{\wftm{\ctx}{\ty}{\kd}}[\projctx{\ctx}{\sctx}]
\end{align*}
\end{theorem}

As a corollary, unrefined equality is invariant under
replacement by refined-equal representatives. In other words, although
the unrefined interpretation of a type may relate more pairs than the
refined interpretation, the unrefined equality test is well-defined on
refined equivalence classes. This follows from the inclusion above and
symmetry and transitivity of the unrefined PER.
\begin{corollary}[Unrefined equality respects refined equality]%
\label{thm:unrefined-equality-respects-refined-equality}
Let $\wfctx{\ctx}$ be a well-formed context, let
$\wftm{\ctx}{\ty}{\kdtype}$ be a well-formed type, and let
$\sctx \in \dom{\sem{\wfctx{\ctx}}}$.

If $\pair{\sval}{\sval'} \in \sem{\wftm{\ctx}{\ty}{\kdtype}}[\sctx]$
and $\pair{\svalalt}{\svalalt'} \in \sem{\wftm{\ctx}{\ty}{\kdtype}}[\sctx]$,
then
\begin{align*}
  \pair{\sval}{\svalalt} \in
  \sembot{\wftm{\ctx}{\ty}{\kdtype}}[\projctx{\ctx}{\sctx}]
  \quad\Longleftrightarrow\quad
  \pair{\sval'}{\svalalt'} \in
  \sembot{\wftm{\ctx}{\ty}{\kdtype}}[\projctx{\ctx}{\sctx}].
\end{align*}
\end{corollary}

\subsubsection{Terms}\label{sec:term-semantics}
Terms are interpreted directly as elements of $\scottd$. The
interpretation is defined on raw terms rather than on typing
derivations and depends only on the assignment of term variables. Type
assignments are therefore erased via a runtime projection. This
exemplifies the fact that the semantics is \emph{annotation
independent} as neither typing judgment, derivation nor type
annotations can influence the interpretation of terms.

To begin, we define a function that projects away all the type
assignments from a context interpretation,
keeping only the term assignments.
\begin{definition}[Runtime Projection]%
\label{def:runtime-projection}
Let $\wfctx{\ctx}$. We define \emph{runtime projection} from refined
environments into runtime environments by induction on $\ctx$:
  \[
    \begin{array}{l @{\;} c @{\;} l@{\quad} l @{\;} c @{\;} l @{\quad} l @{\;} c @{\;} l}
      \multicolumn{9}{c}{\projrun{\ctx}{-} : \dom{\sem{\wfctx{\ctx}}} \to \runctx{\wfctx{\ctx}}} \\% [0.4em]
      \projrun{\ctxemp}{\sunit} & \defeq & \sunit &
      \projrun{\ctx \cons \annty{\var}{\ty}}{\pair{\sctx}{\sval}} & \defeq & \pair{\projrun{\ctx}{\sctx}}{\sval} &
      \projrun{\ctx \cons \annkd{\var}{\kd}}{\pair{\sctx}{\pair{\skd}{\skd\bots}}} & \defeq & \projrun{\ctx}{\sctx}
    \end{array}
  \]
\end{definition}
This projection discards type assignments, as types are erased at runtime.

Literals are then interpreted as their canonical embeddings into $\scottd$.
\[\runtimeDefinition{
\begin{aligned}
  \semtm{n}[\sctx] \defeq \demb{\snat}{n}
  \qquad
  \semtm{b}[\sctx] \defeq \demb{\sbool}{b}
\end{aligned}
}\]

Variables are interpreted by projection from the runtime environment,
$\lambda$-abstractions as the embedding of continuous functions, and application
as application in $\scottd$:
\[\runtimeDefinition{
\begin{aligned}
  \semtm{\var}[\sctx] \defeq \sctx\left(\var\right)
  \qquad
  \semtm{\tmlam{\var}{\tm}}[\sctx]
  \defeq \demb{\to}{\sval \mapsto
    \semtm{\tm}[\pair{\sctx}{\sval}]}
  \qquad
  \semtm{\tm \app \tmalt}[\sctx]
  \defeq \semtm{\tm}[\sctx] \app \semtm{\tmalt}[\sctx]
\end{aligned}
}\]

Type abstraction and application have no computational content, so they are
ignored:
\[\runtimeDefinition{
\begin{aligned}
  \semtm{\tmLam{\var}{\tm}}[\sctx] \defeq \semtm{\tm}[\sctx]
  \qquad
  \semtm{\tm \tmapp \ty}[\sctx] \defeq \semtm{\tm}[\sctx]
\end{aligned}
}\]

As usual we prove that the semantics is PER-respecting.
In particular, the interpretation of a term should not distinguish
between related contexts. For the unrefined semantics, the context
interpretation $\sembot{\wfctx{\ctx}}$ only contains the erased
type-level environment. It does not contain the runtime environment
used by $\semtm{-}$. We therefore need to define $\unrefass{\ctx}{\sctx}$ as
the PER on $\runctx{\wfctx{\ctx}}$ induced by the unrefined types in
$\ctx$.

\begin{definition}[Unrefined runtime assignment]%
\label{def:unrefined-runtime-assignment}
Let $\wfctx{\ctx}$. The \emph{unrefined runtime assignment} over $\sctx \in
\dom{\sembot{\wfctx{\ctx}}}$ is a PER on $\runctx{\wfctx{\ctx}}$, defined as
$\unrefass{\ctx}{\sctx}$:
  \[
    \begin{array}{l @{\;} c @{\;} l}
      \unrefass{\ctx}{-} & : & \dom{\sembot{\wfctx{\ctx}}} \to \per{\runctx{\wfctx{\ctx}}} \\
      \unrefass{\ctxemp}{\sunit} & \defeq & \diag{\ssing} \\
      \unrefass{\ctx \cons \annty{\var}{\ty}}{\sctx} & \defeq &
        \begin{setbuild}{\pair{\pair{\srec}{\sval}}{\pair{\srec'}{\svalalt}}}[l]
          \pair{\srec}{\srec'} \in \unrefass{\ctx}{\sctx},
          \pair{\sval}{\svalalt} \in \sembot{\wfty{\ctx}{\ty}}[\sctx]
        \end{setbuild} \\
      \unrefass{\ctx \cons \annkd{\var}{\kd}}{\pair{\sctx}{\skd}} & \defeq & \unrefass{\ctx}{\sctx}
    \end{array}
  \]
\end{definition}

Related refined runtime projections are related by the unrefined
runtime assignment relation.
\begin{lemma}[Refined context runtime projection is unrefined context
  runtime-related]%
\label[lemma]{thm:refined-runtime-projection-is-unrefined-runtime-interpretation}
Let $\wfctx{\ctx}$. If $\pair{\sctx}{\sctxalt} \in \sem{\wfctx{\ctx}}$, then
%\[
$
  \pair{\projrun{\ctx}{\sctx}}{\projrun{\ctx}{\sctxalt}}
  \in \unrefass{\ctx}{\projctx{\ctx}{\sctx}}
$.
%\]
\end{lemma}

We are now ready to state that both the refined and unrefined term
semantics are PER-respecting.
\begin{theorem}[Semantics is PER-respecting (terms)]%
\label{thm:per-respecting-terms}
Let $\wfctx{\ctx}$ and $\wftm{\ctx}{\tm}{\ty}$. Then:
\begin{enumerate}[label={(\roman*)},topsep=\ListTopSep,itemsep=\ListItemSep,partopsep=\ListPartopSep]
\item{Refined:\quad}\label{thm:per-respecting-terms:refined}
  If $\pair{\sctx}{\sctxalt} \in \sem{\wfctx{\ctx}}$, then
  $\pair{
    \semtm{\tm}[\projrun{\ctx}{\sctx}]
  }{
    \semtm{\tm}[\projrun{\ctx}{\sctxalt}]
  }
  \in
  \sem{\wfty{\ctx}{\ty}}[\sctx]$.
\item{Unrefined:}\label{thm:per-respecting-terms:unrefined}
  If $\sctx \in \dom{\sembot{\wfctx{\ctx}}}$;
  $\pair{\sctxalt}{\sctxaltalt} \in \unrefass{\ctx}{\sctx}$, then
  $\pair{
    \semtm{\tm}[\sctxalt]
  }{
    \semtm{\tm}[\sctxaltalt]
  }
  \in \sembot{\wfty{\ctx}{\ty}}[\sctx]$.
\end{enumerate}
\end{theorem}

Consistency of the syntax then is a corollary of
\Crefsub{thm:per-respecting-terms}{refined}. We represent false as the type
$\tyall{\var}{\kdtype}{\var}$ as from \RuleTmTyApp{label} it can be used to
inhabit every other type.

% : if we had a derivation of
% $\wftm{\ctxemp}{\tm}{\tyall{\var}{\kdtype}{\var}}$, then also every well-formed
% type would be inhabited in the empty context:
% \begin{align*}
% \inference{
%   \inference{
%     \vdots
%   }{
%     \wftm{\ctxemp}{\tm}{\tyall{\var}{\kdtype}{\var}}
%   }
%   \qquad
%   \inference{
%     \vdots
%   }{
%     \wftm{\ctxemp}{\ty}{\kdtype}
%   }
% }{
%   \wftm{\ctxemp}{\tm \tmapp \ty}{\ty}
% }[\RuleTmTyApp{label}]
% \end{align*}

\begin{corollary}[Relative consistency]
\label{thm:consistency}
There is no term $\tm$ such that
$\wftm{\ctxemp}{\tm}{\tyall{\var}{\kdtype}{\var}}$ is derivable.
\end{corollary}
We show that the type is not inhabited as it would imply the existence
of an element of $\emptyset$.

\section{Refinement types}%
\label{sec:model-refinement-types}

In this section we extend \corelan with refinement types. The extension
includes a proposition judgment together with reflection and subtyping rules and
an SMT validity judgment. \Cref{sec:model-refinement-types-syntax} presents the
syntax of the calculus and \Cref{sec:model-refinement-types-semantics} presents
the semantics.

\subsection{Syntax}%
\label{sec:model-refinement-types-syntax}

\begin{figure}
  {\small
  \textbf{Propositions} \hfill \boxed{$\wfprop{\ctx}{\prop}$}
  \begin{gather*}
  \RulePropAnd{full}
  \quad
  \RulePropNot{full}
  \quad
  \RulePropEq{full}
  \end{gather*}
  \textbf{Types continued} \hfill \boxed{$\wftm{\ctx}{\ty}{\kd}$}
  \begin{align*}
    \RuleTyRef{full}
  \end{align*}
  \textbf{Terms continued} \hfill \boxed{$\wftm{\ctx}{\tm}{\ty}$}
  \begin{align*}
  \RuleTmRefl{full}
  \qquad
  \RuleTmSub{full}
  \end{align*}
  \textbf{Subtyping} \hfill \boxed{$\wfsub{\ctx}{\ty}{\tyalt}$}
  \begin{gather*}
  \RuleSubTop{full}
  \quad
  \RuleSubImpl{full}
  \\
  \RuleSubFun{full}
  \quad
  \RuleSubRefine{full}
  \\
  \RuleSubRefl{full}
  \quad
  \RuleSubFlatten{full}
  \\
  \RuleSubForget{full}
  \quad
  \RuleSubAll{full}
  \\
  \RuleSubTrans{full}
  \end{gather*}
  }
  \caption{Typing rules of refined \corelan.}%
  \label{fig:syntax-ref}
\end{figure}

\Cref{fig:syntax-ref} extends the syntax of \corelan with
refinement types. It shows only the new rules.

\sparagraph{Propositions}
 form the syntax of refinements. They are built from term equality
(at any type) and the standard boolean connectives. Since propositions form a
separate layer on top of the programming language
their semantics are not required to be continuous.
This permits forming equalities between terms at
any type (using rule \propEqName). Any boolean-valued term of the language can be embedded as a
proposition by equating it to $\strue$.

This propositional logic is interpreted classically: a proposition is meant to
denote either $\strue$ or $\sfalse$. We can therefore
define all of propositional logic from conjunction and negation, with the other
connectives being defined in terms of those.

\paragraph{Types,} of the previous section,
%The types of the previous section
are extended by the \emph{refinement type}
$\tyref{\var}{\ty}{\prop}$, which can be thought of as the subset of elements of
$\ty$ satisfying $\prop$.

\paragraph{Terms}

Refinements add two new typing rules on terms: \emph{reflection} and
\emph{subtyping}. Reflection (via rule \RuleTmRefl{label}) asserts that a term satisfies the
proposition that it is equal to itself. This is the strongest proposition a term
can satisfy and is therefore
% useful for introducing assumptions in the context.
to increase the precision of type checking~\cite{knowlesHybridTypeChecking2010}.
Subtyping (via rule \RuleTmSub{label}) allows us to use a term at a `larger' type.

\paragraph{Subtyping and SMT}

The judgment $\wfsmt{\ctx}{\prop}$
models the discharge of propositions by an SMT solver.
It is used to give
meaning to the \RuleSubImpl{label} rule, which,
following the subtyping rules of~\cite{jhalaRefinementTypesTutorial2020},
reduces subtyping
to implication between propositions. In addition, subtyping is reflexive and
transitive, a refinement type is a subtype of the type it refines, and any
term of any type satisfies the trivial proposition.
Function types are contravariant in their argument and covariant in their
result, and type quantification is covariant in its body.
Finally, nested refinements
can get flattened
by taking the conjunction of their propositions or
propagate subtyping through the refinement type constructor.

\subsection{Semantics}%
\label{sec:model-refinement-types-semantics}

The semantics extends the core model with interpretations for propositions,
refinement types, and the new typing and subtyping rules.

\sparagraph{Propositions}
have semantics only at the refined level; from the unrefined
point of view they are just annotations. The interpretation of a proposition
 maps refined environments to a boolean value:% \looseness=-1
\begin{align*}
  \semprp{\wfprop{\ctx}{\prop}}[-] : \dom{\sem{\wfctx{\ctx}}} \to \sbool
\end{align*}

Equality is interpreted by checking whether the unrefined interpretation equates
the two terms at their type. As explained in
\Cref{sec:untyped-equality} the notion of equality \emph{cannot} depend on the
refined interpretation of types and
\Cref{thm:unrefined-equality-respects-refined-equality} shows that this
unrefined equality test is invariant under refined-equal replacement. Taking this into account, the interpretation of equality is true when the unrefined interpretations of the two terms are related by the unrefined interpretation of their type:
\[\propDefinition{
\begin{aligned}
  \semprp{\wfprop{\ctx}{\tm \propeq_\ty \tmalt}}[\sctx] \defeq
  % \begin{cases}
    \strue \iff \pair{
      \semtm{\tm}[\projrun{\ctx}{\sctx}]
    }{
      \semtm{\tmalt}[\projrun{\ctx}{\sctx}]
    } \in
    \sembot{\wfty{\ctx}{\ty}}[\projctx{\ctx}{\sctx}]
  %   \\
  %   \sfalse & \text{otherwise}
  % \end{cases}
\end{aligned}
}\]

The remaining connectives are interpreted as expected: conjunction returns the
interpretation of the right conjunct when the left conjunct is true and
$\sfalse$ otherwise; negation is boolean negation.
\[\propDefinition{
\begin{aligned}
\semprp{\wfprop{\ctx}{\prop \propand \propalt}}[\sctx]
&\defeq \semprp{\wfprop{\ctx}{\prop}}[\sctx] \land
  \semprp{\wfprop{\ctx}{\propalt}}[\sctx] \\
  % \begin{cases}
  % \semprp{\wfprop{\ctx}{\propalt}}[\sctx]
  % & \text{if } \semprp{\wfprop{\ctx}{\prop}}[\sctx] = \strue \\
  % \sfalse & \text{otherwise}
% \end{cases} \\
\semprp{\wfprop{\ctx}{\propnot \prop}}[\sctx]
&\defeq \neg \semprp{\wfprop{\ctx}{\prop}}[\sctx]
% &\defeq \begin{cases}
%   \sfalse & \text{if } \semprp{\wfprop{\ctx}{\prop}}[\sctx] = \strue \\
%   \strue & \text{otherwise}
% \end{cases}
\end{aligned}
}\]

As before, the interpretation of propositions respects PERs.
\begin{theorem}[PER-Respecting Propositions]%
\label{thm:refined-per-respecting-propositions}%
Let $\wfctx{\!\ctx\!}$ and $\wfprop{\ctx\!}{\!\prop}$. If
$\pair{\sctx}{\sctxalt}\!\!\in\!\!\sem{\wfctx{\!\ctx\!}}$, then
\begin{align*}
  \semprp{\wfprop{\ctx}{\prop}}[\sctx] =
  \semprp{\wfprop{\ctx}{\prop}}[\sctxalt]
\end{align*}
\end{theorem}

\paragraph{Refinement types}

Refinement types carry no meaning in the unrefined semantics:
\[\botDefinition{
\begin{aligned}
  \sembot{\RuleTyRef{concl}}[\sctx] \defeq \sembot{\wfty{\ctx}{\ty}}[\sctx]
\end{aligned}
}\]
In the refined semantics $\tyref{\var}{\ty}{\prop}$ is the subset of the
interpretation of $\ty$ that satisfies $\prop$. By
\Cref{thm:refined-per-respecting-propositions} propositions do not
distinguish related environments and either component can be used:
\[\refinedDefinition{
\begin{aligned}
  \sem{\RuleTyRef{concl}}[\sctx]
  \defeq
  \begin{setbuild}{
    \pair{\sval}{\svalalt} \in \sem{\wfty{\ctx}{\ty}}[\sctx]
  }
    \semprp{\wfprop{
        \ctx \cons \annty{\var}{\ty}
    }{\prop}}[\pair{\sctx}{\sval}] = \strue
  \end{setbuild}
\end{aligned}
}\]

\paragraph{Subtyping.}
We keep the behaviour of the SMT judgment abstract and require the following
soundness condition: if a proposition is well-formed and valid then it is true
in all environments.
\begin{assumption}[SMT soundness]%
\label{ass:smt-soundness}
If $\wfctx{\ctx}$, $\wfprop{\ctx}{\prop}$, and $\wfsmt{\ctx}{\prop}$,
then $\forall \sctx \in \dom{\sem{\wfctx{\ctx}}}$,
\begin{align*}
  \semprp{\wfprop{\ctx}{\prop}}[\sctx] = \strue
\end{align*}
\end{assumption}

Subtyping judgments are interpreted in the refined semantics as proofs of subset
inclusion between the two interpretations of types. Since the unrefined
semantics ignores refinements a subtype is equal to its supertype with respect
to the unrefined semantics:

\begin{theorem}[Subtyping semantics]%
\label{thm:subtyping-semantics}
If $\wfctx{\ctx}$, $\wfty{\ctx}{\ty}$, $\wfty{\ctx}{\tyalt}$,
and $\wfsub{\ctx}{\ty}{\tyalt}$ then:
\begin{enumerate}[label={(\roman*)},topsep=\ListTopSep,itemsep=\ListItemSep,partopsep=\ListPartopSep]
\item{Refined:\quad}\label{thm:subtyping-semantics:refined}%
  If $\sctx \in \dom{\sem{\wfctx{\ctx}}}$ then
  $\sem{\wfty{\ctx}{\ty}}[\sctx] \subseteq
  \sem{\wfty{\ctx}{\tyalt}}[\sctx]$.
\item{Unrefined:}\label{thm:subtyping-semantics:unrefined}%
  If $\sctx \in \dom{\sembot{\wfctx{\ctx}}}$ then
  $\sembot{\wfty{\ctx}{\ty}}[\sctx] =
  \sembot{\wfty{\ctx}{\tyalt}}[\sctx]$.
\end{enumerate}
\end{theorem}

\paragraph{Terms.} The interpretation of terms is unchanged since no new term
formers are introduced, and the proof of \Cref{thm:per-respecting-terms} covers
the new typing rules.

\section{Inductive types}%
\label{sec:model-inductive}
Finally, we extend the calculus of~\Cref{sec:model-refinement-types}
with inductive types. We add kinds for inductive type constructors and
formation rules for inductive types, reflecting index dependence.

The central question of our design is which recursive occurrences of an
inductive type should be allowed so that its semantic interpretation remains
well defined.
Strict positivity is the familiar discipline from
dependent type theory: a datatype such as lists is accepted because
recursive occurrences appear only on the right of the function
type. Refinement systems also use two broader
patterns. Positive inductive types allow non-strictly-positive shapes
whose overall variance is still covariant. Stratified inductive types
allow recursive occurrences at smaller indices, even in negative
positions.

Strictly positive inductive types admit an elegant canonical syntactic
presentation in terms of W-types, which capture constructions
generated by strictly positive signatures. However, positive and
stratified inductive types allow more general shapes that are not
reducible to standard strictly positive signatures. For this reason
inductive types will be presented via a general axiom schema.

\Cref{sec:model-inductive-syntax} presents the syntax of indexed
kinds, positive and stratified inductive types, and the typing rules
for introduction and elimination, in particular both positive and
stratified types use a single general introduction rule and
elimination schema. \Cref{sec:model-inductive-semantics} presents
the semantics of indexed kinds and inductive types and prove that the
semantics of the eliminator is well defined and satisfies the
semantics of the typing rules.

\subsection{Syntax}%
\label{sec:model-inductive-syntax}
\begin{figure}
  {\small
  \textbf{Kinds continued} \hfill \boxed{$\wfkd{\ctx}{\kd}$}
  \begin{align*}
  \RuleKdFun{full}
  \end{align*}
  \textbf{Types continued} \hfill \boxed{$\wftm{\ctx}{\ty}{\kd}$}
  \begin{align*}
  \RuleTyApp{full}
  \qquad
  \RuleKdIndPos{full}
  \end{align*}
  \begin{align*}
  \RuleKdIndStrat{full}
  \end{align*}
  \textbf{Terms continued} \hfill \boxed{$\wftm{\ctx}{\tm}{\ty}$}
  \begin{align*}
  \RuleTmIntro{full}
  \end{align*}
  \begin{align*}
  \RuleTyElim{full}
  \end{align*}%
  }
  \textbf{Subtyping continued} \hfill \boxed{$\wfsub{\ctx}{\ty}{\tyalt}$}
  \begin{align*}
  \RuleSubIx{full}
  \end{align*}%
  \caption{Typing rules of inductive \corelan.
    Where
    $\stratifiedLong{m}{\ctx}{\ctxalt}{\ctxaltalt}
  {\tyctor}{ts}{\ty} \defeq \stratified{m}{\ctx}{\ctxalt}{\ctxaltalt}
  {\tyctor}{ts}{\ty}$.
  }%
  \label{fig:syntax-ind}
\end{figure}

\Cref{fig:syntax-ind} extends the syntax of \corelan with inductive
types. We only show the new rules.

\paragraph{Kinds} Since in our calculus inductive
types are indexed (\eg the \texttt{Val} type of \Cref{sec:inductive-types}),
rule \RuleKdFun{label} extends kinds with term indexed kinds, the function
space is dependent in the same way as the function space of terms.

\paragraph{Inductive types}
There are two syntactic disciplines to form inductive types: positive
as in rule \RuleKdIndPos{label} and stratified as in rule \RuleKdIndStrat{label}.
Syntactically,
an inductive type $\tyctor$ is defined in the form
$ \defind{\tyctor}{
    \kdfun{\mathcolor{clrBlue}{\ioverline{\mathcolor{black}{k}}{\mathcolor{black}{\ty}}}}{\kdtype}
  }{\variance}{
    \mathcolor{clrMaroon}{\iunderline{\mathcolor{clrMaroon}{l}}{
      \mathcolor{black}{
        \ann{\ctor}{
      \tyfun{
        \mathcolor{clrGreen}{\ioverline{\mathcolor{clrGreen}{j}}{\mathcolor{black}{(\ann{\varalt}{\tyalt})}}}
      }{
        \tyctor \tmapp \mathcolor{clrBlue}{\ioverline{\mathcolor{clrBlue}{k}}{\mathcolor{black}{\tm}}}
      }
    }}}}
  }$,
  where $\variance$ is either $\varpos$ or $\varstratR$ for positive and stratified inductive types respectively.
\tyctor is a type operator taking
$\mathcolor{clrBlue}{k}$ arguments, called the \emph{indices} of the inductive type.
Each type is defined as a collection
of $\mathcolor{clrMaroon}{l}$ constructors and each constructor is specified by a
typing signature which is a, possibly nullary if $\mathcolor{clrGreen}{j} = 0$, dependent
function returning the inductive type
being defined fully applied to
some concrete indices; the argument of the function can be any type,
including types that contain the inductive type being defined.
Both rules at the premises have a predicate
($\mathsf{pol}$ and $\mathsf{strat}$) that ensures the well-formedness of the inductive type and are defined next.

\sparagraph{Positive inductive types} are those where every occurrence of the
type constructor appears under an even number of function domains.
Universal quantification is also treated as variance-flipping binder, since it can
capture occurrences of the type constructors.

\begin{example}[Hofmann’s Routine]\label{ex:hofmann-routine}
As an example of a positive inductive type,
Hofmann’s Routine (from~\Cref{sec:inductive-types}) is defined in \corelan,
to be of kind $\kdtype$, as follows:
\[
  \mathsf{Rou}
  \defeq
  \defind{\mathsf{R}}{\kdtype}{\varpos}{
    \ann{\mathsf{Over}}{\mathsf{R}},
    \ann{\mathsf{Next}}{
      \tyfun{(
        \tyfun{(\tyfun{\mathsf{R}}{\mathsf{ListNat}})}{\mathsf{ListNat}}
      )}{\mathsf{R}}
    }
  }.
\]
The recursive occurrence of $\mathsf{R}$ in the argument of
$\mathsf{Next}$ appears to the left of two function arrows, so its polarity is
positive.
\end{example}

%
%\begingroup
%\setlength{\intextsep}{0.25\baselineskip}
%\setlength{\columnsep}{0.75em}
\begin{wrapfigure}{r}{0.25\textwidth}
%\vspace{-1.5\baselineskip}
\centering
\begin{tikzpicture}[scale=0.78, transform shape]
\node (both) at (0,2)  {$\posboth$};
\node (pos)  at (-1,1) {$\pospos$};
\node (neg)  at (1,1)  {$\posneg$};
\node (none) at (0,0)  {$\posnone$};
\draw (both) -- (pos);
\draw (both) -- (neg);
\draw (pos)  -- (none);
\draw (neg)  -- (none);
\end{tikzpicture}
\vspace{-0.75\baselineskip}
\caption{Polarity lattice.}%
\label{fig:hasse}
\vspace{-0.75\baselineskip}
\end{wrapfigure}
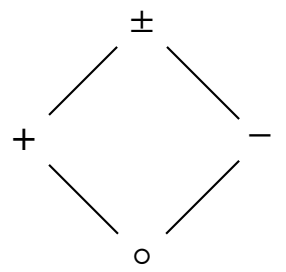
%\endgroup
Formally, the variance of a specific type constructor in a type
is captured by a polarity lattice.%
\begin{definition}[Polarity lattice]%
\label{def:polarity-lattice}
Polarities are elements of a lattice with four elements $\posboth$
denoting both positive and negative occurrences,
$\pospos$ denoting only positive occurrences,
$\posneg$ denoting only negative occurrences,
and $\posnone$ denoting no occurrences.
The Hasse diagram (\Cref{fig:hasse}) uniquely
defines the ordering ($\posleq$), join ($\poslub$) and complement
($\posconv{p}$)
operations on polarities.
\end{definition}

The polarity of a type constructor, %  in a type or kind,
is next defined
as an element of the polarity lattice.
\begin{definition}[Polarity assignment]%
\label{def:positivity}
The polarity of a type constructor $\tyctor$ in a type $\ty$
or kind $\kd$ is an element of the polarity lattice, defined by
structural induction on the type or kind.
\[
\begin{array}{rclcrcl}
\positive{\tyctor}{\tybase} &\defeq& \posnone
&\quad&
\positive{\tyctor}{\tyfun*[\var]{\ty}{\tyalt}} &\defeq& \posconv{\positive{\tyctor}{\ty}} \poslub
  \positive{\tyctor}{\tyalt} \\
\positive{\tyctor}{\tyctor} &\defeq& \pospos
&\quad&
\positive{\tyctor}{\tyref{\var}{\ty}{\prop}} &\defeq& \positive{\tyctor}{\ty} \\
\positive{\tyctor}{\ty \tmapp \ioverline*{k}{\tm}} &\defeq&
\positive{\tyctor}{\ty}
&\quad&
\positive{\tyctor}{\tyall{\var}{\kd}{\ty}} &\defeq& \posconv{\positive{\tyctor}{\kd}} \poslub
\positive{\tyctor}{\ty} \\
\positive{\tyctor}{\kdtype} &\defeq& \posnone
&\quad&
\positive{\tyctor}{\kdfun*[{\var}]{\ty}{\kd}} &\defeq& \posconv{\positive{\tyctor}{\ty}} \poslub
\positive{\tyctor}{\kd}
\end{array}\]
\[\begin{array}{rcl}
\positive{\tyctor}{\declpos[\tyctoralt]} &\defeq& \Lub\limits_{i, j} \positive{\tyctor}{\tyalt_{i, j}} \\
\positive{\tyctor}{\declstrat[\tyctoralt]} &\defeq& \begin{cases}
    \posnone & \text{if } \forall i,j. \positive{\tyctor}{\tyalt_{i,j}} = \posnone \\
    \posboth & \text{otherwise}
  \end{cases}\\
\end{array}
\]
\end{definition}

We say that the occurrence of constructor $\tyctor$ in a type
is positive if $\positive{\tyctor}{\ty} \posleq
\pospos$ and negative if $\positive{\tyctor}{\ty} \posleq \posneg$. The
definitions extend naturally to kinds.

To assign the polarity of a type in a inductive declaration, when
the type is positive we can simply collect the polarities of the type
constructor in each of the constructor argument types.

When the type is stratified, we require that the type constructor does
not appear in any of the argument, not even in positive
positions. This is because stratified types, by self reference in
negative positions, can move a syntactically positive occurrence into a
semantically negative position.\looseness-1

\sparagraph{Stratified inductive types}
permit negative occurrences of the inductive type by enforcing the syntactic discipline of stratification~\cite{jacob-raoIndexStratifiedTypes2018}.
Concretely, given a natural number metric on the indices of the inductive type
we define a structural order on terms.
Now each recursive occurrence of the
inductive type in each of the constructor arguments must be indexed by
terms that are strictly smaller than the indices of the constructor
return type. This ensures that the stratified types are well-founded
as they are generated bottom-up from the well-founded order on the indices.

\begin{example}[Stratified indexed types]\label{ex:stratified-indexed-types}
The indexed value type from \Cref{sec:stratified-types} is  a
stratified type operator
$\mathsf{Val} : \kdfun*[\var]{\mathsf{Typ}}{\kdtype}$, where
$\mathsf{Typ}$ is the inductive type of object-language simple types. The \corelan declaration is the following:
\[
\mathsf{Val} \defeq
  \defind{\mathsf{V}}{\kdfun*[\var]{\mathsf{Typ}}{\kdtype}}{\varstratR}{
    \ann{\mathsf{VIota}}{\tyfun{\tynat}{\!\mathsf{V}\tmapp\mathsf{TIota}}},
    \ann{\mathsf{VArr}}{
      \tyfun*[t]{\mathsf{Typ}}{\!
      \tyfun*[u]{\mathsf{Typ}}{\!
      \tyfun{
        (\tyfun{\mathsf{V}\tmapp t}{\mathsf{V}\tmapp u})\!
      }{\!\mathsf{V}\tmapp(\mathsf{TArr}\tmapp t\tmapp u)}}}
    }
  }
\]
The function argument of $\mathsf{VArr}$ contains two recursive occurrences, one
of them, $\mathsf{V}\tmapp t$, in a negative position, but both occurrences are indexed by the subcomponents
$t$ and $u$ of the result index $\mathsf{TArr}\tmapp t\tmapp u$. Taking the
metric $\varstratR$ to be the size of
%a simple type
$\mathsf{Typ}$, both recursive indices are strictly
smaller than the result index, so
 $\mathsf{Val}$ is well defined.
\end{example}

More formally, given a natural number metric on the indices,
the stratification predicate checks that all recursive occurrences of the
inductive type in a type are indexed by terms that are strictly
smaller than the indices of the type being defined.

\begin{definition}[Stratification predicate]%
\label{def:stratification}
The stratification of a type constructor $\tyctor$ in a type $\ty$
or kind $\kd$ is defined as a predicate by structural induction on the
type or kind.
\begin{align*}
\stratified{m}{\ctx}{\ctxalt}{\ctxaltalt}{\tyctor}{ts}{\tybase}
  &\defeq \strue \\
\stratified{m}{\ctx}{\ctxalt}{\ctxaltalt}{\tyctor}{ts}{\ty \tmapp \ioverline*{k'}{\tmalt}}
  &\defeq \stratified{m}{\ctx}{\ctxalt}{\ctxaltalt}{\tyctor}{ts}{\ty}
    \text{ where } \ty \neq \tyctor \\
\stratified{m}{\ctx}{\ctxalt}{\ctxaltalt}{\tyctor}{ts}{\tyfun*[\var]{\ty}{\tyalt}}
  &\defeq \stratified{m}{\ctx}{\ctxalt}{\ctxaltalt}{\tyctor}{ts}{\ty} \land
    \stratified{m}{\ctx}{(\ctxalt \cons \annty{\var}{\ty})}{\ctxaltalt}{\tyctor}{ts}{\tyalt} \\
\stratified{m}{\ctx}{\ctxalt}{\ctxaltalt}{\tyctor}{ts}{\tyref{\var}{\ty}{\prop}}
  &\defeq \stratified{m}{\ctx}{\ctxalt}{\ctxaltalt}{\tyctor}{ts}{\ty} \\
\stratified{m}{\ctx}{\ctxalt}{\ctxaltalt}{\tyctor}{ts}{\tyall{\var}{\kd}{\ty}}
  &\defeq \stratified{m}{\ctx}{\ctxalt}{\ctxaltalt}{\tyctor}{ts}{\kd} \land
    \stratified{m}{\ctx}{(\ctxalt \cons \annkd{\var}{\kd})}{\ctxaltalt}{\tyctor}{ts}{\ty} \\
\stratified{m}{\ctx}{\ctxalt}{\ctxaltalt}{\tyctor}{ts}{\kdtype}
  &\defeq \strue \\
\stratified{m}{\ctx}{\ctxalt}{\ctxaltalt}{\tyctor}{ts}{\kdfun*[{\var}]{\ty}{\kd}}
  &\defeq \stratified{m}{\ctx}{\ctxalt}{\ctxaltalt}{\tyctor}{ts}{\ty} \land
    \stratified{m}{\ctx}{(\ctxalt \cons \annty{\var}{\ty})}{\ctxaltalt}{\tyctor}{ts}{\kd}
\end{align*}
When the type is an inductive type, we check that all the constructor
argument types are stratified with respect to the metric.
\[
\resizebox{\linewidth}{!}{$
%\begin{multline*}
\stratified{m}{\ctx}{\ctxalt}{\ctxaltalt}{\tyctor}{ts}{
  \declstrat[\tyctoralt]
} \defeq
\bigwedge_{i, j} \stratified{m}{\ctx}{
(\ctxalt \cons \annkd{\tyctoralt}{\kdfun{\ioverline*{k}{\ty}}{\kdtype}})
}{\ctxaltalt}{\tyctor}{ts}{\tyalt_{i,j}}
%\end{multline*}
$}
\]
The interesting case is when the type is a recursive occurrence
of the type being defined:
\begin{multline*}
\stratified{m}{\ctx}{\ctxalt}{\ctxaltalt}{\tyctor}{\ioverline*{k}{\tm}}{\tyctor
  \tmapp \ioverline*{j}{\tmalt}} \\
\defeq \left\{\begin{array}{l}
    j = k, \forall \sctx \in \dom{\sem{\wfctx{\ctx}}},
    \forall \pair{\sctx}{\sctxalt} \in \dom{\sem{\wfctx{\ctx \cons \ctxalt}}},
    \forall \pair{\sctx}{\sctxaltalt} \in \dom{\sem{\wfctx{\ctx \cons \ctxaltalt}}}, \\
    \dret{\snat}{\semtm{m}[\projrun{\ctx}{\sctx}] \app \ioverline{k}{\semtm{\tmalt}[\projrun{\ctx \cons \ctxalt}{\pair{\sctx}{\sctxalt}}]}}
    <
    \dret{\snat}{\semtm{m}[\projrun{\ctx}{\sctx}] \app \ioverline{k}{\semtm{\tm}[\projrun{\ctx \cons \ctxaltalt}{\pair{\sctx}{\sctxaltalt}}]}}
    \end{array}\right.
\end{multline*}

\end{definition}
A recursive occurrence of a type constructor $\tyctor$ in a type $\ty$
is stratified with respect to a sequence of indices
$\ioverline*{k}{\tm}$ if
$\stratified{m}{\ctx}{\ctxalt}{\ctxaltalt}{\tyctor}{\ioverline*{k}{\tm}}{\ty}$ holds and the
definition extends naturally to kinds.

While the condition requires checking the stratification predicate for
all environments, in practice it can be checked syntactically by
ensuring that the indices of the recursive occurrences are
syntactically smaller than those of the constructor return type,
just as structural termination is checked.\looseness-1

\paragraph{Introduction and elimination,} are neutral with respect to
the formation discipline, as they are semantically justified by the
fixed point construction of inductive types. Thus, the same introduction and
elimination rules apply to both positive and stratified inductive types; the
formation discipline is handled entirely by the corresponding type formation
rule.

The introduction rule \RuleTmIntro{label} simply unfolds the inductive type in the
constructor types.
The elimination rule \RuleTyElim{label}
instead is more complicated: it takes a target
type $\tyaltalt$ where the inductive type is eliminated, and a case
$c_i$ for each constructor $\tyctor_i$ of the inductive type. Each
case $c_i$ has access both to the unfolding of the constructor
arguments $\tyalt_j\subst{D}{\tyctor}$ and to the inductive
hypothesis for each strictly positive occurrence of the inductive type
in the constructor arguments
$\tyel{\tyctor}{\tyaltalt}{\varaltalt_j}{\tyalt_j}$.
Introducing the
inductive hypothesis in strictly positive positions corresponds to the permitted
recursive calls of structurally recursive
functions over inductive types with pattern
matching~\cite{abelSemanticAnalyisStructural}.
\begin{definition}[Inductive hypothesis]%
Given a type constructor $\tyctor$ and a type $\ty$,  the
inductive hypothesis $\tyel{\tyctor}{\tyaltalt}{\tmalt}{\ty}$ is defined
by structural induction over $\ty$.
\[\begin{array}{rclcrcl}
\tyel{\tyctor}{\tyaltalt}{\tmalt}{
  \ty \tmapp \ioverline*{n}{\tm}
} & \defeq & \tyunit &\quad&
\tyel{\tyctor}{\tyaltalt}{\tmalt}{
  \tyctor \tmapp \ioverline*{k}{\tm}
} & \defeq & \tyaltalt\subst{\ioverline*{k}{
  \tm
}}{\ioverline*{k}{\var}}%
  \subst{\tmalt}{\varalt} \\
\tyel{\tyctor}{\tyaltalt}{\tmalt}{
  \tybase
} & \defeq & \tyunit &\quad&
\tyel{\tyctor}{\tyaltalt}{\tmalt}{
  \tyfun[\var]{\ty}{\tyalt}
} & \defeq & \tyfun[\var]{\ty}{
  \tyel{\tyctor}{\tyaltalt}{\tmalt \app \var}{\tyalt}
} \\
\tyel{\tyctor}{\tyaltalt}{\tmalt}{
  \tyref{\var}{\ty}{\prop}
} & \defeq & \tyel{\tyctor}{\tyaltalt}{\tmalt}{\ty} &\quad&
\tyel{\tyctor}{\tyaltalt}{\tmalt}{
  \tyall{\var}{\kd}{\ty}
} & \defeq & \tyall{\var}{\kd}{
  \tyel{\tyctor}{\tyaltalt}{\tmalt \tmapp \var}{\ty}
}\\
\multicolumn{7}{l}{
\text{where unit can be
encoded as } \tyunit \defeq
\tyref{\var}{\tybool}{\var \propeq \tmtrue} \text{ and } \sunit \defeq
\strue
}
\end{array}\]
\end{definition}
\noindent

Next, we exemplify the elimination rule
with the types defined in examples~\ref{ex:hofmann-routine} and~\ref{ex:stratified-indexed-types}.

\begin{example}[Eliminating Hofmann’s Routine]\label{ex:hofmann-elim}
For the routine type $\mathsf{Rou}$, an eliminator into a target type
$\rho$ depending on $\mathsf{r} : \mathsf{Rou}$ has the shape
$ \tmelim{\mathsf{Rou}}{c_{\mathsf{Over}}, c_{\mathsf{Next}}}
  :
  \tyfun*[{\mathsf{r}}]{\mathsf{Rou}}{\rho(\mathsf{r})}.
$
The two cases correspond to the two constructors:
\[
\begin{array}{rcl}
  c_{\mathsf{Over}}
  &:&
  \rho(\mathsf{Over}_{\mathsf{Rou}})
  \\
  c_{\mathsf{Next}}
  &:&
  \tyfun[g]{(\tyfun{(\tyfun{\mathsf{Rou}}{\mathsf{ListNat}})}{\mathsf{ListNat}})}{
  \tyfun[h_g]{(\tyfun{(\tyfun{\mathsf{Rou}}{\mathsf{ListNat}})}{\tyunit})}{
    \rho(\mathsf{Next}_{\mathsf{Rou}}\tmapp g)
  }} .
\end{array}
\]
The recursive occurrence of $\mathsf{R}$ in the constructor argument is positive
but not strictly positive, so the $h_g$ carries no
information.
Using this we can write all the structurally recursive functions
(all the functions except ``extract'')
in \citet{HofmannNonStrictlyPositive}'s original presentation of the
Routine type.
\end{example}

\begin{example}[Eliminating indexed values]\label{ex:stratified-elim}
For the stratified type $\mathsf{Val}$, an eliminator into a target type
$\rho(t, v)$ depending on an index $t : \mathsf{Typ}$ and a value
$v : \mathsf{Val}\tmapp t$ has the shape
$
  \tmelim{\mathsf{Val}}{c_{\mathsf{I}}, c_{\mathsf{A}}}
  :
  \tyfun*[t]{\mathsf{Typ}}{
  \tyfun*[v]{\mathsf{Val}\tmapp t}{\rho(t, v)}}.
$
The base case proves the target for values built by
$\mathsf{VIota}$ while the arrow case receives a function argument together with the induction
hypothesis generated for its codomain:
\[
\begin{array}{r@{\;}c@{\;}c@{\;}c@{\;}c@{\;}c@{\;}c@{\;}c}
  c_{\mathsf{I}}
  &:&
  (\ann{n}{\tynat}) &\to& \tyunit &\to&
  \rho(\mathsf{TIota}, \mathsf{VIota}_{\mathsf{Val}}\tmapp n)\\% [0.3em]
  c_{\mathsf{A}}
  &:&
  (\ann{t}{\mathsf{Typ}}) &\to&
  (\ann{u}{\mathsf{Typ}}) &\to&
  (\ann{f}{(\mathsf{Val}\tmapp t \to \mathsf{Val}\tmapp u)}) \\
  &\to&
  \tyunit &\to&
  \tyunit &\to&
  ((\ann{x}{\mathsf{Val}\tmapp t}) \to \rho(u, f \tmapp x)) &\to
  \rho(
    \mathsf{TArr}\tmapp t\tmapp u,
    \mathsf{VArr}_{\mathsf{Val}}\tmapp t\tmapp u\tmapp f
  )
\end{array}
\]
Thus, to eliminate a function value, the case for $\mathsf{VArr}$ may use the
induction hypothesis on every result $f \tmapp x$ at the smaller index $u$.
\end{example}

\paragraph{Subtyping,} is extended with the rule \RuleSubIx{label},
which allows equal indices to be substituted in the type of an
inductive datatype. Uniform parametric datatypes do not require any
additional subtyping rules, since they can be encoded using universal
and existential quantification~\cite{girardProofsTypes1989}.
Since the encoding is expressed entirely using existing type formers,
subtyping for parametric datatypes is inherited from the subtyping
rules of the basic type formers, without requiring any
datatype-specific subtyping rule.

\subsection{Semantics}%
\label{sec:model-inductive-semantics}

The semantics of inductive types extends both interpretations with indexed type
operators. The unrefined semantics forgets the indices and uses a guarded fixed
point construction, while the refined semantics keeps the indices and justifies
positive and stratified declarations by separate fixed point arguments.

\subsubsection{Type operators and application.}
The unrefined interpretation of kinds just ignores all the index information,
since the unrefined semantics ignores every term dependency:
\[\botDefinition{
\begin{aligned}
\sembot{\RuleKdFun{concl}}[\sctx] &\defeq
\sembot{\wfkd{\ctx \cons \annty{\var}{\ty}}{\kd}}[\sctx]
\\
\sembot{\RuleTyApp{concl}}[\sctx] &\defeq
\sembot{\wftm{\ctx}{\tyalt}{\kdfun*[{\var}]{\ty}{\kd}}}[\sctx]
\end{aligned}
}\]
These definitions make $\sembot{\wfkd{\ctx}{\kd}}[\sctx] =
\pper{\dinf}$ for all kinds $\kd$.

Instead, in the refined semantics, function kinds are interpreted as
set-theoretic functions, which take as arguments as many elements of
$\dinf$ as the arity of the kind and return an element of
$\per{\dinf}$; as a side condition, the function respects the
signature given by the kind and respects the equivalence relation of
the domain of the function.
\[\refinedDefinition{
\begin{aligned}
&\sem{\RuleKdFun{concl}}[\sctx] \\ &\quad \defeq
\begin{setbuild}{\sfun}[c@{\quad}c]
  \sfun \in \dinf \to \left(\dinf \to \right)^{\arity{\kd}}
  \per{\dinf} &\textcolor{gray}{\text{(i)}} \\
  \forall \sval \in \dom{\sem{\wfty{\ctx}{\ty}}[\sctx]} \;
  \sfun(\sval) \in \sem{\wfkd{\ctx \cons \annty{\var}{\ty}}{\kd}}[\pair{\sctx}{\sval}]
  & \textcolor{gray}{\text{(ii)}}\\
  \forall \pair{\sval}{\svalalt} \in \left(
  \sembot{\wfty{\ctx}{\ty}}[\projctx{\ctx}{\sctx}] \cap \dom{\sem{\wfty{\ctx}{\ty}}[\sctx]}^2
  \right)\;
  \sfun(\sval) = \sfun(\svalalt) & \textcolor{gray}{\text{(iii)}}
\end{setbuild}
\end{aligned}
}\]
where the arity of a kind is defined
as

$\arity{\kdtype} \defeq 0$
and
$\arity{\kdfun*[{\var}]{\ty}{\kd}} \defeq 1 + \arity{\kd}$.

One might wonder why the semantics of the function kind is not defined
directly as functions $\sval \in \dom{\sem{\wfty{\ctx}{\ty}}[\sctx]}
\to \sem{\wfkd{\ctx \cons
\annty{\var}{\ty}}{\kd}}[\pair{\sctx}{\sval}]$. But, since set
theoretically functions are defined as sets of pairs, the
interpretation of kinds cannot be related by set-theoretic subsets.
Conditions (i) and (ii) allow us to define the interpretation of
function kinds as a direct definition, while not constraining the
domain to be exactly $\dom{\sem{\wfty{\ctx}{\ty}}[\sctx]}$.
Condition (iii) ensures that the interpretation of indexed types cannot
distinguish between related, hence equal, terms; as inductive types
can witness equalities between them. As an example, the subtyping
rule \RuleSubIx{label} allows substitution of equal indices in the
type of an inductive type.

Application instead can be interpreted as set-theoretic application; the side condition in the definition of the
interpretation of function kinds ensures that the application is
well-defined.
\[\refinedDefinition{
\begin{aligned}
\sem{\RuleTyApp{concl}}[\sctx] \defeq
  \sem{\wftm{\ctx}{\tyalt}{\kdfun*[{\var}]{\ty}{\kd}}}[\sctx]\left(
    \semtm{\tm}[\projrun{\ctx}{\sctx}]
  \right)
\end{aligned}
}\]

\paragraph{Extending type projections}
Indexed kinds are the reason type projections
(\Cref{def:type-projection}) are required, so that we can
check if the refined interpretation of a type
in $\sem*{\wfkd{\ctx}{\kdfun*[{\var}]{\ty}{\kd}}}[\sctx]$ (which is a
function) is a subset of the unrefined interpretation of a type in
$\sembot{\wfkd{\ctx}{\kdfun*[{\var}]{\ty}{\kd}}}[\sctx]$ (which is a
CUPER).
Thus instead of looking directly at the function for comparison we
look at its image, since kinds can be defined outside the
interpretation of the argument, we cannot take the simple
set-theoretic image, but we need to take the image of the function
restricted to the domain of the interpretation of the domain type.
More formally:
\begingroup
\renewcommand{\thedefinition}{\ref{def:type-projection} - Extended}
\begin{definition}[Type projection]
The
definition of type projection is extended to indexed kinds as follows:
$
  \projty{\ctx}{\sctx}{\kdfun*[{\var}]{\ty}{\kd}}{\sfun} \defeq
  \cup \ \{
  \projty{
    \ctx \cons \annty{\var}{\ty}
  }{\pair{\sctx}{\sval}}{\kd}{\sfun \app \sval}
 \mid
  \sval \in \dom{\sem{\wfty{\ctx}{\ty}}[\sctx]}
  \}
$.
% \begin{align*}
%   \projty{\ctx}{\sctx}{\kdfun*[{\var}]{\ty}{\kd}}{\sfun} \defeq
%   \negspace{22}
%   \bigcup_{\subscript{\sval \in \dom{\sem{\wfty{\ctx}{\ty}}[\sctx]}}}
%   \negspace{22}
%   \projty{
%     \ctx \cons \annty{\var}{\ty}
%   }{\pair{\sctx}{\sval}}{\kd}{\sfun \app \sval}
% \end{align*}
\end{definition}
\endgroup

\subsubsection{Unrefined inductive types.}
In the unrefined semantics inductive types are interpreted as
fixed-point of a contractive operator on CUPERs.
The operator acts on the $n$-th approximations (per \cref{def:rel-approx}) and the
determinacy of a CUPER by its approximations (\cref{lem:cuper-det}) then yields
its fixed point. We show that the unrefined
interpretation of a type with a free type variable is either contractive
(\cref{def:contractive-op}) or the identity function, which is sufficient
for the existence of a fixed point.\looseness-1
\begin{theorem}[Unrefined type semantics is contractive or identity]%
\label{thm:unrefined-semantics-contractive}
Let $\wfctx{\ctx \cons \annkd{\var}{\kd} \cons \ctxalt}$ be a
well-formed context and
$\wftm{\ctx \cons \annkd{\var}{\kd} \cons \ctxalt}{\ty}{\kdalt}$ a
well-formed type. For every context interpretation
$\pair{\pair{\sctx}{\skd}}{\sctxalt} \in \dom{\sembot{\wfctx{\ctx \cons
\annkd{\var}{\kd} \cons \ctxalt}}}$, the following operator, which
varies the interpretation assigned to $\var$, is either contractive
or the identity function:
\[
\sembot{\wftm{\ctx \cons \annkd{\var}{\kd} \cons
\ctxalt}{\ty}{\kdalt}}[\pair{\pair{\sctx}{-}}{\sctxalt}]
: \pper{\dinf} \to \pper{\dinf}
\]
\end{theorem}

Because each contractive operator has a unique fixed point (\Cref{lem:guarded-fixpoint}); when the operator is the identity, every CUPER is a fixed point. In
either case a fixed point exists.

The unrefined semantics of inductive types are defined as the
fixed point of all the constructors:
\[\botDefinition{
\begin{array}{c}
\sembot{\wftm{\ctx}{\decl}{
  \kdfun{\ioverline{k}{\ty}}{\kdtype}
}}[\sctx] \defeq \mathsf{fix}_X \; \mathcal{F}\bots(X)
\\
\multicolumn{1}{l}{\textit{where,}}
\\
\mathcal{F}\bots(X) \defeq
\scalebox{0.89}{$
\bigcup\limits_i
\begin{setbuild}{\pair{
  \pair{\ctor_i}{\ioverline{j}{\sval}}
}{
  \pair{\ctor_i}{\ioverline{j}{\svalalt}}
}}
  \pair{\sval_j}{\svalalt_j} \in \sembot{\wfty{
    \ctx \cons
    \annkd{\tyctor}{\kdfun{\ioverline*{k}{\ty}}{\kdtype}} \cons
    \ioverline*{j - 1}{\annty{\var}{\tyalt}}
  }{\tyalt_j}}[\pair{\sctx}{X}]
\end{setbuild} \cup \left\{ \pair{\bot}{\bot} \right\}$}
\end{array}
}\]

From \Cref{thm:unrefined-semantics-contractive} each element of the union is
either contractive or the identity function, and in either case non-expansive.
Products and tagged unions act on approximations componentwise and are therefore
non-expansive---$(A \times B)_n$ is determined by $A_n$ and $B_n$---while tagging
a component by a constructor is contractive, as its $(n+1)$-st approximation
depends only on the $n$-th approximation of the component. Every branch of
$\mathcal{F}\bots$ tags its arguments by a constructor $\ctor_i$,
so each branch is contractive; taking the componentwise product of the arguments
and the union over constructors then preserves contractivity. Hence the
$\mathcal{F}\bots$ operator is contractive, and by
\Cref{lem:guarded-fixpoint} it has a unique fixed point---so the fixed point
$\mathsf{fix}_X \; \mathcal{F}\bots(X)$ defined above is
well-posed.

\subsubsection{Refined inductive types.}
The refined semantics instead uses more traditional techniques from
the semantics of inductive types: for positive inductive types the
type operator is monotone and the Knaster-Tarski theorem yields a
least fixed point; for stratified inductive types the well-founded
order over the indices yields a unique, hence least, fixed point by a
bottom-up construction. These arguments discharge the new induction
cases of \cref{thm:type-interpretation-well-formed}.
For \emph{least} to be meaningful we order the semantics of types by
the pointwise lift of PER inclusion and the semantics of kinds by
inclusion.
\begin{definition}[Lattice structure of types]%
\label{def:type-operator-lattice}
  Given $\wfctx{\ctx}$ and $\wfkd{\ctx}{\kd}$, we define the following
  order on $\skd, \skdalt \in \sem{\wfkd{\ctx}{\kd}}[\sctx]$:
  \begin{itemize}[topsep=\ListTopSep,itemsep=\ListItemSep,partopsep=\ListPartopSep]
    \item For $\kd = \kdtype$, $\skd \leq \skdalt$ if $\skd \subseteq
      \skdalt$.
    \item For $\kd = \kdfun*[{\var}]{\ty}{\kdalt}$, $\skd \leq
      \skdalt$
      if for all $\sval \in \dom{\sem{\wfty{\ctx}{\ty}}[\sctx]}$,
      $\skd \app \sval \leq \skdalt \app \sval$.
  \end{itemize}
\end{definition}
This order is a complete lattice: meets are intersections, and the
join of $A$ and $B$ is the least PER containing $A \cup B$, which
exists because PERs are closed under intersection and $\dinf \times
\dinf$ is the top PER; the pointwise lift of a complete lattice is
again a complete lattice. Here we identify functions that agree on
the domain of the interpretation of the domain type, as only that
restriction is ever used.\looseness-1

Each inductive type is then interpreted as the least fixed point of
the operator $\mathcal{F}$, which unfolds a candidate interpretation
$X$ into the union of all the possible constructor assignments.
\[\refinedDefinition{
\begin{array}{c}
\sem{\wftm{\ctx}{\decl}{
  \kdfun{\ioverline*{k}{\ty}}{\kdtype}
  }}[\sctx] \defeq \mathsf{lfp}_X \mathcal{F}(X)
\\
\multicolumn{1}{l}{
\textit{where,}}\\
\multicolumn{1}{l}{
\quad
D \defeq \decl; \quad
\kind_D \defeq \kdfun{\ioverline*{k}{\ty}}{\kdtype}} \\
\multicolumn{1}{l}{
  \quad
\semtypem \defeq \sem{\wfty{
      \ctx \cons
      \annkd{\tyctor}{\kind_D} \cons
      \ioverline*{j - 1}{\annty{\var}{\tyalt}}
    }{\tyalt_j}}; \quad
\semdeclm \defeq \sembot{\wftm{\ctx}{D}{\kind_D}} ;\quad
\semidxm \defeq \sembot{\wfty{
      \ctx \cons \ioverline*{k - 1}{\annty{\var}{\ty}}
    }{\ty_k}}}
\\
\mathcal{F}(X) \defeq
\displaystyle
  \ioverline{k}{\svalaltalt} \mapsto
  \Bigcup_i \begin{setbuild}{\pair{
    \pair{\ctor_i}{\ioverline{j}{\sval}}
  }{
    \pair{\ctor_i}{\ioverline{j}{\svalalt}}
  }}
    \pair{\sval_j}{\svalalt_j} \in \semtypem[\pair{\pair{\sctx}{\pair{X}{
        \semdeclm [\projctx{\ctx}{\sctx}]
      }}}{\ioverline*{j - 1}{\sval}}] \\
    \pair{\svalaltalt_k}{\semtm{\tm_k}[\pair{\projrun{\ctx}{\sctx}}{\ioverline{j}{\sval}}]}
    \in \semidxm [\projctx{\ctx}{\sctx}]
  \end{setbuild}
\end{array}
}\]

In $\mathcal{F}(X)$ the first condition relates the constructor
arguments at their declared types, with the candidate $X$
interpreting the recursive occurrences of the type; the second
requires the input indices to be related to the values of the
constructor's index terms.
The environment used in $\mathcal{F}(X)$ is well formed only when $X$ is
included in $\semdeclm$, in the sense required of type variables in contexts,
so the least fixed point is taken in the sublattice of such $X$. This is again
a complete lattice, and $\mathcal{F}$ preserves it.
Importantly, $\mathcal{F}$ has a least fixed point, both for positive and stratified types, as we show next.

\paragraph{Positive refined inductive types.}
For positive inductive types the fixed point exists by the
Knaster-Tarski (\cref{thm:knaster-tarski}), so it suffices to
show that $\mathcal{F}$ is monotone. Since kinds may quantify over
types indexed by the type being defined, we  show that the
semantics of kinds is (anti-)monotone; there the order is plain
set-theoretic inclusion, as kinds are interpreted as sets of
set-theoretic functions.\looseness-1
\begin{lemma}[(Anti-)Monotonicity of types]%
\label[lemma]{thm:monotonicity-types}
Let $\ctxalt \defeq \ctx \cons \annkd{\tyctor}{\kdalt} \cons
\ioverline*{i}{\ann{\var}{\tyalt}}$, where $\var_i$ can bind
both type and term variables, and let $\wfctx{\ctxalt}$,
$\wfkd{\ctxalt}{\kdalt}$, and $\wftm{\ctxalt}{\ty}{\kdalt}$ be
well-formed judgments. Given two context interpretations
$\pair*{\pair*{\sctx}{\pair*{\skd}{\skd_{\bot}}}}{\ioverline*{i}{\sval}},
\pair*{\pair*{\sctx}{\pair*{\skdalt}{\skd_{\bot}}}}{\ioverline*{i}{\sval}}
\in{} \dom{\sem*{\wfctx{\ctxalt}}}$ that differ only in the
assignment of $\tyctor$, with $\skd \leq \skdalt$, then
\begin{enumerate}[label={(\roman*)},topsep=\ListTopSep,itemsep=\ListItemSep,partopsep=\ListPartopSep]
  \item If $\positive{\tyctor}{\ty} \posleq \pospos$, then
    $\sem{\wftm{\ctxalt}{\ty}{\kdalt}}[\pair{\pair{\sctx}{\pair{\skd}{\skd_{\bot}}}}{\ioverline*{i}{\sval}}]
    \leq
    \sem{\wftm{\ctxalt}{\ty}{\kdalt}}[\pair{\pair{\sctx}{\pair{\skdalt}{\skd_{\bot}}}}{\ioverline*{i}{\sval}}]$; and
  \item If $\positive{\tyctor}{\ty} \posleq \posneg$, then
    $\sem{\wftm{\ctxalt}{\ty}{\kdalt}}[\pair{\pair{\sctx}{\pair{\skd}{\skd_{\bot}}}}{\ioverline*{i}{\sval}}]
    \geq
    \sem{\wftm{\ctxalt}{\ty}{\kdalt}}[\pair{\pair{\sctx}{\pair{\skdalt}{\skd_{\bot}}}}{\ioverline*{i}{\sval}}]$.
\end{enumerate}
\end{lemma}
The semantics of kinds satisfy a similar property, with set inclusion instead of the order on types.

Since $\mathcal{F}$ only builds tuples out of the interpretations of
the constructor argument types, in which $\tyctor$ occurs positively,
it is monotone by \cref{thm:monotonicity-types}, and by
\cref{thm:knaster-tarski} it has a least fixed point.

\paragraph{Stratified refined inductive types.}
For stratified inductive types the metric $m$ over the
indices justifies a bottom-up construction. The stratification
condition ensures that the interpretation at an index
$\ioverline*{k}{\sval}$ instantiates every recursive occurrence of
the type constructor only at indices $\ioverline*{k'}{\svalalt}$ with
$m(\ioverline*{k'}{\svalalt}) < m(\ioverline*{k}{\sval})$, and hence
depends only on interpretations at strictly $m$-smaller indices. The
semantics is therefore defined by well-founded recursion on the order
induced by $m$, which yields a unique solution of the fixed-point
equation, hence it is also the least.

\subsubsection{Introduction and elimination.}
The introduction simply constructs a pair of the constructor tag and
the tuple of the arguments while
% \[\runtimeDefinition{
% \begin{aligned}
% \semtm{\ctor(i)_{\decl}}[\sctx] \defeq
% \demb{\to}{\ioverline*{j}{\sval} \mapsto
% \pair{\ctor_i}{\ioverline*{j}{\sval}}}
% \end{aligned}
% }\]
% \nv{What is this fancy i? I believe it is somewhere in the early sections, but please remind the reader here.}
the elimination interpretation is the least fixed point of semantic
case analysis:
\[\runtimeDefinition{
\vcenter{\hbox{\resizebox{0.97\textwidth}{!}{$\displaystyle
\begin{array}{r@{}c@{}l@{}}
\semtm{\ctor(i)_{\decl}}[\sctx] &\defeq&
\demb{\to}{\ioverline*{j}{\sval} \mapsto
\pair{\ctor_i}{\ioverline*{j}{\sval}}}\\
 \semtm{\tmelim{\decl}{\ioverline*{i}{c}}}[\sctx]  &\defeq&
\demb{\to}{\mathsf{lfp}_r
\ioverline{k}{\sval} \cons   % indexes
\svalalt             \mapsto % value
\begin{cases}
  \semtm{c_i}[\sctx] \app
  \ioverline*{j}{\svalaltalt} \app
  \ioverline*{j}{
    \tmel{\tyctor}{\tyalt}{
      \pair{\sctx}{\ioverline*{j-1}{\svalaltalt}}
    }{
      \svalaltalt_j
    }{r}
  }
  & \text{if } \svalalt = \pair{
      \ctor_i
    }{
      \ioverline*{j}{\svalaltalt}
    } \\
  \bot & \text{otherwise}
\end{cases}}
\end{array}
$}}}
}\]
\noindent
Where the semantics of the actual recursive calls follow
the same pattern as the typing rules for the eliminator.
Namely, the semantics is inductively defined on the type of the constructor argument by
invoking $r$\nv{What is r?} on all the strictly positive recursive occurrences of
the type constructor:
\begin{definition}[Semantics of recursive calls]
Given a type constructor $\tyctor$, a type $\ty$, the interpretation of
the motive $r$, an environment $\sctx$ recording the interpretations
of the preceding constructor arguments, and the current argument
$\sval$,the semantics of the recursive calls is defined as follows:
\[
  \begin{array}{r@{\;}c@{\;}l@{\;}c@{\;}r@{\;}c@{\;}l}
    \tmel{\tyctor}{
    \tybase
  }{\sctx}{\sval}{r} & \defeq &
  \sunit &\qquad\qquad\qquad&
  \tmel{\tyctor}{
    \tyctor \app \ioverline*{k}{\tm}
  }{\sctx}{\sval}{r} & \defeq &
  r \app \ioverline{k}{\semtm{\tm}[\sctx]} \app \sval \\
  \tmel{\tyctor}{
    \ty \app \ioverline*{n}{\tm}
  }{\sctx}{\sval}{r} & \defeq &
  \sunit &\quad&
  \tmel{\tyctor}{
    \tyref{\var}{\ty}{\prop}
  }{\sctx}{\sval}{r} & \defeq &
  \tmel{\tyctor}{
    \ty
  }{\sctx}{\sval}{r}
\\
  \tmel{\tyctor}{
    \tyall{\var}{\kd}{\ty}
  }{\sctx}{\sval}{r} &\defeq &
  \multicolumn{5}{@{}l}{
  \tmel{\tyctor}{
    \ty
  }{\sctx}{\sval}{r}}
  \\
  \tmel{\tyctor}{
    \tyfun[{\var}]{\ty}{\tyalt}
  }{\sctx}{\sval}{r} &\defeq &
  \multicolumn{5}{@{}l}{
  \demb{\to}{\svalalt \mapsto \tmel{\tyctor}{
    \tyalt
  }{\pair{\sctx}{\svalalt}}{\sval \app \svalalt}{r}}}
  \end{array}
\]
\end{definition}
Notice that the whole construction branches over the syntax of
the type in the first argument and
it does not depend on the semantics.

We proved that the eliminator is PER-respecting and it
realizes the type of the inductive hypotheses. For the refined
semantics we prove the following stronger statement that ensures that the eliminator works for
any under approximation $\skd$ of the inductive type: if the
realizers $r$ and $r'$ map related indices and $\skd$-related values
to related results (hypothesis (i)) and are moreover related at the
unrefined type of the eliminator (hypothesis (ii)), then the
recursive-call packages computed by $\mathsf{tmEl}$ inhabit the
corresponding inductive-hypothesis types $\mathsf{El}$. The
generality in $\skd$ is forced on us: the refined interpretation of
the inductive type is defined as a fixed point, so while proving the
eliminator correct the inductive type is only available up to an
approximation.
\begin{lemma}[Refined semantics is PER-respecting (Eliminator)]%
\label[lemma]{lem:per-respecting-eliminator} $\ $\\
Let $D \defeq \decl$, $\ctxalt(i) \defeq
\annkd{\tyctor}{\kdfun*{({\ioverline*{k}{\ann{\var}{\ty}}})}{\kdtype}}
\cons \ioverline*{i}{\ann{\varalt}{\tyalt}}$ and $\skd\bots(\sctx)
\defeq
\sembot*{\wftm{\ctx}{D}{\kdfun*{({\ioverline*{k}{\ann{\var}{\ty}}})}{\kdtype}}}[\projctx{\ctx}{\sctx}]$.
Assume the following judgments are well-formed: $\wfctx{\ctx \cons
\ctxalt(j)}$ and $\wfty{\ctx \cons \ioverline{k}{\ann{\var}{\ty}}
\cons \annty{\varalt}{D \tmapp \ioverline{k}{\var}}}{\tyaltalt}$.
Given an under approximation of $D$: $\skd \leq
\sem{\wftm{\ctx}{D}{\kdfun*{({\ioverline*{k}{\ann{\var}{\ty}}})}{\kdtype}}}[\sctx]$,
in two related context interpretations $\pair{
  \pair{\pair{\sctx}{\pair{\skd}{\skd\bots(\sctx)}}}{\ioverline*{j-1}{\sval}}
}{
  \pair{\pair{\sctxalt}{\pair{\skd}{\skd\bots(\sctxalt)}}}{\ioverline*{j-1}{\svalalt}}
} \in \sem*{\wfctx{\ctx \cons \ctxalt(j-1)}}$ and
$r$ and $r'$ related realizers for the restricted $\skd$,
\ie{}
\begin{enumerate}[label={(\roman*)},topsep=\ListTopSep,itemsep=\ListItemSep,partopsep=\ListPartopSep]
  \item $\forall\; \ioverline*{k}{\pair{t}{t'}} \in \sem*{\wfty{\ctx \cons
    \ioverline*{k - 1}{\annty{\var}{\ty}}
  }{\ty_k}}[\pair{\sctx}{\ioverline*{k - 1}{t}}]$,
  $\forall \pair{\sval}{\svalalt} \in \skd \app \ioverline*{k}{t}$,
    $\pair{
      r \app \ioverline*{k}{t} \app \sval
    }{
      r' \app \ioverline*{k}{t'} \app \svalalt
    } \in
    \sem{\wfty{
      \ctx \cons
      \ioverline*{k}{\annty{\var}{\ty}} \cons
      \annty{\varalt}{D \tmapp \ioverline*{k}{\var}}
    }{\tyaltalt}}[\pair{\pair{\sctx}{\ioverline*{k}{t}}}{\sval}]$, and
  \item $\pair*{r}{r'} \in
    \sembot*{\wfty{\ctx}{\tyfun{\left(\ioverline*{k}{\annty{\var}{\ty}}\right)}{\tyfun{{D \app
    \ioverline*{k}{\var}}}{\tyaltalt}}}}[\projctx{\ctx}{\sctx}]$
\end{enumerate}
and some term $\wftm{\ctx \cons \ctxalt(j -
1)}{\tmaltaltalt}{\tyalt_j}$, then the restricted eliminators realize
the type of the induction hypothesis over argument $\tmaltaltalt$:
{% \setlength{\abovedisplayskip}{\AlignTopSep}
\begin{align*}
  \vpair{
    \tmel{\tyctor}{\tyalt_j}{\sctx_\rho}{\semtm{\tmaltaltalt}[\sctx_\rho]}{r}
  }{
    \tmel{\tyctor}{\tyalt_j}{\sctxalt_\rho}{\semtm{\tmaltaltalt}[\sctxalt_\rho]}{r'}
  } \in \sem{\wfty{\ctx \cons \ctxalt(j - 1)}{
    \tyel{\tyctor}{\tyaltalt}{\tmaltaltalt}{\tyalt_j}
  }}[\pair{\pair{\sctx}{\pair{\skd}{\skd\bots(\sctx)}}}{\ioverline*{j - 1}{\sval}}]\\
\text{where } \sctx_\rho \defeq
\projrun{
  \ctx \cons \ctxalt(j - 1)
}{
  \pair{
    \pair{\sctx}{\pair{\skd}{\skd\bots(\sctx)}}
  }{
    \ioverline*{j-1}{\sval}
  }
}\text{ and }\sctxalt_\rho \defeq
\projrun{
  \ctx \cons \ctxalt(j - 1)
}{
  \pair{
    \pair{\sctxalt}{\pair{\skd}{\skd\bots(\sctxalt)}}
  }{
    \ioverline*{j-1}{\svalalt}
  }
}
\end{align*}
}
\end{lemma}
The same holds for the unrefined interpretation, but there the
stronger statement is unnecessary: every unrefined type interpretation
is pointed and complete, so it contains the least fixed point.

In the refined case completeness is not for free, so we argue
differently, depending on the formation discipline. For positive
types, write $T$ for the refined interpretation of the inductive
type, \ie the least fixed point of the operator $\mathcal{F}$, and
let $P \leq T$ be the restriction of $T$ to the pairs on which the
eliminator respects the typing of the motive. We show that $P$ is a
pre-fixed point of $\mathcal{F}$, so that $T \leq P$, and hence $P =
T$. The inclusion $\mathcal{F}(P) \leq P$ amounts to two conditions:
membership in $T$, which holds as $\mathcal{F}(P) \leq \mathcal{F}(T)
= T$ by monotonicity of $\mathcal{F}$; and \Cref{lem:per-respecting-eliminator}.

For stratified types $\mathcal{F}$ need not be monotone; instead
we define a family of truncations $P(n)$, where $P(n)$ agrees with
$T$ at all indices of stratum smaller than $n$ and is empty
elsewhere.
By \Cref{lem:per-respecting-eliminator}
and induction on $n$, we show that the eliminator
respects the typing of the motive on each $P(n)$.\looseness-1

This completes the treatment of inductive types. As for refinement types, the
statements of the theorems of \Cref{sec:model-basic} are unchanged, and their
proofs are extended with the cases for the rules of \Cref{fig:syntax-ind}. Hence
soundness (\Cref{thm:per-respecting-terms}) and consistency
(\Cref{thm:consistency}) hold for the complete \corelan calculus, with
refinement types, subtyping, and both positive and stratified inductive types.

\section{Compiling indexed types}%
\label{sec:compilation}
We implemented the positive and stratified inductive types of \corelan in Liquid Haskell
and use it to compile the examples of \cref{sec:inductive-types}.
To permit positive types, we simply relaxed
the existing strictly positivity check.
The encoding of stratified inductive types is more involved, as Liquid Haskell does not have a direct support of indexed inductive types.
Yet, following the technique of \emph{data propositions}~\cite{borkowskiMechanizingRefinementTypes2024},
stratified types can compile to ordinary inductive types, by encoding the index information in the refinement logic,
via an uninterpreted function.

Finally, we implemented a stratification condition that ensures that the well-formedness
condition of the declaration is satisfied.\looseness-1

Stratification is checked when a datatype
is declared as \lstinline{stratified}.
For example, the \lstinline{Val} type of \cref{sec:inductive-types} needs to be declared as
\lstinline{stratified}, otherwise the positivity check of Liquid Haskell would reject it.
Yet, this implementation is direct:
in Liquid Haskell, it required
$\sim$400 lines of code,
thus we believe that it can be easily adapted to other refinement type systems.

\section{Related work}%
\label{sec:related-work}

\paragraph{Models of refinement type systems.}
Refinement types were introduced by Freeman and
Pfenning~\cite{freemanRefinementTypesML1991} as a discipline for
carving out subsets of ML datatypes. Stating properties of data by
predicates over unindexed types, as refinement types do, has a long history in
logic-enriched type theories~\cite{gambino_2006,adams_2010}. In Sections~\ref{sec:features} and
\ref{sec:existing-formalizations} we already discussed in detail what
is and what is not captured by existing formalizations of refinement
type systems. Other works, such
as~\cite{bovelFirstClassRefinementTypes2026}, build
on~\cite{hamzaSystemFRFormalized2019} by reintroducing a subtyping
judgment, validating its necessity. However, they establish only a
partial correctness result relative to termination and do not
guarantee consistency. \citet{lepigre_2016} gives a classical realizability
model of a second-order language with impredicative polymorphism and an
equality type on untyped terms, interpreted as observational equivalence of
programs rather than by the unrefined interpretation of a type; it has neither
subtyping nor inductive types. Closest to ours in spirit is the work of~\citet{economouFocusingRefinementTyping2023}, who
likewise give an elementary domain-theoretic denotational semantics
for a refinement type system with subtyping and use it to derive
logical consistency. Their system, however, belongs to the
index-refinement tradition of Dependent ML and liquid
types~\cite{rondonLiquidTypes2008}: it
supports neither impredicative polymorphism nor the non-standard
inductive types that are our central concern.
The view of refinements as an annotation layer
over an unrefined language is made precise by~\citet{melliesFunctorsAreType2015}, who axiomatise a
refinement type system as a functor from a category of refined typing
derivations down to its unrefined base. Our dual-layer
model---refined PERs sitting above a layer of CUPERs that models
computation---is a semantic instance of this picture: a type denotes a PER
over $\dinf$ included in a CUPER over $\dinf$, a term denotes an element of
$\dinf$ preserving both relations, and annotation independence is its
concrete manifestation. Methodologically, our
interpretation of types as PERs follows the recipe
of~\citet{harperConstructingTypeSystems1992} for constructing type systems over
an operational semantics, with the closed terms replaced by the elements of
$\dinf$.

\paragraph{Models of System F}
Interpreting types as PERs over a universal domain is a standard technique of
realizability~\citep{abadiModelPolymorphismRecursive1990,amadioRecursionRealizabilityStructures1991,bruce_1992}.
We build on the model of System~F with general inductive types given
by~\citet{amadioRecursionRealizabilityStructures1991}, which
interprets types as CUPERs over a domain and itself descends from the
ideal model of~\citet{macqueenIdealModelRecursive1986}.  Those models solve
recursive type equations by a \citet{banach_1922} contraction on a
metric of relations; we adapt the same contractive argument to the
minimal-invariant projections of the domain~\cite{pitts_1996}, avoiding
the metric. \citet{birkedal_2007} obtain recursive types in admissible PERs
from parametricity instead.

\paragraph{Stratified inductive types.}
\citet{jacob-raoIndexStratifiedTypes2018} introduce
stratified types as a way to circumvent the absence of large
elimination in Tores. Their system omits
impredicative polymorphism,
subtyping, and refinement types, and instead supports general
recursive types via \citet{mendlerInductiveTypesType1991}-style
recursion. Their development also
includes a normalization proof. Our work can be seen as extending this
setting with impredicative polymorphism and refinement structure. The
strategy of relaxing syntactic strict positivity in an impredicative
setting goes back to~\citet{matthesMonotoneCoinductiveTypes1999},
who replaces positivity with a monotonicity witness while retaining
strong normalization for System~F extended with fixed-point types.
Sized types~\citep{abel_2008} index an inductive type by an ordinal that
bounds the height of its elements, in order to check termination. Sizes
form a dedicated kind, the operator must be monotone, and types do not depend
on terms.
\citet{abel_2012} notes that the size-indexed approximants remain well defined
for a non-monotone operator, but does not study the metatheory of that case.

\paragraph{Dependent type theory.}
In contrast to refinement types, standard dependent type theories,
such as the Calculus of Inductive Constructions enforce strict
positivity requirements for inductive definitions. This restriction is
fundamentally necessary in standard type theory to preserve logical
consistency, particularly when combined with impredicative
polymorphism, as demonstrated by the~\citet{coquandInductivelyDefinedTypes1990} paradox. Furthermore,
systems with large elimination allow for the construction of power
sets from inductive types, making non-strictly positive definitions
inherently dangerous. Our PER-based semantics demonstrates that, in a
refinement setting without large elimination, this strict requirement
can be safely relaxed.

\section{Conclusion \& Future Work}%
\label{sec:conclusion}
We presented a semantic model for refinement type systems based on
partial equivalence relations over a Scott domain. The model accounts
for all subtyping, unrefined equality, impredicative
polymorphism, and non-standard inductive types, and
comes with PER-respecting interpretations.

This realizability model shows
that the core refinement type calculus is consistent, even in the presence of positive and
stratified inductive types.
These two inductive type schemes are currently
permitted by Liquid Haskell and we believe that the same design can guide future
extensions of more expressive type systems.

Our refinements express total correctness: a term of a refined type
terminates and satisfies its predicate. Refined types are therefore
ordinary PERs and need not contain a bottom element, whose presence
under lazy evaluation makes the refinement logic
inconsistent~\citep{vazouRefinementTypesHaskell2014}.
Partial correctness and quotient types would require a different
model, in both cases because equality is unrefined. Partial correctness
requires the refined semantics of types to be complete~\citep{crary_1998},
which is not the case in general. Quotient types~\citep{hewer_2024}
identify more elements than unrefined equality does, which contradicts
\Cref{thm:refined-sub-unrefined}.

Our proofs are not mechanically verified. A complete mechanization
would require formalizing the underlying domain theory, the
PER/CUPER constructions, and the semantic interpretation of the
inductive types introduced in this work. While mechanized
developments of syntactic metatheory are more common, mechanizations
of semantic models at a comparable level of expressiveness remain
an active research field. Developing the semantic
infrastructure and carrying out the corresponding soundness proofs
would constitute a substantial project in its own right.

\begin{acks}
We would like to thank the anonymous reviewers for their valuable feedback and suggestions.
This work was funded by
the Excellence Seal Mar\'{i}a de Maeztu Grant
(CEX2024-001471-M funded by MICIU/AEI/10.13039/501100011033)
and the ERC Grant CRETE (GA 101039196).
Views and opinions expressed are however those of the
authors only and do not necessarily reflect those of the European
Union or the European Research Council; neither of which can be held responsible for them.
\end{acks}

\section*{Disclosure of Use of AI in Research}
No AI tools were used to conduct the research presented in this work.

\bibliography{references}

@inproceedings{abadiModelPolymorphismRecursive1990,
  title = {A {{PER}} Model of Polymorphism and Recursive Types},
  booktitle = {Proceedings of the Fifth Annual {IEEE} Symposium on Logic in Computer Science ({LICS} '90)},
  author = {Abadi, Mart{\'\i}n and Plotkin, Gordon D.},
  date = {1990-06},
  pages = {355--365},
  doi = {10.1109/LICS.1990.113761},
  url = {https://ieeexplore.ieee.org/document/113761},
  urldate = {2026-02-05},
  eventtitle = {[1990]. {{Fifth Annual IEEE Symposium}} on {{Logic}} in {{Computer Science}}}
}

@mastersthesis{abelSemanticAnalyisStructural,
  title = {A {{Semantic Analysis}} of {{Structural Recursion}}},
  author = {Abel, Andreas},
  date = {1999-02-26},
  school = {Ludwig-Maximilians-Universität München},
  langid = {english},
  type = {Diplomarbeit},
}

@inproceedings{cardone_1989,
  title = {Relational Semantics for Recursive Types and Bounded Quantification},
  booktitle = {Automata, {{Languages}} and {{Programming}}},
  author = {Cardone, Felice},
  editor = {Ausiello, Giorgio and Dezani-Ciancaglini, Mariangiola and {Ronchi Della Rocca}, Simona},
  date = {1989},
  series = {Lecture {{Notes}} in {{Computer Science}}},
  volume = {372},
  pages = {164--178},
  publisher = {Springer},
  location = {Berlin, Heidelberg},
  doi = {10.1007/BFb0035759}
}

@article{amadioRecursionRealizabilityStructures1991,
  title = {Recursion over Realizability Structures},
  author = {Amadio, Roberto M.},
  date = {1991-03-01},
  journaltitle = {Information and Computation},
  journal = {Information and Computation},
  shortjournal = {Information and Computation},
  volume = {91},
  number = {1},
  pages = {55--85},
  issn = {0890-5401},
  doi = {10.1016/0890-5401(91)90074-C},
  url = {https://www.sciencedirect.com/science/article/pii/089054019190074C},
  urldate = {2026-02-10}
}

@book{girardProofsTypes1989,
  title = {Proofs and Types},
  author = {Girard, Jean-Yves},
  date = {1989},
  series = {Cambridge Tracts in Theoretical Computer Science},
  number = {7},
  publisher = {Cambridge University Press},
  location = {Cambridge [England] ; New York},
  isbn = {978-0-521-37181-0},
  langid = {english},
  pagetotal = {176}
}

@inproceedings{bergerMartinHofmannsCase2019,
  title = {Martin {{Hofmann}}’s {{Case}} for {{Non-Strictly Positive Data Types}}},
  author = {Berger, Ulrich and Matthes, Ralph and Setzer, Anton},
  editor = {Dybjer, Peter and Espírito Santo, José and Pinto, Luís},
  date = {2019},
  volume = {130},
  publisher = {Schloss Dagstuhl – Leibniz-Zentrum für Informatik},
  issn = {1868-8969},
  doi = {10.4230/LIPICS.TYPES.2018.1},
  url = {https://drops.dagstuhl.de/entities/document/10.4230/LIPIcs.TYPES.2018.1},
  urldate = {2025-10-28},
  isbn = {9783959771061},
  langid = {english},
  version = {1.0},
  booktitle = {24th International Conference on Types for Proofs and Programs ({TYPES} 2018)},
  series = {Leibniz International Proceedings in Informatics ({LIPIcs})},
  pages = {1:1--1:22},
}

@article{borkowskiMechanizingRefinementTypes2024,
  title = {Mechanizing {{Refinement Types}}},
  author = {Borkowski, Michael H. and Vazou, Niki and Jhala, Ranjit},
  date = {2024-01-05},
  journaltitle = {Proc. ACM Program. Lang.},
  journal = {Proc. ACM Program. Lang.},
  shortjournal = {Proc. ACM Program. Lang.},
  volume = {8},
  pages = {70:2099--70:2128},
  doi = {10.1145/3632912},
  url = {https://dl.acm.org/doi/10.1145/3632912},
  urldate = {2025-09-19},
  issue = {POPL}
}

@online{bovelFirstClassRefinementTypes2026,
  title = {First-{{Class Refinement Types}} for {{Scala}}},
  author = {Bovel, Matt and Kunčak, Viktor and Odersky, Martin},
  date = {2026},
  doi = {10.48550/ARXIV.2605.08369},
  url = {https://arxiv.org/abs/2605.08369},
  urldate = {2026-06-04},
  pubstate = {prepublished},
  version = {1}
}

@article{breazutannen_1991,
  title = {Inheritance as Implicit Coercion},
  author = {Breazu-Tannen, Val and Coquand, Thierry and Gunter, Carl A. and Scedrov, Andre},
  date = {1991-07},
  journaltitle = {Information and Computation},
  journal = {Information and Computation},
  volume = {93},
  number = {1},
  pages = {172--221},
  doi = {10.1016/0890-5401(91)90055-7}
}

@inproceedings{cardelli_1984,
  title = {A Semantics of Multiple Inheritance},
  booktitle = {Semantics of {{Data Types}}},
  author = {Cardelli, Luca},
  editor = {Kahn, Gilles and MacQueen, David B. and Plotkin, Gordon},
  date = {1984},
  series = {Lecture {{Notes}} in {{Computer Science}}},
  volume = {173},
  pages = {51--67},
  publisher = {Springer},
  location = {Berlin, Heidelberg},
  doi = {10.1007/3-540-13346-1_2}
}

@inproceedings{coquandInductivelyDefinedTypes1990,
  title = {Inductively Defined Types},
  booktitle = {{COLOG-88}: International Conference on Computer Logic},
  author = {Coquand, Thierry and Paulin, Christine},
  editor = {Martin-Löf, Per and Mints, Grigori},
  date = {1990},
  pages = {50--66},
  publisher = {Springer},
  location = {Berlin, Heidelberg},
  doi = {10.1007/3-540-52335-9_47},
  isbn = {978-3-540-46963-6},
  langid = {english}
}

@article{economouFocusingRefinementTyping2023,
  title = {Focusing on {{Refinement Typing}}},
  author = {Economou, Dimitrios J. and Krishnaswami, Neel and Dunfield, Jana},
  date = {2023-12-20},
  journaltitle = {ACM Trans. Program. Lang. Syst.},
  journal = {ACM Trans. Program. Lang. Syst.},
  volume = {45},
  number = {4},
  pages = {22:1--22:62},
  doi = {10.1145/3610408},
  url = {https://dl.acm.org/doi/10.1145/3610408},
  urldate = {2026-06-16}
}

@inproceedings{ferrariniLargeEliminationIndexed,
  title = {Large {{Elimination}} and {{Indexed Types}} in {{Refinement Types}}},
  booktitle = {31st {{International Conference}} on {{Types}} for {{Proofs}} and {{Programs}} ({{TYPES}} 2025), {{Book}} of {{Abstracts}}},
  author = {Ferrarini, Alessio and Vazou, Niki},
  date = {2025},
  location = {Glasgow, Scotland},
  url = {https://msp.cis.strath.ac.uk/types2025/abstracts/TYPES2025_paper33.pdf},
  eventtitle = {{{TYPES}} 2025},
  langid = {english}
}

@article{ferrariniPLEXNormalizationRefinement2026,
  title = {{{PLEX}}: {{Normalization}} for {{Refinement Types}}},
  shorttitle = {{{PLEX}}},
  author = {Ferrarini, Alessio and Vazou, Niki and Swierstra, Wouter},
  date = {2026-04-10},
  journaltitle = {Proc. ACM Program. Lang.},
  journal = {Proc. ACM Program. Lang.},
  shortjournal = {Proc. ACM Program. Lang.},
  volume = {10},
  pages = {140:1321--140:1347},
  doi = {10.1145/3798248},
  url = {https://dl.acm.org/doi/10.1145/3798248},
  urldate = {2026-04-16},
  issue = {OOPSLA1}
}

@article{freemanRefinementTypesML1991,
  title = {Refinement Types for {ML}},
  author = {Freeman, Tim and Pfenning, Frank},
  date = {1991},
  journaltitle = {SIGPLAN Not.},
  journal = {SIGPLAN Not.},
  volume = {26},
  number = {6},
  pages = {268--277},
  doi = {10.1145/113446.113468}
}

@article{ghalayiniExplicitRefinementTypes2023,
  title = {Explicit {{Refinement Types}}},
  author = {Ghalayini, Jad Elkhaleq and Krishnaswami, Neel},
  date = {2023-08-31},
  journaltitle = {Proc. ACM Program. Lang.},
  journal = {Proc. ACM Program. Lang.},
  volume = {7},
  pages = {195:187--195:214},
  doi = {10.1145/3607837},
  url = {https://dl.acm.org/doi/10.1145/3607837},
  urldate = {2026-03-10},
  issue = {ICFP}
}

@article{hamzaSystemFRFormalized2019,
  title = {System {{FR}}: Formalized Foundations for the Stainless Verifier},
  shorttitle = {System {{FR}}},
  author = {Hamza, Jad and Voirol, Nicolas and Kunčak, Viktor},
  date = {2019-10-10},
  journaltitle = {Proc. ACM Program. Lang.},
  journal = {Proc. ACM Program. Lang.},
  volume = {3},
  pages = {166:1--166:30},
  doi = {10.1145/3360592},
  url = {https://dl.acm.org/doi/10.1145/3360592},
  urldate = {2026-03-10},
  issue = {OOPSLA}
}

@article{harperConstructingTypeSystems1992,
  title = {Constructing Type Systems over an Operational Semantics},
  author = {Harper, Robert},
  date = {1992-07},
  journaltitle = {Journal of Symbolic Computation},
  journal = {Journal of Symbolic Computation},
  volume = {14},
  number = {1},
  pages = {71--84},
  doi = {10.1016/0747-7171(92)90026-Z}
}

@online{HofmannNonStrictlyPositive,
  title = {Non {{Strictly Positive Datatypes}} in {{System F}}},
  author = {Hofmann, Martin},
  date = {1993-02-15},
  url = {https://www.seas.upenn.edu/~sweirich/types/archive/1993/msg00027.html},
  urldate = {2026-05-21}
}

@online{IncompatibilitiesFunctionalExtensionality,
  title = {Incompatibilities between Functional Extensionality, Subtyping, and Untyped Equality · {{Issue}} \#1542 · {{FStarLang}}/{{FStar}}},
  author = {Maillard, Kenji},
  date = {2018-11-28},
  url = {https://github.com/FStarLang/FStar/issues/1542},
  urldate = {2026-04-16},
  langid = {english},
  organization = {GitHub}
}

@inproceedings{jacob-raoIndexStratifiedTypes2018,
  title = {Index-{{Stratified Types}}},
  author = {Jacob-Rao, Rohan and Pientka, Brigitte and Thibodeau, David},
  editor = {Kirchner, Hélène},
  date = {2018},
  volume = {108},
  publisher = {Schloss Dagstuhl – Leibniz-Zentrum für Informatik},
  issn = {1868-8969},
  doi = {10.4230/LIPICS.FSCD.2018.19},
  url = {https://drops.dagstuhl.de/entities/document/10.4230/LIPIcs.FSCD.2018.19},
  urldate = {2025-09-19},
  isbn = {9783959770774},
  langid = {english},
  booktitle = {3rd International Conference on Formal Structures for Computation and Deduction ({FSCD} 2018)},
  series = {Leibniz International Proceedings in Informatics ({LIPIcs})},
  pages = {19:1--19:17},
}

@online{jhalaRefinementTypesTutorial2020,
  title = {Refinement {{Types}}: {{A Tutorial}}},
  shorttitle = {Refinement {{Types}}},
  author = {Jhala, Ranjit and Vazou, Niki},
  date = {2020-10-15},
  eprint = {2010.07763},
  eprinttype = {arXiv},
  eprintclass = {cs},
  doi = {10.48550/arXiv.2010.07763},
  url = {http://arxiv.org/abs/2010.07763},
  urldate = {2025-09-19},
  pubstate = {prepublished}
}

@article{knowlesHybridTypeChecking2010,
  title = {Hybrid Type Checking},
  author = {Knowles, Kenneth and Flanagan, Cormac},
  date = {2010-02-08},
  journaltitle = {ACM Trans. Program. Lang. Syst.},
  journal = {ACM Trans. Program. Lang. Syst.},
  shortjournal = {ACM Trans. Program. Lang. Syst.},
  volume = {32},
  number = {2},
  pages = {6:1--6:34},
  issn = {0164-0925},
  doi = {10.1145/1667048.1667051},
  url = {https://dl.acm.org/doi/10.1145/1667048.1667051},
  urldate = {2026-06-11}
}

@article{lehmannFluxLiquidTypes2023,
  title = {Flux: {{Liquid Types}} for {{Rust}}},
  shorttitle = {Flux},
  author = {Lehmann, Nico and Geller, Adam T. and Vazou, Niki and Jhala, Ranjit},
  date = {2023-06-06},
  journaltitle = {Proc. ACM Program. Lang.},
  journal = {Proc. ACM Program. Lang.},
  volume = {7},
  pages = {169:1533--169:1557},
  doi = {10.1145/3591283},
  url = {https://dl.acm.org/doi/10.1145/3591283},
  urldate = {2025-09-19},
  issue = {PLDI}
}

@article{livshitsDefenseSoundinessManifesto2015,
  title = {In Defense of Soundiness: A Manifesto},
  shorttitle = {In Defense of Soundiness},
  author = {Livshits, Benjamin and Sridharan, Manu and Smaragdakis, Yannis and Lhoták, Ondřej and Amaral, J. Nelson and Chang, Bor-Yuh Evan and Guyer, Samuel Z. and Khedker, Uday P. and Møller, Anders and Vardoulakis, Dimitrios},
  date = {2015-01-28},
  journaltitle = {Communications of the ACM},
  journal = {Communications of the ACM},
  shortjournal = {Commun. ACM},
  volume = {58},
  number = {2},
  pages = {44--46},
  issn = {0001-0782},
  doi = {10.1145/2644805},
  url = {https://dl.acm.org/doi/10.1145/2644805},
  urldate = {2026-06-02}
}

@article{macqueenIdealModelRecursive1986,
  title = {An Ideal Model for Recursive Polymorphic Types},
  author = {MacQueen, David and Plotkin, Gordon and Sethi, Ravi},
  date = {1986-10},
  journaltitle = {Information and Control},
  journal = {Information and Control},
  volume = {71},
  number = {1-2},
  pages = {95--130},
  doi = {10.1016/S0019-9958(86)80019-5}
}

@article{matthesMonotoneCoinductiveTypes1999,
  title = {Monotone (Co)Inductive Types and Positive Fixed-Point Types},
  author = {Matthes, Ralph},
  date = {1999},
  journaltitle = {RAIRO - Theoretical Informatics and Applications},
  journal = {RAIRO - Theoretical Informatics and Applications},
  volume = {33},
  number = {4-5},
  pages = {309--328},
  doi = {10.1051/ita:1999120}
}

@inproceedings{melliesFunctorsAreType2015,
  series = {{POPL} '15},
  title = {Functors Are Type Refinement Systems},
  author = {Melliès, Paul-André and Zeilberger, Noam},
  date = {2015},
  booktitle = {Proceedings of the 42nd Annual {ACM} {SIGPLAN}-{SIGACT} Symposium on Principles of Programming Languages},
  pages = {3--16},
  publisher = {Association for Computing Machinery},
  doi = {10.1145/2676726.2676970}
}

@article{mendlerInductiveTypesType1991,
  title = {Inductive Types and Type Constraints in the Second-Order Lambda Calculus},
  author = {Mendler, Nax Paul},
  date = {1991},
  journaltitle = {Annals of Pure and Applied Logic},
  journal = {Annals of Pure and Applied Logic},
  volume = {51},
  number = {1-2},
  pages = {159--172},
  doi = {10.1016/0168-0072(91)90069-X}
}

@inproceedings{reynoldsPolymorphismNotSettheoretic1984,
  title = {Polymorphism Is Not Set-Theoretic},
  booktitle = {Semantics of {{Data Types}}},
  author = {Reynolds, John C.},
  editor = {Kahn, Gilles and MacQueen, David B. and Plotkin, Gordon},
  date = {1984},
  pages = {145--156},
  publisher = {Springer},
  location = {Berlin, Heidelberg},
  doi = {10.1007/3-540-13346-1_7},
  isbn = {978-3-540-38891-3},
  langid = {english}
}

@article{rondonLiquidTypes2008,
  title = {Liquid Types},
  author = {Rondon, Patrick M. and Kawaguchi, Ming and Jhala, Ranjit},
  date = {2008-06-07},
  journaltitle = {SIGPLAN Not.},
  journal = {SIGPLAN Not.},
  volume = {43},
  number = {6},
  pages = {159--169},
  issn = {0362-1340},
  doi = {10.1145/1379022.1375602},
  url = {https://dl.acm.org/doi/10.1145/1379022.1375602},
  urldate = {2025-09-19}
}

@inproceedings{swamyDependentTypesMultimonadic2016,
  title = {Dependent Types and Multi-Monadic Effects in {{F}}*},
  booktitle = {Proceedings of the 43rd {{Annual ACM SIGPLAN-SIGACT Symposium}} on {{Principles}} of {{Programming Languages}}},
  author = {Swamy, Nikhil and Hriţcu, Cătălin and Keller, Chantal and Rastogi, Aseem and Delignat-Lavaud, Antoine and Forest, Simon and Bhargavan, Karthikeyan and Fournet, Cédric and Strub, Pierre-Yves and Kohlweiss, Markulf and Zinzindohoue, Jean-Karim and Zanella-Béguelin, Santiago},
  date = {2016-01-11},
  series = {{{POPL}} '16},
  pages = {256--270},
  publisher = {Association for Computing Machinery},
  location = {New York, NY, USA},
  doi = {10.1145/2837614.2837655},
  url = {https://dl.acm.org/doi/10.1145/2837614.2837655},
  urldate = {2026-03-12},
  isbn = {978-1-4503-3549-2}
}

@inproceedings{vazouHowSafelyUse2022,
  title = {How to Safely Use Extensionality in {{Liquid Haskell}}},
  booktitle = {Proceedings of the 15th {{ACM SIGPLAN International Haskell Symposium}}},
  author = {Vazou, Niki and Greenberg, Michael},
  date = {2022-09-06},
  series = {Haskell 2022},
  pages = {13--26},
  publisher = {Association for Computing Machinery},
  location = {New York, NY, USA},
  doi = {10.1145/3546189.3549919},
  url = {https://dl.acm.org/doi/10.1145/3546189.3549919},
  urldate = {2025-09-19},
  isbn = {978-1-4503-9438-3}
}

@article{vazouRefinementReflectionComplete2017,
  title = {Refinement Reflection: Complete Verification with {{SMT}}},
  shorttitle = {Refinement Reflection},
  author = {Vazou, Niki and Tondwalkar, Anish and Choudhury, Vikraman and Scott, Ryan G. and Newton, Ryan R. and Wadler, Philip and Jhala, Ranjit},
  date = {2017-12-27},
  journaltitle = {Proc. ACM Program. Lang.},
  journal = {Proc. ACM Program. Lang.},
  volume = {2},
  pages = {53:1--53:31},
  doi = {10.1145/3158141},
  url = {https://dl.acm.org/doi/10.1145/3158141},
  urldate = {2025-09-19},
  issue = {POPL}
}

@article{vazouRefinementTypesHaskell2014,
  title = {Refinement Types for {{Haskell}}},
  author = {Vazou, Niki and Seidel, Eric L. and Jhala, Ranjit and Vytiniotis, Dimitrios and {Peyton Jones}, Simon},
  date = {2014-08-19},
  journaltitle = {SIGPLAN Not.},
  journal = {SIGPLAN Not.},
  volume = {49},
  number = {9},
  pages = {269--282},
  issn = {0362-1340},
  doi = {10.1145/2692915.2628161},
  url = {https://dl.acm.org/doi/10.1145/2692915.2628161},
  urldate = {2025-09-19}
}

@inproceedings{reynolds_1980,
  title = {Using Category Theory to Design Implicit Conversions and Generic Operators},
  booktitle = {Semantics-Directed Compiler Generation},
  author = {Reynolds, John C.},
  editor = {Jones, Neil D.},
  date = {1980},
  series = {Lecture Notes in Computer Science},
  volume = {94},
  pages = {211--258},
  publisher = {Springer},
  location = {Berlin, Heidelberg},
  doi = {10.1007/3-540-10250-7_24}
}

@inproceedings{reynolds_1984,
  location = {Amsterdam},
  title = {Types, Abstraction, and Parametric Polymorphism},
  pages = {513--523},
  booktitle = {Information Processing 83: Proceedings of the {IFIP} 9th World Computer Congress, Paris, France, September 19-23, 1983},
  publisher = {Elsevier Science Publishers B. V. (North-Holland)},
  author = {Reynolds, John C.},
  editor = {Mason, R. E. A},
  date = {1984}
}

@phdthesis{girard_1972,
  title = {Interprétation fonctionelle et élimination des coupures de l'arithmétique d'ordre supérieur},
  school = {Université Paris {VII}},
  type = {Thèse d'État},
  author = {Girard, Jean-Yves},
  date = {1972},
}

@article{cardelli_1985,
  title = {On understanding types, data abstraction, and polymorphism},
  volume = {17},
  doi = {10.1145/6041.6042},
  pages = {471--523},
  number = {4},
  journaltitle = {{ACM} Computing Surveys},
  journal = {{ACM} Computing Surveys},
  author = {Cardelli, Luca and Wegner, Peter},
  date = {1985},
}

@article{leivant_1991,
  title = {Finitely stratified polymorphism},
  volume = {93},
  doi = {10.1016/0890-5401(91)90053-5},
  pages = {93--113},
  number = {1},
  journaltitle = {Information and Computation},
  journal = {Information and Computation},
  author = {Leivant, Daniel},
  date = {1991},
}

@inproceedings{mendler_1987,
  title = {Recursive Types and Type Constraints in Second-Order Lambda Calculus},
  booktitle = {Proceedings of the Symposium on Logic in Computer Science ({LICS} '87)},
  author = {Mendler, Nax Paul},
  date = {1987},
  pages = {30--36},
  publisher = {IEEE Computer Society Press},
  location = {Ithaca, NY, USA},
}

@book{pierce_2002,
  title = {Types and Programming Languages},
  author = {Pierce, Benjamin C.},
  date = {2002},
  publisher = {MIT Press},
  location = {Cambridge, MA, USA},
  isbn = {978-0-262-16209-8},
}

@book{harper_2016,
  title = {Practical Foundations for Programming Languages},
  edition = {2},
  author = {Harper, Robert},
  date = {2016},
  publisher = {Cambridge University Press},
  doi = {10.1017/CBO9781316576892},
  isbn = {978-1-107-15030-0},
}

@incollection{pitts_1987,
  title = {Polymorphism is set theoretic, constructively},
  volume = {283},
  issn = {16113349},
  url = {https://www.cl.cam.ac.uk/~amp12/papers/polist/polist.pdf},
  doi = {10.1007/3-540-18508-9_18},
  pages = {12--39},
  booktitle = {Category Theory and Computer Science},
  publisher = {Springer, Berlin, Heidelberg},
  author = {Pitts, Andrew M.},
  editor = {Pitt, David H. and Poign{\'e}, Axel and Rydeheard, David E.},
  date = {1987},
}

@article{pitts_1996,
  title = {Relational Properties of Domains},
  author = {Pitts, Andrew M.},
  date = {1996},
  journaltitle = {Information and Computation},
  journal = {Information and Computation},
  volume = {127},
  number = {2},
  pages = {66--90},
  doi = {10.1006/inco.1996.0052},
}

@article{scott_1976,
  title = {Data Types as Lattices},
  author = {Scott, Dana S.},
  date = {1976-09},
  journaltitle = {SIAM Journal on Computing},
  journal = {SIAM Journal on Computing},
  volume = {5},
  number = {3},
  pages = {522--587},
  doi = {10.1137/0205037},
}

@article{smyth_1982,
  title = {The Category-Theoretic Solution of Recursive Domain Equations},
  author = {Smyth, Michael B. and Plotkin, Gordon D.},
  date = {1982-11},
  journaltitle = {SIAM Journal on Computing},
  journal = {SIAM Journal on Computing},
  volume = {11},
  number = {4},
  pages = {761--783},
  doi = {10.1137/0211062},
}

@incollection{abramsky_1994,
  title = {Domain Theory},
  author = {Abramsky, Samson and Jung, Achim},
  booktitle = {Handbook of Logic in Computer Science (Vol. 3): Semantic Structures},
  editor = {Abramsky, S. and Gabbay, Dov M. and Maibaum, T. S. E.},
  date = {1994},
  pages = {1--168},
  publisher = {Oxford University Press},
  location = {Oxford, UK},
  isbn = {978-0-19-853762-5},
}

@book{gunter_1992,
  title = {Semantics of Programming Languages: Structures and Techniques},
  author = {Gunter, Carl A.},
  date = {1992},
  series = {Foundations of Computing},
  publisher = {MIT Press},
  location = {Cambridge, MA},
  isbn = {978-0-262-07143-7},
}

@article{barendregt_1983,
  title = {A {{Filter Lambda Model}} and the {{Completeness}} of {{Type Assignment}}},
  author = {Barendregt, Henk and Coppo, Mario and Dezani-Ciancaglini, Mariangiola},
  date = {1983},
  journaltitle = {Journal of Symbolic Logic},
  journal = {Journal of Symbolic Logic},
  shortjournal = {J. Symb. Log.},
  volume = {48},
  number = {4},
  eprint = {2273659},
  eprinttype = {jstor},
  pages = {931--940},
  publisher = {Association for Symbolic Logic},
  issn = {0022-4812},
  doi = {10.2307/2273659},
  url = {https://www.jstor.org/stable/2273659},
  urldate = {2026-10-15}
}

@article{frisch_2008,
  title = {Semantic Subtyping: {{Dealing}} Set-Theoretically with Function, Union, Intersection, and Negation Types},
  shorttitle = {Semantic Subtyping},
  author = {Frisch, Alain and Castagna, Giuseppe and Benzaken, Véronique},
  date = {2008-09-18},
  journaltitle = {J. ACM},
  journal = {J. ACM},
  shortjournal = {J. ACM JACM},
  volume = {55},
  number = {4},
  pages = {19:1--19:64},
  issn = {0004-5411},
  doi = {10.1145/1391289.1391293},
  url = {https://dl.acm.org/doi/10.1145/1391289.1391293},
  urldate = {2026-10-15}
}

@article{abel_2008,
  title = {Semi-Continuous {{Sized Types}} and {{Termination}}},
  author = {Abel, Andreas},
  date = {2008-04-10},
  journaltitle = {Logical Methods in Computer Science},
  journal = {Logical Methods in Computer Science},
  shortjournal = {Log. Methods Comput. Sci.},
  volume = {4},
  number = {2},
  publisher = {Episciences.org},
  issn = {1860-5974},
  doi = {10.2168/LMCS-4(2:3)2008},
  url = {https://lmcs.episciences.org/1236},
  urldate = {2026-10-13},
  langid = {english}
}

@article{abel_2012,
  title = {Type-{{Based Termination}}, {{Inflationary Fixed-Points}}, and {{Mixed Inductive-Coinductive Types}}},
  author = {Abel, Andreas},
  date = {2012-02-14},
  journaltitle = {Electronic Proceedings in Theoretical Computer Science},
  journal = {Electronic Proceedings in Theoretical Computer Science},
  shortjournal = {Electron. Proc. Theor. Comput. Sci.},
  volume = {77},
  eprint = {1202.3496},
  eprinttype = {arXiv},
  eprintclass = {cs.LO},
  pages = {1--11},
  issn = {2075-2180},
  doi = {10.4204/EPTCS.77.1},
  url = {http://arxiv.org/abs/1202.3496},
  urldate = {2026-10-12}
}

@inproceedings{crary_1998,
  title = {Admissibility of Fixpoint Induction over Partial Types},
  booktitle = {Automated {{Deduction}} — {{CADE-15}}},
  author = {Crary, Karl},
  editor = {Kirchner, Claude and Kirchner, Hélène},
  date = {1998},
  pages = {270--285},
  publisher = {Springer},
  location = {Berlin, Heidelberg},
  doi = {10.1007/BFb0054265},
  isbn = {978-3-540-69110-5},
  langid = {english}
}

@book{amadio_1998,
  title = {Domains and Lambda-Calculi},
  author = {Amadio, Roberto M. and Curien, Pierre-Louis},
  date = {1998},
  series = {Cambridge Tracts in Theoretical Computer Science},
  number = {46},
  publisher = {Cambridge University Press},
  doi = {10.1017/CBO9780511983504},
  isbn = {978-0-521-62277-8},
}

@book{streicher_2006,
  title = {Domain-Theoretic Foundations of Functional Programming},
  author = {Streicher, Thomas},
  date = {2006},
  publisher = {World Scientific},
  location = {Singapore},
  doi = {10.1142/6284},
  isbn = {978-981-270-142-8},
}

@book{davey_2002,
  title = {Introduction to Lattices and Order},
  author = {Davey, Brian A. and Priestley, Hilary A.},
  date = {2002},
  edition = {2},
  publisher = {Cambridge University Press},
  doi = {10.1017/CBO9780511809088},
  isbn = {978-0-521-78451-0},
}

@article{tarski_1955,
  title = {A Lattice-Theoretical Fixpoint Theorem and Its Applications},
  author = {Tarski, Alfred},
  date = {1955},
  journaltitle = {Pacific Journal of Mathematics},
  journal = {Pacific Journal of Mathematics},
  volume = {5},
  number = {2},
  pages = {285--309},
  doi = {10.2140/pjm.1955.5.285},
}

@article{banach_1922,
  title = {Sur les opérations dans les ensembles abstraits et leur application aux équations intégrales},
  author = {Banach, Stefan},
  date = {1922},
  journaltitle = {Fundamenta Mathematicae},
  journal = {Fundamenta Mathematicae},
  volume = {3},
  number = {1},
  pages = {133--181},
  doi = {10.4064/fm-3-1-133-181},
}

@article{henkin_1950,
  title = {Completeness in the theory of types},
  volume = {15},
  doi = {10.2307/2266967},
  pages = {81--91},
  number = {2},
  journaltitle = {Journal of Symbolic Logic},
  journal = {Journal of Symbolic Logic},
  author = {Henkin, Leon},
  date = {1950},
}

@article{barendregt_1991,
  title = {Introduction to generalized type systems},
  volume = {1},
  doi = {10.1017/S0956796800020025},
  pages = {125--154},
  number = {2},
  journaltitle = {Journal of Functional Programming},
  journal = {Journal of Functional Programming},
  author = {Barendregt, Henk},
  date = {1991},
}

@article{longo_1991,
  title = {Constructive natural deduction and its $\omega$-set interpretation},
  volume = {1},
  doi = {10.1017/S0960129500001298},
  pages = {215--254},
  number = {2},
  journaltitle = {Mathematical Structures in Computer Science},
  journal = {Mathematical Structures in Computer Science},
  author = {Longo, Giuseppe and Moggi, Eugenio},
  date = {1991},
}

@inproceedings{lepigre_2016,
  title = {A Classical Realizability Model for a Semantical Value Restriction},
  author = {Lepigre, Rodolphe},
  editor = {Thiemann, Peter},
  booktitle = {Programming Languages and Systems: 25th European Symposium on Programming, {ESOP} 2016},
  series = {Lecture Notes in Computer Science},
  volume = {9632},
  pages = {476--502},
  publisher = {Springer},
  location = {Berlin, Heidelberg},
  date = {2016},
  doi = {10.1007/978-3-662-49498-1_19},
}

@article{gambino_2006,
  title = {The Generalised Type-Theoretic Interpretation of Constructive Set Theory},
  author = {Gambino, Nicola and Aczel, Peter},
  journaltitle = {Journal of Symbolic Logic},
  journal = {Journal of Symbolic Logic},
  volume = {71},
  number = {1},
  pages = {67--103},
  date = {2006},
  doi = {10.2178/jsl/1140641163},
}

@article{adams_2010,
  title = {Classical Predicative Logic-Enriched Type Theories},
  author = {Adams, Robin and Luo, Zhaohui},
  journaltitle = {Annals of Pure and Applied Logic},
  journal = {Annals of Pure and Applied Logic},
  volume = {161},
  number = {11},
  pages = {1315--1345},
  date = {2010},
  doi = {10.1016/j.apal.2010.04.005},
}

@inproceedings{bruce_1992,
  title = {{PER} Models of Subtyping, Recursive Types and Higher-Order Polymorphism},
  author = {Bruce, Kim B. and Mitchell, John C.},
  booktitle = {Proceedings of the 19th {ACM} {SIGPLAN}-{SIGACT} Symposium on Principles of Programming Languages},
  series = {{POPL} '92},
  pages = {316--327},
  publisher = {ACM},
  location = {New York, NY, USA},
  date = {1992},
  doi = {10.1145/143165.143230},
}

@article{birkedal_2007,
  title = {Domain-Theoretical Models of Parametric Polymorphism},
  author = {Birkedal, Lars and M{\o}gelberg, Rasmus Ejlers and Petersen, Rasmus Lerchedahl},
  date = {2007-12},
  journaltitle = {Theoretical Computer Science},
  journal = {Theoretical Computer Science},
  volume = {388},
  number = {1-3},
  pages = {152--172},
  issn = {0304-3975},
  doi = {10.1016/j.tcs.2007.06.016},
  url = {https://doi.org/10.1016/j.tcs.2007.06.016},
}

@article{hewer_2024,
  title = {Quotient {{Haskell}}: {{Lightweight Quotient Types}} for {{All}}},
  shorttitle = {Quotient {{Haskell}}},
  author = {Hewer, Brandon and Hutton, Graham},
  date = {2024-01},
  journaltitle = {Proceedings of the ACM on Programming Languages},
  journal = {Proceedings of the ACM on Programming Languages},
  volume = {8},
  number = {POPL},
  pages = {27:785--27:815},
  doi = {10.1145/3632869},
  url = {https://dl.acm.org/doi/10.1145/3632869},
}

\newpage
\appendix
\allowdisplaybreaks

\section{Compilation examples}

The example of simply typed lambda calculus
values (\Cref{sec:stratified-types}) can be compiled to the following
non indexed type
\begin{lstlisting}
data Val where
{-@ VIota :: Int -> {v:Val | prop v = TIota} @-}
    VIota :: Int -> Val
{-@ VArr :: t:Typ -> u:Typ -> ({v: Val | prop v = t} -> {v: Val | prop v = u})
         -> {v: Val | prop v = TArr t u} @-}
    VArr :: Typ -> Typ -> (Val -> Val) -> Val
\end{lstlisting}

\section{Elimination examples}

\subsection{Negative inductive types are still inconsistent}

If the type declaration of \texttt{Bad} (\Cref{sec:inductive-types})
were allowed, where $\bot$ is an empty type:
\[
  \mathsf{Bad} : \kdtype
  \defeq
  \defind{\mathsf{B}}{\kdtype}{\varpos}{
    \ann{\mathsf{MkBad}}{(\mathsf{B} \to \bot) \to \mathsf{B}}
  }.
\]

The eliminator for \texttt{Bad} would have type
\[
  \tmelim{\mathsf{Bad}}{c_{\mathsf{MkBad}}}
  : (\ann{b}{\mathsf{Bad}}) \to \rho(b), \text{where}
\]
\[
  c_{\mathsf{MkBad}} :
    (\ann{f}{\mathsf{Bad} \to \bot}) \to
    (\mathsf{Bad} \to \tyunit) \to
    \rho(\mathsf{MkBad}\ f)
\]

We can show that $\mathsf{Bad}$ must be empty: instantiating the
motive at the constant type $\rho(b) \defeq \mathsf{Bad} \to \bot$,
the eliminator has type $\mathsf{Bad} \to \mathsf{Bad} \to \bot$, and
diagonalizing gives
\[
  notBad : \mathsf{Bad} \to \bot \defeq \lambda\; b.\; \tmelim{\mathsf{Bad}}{
    \lambda \; f \; g .\; f
  } \tmapp b \tmapp b
\]

But, at the same time, show that it is inhabited
\[
  bad : \mathsf{Bad} \defeq \mathsf{MkBad}\ notBad
\]

Hence, deriving a contradiction.
\[
  contra : \bot \defeq notBad \tmapp bad
\]

The inconsistency arises because the eliminator exposes recursive
occurrences of Bad directly. Mendler-style recursion, by contrast, hides
those occurrences, preventing such self-application arguments and
preserving normalization even for datatypes that are not positive.

\subsection{Using Hofmann's routine}

We can use Hofmann's routine type that we defined in \corelan in
\Cref{ex:hofmann-elim} to define the structural recursive definition
of the \texttt{unfold} and \texttt{br} function
from~\cite{bergerMartinHofmannsCase2019}:
\[
  unfold : \mathsf{Rou} \to (\mathsf{Rou} \to \mathsf{ListNat}) \to \mathsf{ListNat}
  \defeq \tmelim{\mathsf{Rou}}{
    \lambda \; k.\; k \tmapp \mathsf{Over},
    \lambda \; f \; g.\; f
  }
\]

\[
  \mathsf{TreeNat} : \kdtype \defeq \defind{\mathsf{T}}{\kdtype}{\varpos}{
    \ann{\mathsf{Leaf}}{\tynat \to \mathsf{T}},
    \ann{\mathsf{Node}}{\mathsf{T} \to \tynat \to \mathsf{T} \to
      \mathsf{T}}
  },
\]
\begin{multline*}
  br : \mathsf{TreeNat} \to \mathsf{Rou} \to \mathsf{Rou} \\ \defeq
  \tmelim{\mathsf{TreeNat}}{\begin{array}{l}
    \lambda \; n \; r.\; \lambda\; c.\;
    \mathsf{Next} (\lambda\; k.\; \mathsf{Cons} \tmapp n \tmapp (unfold \tmapp c \tmapp k)), \\
    \lambda \; tl \; n \; tr \; r_1\; r_2\; r_3. \; \lambda\; c.\;
    \mathsf{Next} (\lambda\; k.\; \mathsf{Cons} \tmapp n \tmapp (unfold \tmapp c \tmapp (k \circ r_1 \circ r_3)))
  \end{array}}
\end{multline*}
where the motive is the constant type $\mathsf{Rou} \to \mathsf{Rou}$,
so that the induction hypotheses $r_1, r_3 : \mathsf{Rou} \to
\mathsf{Rou}$ for the two subtrees can be composed with the
continuation $k$ ($r_2 : \tyunit$ is the trivial hypothesis for the
$\tynat$ argument).

The \texttt{extract} function is non structural, but by allowing
non-structural recursion in \corelan, we can define it as follows:
\[
  extract : \mathsf{Rou} \to \mathsf{ListNat} \defeq
  \tmelim{\mathsf{Rou}}{
    \mathsf{Nil},
    \lambda \; f \; g.\; f \tmapp extract
  }
\]

Finally, we can define \texttt{breadthfirst} and complete
Hofmann's program:
\[
  breadthfirst : \mathsf{TreeNat} \to \mathsf{ListNat} \defeq
  \lambda\; t.\; extract \tmapp (br \tmapp t \tmapp \mathsf{Over})
\]

\section{Weakening, substitution and narrowing}

We use the notation $\ctx \vdash J$ to mean any of the judgments
$\wfkd{\ctx}{\kd}$, $\wftm{\ctx}{\ty}{\kd}$, $\wftm{\ctx}{\tm}{\ty}$,
$\wfprop{\ctx}{\prop}$, $\wfsub{\ctx}{\ty}{\tyalt}$.

As usual, we have weakening and substitution lemmas for the syntactic
judgments, which are proved by induction on the derivation. The
weakening lemma is used in the proof of the substitution lemma.

\begin{lemma}[Weakening]%
\label[lemma]{thm:weakening}
Let $\ctx \cons \annty{\var}{\ty} \cons \ctxalt$ (resp.\
$\ctx \cons \annkd{\var}{\kd} \cons \ctxalt$) be a well-formed context,
with $\var$ fresh. If $\ctx \cons \ctxalt \vdash J$
is derivable, then so is the
corresponding judgment over the enlarged context
$\ctx \cons \annty{\var}{\ty} \cons \ctxalt \vdash J$ (resp.\
$\ctx \cons \annkd{\var}{\kd} \cons \ctxalt \vdash J$).
\end{lemma}
\begin{proof}
By induction on the derivation. Every rule is closed under inserting
an unused binding: extend $\ctxalt$ every time a rule introduces a
binder.
\end{proof}

\begin{lemma}[Syntactic substitution]%
\label[lemma]{thm:syntactic-substitution}
Let $\wftm{\ctx}{\tm}{\ty}$ (resp.\ $\wftm{\ctx}{\ty}{\kd}$). If the
judgment $\ctx \cons \annty{\var}{\ty} \cons \ctxalt \vdash J$
(resp.\ $\ctx \cons \annkd{\var}{\kd} \cons \ctxalt \vdash J$)
is derivable, then so is the judgment obtained by applying the
substitution $\subst{\tm}{\var}$ (resp.\ $\subst{\ty}{\var}$) to all its
components, $\ctx \cons \ctxalt\subst{\tm}{\var} \vdash J\subst{\tm}{\var}$
(resp.\ $\ctx \cons \ctxalt\subst{\ty}{\var} \vdash J\subst{\ty}{\var}$).
\end{lemma}
\begin{proof}
By induction on the derivation. In the variable case for $\var$
the use of $\var$ is replaced by the assumed derivation of
$\wftm{\ctx}{\tm}{\ty}$ (resp.\ $\wftm{\ctx}{\ty}{\kd}$), weakened by
\Cref{thm:weakening}. The remaining rules commute with the
substitution, which distributes over every former and passes under
binders.
\end{proof}

Terms are interpreted by induction on their raw syntax, so substitution
of a term commutes with the interpretation independently of any typing
derivation.

\begin{lemma}[Term substitution]%
\label[lemma]{thm:term-substitution}
For all terms $\tm$, $\tmalt$, every variable $\var$ and every runtime
assignment $\sctx$ defined on the free variables of $\tm$ and of
$\tmalt$ other than $\var$,
\[
  \semtm{\tmalt\subst{\tm}{\var}}[\sctx]
  = \semtm{\tmalt}[\pair{\sctx}{\semtm{\tm}[\sctx]}].
\]
\end{lemma}
\begin{proof}
By induction on the raw syntax of $\tmalt$.
\end{proof}

The four semantic substitution lemmas below all assert that extending
an environment by the denotation of the substituted term or type
coincides with first performing the substitution syntactically.
\begin{lemma}[Semantic substitution: term into a kind, refined]%
\label[lemma]{thm:substitution-refined-kind-ty}
If $\wfctx{\ctx \cons \annty{\var}{\tyalt} \cons \ctxalt}$ and
$\wfkd{\ctx \cons \annty{\var}{\tyalt} \cons \ctxalt}{\kd}$, then for
every
$\pair{\pair{\sctx}{\semtm{\tm}[\projrun{\ctx}{\sctx}]}}{\sctxalt} \in
\dom{\sem{\wfctx{\ctx \cons \annty{\var}{\tyalt} \cons \ctxalt}}}$,
\[
  \sem{\wfkd{\ctx \cons \annty{\var}{\tyalt} \cons \ctxalt}{\kd}}[\pair{\pair{\sctx}{\semtm{\tm}[\projrun{\ctx}{\sctx}]}}{\sctxalt}]
  = \sem{\wfkd{\ctx \cons \ctxalt\subst{\tm}{\var}}{\kd\subst{\tm}{\var}}}[\pair{\sctx}{\sctxalt}].
\]
\end{lemma}

\begin{lemma}[Semantic substitution: term into a type, refined]%
\label[lemma]{thm:substitution-refined-ty-tm}
If $\wfctx{\ctx \cons \annty{\var}{\tyalt} \cons \ctxalt}$ and
$\wftm{\ctx \cons \annty{\var}{\tyalt} \cons \ctxalt}{\tyaltalt}{\kd}$, then
for every
$\pair{\pair{\sctx}{\semtm{\tm}[\projrun{\ctx}{\sctx}]}}{\sctxalt} \in
\dom{\sem{\wfctx{\ctx \cons \annty{\var}{\tyalt} \cons \ctxalt}}}$,
\begin{multline*}
  \sem{\wftm{\ctx \cons \annty{\var}{\tyalt} \cons \ctxalt}{\tyaltalt}{\kd}}[\pair{\pair{\sctx}{\semtm{\tm}[\projrun{\ctx}{\sctx}]}}{\sctxalt}]
  \\ = \sem{\wftm{\ctx \cons \ctxalt\subst{\tm}{\var}}{\tyaltalt\subst{\tm}{\var}}{\kd\subst{\tm}{\var}}}[\pair{\sctx}{\sctxalt}].
\end{multline*}
\end{lemma}

\begin{lemma}[Semantic substitution: type into a kind, refined]%
\label[lemma]{thm:substitution-refined-kindtype}
If $\wfctx{\ctx \cons \annkd{\var}{\kd} \cons \ctxalt}$ and
$\wfkd{\ctx \cons \annkd{\var}{\kd} \cons \ctxalt}{\tyalt}$, then
for every
$\pair{\pair{\sctx}{
  \pair{\sem{\wftm{\ctx}{\ty}{\kd}}[\sctx]}{\sembot{\wftm{\ctx}{\ty}{\kd}}[\projctx{\ctx}{\sctx}]}
}}{\sctxalt} \in
\dom*{\sem*{\wfctx{\ctx \cons \annkd{\var}{\kd} \cons \ctxalt \\}}}$,
\begin{multline*}
  \sem{\wfkd{\ctx \cons \annkd{\var}{\kd} \cons
      \ctxalt}{\tyalt}}[\pair{\pair{\sctx}{
\pair{\sem{\wftm{\ctx}{\ty}{\kd}}[\sctx]}{\sembot{\wftm{\ctx}{\ty}{\kd}}[\projctx{\ctx}{\sctx}]}
    }}{\sctxalt}]
  \\ = \sem{\wfkd{\ctx \cons \ctxalt\subst{\ty}{\var}}{\tyalt\subst{\ty}{\var}}}[\pair{\sctx}{\sctxalt}].
\end{multline*}
\end{lemma}

\begin{lemma}[Semantic substitution: type into a type, refined]%
\label[lemma]{thm:substitution-refined-ty-ty}
If $\wfctx{\ctx \cons \annkd{\var}{\kd} \cons \ctxalt}$ and
$\wftm{\ctx \cons \annkd{\var}{\kd} \cons \ctxalt}{\tyalt}{\kdalt}$, then
for every
$\pair{\pair{\sctx}{
  \pair{\sem{\wftm{\ctx}{\ty}{\kd}}[\sctx]}{\sembot{\wftm{\ctx}{\ty}{\kd}}[\projctx{\ctx}{\sctx}]}
}}{\sctxalt} \in
\dom*{\sem*{\wfctx{\ctx \cons \annkd{\var}{\kd} \cons \\ \ctxalt}}}$,
\begin{multline*}
  \sem{\wftm{\ctx \cons \annkd{\var}{\kd} \cons
      \ctxalt}{\tyalt}{\kdalt}}[\pair{\pair{\sctx}{
\pair{\sem{\wftm{\ctx}{\ty}{\kd}}[\sctx]}{\sembot{\wftm{\ctx}{\ty}{\kd}}[\projctx{\ctx}{\sctx}]}
    }}{\sctxalt}]
  \\ = \sem{\wftm{\ctx \cons \ctxalt\subst{\ty}{\var}}{\tyalt\subst{\ty}{\var}}{\kdalt\subst{\ty}{\var}}}[\pair{\sctx}{\sctxalt}].
\end{multline*}
\end{lemma}

In the unrefined semantics substitution into a kind, and substitution
of a term, are degenerate: the unrefined interpretation of every kind
is $\pper{\dinf}$ and that of a type never inspects the term
assignment.

\begin{lemma}[Semantic substitution: type into a type, unrefined]%
\label[lemma]{thm:substitution-unrefined-ty-ty}
If $\wfctx{\ctx \cons \annkd{\var}{\kd} \cons \ctxalt}$ and
$\wftm{\ctx \cons \annkd{\var}{\kd} \cons \ctxalt}{\tyalt}{\kdalt}$, then
for every
$\pair{\pair{\sctx}{
  {\sembot{\wftm{\ctx}{\ty}{\kd}}[\projctx{\ctx}{\sctx}]}
}}{\sctxalt} \in
\dom{\sembot{\wfctx{\ctx \cons \annkd{\var}{\kd} \cons \ctxalt}}}$,
\begin{multline*}
  \sembot{\wftm{\ctx \cons \annkd{\var}{\kd} \cons
      \ctxalt}{\tyalt}{\kdalt}}[\pair{\pair{\sctx}{
{\sembot{\wftm{\ctx}{\ty}{\kd}}[\projctx{\ctx}{\sctx}]}
    }}{\sctxalt}] \\
  = \sembot{\wftm{\ctx \cons \ctxalt\subst{\ty}{\var}}{\tyalt\subst{\ty}{\var}}{\kdalt\subst{\ty}{\var}}}[\pair{\sctx}{\sctxalt}].
\end{multline*}
\end{lemma}

\begin{lemma}[Semantic substitution: term into a proposition]%
\label[lemma]{thm:substitution-refined-prop-tm}
If $\wfctx{\ctx \cons \annty{\var}{\tyalt} \cons \ctxalt}$ and
$\wfprop{\ctx \cons \annty{\var}{\tyalt} \cons \ctxalt}{\prop}$, then
for every
$\pair{\pair{\sctx}{
  \semtm{\tm}[\projrun{\ctx}{\sctx}]
}}{\sctxalt} \in
\dom{\sem{\wfctx{\ctx \cons \annty{\var}{\tyalt} \cons \ctxalt}}}$,
\begin{align*}
  \semprp{\wfprop{\ctx \cons \annty{\var}{\tyalt} \cons \ctxalt}{\prop}}[\pair{\pair{\sctx}{\semtm{\tm}[\projrun{\ctx}{\sctx}]}}{\sctxalt}]
  = \semprp{\wfprop{\ctx \cons \ctxalt\subst{\tm}{\var}}{\prop\subst{\tm}{\var}}}[\pair{\sctx}{\sctxalt}],
\end{align*}
\end{lemma}

\begin{lemma}[Semantic substitution: type into a proposition]%
\label[lemma]{thm:substitution-refined-prop-ty}
If $\wfctx{\ctx \cons \annkd{\var}{\kd} \cons \ctxalt}$ and
$\wfprop{\ctx \cons \annkd{\var}{\kd} \cons \ctxalt}{\prop}$, then
for every
$\pair{\pair{\sctx}{
\pair{\sem{\wftm{\ctx}{\ty}{\kd}}[\sctx]}{\sembot{\wftm{\ctx}{\ty}{\kd}}[\projctx{\ctx}{\sctx}]}
}}{\sctxalt} \in
\dom*{\sem*{\wfctx{\ctx \cons \annkd{\var}{\kd} \cons \\ \ctxalt}}}$,
\begin{multline*}
  \semprp{\wfprop{\ctx \cons \annkd{\var}{\kd} \cons
  \ctxalt}{\prop}}[\pair{\pair{\sctx}{
\pair{\sem{\wftm{\ctx}{\ty}{\kd}}[\sctx]}{\sembot{\wftm{\ctx}{\ty}{\kd}}[\projctx{\ctx}{\sctx}]}
  }}{\sctxalt}] \\
  = \semprp{\wfprop{\ctx \cons \ctxalt\subst{\ty}{\var}}{\prop\subst{\ty}{\var}}}[\pair{\sctx}{\sctxalt}],
\end{multline*}
\end{lemma}

\begin{proof}[Proof of \Cref{%
    thm:substitution-refined-ty-tm,%
    thm:substitution-refined-kind-ty,%
    thm:substitution-refined-kindtype,%
    thm:substitution-refined-ty-ty,%
    thm:substitution-unrefined-ty-ty,%
    thm:substitution-refined-prop-tm,%
    thm:substitution-refined-prop-ty%
}]
The statements are proved by induction on the derivation of the
substituted judgment, the cases that reach a term are discharged by
\Cref{thm:term-substitution}.
The right-hand sides are well-formed by \Cref{thm:syntactic-substitution}.
\end{proof}

\begin{lemma}[Syntactic narrowing]%
\label[lemma]{thm:syntactic-narrowing}
Let $\wfsub{\ctx}{\ty}{\tyalt}$ and $\wfty{\ctx}{\ty}$. If
$\ctx \cons \annty{\var}{\tyalt} \cons \ctxalt \vdash J$ is derivable,
then so is $\ctx \cons \annty{\var}{\ty} \cons \ctxalt \vdash J$.
\end{lemma}
\begin{proof}
By induction on the derivation. Apply the \RuleTmSub{label} and
\Cref{thm:weakening} in the \RuleTmVar{label} case.
\end{proof}

\begin{lemma}[Semantic narrowing]%
\label[lemma]{thm:semantic-narrowing}
Let $\wfsub{\ctx}{\ty}{\tyalt}$, and suppose both
$\ctx \cons \annty{\var}{\ty} \cons \ctxalt \vdash J$ and
$\ctx \cons \annty{\var}{\tyalt} \cons \ctxalt \vdash J$ are derivable.
\begin{enumerate}[label={(\roman*)}]
\item{Refined:}\label{thm:semantic-narrowing:refined}
  If $J$ is a kind or type judgment and
  $\sctx \in \dom{\sem{\wfctx{\ctx \cons \annty{\var}{\ty} \cons \ctxalt}}}$ and
  $\sctx \in \dom*{\sem*{\wfctx{\ctx \cons \annty{\var}{\tyalt} \cons \ctxalt}}}$, then
  \[
    \sem{\ctx \cons \annty{\var}{\ty} \cons \ctxalt \vdash J}[\sctx]
    =
    \sem{\ctx \cons \annty{\var}{\tyalt} \cons \ctxalt \vdash J}[\sctx],
  \]
\item{Unrefined:}\label{thm:semantic-narrowing:unrefined}
  If $J$ is a kind or type judgment and
  $\sctx \in \dom{\sembot{\wfctx{\ctx \cons \annty{\var}{\ty} \cons \ctxalt}}}$ and
  $\sctx \in \dom*{\sembot*{\wfctx{\ctx \cons \annty{\var}{\tyalt} \cons \ctxalt}}}$, then
  \[
    \sembot{\ctx \cons \annty{\var}{\ty} \cons \ctxalt \vdash J}[\sctx]
    =
    \sembot{\ctx \cons \annty{\var}{\tyalt} \cons \ctxalt \vdash J}[\sctx],
  \]
\item{Propositions:}\label{thm:semantic-narrowing:prop}
  If $\sctx \in \dom{\sem{\wfctx{\ctx \cons \annty{\var}{\ty} \cons
\ctxalt}}}$ and
  $\sctx \in \dom{\sem{\wfctx{\ctx \cons \annty{\var}{\tyalt} \cons \ctxalt}}}$, then
  \[
    \semprp{\wfprop{\ctx \cons \annty{\var}{\ty} \cons \ctxalt}{\prop}}[\sctx]
    =
    \semprp{\wfprop{\ctx \cons \annty{\var}{\tyalt} \cons \ctxalt}{\prop}}[\sctx].
  \]
\end{enumerate}
\end{lemma}
\begin{proof}
By structural induction on $J$. By \Cref{thm:syntactic-narrowing} the
two derivations differ only by \RuleTmSub{label} steps inserted at
occurrences of $\var$; every kind, type, and proposition formation
node is shared, and no such node inspects the type annotation of a
context entry. The projections $\projctx{-}{-}$ and $\projrun{-}{-}$
either drop the entry for $\var$ or forward its value without reading
its annotation, and terms enter the interpretations only through
$\semtm{-}$, which is defined on raw syntax and depends only on
$\projrun{-}{-}$; hence the inserted \RuleTmSub{label} steps are
invisible to the semantics.
\end{proof}

\section{Domain theory}

\begin{definition}[Complete lattice]
A complete lattice is a partially ordered set $(L, \leq)$ such that
every subset $S \subseteq L$ has a least upper bound $\Lub S$ and
a greatest lower bound $\Glb S$ in $L$.
\end{definition}

\begin{definition}[Complete partial order]
A complete partial order (CPO) is a partially ordered set $(P, \leq)$
such that every directed subset $D \subseteq P$ has a least upper
bound $\Lub D$ in $P$.

A CPO is \emph{pointed} if it has a least element $\bot \in P$.
\end{definition}

\begin{definition}[Monotone function]
  Given two partially ordered sets $(P, \leq_P)$ and $(Q, \leq_Q)$, a
  function $f: P \to Q$ is monotone if for all $p_1, p_2 \in P$, if
  $p_1 \leq_P p_2$ then $f(p_1) \leq_Q f(p_2)$.
\end{definition}

\begin{definition}[Continuous function]
  Given two complete partial orders $(P, \leq_P)$ and $(Q, \leq_Q)$, a monotone
  $f: P \to Q$ is continuous if for every directed subset $D \subseteq P$ $f(\Lub
  D) = \Lub f(D)$.
\end{definition}

\begin{lemma}[Projection of tagged tuples]
\label[lemma]{lem:tagged-tuple-projection}
For every $n$ there exists $n_j \le n$ such that
\[ \pi_{n+1}(\pair{\ctor}{\ioverline*{m}{\sval_j}}) = \pair{\pi_n(\ctor)}
{\ioverline*{m}{\pi_{n_j}(\sval_j)}} \]
\end{lemma}

\section{Supporting lemmas}

The equivalent theorem to \Cref{thm:monotonicity-types} for kinds.
\begin{lemma}[(Anti-)Monotonicity of kinds]%
\label[lemma]{thm:monotonicity-kinds}
Let $\ctxalt \defeq \ctx \cons \annkd{\tyctor}{\kdalt} \cons
\ioverline*{i}{\ann{\var}{\tyalt}}$, where $\var_i$ can bind
both type and term variables; given two context interpretations
different only on the assignment of $\tyctor$
$\pair*{\pair*{\sctx}{\pair*{\skd}{\skd_{\bot}}}}{\ioverline*{i}{\sval}},
\pair*{\pair*{\sctx}{\pair*{\skdalt}{\skd_{\bot}}}}{\ioverline*{i}{\sval}}
\in{} \dom{\sem*{\wfctx{\ctxalt}}}$, where $\skd \leq \skdalt$.
Let $\wfctx{\ctxalt}$ and $\wfkd{\ctxalt}{\kd}$ be
well-formed judgments, then
\begin{enumerate}[label={(\roman*)}]
\item If $\positive{\tyctor}{\kd} \posleq \pospos$, then
  $\sem{\wfkd{\ctxalt}{\kd}}[\pair{\pair{\sctx}{\pair{\skd}{\skd_{\bot}}}}{\ioverline*{i}{\sval}}]
  \subseteq
  \sem{\wfkd{\ctxalt}{\kd}}[\pair{\pair{\sctx}{\pair{\skdalt}{\skd_{\bot}}}}{\ioverline*{i}{\sval}}]$,
\item If $\positive{\tyctor}{\kd} \posleq \posneg$, then
  $\sem{\wfkd{\ctxalt}{\kd}}[\pair{\pair{\sctx}{\pair{\skd}{\skd_{\bot}}}}{\ioverline*{i}{\sval}}]
  \supseteq
  \sem{\wfkd{\ctxalt}{\kd}}[\pair{\pair{\sctx}{\pair{\skdalt}{\skd_{\bot}}}}{\ioverline*{i}{\sval}}]$.
\end{enumerate}
\end{lemma}

The following is the analogue of \Cref{lem:per-respecting-eliminator}
for the unrefined semantics; as discussed in
\cref{sec:model-inductive-semantics}, a less general statement
suffices here, since no under-approximation of the inductive type is
needed.

\begin{lemma}[Unrefined semantics is PER-respecting (Eliminator)]%
\label[lemma]{lem:unref-per-respecting-eliminator}$\ $\\
Let $D \defeq \decl$, $\ctxalt(i) \defeq
\annkd{\tyctor}{\kdfun*{({\ioverline*{k}{\ann{\var}{\ty}}})}{\kdtype}}
\cons \ioverline*{i}{\ann{\varalt}{\tyalt}}$ and $\skd\bots(\sctx)
\defeq
\sembot*{\wftm{\ctx}{D}{\kdfun*{({\ioverline*{k}{\ann{\var}{\ty}}})}{\kdtype}}}[\projctx{\ctx}{\sctx}]$.
If the following judgments are well-formed: $\wfctx{\ctx \cons
\ctxalt(j)}$ and $\wfty{\ctx \cons \ioverline{k}{\ann{\var}{\ty}}
\cons \annty{\varalt}{D \tmapp \ioverline{k}{\var}}}{\tyaltalt}$.
Given two related context interpretations $\pair{
  \pair{\pair{\sctx}{\skd\bots(\sctx)}}{\sctxaltalt}
}{
  \pair{\pair{\sctxalt}{\skd\bots(\sctxalt)}}{\sctxaltaltalt}
} \in \sembot*{\wfctx{\ctx \cons \ctxalt(j-1)}}$, two related
realizers of the eliminator $\pair{r}{r'} \in
\sembot{\wfty{\ctx}{\tyfun{\left(\ioverline*{k}{\annty{\var}{\ty}}\right)}{\tyfun{{D \app
\ioverline*{k}{\var}}}{\tyaltalt}}}}[\sctx]$ and
some term $\wftm{\ctx \cons \ctxalt(j - 1)}{\tmaltaltalt}{\tyalt_j}$,
then the eliminators realize induction hypothesis over argument
$\tmaltaltalt$:
\begin{align*}
  \vpair{
    \tmel{\tyctor}{\tyalt_j}{\sctx_\rho}{\semtm{\tmaltaltalt}[\sctx_\rho]}{r}
  }{
    \tmel{\tyctor}{\tyalt_j}{\sctxalt_\rho}{\semtm{\tmaltaltalt}[\sctxalt_\rho]}{r'}
  } \in \sembot{\wfty{\ctx \cons \ctxalt(j - 1)}{
    \tyel{\tyctor}{\tyaltalt}{\tmaltaltalt}{\tyalt_j}
  }}[\pair{\pair{\sctx}{\skd\bots(\sctx)}}{\sctxaltalt}]
\end{align*}
Where $\pair{\sctx_\rho}{\sctxalt_\rho} \in
\unrefass{\ctx \cons \ctxalt(j-1)}{\pair{\pair{\sctx}{\skd\bots(\sctx)}}{\sctxaltalt}}$
\end{lemma}

\section{Main Proofs}

The unrefined statements
(\Crefsub{thm:semantics-per-respecting-kinds}{unrefined},
\Crefsub{thm:type-interpretation-well-formed}{unrefined},
\Crefsub{thm:per-respecting-types}{unrefined},
\Crefsub{thm:per-respecting-terms}{unrefined},
\Crefsub{thm:subtyping-semantics}{unrefined},
\Cref{thm:unrefined-semantics-contractive}, and
\Cref{lem:unref-per-respecting-eliminator}) never mention the refined
interpretations and are proved by mutual induction on typing
derivations.
Monotonicity
(\Cref{thm:monotonicity-types,thm:monotonicity-kinds}) and
\Cref{thm:refined-sub-unrefined}, together with
\Cref{thm:unrefined-equality-respects-refined-equality} and
\Cref{thm:refined-runtime-projection-is-unrefined-runtime-interpretation},
follow by induction on derivations, using only the statements above.
These lemmas are statements \emph{about} the refined interpretations
but consume only their definitions, never the refined statements
below; the definitions themselves are given by induction on
derivations, simultaneously with
\Cref{thm:unrefined-semantics-contractive} and
\Cref{thm:monotonicity-types}, which justify the fixed points taken
in the interpretation of inductive types at strictly smaller
derivations.
Finally, the refined statements
(\Crefsub{thm:semantics-per-respecting-kinds}{refined},
\Crefsub{thm:type-interpretation-well-formed}{refined},
\Crefsub{thm:per-respecting-types}{refined},
\Crefsub{thm:per-respecting-terms}{refined},
\Crefsub{thm:subtyping-semantics}{refined},
\Cref{thm:refined-per-respecting-propositions}, and
\Cref{lem:per-respecting-eliminator}) are proved by one further mutual
induction on typing derivations; in particular, a refined statement
may invoke its unrefined counterpart at the \emph{same} judgment.

\begin{proof}[Proof of \Cref{lem:cuper-restriction}]
If $\pair{\pi_n(a)}{\pi_n(b)} \in S_n$ with $\pair{a}{b} \in S$, then
$\pair{\pi_n(a)}{\pi_n(b)} \in S$ by uniformity, and $\pi_n(a), \pi_n(b) \in
D_n$. Conversely, if $\pair{x}{y} \in S$ with $x, y \in D_n$ then $x = \pi_n(x)$
and $y = \pi_n(y)$, so $\pair{x}{y} \in S_n$. For the inclusion, $D_j \subseteq
D_n$ when $j \leq n$ because $\pi_n \circ \pi_j = \pi_j$, whence $S \cap (D_j
\times D_j) \subseteq S \cap (D_n \times D_n)$.
\end{proof}

\begin{proof}[Proof of \Cref{lem:cuper-det}]
Let $\pair{a}{b} \in A$. By uniformity and \cref{lem:cuper-restriction},
$\pair{\pi_n(a)}{\pi_n(b)} \in A_n = B_n \subseteq B$ for every $n$. The family
${\{\pair{\pi_n(a)}{\pi_n(b)}\}}_n$ is directed with least upper bound
$\pair{a}{b}$, since $\Lub_n \pi_n = id$; by completeness of $B$, $\pair{a}{b}
\in B$. Hence $A \subseteq B$, and symmetrically $B \subseteq A$.
\end{proof}

\begin{proof}[Proof of \Cref{lem:guarded-fixpoint}]
\emph{Uniqueness.} Suppose $\Phi(R) = R$ and $\Phi(S) = S$. We show $R_n = S_n$
by induction on $n$. Since $\pi_0$ is the constant map $\bot$, both
approximations equal $\{\pair{\bot}{\bot}\}$ at $n = 0$; and if $R_n = S_n$ then contractivity gives
$R_{n+1} = \Phi(R)_{n+1} = \Phi(S)_{n+1} = S_{n+1}$. By \cref{lem:cuper-det}, $R
= S$.

\emph{Existence.} Put $R^0 \defeq D \times D$ and $R^{k+1} \defeq \Phi(R^k)$;
each $R^k$ is a CUPER. Using contractivity, a routine induction shows the
iterates stabilise, $(R^m)_k = (R^k)_k$ whenever $m \geq k$, the base case
holding because all CUPERs agree at level $0$. Writing $R_n \defeq (R^n)_n$ for
the stabilised level-$n$ data---not yet known to coincide with the approximation
$(R)_n$ of the limit $R$ defined below---the family is increasing and coherent,
$(R_{n+1})_n = R_n$, by \cref{lem:cuper-restriction} and stabilisation. Define
\begin{align*}
  R \defeq
  \begin{setbuild}{\pair{a}{b} \in D \times D}
    \forall n.\ \pair{\pi_n(a)}{\pi_n(b)} \in R_n
  \end{setbuild}.
\end{align*}
Then $R$ is a CUPER: symmetry and transitivity hold because each $R_n = R^n \cap
(D_n \times D_n)$ is a PER (\cref{lem:cuper-restriction}); pointedness is
immediate, and uniformity follows from the monotonicity $R_j \subseteq R_n$ ($j
\leq n$); and for completeness, if ${\{\pair{a_i}{b_i}\}}_i \subseteq R$ is
directed with least upper bound $\pair{a}{b}$, then continuity of each $\pi_n$
makes $\pair{\pi_n(a)}{\pi_n(b)}$ the least upper bound of the directed family
${\{\pair{\pi_n(a_i)}{\pi_n(b_i)}\}}_i \subseteq R_n$; since $R_n = R^n \cap (D_n
\times D_n)$ is closed under directed least upper bounds---$R^n$ being complete
and $D_n \times D_n$ Scott-closed---this bound lies in $R_n$, so $\pair{a}{b} \in
R$. Moreover
$(R)_n = R_n$: the inclusion $(R)_n \subseteq R_n$ is immediate, and conversely,
given $\pair{x}{y} \in R_n \subseteq D_n \times D_n$, coherence and monotonicity
give $\pair{\pi_m(x)}{\pi_m(y)} \in R_m$ for every $m$, so $\pair{x}{y} \in R$;
as $x = \pi_n(x)$ and $y = \pi_n(y)$, this places $\pair{x}{y}$ in $(R)_n$.
Finally $R$ is a fixed point: $R$ and $R^n$
share the $n$-th approximation $R_n$, so contractivity gives $\Phi(R)_{n+1} =
\Phi(R^n)_{n+1} = (R^{n+1})_{n+1} = R_{n+1}$, while $\Phi(R)_0 =
\{\pair{\bot}{\bot}\} = R_0$; hence $\Phi(R) = R$ by \cref{lem:cuper-det}.
\end{proof}

\begin{proof}[Proof of \Cref{lem:tagged-tuple-projection}]
In the bilimit, $\pi_{n+1}\pair{a}{b} = \pair{\pi_n(a)}{\pi_n(b)}$
hence we can apply the identity recursively inside the nested pairs.
\end{proof}

\begin{proof}[Proof of \Crefsub{thm:semantics-per-respecting-kinds}{refined}]
By induction on the structure of $\wfkd{\ctx}{\kd}$.
\begin{itemize}
\item \RuleKdType{label}
\[
  \sem{\RuleKdType{concl}}[\sctx] = \per{\dinf} =
  \sem{\RuleKdType{concl}}[\sctxalt]
\]

\item \RuleKdFun{label}
\begin{align*}
  & \sem{\RuleKdFun{concl}}[\sctx] \\
= & \begin{setbuild}{\sfun}
  \sfun \in \dinf \to \left(\dinf \to \right)^{\arity{\kd}} \per{\dinf} \\
  \forall \sval \in \dom{\sem{\wfty{\ctx}{\ty}}[\sctx]} \;
  \sfun(\sval) \in \sem{\wfkd{\ctx \cons \annty{\var}{\ty}}{\kd}}[\pair{\sctx}{\sval}] \\
  \forall \pair{\sval}{\svalalt} \in \left(\sembot{\wfty{\ctx}{\ty}}[\projctx{\ctx}{\sctx}] \cap \dom{\sem{\wfty{\ctx}{\ty}}[\sctx]}^2\right) \;
  \sfun(\sval) = \sfun(\svalalt)
\end{setbuild} \\
  & \text{By \ih{} and \Cref{thm:per-respecting-types}} \\
= & \begin{setbuild}{\sfun}
  \sfun \in \dinf \to \left(\dinf \to \right)^{\arity{\kd}} \per{\dinf} \\
  \forall \sval \in \dom{\sem{\wfty{\ctx}{\ty}}[\sctxalt]} \;
  \sfun(\sval) \in \sem{\wfkd{\ctx \cons \annty{\var}{\ty}}{\kd}}[\pair{\sctxalt}{\sval}] \\
  \forall \pair{\sval}{\svalalt} \in \left(\sembot{\wfty{\ctx}{\ty}}[\projctx{\ctx}{\sctxalt}] \cap \dom{\sem{\wfty{\ctx}{\ty}}[\sctxalt]}^2\right) \;
  \sfun(\sval) = \sfun(\svalalt)
\end{setbuild} \\
= & \sem{\RuleKdFun{concl}}[\sctxalt]
\end{align*}
\end{itemize}
\end{proof}

\begin{proof}[Proof of \Crefsub{thm:semantics-per-respecting-kinds}{unrefined}]
By induction on the structure of $\wfkd{\ctx}{\kd}$.
\begin{itemize}
\item \RuleKdType{label}
\[
  \sembot{\RuleKdType{concl}}[\sctx] = \pper{\dinf} =
  \sembot{\RuleKdType{concl}}[\sctxalt]
\]

\item \RuleKdFun{label}
\begin{align*}
  & \sembot{\RuleKdFun{concl}}[\sctx] \\
= & \sembot{\wfkd{\ctx \cons \annty{\var}{\ty}}{\kd}}[\sctx] \\
  & \text{By \ih{}} \\
= & \sembot{\wfkd{\ctx \cons \annty{\var}{\ty}}{\kd}}[\sctxalt] \\
= & \sembot{\RuleKdFun{concl}}[\sctxalt]
\end{align*}
\end{itemize}
\end{proof}

\begin{proof}[Proof of \Crefsub{thm:type-interpretation-well-formed}{refined}]
By induction on the structure of $\wftm{\ctx}{\ty}{\kd}$.
\begin{itemize}
\item \RuleTyNat{label} and \RuleTyBool{label}
\[
  \sem{\wftm{\ctx}{\tybase}{\kd}}[\sctx] =
  \begin{setbuild}{\pair{\demb{\tybase}{\sval}}{\demb{\tybase}{\sval}}}
    \sval \in \tybase
  \end{setbuild}
  \in
  \sem{\wfkd{\ctx}{\kd}}[\sctx] = \per{\dinf}
\]
The diagonal relation over a subset of $\dinf$ is a PER.

\item \RuleTyFun{label}
\begin{align*}
  & \sem{\RuleTyFun{concl}}[\sctx]
  \\
  & =
  \begin{setbuild}{\pair{\sfun}{\sfunalt}}[lc@{\;}llc@{\;}lll]
      \forall&\pair{\sval}{\svalalt}
      &\in{}& \sem{\wfty{\ctx}{\ty}}[\sctx].\;
      &\pair{\sfun \app \sval}{\sfunalt \app \svalalt} &\in{}&
      \sem{\wfty{
        \ctx \cons \annty{\var}{\ty}
      }{\tyalt}}[\pair{\sctx}{\sval}]
    \\
      \forall&\pair{\sval\bots}{\svalalt\bots}
      &\in{}& \sembot{\wfty{\ctx}{\ty}}[\projctx{\ctx}{\sctx}].\;
      &\pair{\sfun \app \sval\bots}{\sfunalt \app \svalalt\bots} &\in{}&
      \sembot{\wfty{
        \ctx \cons \annty{\var}{\ty}
      }{\tyalt}}[\projctx{\ctx}{\sctx}]
  \end{setbuild}
  \\
  & \in \sem{\wfkd{\ctx}{\kdtype}}[\sctx] = \per{\dinf}
\end{align*}

It is not enough to show that each of the two defining conditions
determines a PER on its own, since neither is in general closed
under symmetry by itself. Instead, we show directly that the whole
set is a PER.
\begin{enumerate}
  \item \textbf{Symmetry}:
    Fix $\pair{\sfun}{\sfunalt}$ in the set. We must show
    that $\pair{\sfunalt}{\sfun}$ satisfies the defining condition for
    every
    $\pair{\sval}{\svalalt} \in \sem{\wfty{\ctx}{\ty}}[\sctx]$ and
    $\pair{\sval\bots}{\svalalt\bots} \in
    \sembot{\wfty{\ctx}{\ty}}[\projctx{\ctx}{\sctx}]$.
    By symmetry of the domain PERs,
    $\pair{\svalalt}{\sval} \in \sem{\wfty{\ctx}{\ty}}[\sctx]$ and
    $\pair{\svalalt\bots}{\sval\bots} \in
    \sembot*{\wfty{\ctx}{\ty}}[\projctx{\ctx}{\sctx}]$. Instantiating
    the assumption for $\pair{\sfun}{\sfunalt}$ at these swapped
    arguments gives
    \[
      \pair{\sfun \app \svalalt}{\sfunalt \app \sval}
      \in
      \sem{\wfty{\ctx \cons \annty{\var}{\ty}}{\tyalt}}
      [\pair{\sctx}{\svalalt}]
    \]
    and
    \[
      \pair{\sfun \app \svalalt\bots}
            {\sfunalt \app \sval\bots}
      \in
      \sembot{\wfty{\ctx \cons \annty{\var}{\ty}}{\tyalt}}
      [\projctx{\ctx}{\sctx}].
    \]
    By symmetry of these PERs we obtain the same pairs in the opposite
    order. Finally,
    $\pair{\pair{\sctx}{\svalalt}}{\pair{\sctx}{\sval}}
    \in \sem{\wfctx{\ctx \cons \annty{\var}{\ty}}}$, so
    \Crefsub{thm:per-respecting-types}{refined} transports the refined
    codomain relation from $[\pair{\sctx}{\svalalt}]$ to
    $[\pair{\sctx}{\sval}]$. Hence $\pair{\sfunalt}{\sfun}$ is in the
    set.
  \item \textbf{Transitivity}:
    Fix $\pair{\sfun}{\sfunalt}$ and
    $\pair{\sfunalt}{\sfunaltalt}$ in the set. We again fix
    arbitrary
    $\pair{\sval}{\svalalt} \in \sem*{\wfty{\ctx}{\ty}}[\sctx]$ and
    $\pair{\sval\bots}{\svalalt\bots} \in
    \sembot{\wfty{\ctx}{\ty}}[\projctx{\ctx}{\sctx}]$. By partial
    reflexivity,
    $\pair{\sval}{\sval} \in \sem{\wfty{\ctx}{\ty}}[\sctx]$ and
    $\pair{\sval\bots}{\sval\bots} \in
    \sembot{\wfty{\ctx}{\ty}}[\projctx{\ctx}{\sctx}]$. Instantiating
    the first assumption at these reflexive pairs gives
    \[
      \pair{\sfun \app \sval}{\sfunalt \app \sval}
      \in
      \sem{\wfty{\ctx \cons \annty{\var}{\ty}}{\tyalt}}
      [\pair{\sctx}{\sval}]
    \]
    and
    \[
      \pair{\sfun \app \sval\bots}{\sfunalt \app \sval\bots}
      \in
      \sembot{\wfty{\ctx \cons \annty{\var}{\ty}}{\tyalt}}
      [\projctx{\ctx}{\sctx}].
    \]
    Instantiating the second assumption at
    $\pair{\sval}{\svalalt}$ and
    $\pair{\sval\bots}{\svalalt\bots}$ gives the corresponding pairs
    from $\sfunalt$ to $\sfunaltalt$. Transitivity of the codomain
    PERs therefore yields
    \[
      \pair{\sfun \app \sval}{\sfunaltalt \app \svalalt}
      \in
      \sem{\wfty{\ctx \cons \annty{\var}{\ty}}{\tyalt}}
      [\pair{\sctx}{\sval}]
    \]
    and
    \[
      \pair{\sfun \app \sval\bots}
            {\sfunaltalt \app \svalalt\bots}
      \in
      \sembot{\wfty{\ctx \cons \annty{\var}{\ty}}{\tyalt}}
      [\projctx{\ctx}{\sctx}].
    \]
    Hence $\pair{\sfun}{\sfunaltalt}$ is in the set.
\end{enumerate}

\item \RuleTyAll{label}
\begin{align*}
   & \sem{\RuleTyAll{concl}}[\sctx]
  \\
  & = \begin{setbuild}{\pair{\sval}{\svalalt}}[c]
    \forall \skd \in \sem{\wfkd{\ctx}{\kd}}[\sctx],\;
    \forall \skd\bots \in
    \sembot{\wfkd{\ctx}{\kd}}[\projctx{\ctx}{\sctx}] \\
    \projty{\ctx}{\sctx}{\kd}{\skd} \subseteq \skd\bots
    \Rightarrow
    \pair{\sval}{\svalalt} \in
    \sem{\wfty{
      \ctx \cons \annkd{\var}{\kd}
    }{\ty}}[\pair{\sctx}{\pair{\skd}{\skd\bots}}]
  \end{setbuild}
  \\
  & \in \sem{\wfkd{\ctx}{\kdtype}}[\sctx] = \per{\dinf}
\end{align*}
Since PERs are closed under intersection, it suffices to show that
for each fixed $\skd \in \sem{\wfkd{\ctx}{\kd}}[\sctx]$ and
$\skd\bots \in \sembot{\wfkd{\ctx}{\kd}}[\projctx{\ctx}{\sctx}]$ the
set
\[
  \begin{setbuild}{\pair{\sval}{\svalalt}}
    \projty{\ctx}{\sctx}{\kd}{\skd} \subseteq \skd\bots
    \Rightarrow
    \pair{\sval}{\svalalt} \in
    \sem{\wfty{\ctx \cons \annkd{\var}{\kd}}{\ty}}
    [\pair{\sctx}{\pair{\skd}{\skd\bots}}]
  \end{setbuild}
\]
is a PER. If $\projty{\ctx}{\sctx}{\kd}{\skd} \not\subseteq \skd\bots$
the implication holds vacuously for every pair, so the set is all of
$\dinf \times \dinf$, which is trivially a PER. Otherwise the set
equals
$\sem{\wfty{\ctx \cons \annkd{\var}{\kd}}{\ty}}[\pair{\sctx}{\pair{\skd}{\skd\bots}}]$,
which we show is a PER.
\begin{enumerate}
\item \textbf{Symmetry}:
  Fix $\pair{\sval}{\svalalt}$ in the set, by \ih{} and symmetry of PERs
  we have $\pair{\svalalt}{\sval} \in
  \sem{\wfty{\ctx \cons \annkd{\var}{\kd}}{\ty}}[\pair{\sctx}{\pair{\skd}{\skd\bots}}]$
  thus $\pair{\svalalt}{\sval}$ is in the set.
\item \textbf{Transitivity}:
  Fix $\pair{\sval}{\svalalt}$ and $\pair{\svalalt}{\svalaltalt}$ in
  the set, by \ih{} and transitivity of PERs we have
  $\pair{\sval}{\svalaltalt} \in
  \sem{
    \wfty{\ctx \cons \annkd{\var}{\kd}}{\ty}
  }[\pair{\sctx}{\pair{\skd}{\skd\bots}}]$
  thus $\pair{\sval}{\svalaltalt}$ is in the set.
\end{enumerate}

\item \RuleTyVar{label}
  \[
    \sem{\RuleTyVar{concl}}[\sctx] =
    \sctx\left(\var\right)_1
    \in \sem{\wfkd{\ctx}{\kd}}[\sctx]
  \] by assumption.

\item \RuleTyRef{label}
\begin{multline*}
  \sem{\RuleTyRef{concl}}[\sctx]
  = \begin{setbuild}{\pair{\sval}{\svalalt}}
      \pair{\sval}{\svalalt} \in \sem{\wfty{\ctx}{\ty}}[\sctx] \\
      \semprp{
        \wfprop{\ctx \cons \annty{\var}{\ty}}{\prop}
      }[\pair{\sctx}{\sval}] = \strue
    \end{setbuild} \\
  \in \sem{\wfkd{\ctx}{\kd}}[\sctx] = \per{\dinf}
\end{multline*}
we show that it is a PER
\begin{enumerate}
\item \textbf{Symmetry}:
  Fix $\pair{\sval}{\svalalt}$ in the set, by \ih{} and symmetry of PERs
  we have $\pair{\svalalt}{\sval} \in \sem{\wfty{\ctx}{\ty}}[\sctx]$
  and by \Cref{thm:refined-per-respecting-propositions} we have
  $\semprp{
    \wfprop{\ctx \cons \annty{\var}{\ty}}{\prop}
  }[\pair{\sctx}{\svalalt}] = \strue$ thus
  $\pair{\svalalt}{\sval}$ is in the set.
\item \textbf{Transitivity}:
  Fix $\pair{\sval}{\svalalt}$ and $\pair{\svalalt}{\svalaltalt}$ in
  the set, by \ih{} and transitivity of PERs we have
  $\pair{\sval}{\svalaltalt} \in \sem{\wfty{\ctx}{\ty}}[\sctx]$ and
  thus $\pair{\sval}{\svalaltalt}$ is in the set.
\end{enumerate}

\item \RuleTyApp{label}
\[
  \sem{\RuleTyApp{concl}}[\sctx]
  = \sem{\wftm{\ctx}{\tyalt}{\tyfun*[\var]{\ty}{\kd}}}[\sctx]
    \left(\semtm{\tm}[\projrun{\ctx}{\sctx}]\right)
  \in \sem{\wfkd{\ctx}{\kd \subst{\tm}{\var}}}[\sctx]
\]
by \ih{} $f = \sem{\wftm{\ctx}{\tyalt}{\tyfun*[\var]{\ty}{\kd}}}[\sctx]
\in \sem{\wfkd{\ctx}{\tyfun*[\var]{\ty}{\kd}}}[\sctx]$ thus
$f \left(\semtm{\tm}[\projrun{\ctx}{\sctx}]\right) \in \sem{
  \wfkd{\ctx \cons \annty{\var}{\ty}}{\kd}
}[\pair{\sctx}{\semtm{\tm}[\projrun{\ctx}{\sctx}]}]$
by \Cref{thm:substitution-refined-kind-ty}
$= \sem{\wfkd{\ctx}{\kd \subst{\tm}{\var}}}[\sctx]$.

\item \RuleKdIndPos{label} and \RuleKdIndStrat{label}
\begin{multline*}
  \sem{\wftm{\ctx}{\decl}{\ioverline*{k}{\ty} \to \kdtype}}[\sctx] \\
= \;\mathsf{lfp}_X \left(\scalebox{0.5}{$\displaystyle
    \underbrace{\ioverline*{k}{\svalaltalt} \mapsto
      % \underbrace{
        \Bigcup_i \begin{setbuild}{\pair{
      \pair{\ctor_i}{\ioverline*{j}{\sval}}
    }{
      \pair{\ctor_i}{\ioverline*{j}{\svalalt}}
    }}
      \pair{\sval_j}{\svalalt_j} \in \sem{\wfty{
        \ctx \cons
        \annkd{\tyctor}{\kdfun{\ioverline*{k}{\ty}}{\kdtype}} \cons
        \ioverline*{j - 1}{\annty{\var}{\tyalt}}
      }{\tyalt_j}}[\pair{\pair{\sctx}{\pair{X}{
          \sembot{\wftm{\ctx}{\decl}{\kdfun{\ioverline*{k}{\ty}}{\kdtype}}}[\projctx{\ctx}{\sctx}]
        }}}{\ioverline*{j - 1}{\sval}}] \\
      \pair{\svalaltalt_k}{\semtm{\tm_k}[\pair{\projrun{\ctx}{\sctx}}{\ioverline*{j}{\sval}}]}
      \in \sembot{\wfty{
        \ctx \cons \ioverline*{k - 1}{\annty{\var}{\ty}}
      }{\ty_k}}[\projctx{\ctx}{\sctx}]
  \end{setbuild}
% }_{\beta}
  }_{\alpha}$}\right) \\
  \in \sem{\wfkd{\ctx}{\ioverline*{k}{\ty} \to \kdtype}}[\sctx]
\end{multline*}

First, for the shared part of the two rules, we show that if
$X \in \sem{\wfkd{\ctx}{\ioverline*{k}{\ty} \to \kdtype}}[\sctx]$ lies
below the unrefined interpretation of the declaration, so that the
environment in $\alpha$ is well formed, then
$\alpha \in \sem{\wfkd{\ctx}{\ioverline*{k}{\ty} \to
\kdtype}}[\sctx]$; we have to show two things:
Fixed $\pair{\ioverline*{k}{\svalaltalt}}{\ioverline*{k}{\svalaltaltalt}} \in
\sembot*{\wfty{\ctx}{\ioverline*{k}{\ty}}}[\projctx{\ctx}{\sctx}]$ we show that:
\begin{itemize}
\item Fixed $\ioverline*{k}{\svalaltalt} \in \dom{\sem{\wfty{\ctx}{\ioverline*{k}{\ty}}}[\sctx]}$,
    $\alpha(\ioverline*{k}{\svalaltalt})$ is a PER, since it is the
    union of disjoint $C_i$ we just need to show that each $C_i$ is a
    PER, by \Crefsub{thm:per-respecting-types}{refined} the semantics
    is invariant under related contexts and hence it is a PER.
\item Fixed
  $\pair{\ioverline*{k}{\svalaltalt}}{\ioverline*{k}{\svalaltaltalt}} \in
  \sembot{\wfty{\ctx}{\ioverline*{k}{\ty}}}[\projctx{\ctx}{\sctx}] \cap
  \dom{\sem{\wfty{\ctx}{\ioverline*{k}{\ty}}}[\sctx]}^2$ we show that
    $\alpha(\ioverline*{k}{\svalaltalt}) =
    \alpha(\ioverline*{k}{\svalaltaltalt})$;
    it follows that
    $\pair{\ioverline*{k}{\svalaltalt}}{\ioverline*{k}{\svalaltaltalt}} \in
    \sembot{\wfty{\ctx}{\ioverline*{k}{\ty}}}[\projctx{\ctx}{\sctx}]$ hence
    either $\forall k\;
    \pair*{\svalaltalt_k}{\semtm*{\tm_k}[\pair*{\projrun{\ctx}{\sctx}}{\\ \ioverline*{j}{\sval}}]}
    \in \sembot{\wfty{
      \ctx \cons \ioverline*{k - 1}{\annty{\var}{\ty}}
    }{\ty_k}}[\projctx{\ctx}{\sctx}]$ thus by transitivity of the PERs
    $\pair*{\svalaltaltalt_k}{\semtm*{\tm_k}[\pair*{\projrun{\ctx}{\sctx}}{\\\ioverline*{j}{\sval}}]}
    \in \sembot{\wfty{
      \ctx \cons \ioverline*{k - 1}{\annty{\var}{\ty}}
    }{\ty_k}}[\projctx{\ctx}{\sctx}]$
    and hence the corresponding $C_i$ are equal, as the semantics of each
    $\tyalt_j$ does not depend on $\svalaltalt_k$; or if $\exists k\;
    \pair*{\svalaltalt_k}{\semtm{\tm_k}[\pair*{\projrun{\ctx}{\sctx}}{\ioverline*{j}{\sval}}]}
    \not \in
    \sembot*{\wfty{
      \ctx \cons \ioverline*{k - 1}{\annty{\var}{\ty}}
    }{\ty_k}}[\projctx{\ctx}{\sctx}]$ then by transitivity of the PERs
    $\pair{\svalaltaltalt_k}{\semtm{\tm_k}[\pair{\projrun{\ctx}{\sctx}}{\ioverline*{j}{\sval}}]}
    \not \in \sembot*{\wfty{
      \ctx \cons \ioverline*{k - 1}{\annty{\var}{\ty}}
    }{\ty_k}}[\projctx{\ctx}{\sctx}]$ and hence the corresponding
    sets $C_i$ are empty and hence equal.
\end{itemize}

Now given this we can show that the least fixed point of $\alpha$
exists, we consider the two cases separately:
\begin{itemize}
\item \RuleKdIndPos{label}
By assumption
$\iforall{l, j} \positive{\tyctor}{\ty_{l, j}} \posleq \pospos$
and by \Cref{thm:monotonicity-types} each \\
$\sem{\wfty{
  \ctx \cons
  \annkd{\tyctor}{\kdfun{\ioverline*{k}{\ty}}{\kdtype}} \cons
  \ioverline*{j - 1}{\annty{\var}{\tyalt}}
}{\tyalt_j}}[\pair{\pair{\sctx}{\pair{X}{
  \dots
}}}{\ioverline*{j - 1}{\sval}}]$ is monotone in $X$, thus $\alpha$
is monotone on such $X$, which form a complete lattice, and
by \Cref{thm:knaster-tarski} the least fixpoint of $\alpha$ exists.
\item \RuleKdIndStrat{label}
By assumption each self reference in $\ty_{i, j}$ is on a strictly
smaller index with respect to the stratification function $m$,
hence we can give the alternative definition of the functional
\[
\mathcal{F}_{rec}\left(\ioverline*{k}{\sval}\right) =
  (\alpha^{\left(1 + n\right)}(\ioverline*{k}{\svalaltalt} \mapsto
  \emptyset))(\ioverline*{k}{\sval})
  \text{ where } n \defeq \dret{\snat}{\semtm{m}[\projrun{\ctx}{\sctx}] \app
  \ioverline*{k}{\sval}}
\]
where $\alpha^{(n)}$ is the $n$-fold composition of
$\alpha$.
\\

Since all the recursive occurrences are on strictly
smaller indices, this is well defined.
This follows from the fact that fixed ${\sval_k} \in
\dom*{\sem*{\wfty{
  \ctx \cons \annty{\var}{\ioverline*{k-1}{\ty_k}}
}{\kdtype}}[\pair{\sctx}{\ioverline*{k-1}{\sval}}]}$
$\sem{\wftm{\ctx}{\decl}{\ioverline*{k}{\ty} \to \kdtype}}[\sctx] \app
\ioverline*{k}{\sval}$ cannot depend on any recursive occurrences of
$\ioverline*{k}{\svalalt}$ where
$\dret{\snat}{\semtm{m}[\projrun{\ctx}{\sctx}] \app
\ioverline*{k}{\svalalt}} \geq
\dret{\snat}{\semtm{m}[\projrun{\ctx}{\sctx}] \app
\ioverline*{k}{\sval}}$ by the assumption of the stratification
condition.
\\

This definition is stable, meaning that forall
$\ioverline*{k}{\svalaltalt} \in \dom*{\sem*{\wfty{
  \ctx \cons \annty{\var}{\ioverline*{k-1}{\ty}}
}{\kdtype}}[\pair{\sctx}{\ioverline*{k-1}{\sval}}]}$
where $n = \dret{\snat}{\semtm{m}[\projrun{\ctx}{\sctx}] \app
\ioverline*{k}{\svalaltalt}}$ $(\alpha^{(1 +
n)}(\ioverline*{k}{\svalaltalt} \mapsto
\emptyset))(\ioverline*{k}{\svalaltalt}) = (\alpha^{(1 +
  m)}(\ioverline*{k}{\svalaltalt} \mapsto
\emptyset))(\ioverline*{k}{\svalaltalt})$ for all $m \geq n$.
By strong induction on $n$: $n = 0$ then there are no recursive
occurrences and hence the definition is stable by definition as the
argument to $\alpha$ does not matter. For $n > 0$ the semantics
depends only on the recursive occurrences of strictly smaller indices,
for which by the \ih{} the definition is stable, hence the whole
definition is stable.
\\

Now we show that it is a fixed point, hence $\mathcal{F}_{rec} =
\alpha(\mathcal{F}_{rec})$; fixed ${\sval_k} \in
\dom*{\sem*{\wfty{
  \ctx \cons \annty{\var}{\ioverline*{k-1}{\ty}}
}{\kdtype}}[\pair{\sctx}{\ioverline*{k-1}{\sval}}]}$
and let $n \defeq \dret{\snat}{\semtm{m}[\projrun{\ctx}{\sctx}] \app
\ioverline*{k}{\sval}}$ we have
\begin{align*}
  & \mathcal{F}_{rec}(\ioverline*{k}{\sval}) \\
={} & (\alpha^{(1 + n)}(\ioverline*{k}{\svalaltalt} \mapsto
    \emptyset))(\ioverline*{k}{\sval}) \\
& \text{By stability} \\
={} & (\alpha^{(2 + n)}(\ioverline*{k}{\svalaltalt} \mapsto
    \emptyset))(\ioverline*{k}{\sval}) \\
={}  & \alpha(\mathcal{F}_{rec})(\ioverline*{k}{\sval}) \\
\end{align*}

Finally we show that it is unique and hence trivially the least fixed
point.  Let $G$ be another fixed point of $\alpha$ thus $G =
\alpha(G)$, fixed ${\sval_k} \in
\dom*{\sem*{\wfty{
  \ctx \cons \annty{\var}{\ioverline*{k-1}{\ty}}
}{\kdtype}}[\pair{\sctx}{\ioverline*{k-1}{\sval}}]}$
and let $n \defeq \dret{\snat}{\semtm{m}[\projrun{\ctx}{\sctx}] \app
\ioverline*{k}{\sval}}$ by induction on $n$: if $n = 0$ then
\begin{align*}
    & \mathcal{F}_{rec}(\ioverline*{k}{\sval}) \\
={} & \alpha(\ioverline*{k}{\svalaltalt} \mapsto \emptyset)(\ioverline*{k}{\sval}) \\
    & \text{since there are no recursive occurrences} \\
    & \text{does not depend on the argument} \\
={} & \alpha(G)(\ioverline*{k}{\sval}) \\
    & \text{since $G$ is a fixed point of $\alpha$} \\
={} & G(\ioverline*{k}{\sval})
\end{align*}
if $n > 0$ then
\begin{align*}
    & \mathcal{F}_{rec}(\ioverline*{k}{\sval}) \\
={} & (\alpha^{(2 + n)}(\ioverline*{k}{\svalaltalt} \mapsto \emptyset))(\ioverline*{k}{\sval}) \\
    & \text{By \ih{}} \\
={} & (\alpha(G))(\ioverline*{k}{\sval}) \\
    & \text{since $G$ is a fixed point of $\alpha$} \\
={} & G(\ioverline*{k}{\sval})
\end{align*}
Note that the proof does not rely on the fact that $\alpha$ is
monotone but only on the uniqueness of the fixed point.
\end{itemize}
\end{itemize}
\end{proof}

\begin{proof}[Proof of \Crefsub{thm:type-interpretation-well-formed}{unrefined}]
By induction on the structure of $\wftm{\ctx}{\ty}{\kd}$.
\begin{itemize}
\item \RuleTyNat{label} and \RuleTyBool{label}
\[
  \sembot{\wftm{\ctx}{\tybase}{\kd}}[\sctx] =
  \begin{setbuild}{\pair{\demb{\tybase}{\sval}}{\demb{\tybase}{\sval}}}
    \sval \in \tybase
  \end{setbuild} \cup \left\{\pair{\bot}{\bot}\right\}
  \in
  \sembot{\wfkd{\ctx}{\kdtype}}[\sctx] = \pper{\dinf}
\]
The diagonal relation over a subset of $\dinf$ is a PER.  Since
$\pair{\bot}{\bot}$ is in the set, it is clearly pointed, and since
$\tybase$ is a flat domain, completeness and uniformity follow.

\item \RuleTyFun{label}
\begin{multline*}
\sembot{\RuleTyFun{concl}}[\sctx]
  = \\ \begin{setbuild}{\pair{\sfun}{\sfunalt}}[l]
    \forall \pair{\sval}{\svalalt} \in \sembot{\wfty{\ctx}{\ty}}[\sctx].\;
    \pair{\sfun \app \sval}{\sfunalt \app \svalalt} \in
    \sembot{\wfty{
      \ctx \cons \annty{\var}{\ty}
    }{\tyalt}}[\sctx]
  \end{setbuild} \\
  \in
  \sembot{\wfkd{\ctx}{\kdtype}}[\sctx] = \pper{\dinf}
\end{multline*}
Since CUPERs are closed under intersection we just show that fixed
$\pair{\sval}{\svalalt} \in \sembot{\wfty{\ctx}{\ty}}[\sctx]$ the set
is a CUPER.
\begin{enumerate}
  \item \textbf{Symmetry}:
    Fix $\pair{\sfun}{\sfunalt}$ in the set,
    we have
    $\pair{\sfunalt \app \sval}{\sfun \app \svalalt} \in
    \sembot{\wfty{\ctx \cons \annty{\var}{\ty}}{\tyalt}}[\sctx]$
    by \ih{} and symmetry of CUPERs thus $\pair{\sfunalt}{\sfun}$ is in
    the set.
  \item \textbf{Transitivity}:
    Fix $\pair{\sfun}{\sfunalt}$ and $\pair{\sfunalt}{\sfunaltalt}$ in
    the set by \ih{} and transitivity of CUPERs we have
    $\pair{\sfun \app \sval}{\sfunaltalt \app \svalalt} \in
    \sembot{\wfty{\ctx \cons \annty{\var}{\ty}}{\tyalt}
    }[\sctx]$ thus $\pair{\sfun}{\sfunaltalt}$ is in the
    set.
  \item \textbf{Pointedness}:
    By \ihs{} $\pair{\bot}{\bot} \in \sembot{
      \wfty{\ctx}{\ty}
    }[\sctx]$, and
    $\pair{\bot}{\bot} \in \sembot{
      \wfty{\ctx \cons \annty{\var}{\ty}}{\tyalt}
    }[\sctx]$. Since $\bot \app \sval = \bot$ we have
    $\pair{\bot}{\bot}$ is in the set.
  \item \textbf{Completeness}:
    Let ${\{ \pair{\sfun_i}{\sfunalt_i} \}_{i \in I}}$ be a directed
    subset of the set. Fix $\pair{\sval}{\svalalt} \in
    \sembot*{\wfty{\ctx}{\ty}}[\sctx]$.  By the definition of the set, for
    every $i \in I$ by \ih{} $\pair*{
      \sfun_i \app \sval
    }{
      \sfunalt_i \app \svalalt
    } \in \sembot*{
      \wfty{\ctx \cons \annty{\var}{\ty}}{\tyalt}
    }[\sctx]$, since application is continuous, $\pair{
      (\Lub_{i \in I} \sfun_i)\app \sval
    }{
      (\Lub_{i \in I} \sfunalt_i)\app \svalalt
    } = \Lub_{i \in I} \pair{
      \sfun_i \app \sval
    }{
      \sfunalt_i \app \svalalt
    }$. By \ih{} $\sembot{
      \wfty{\ctx \cons \annty{\var}{\ty}}{\tyalt}
    }[\sctx]$ is complete, hence the lub pair is in the
    relation, so $\pair{\Lub_i \sfun_i}{\Lub_i \sfunalt_i}$ is in the
    set.
  \item \textbf{Uniformity}:
    Let $\pair{\sfun}{\sfunalt}$ be in the set, and fix an arbitrary
    $\pair{\sval}{\svalalt} \in \sembot{\wfty{\ctx}{\ty}}[\sctx]$. By
    \ih{} the domain PER is uniform, so
    $\pair{\pi_n(\sval)}{\pi_n(\svalalt)} \in
    \sembot{\wfty{\ctx}{\ty}}[\sctx]$, hence $\pair*{
      \sfun \app \pi_n(\sval)
    }{
      \sfunalt \app \pi_n(\svalalt)
    } \in \sembot{
      \wfty{\ctx \cons \annty{\var}{\ty}}{\tyalt}
    }[\sctx]$. By \ih{} the codomain PER is uniform as well, so
    $\pair{
      \pi_n(\sfun \app \pi_n(\sval))
    }{
      \pi_n(\sfunalt \app \pi_n(\svalalt))
    }$ is in it. The projections of the bilimit satisfy
    $\pi_{n + 1}(\sfun) \app \sval = \pi_n(\sfun \app \pi_n(\sval))$,
    hence $\pair{
      \pi_{n + 1}(\sfun) \app \sval
    }{
      \pi_{n + 1}(\sfunalt) \app \svalalt
    } \in \sembot{
      \wfty{\ctx \cons \annty{\var}{\ty}}{\tyalt}
    }[\sctx]$. Since $\pair{\sval}{\svalalt}$ was arbitrary,
    $\pair{\pi_{n + 1}(\sfun)}{\pi_{n + 1}(\sfunalt)}$ is in the set;
    for $n = 0$ we have $\pi_0(\sfun) = \bot = \pi_0(\sfunalt)$ and
    the claim follows from pointedness.
\end{enumerate}

\item \RuleTyAll{label}
\begin{multline*}
  \sembot{\RuleTyAll{concl}}[\sctx]
  \\ = \begin{setbuild}{\pair{\sval}{\svalalt}}[l]
    \forall \skd \in \sembot{\wfkd{\ctx}{\kd}}[\sctx].\;
    \pair{\sval}{\svalalt} \in
    \sembot{\wfty{
      \ctx \cons \annkd{\var}{\kd}
    }{\ty}}[\pair{\sctx}{\skd}]
  \end{setbuild} \\
  \in \sembot{\wfkd{\ctx}{\kdtype}}[\sctx] = \pper{\dinf}
\end{multline*}
Since CUPERs are closed under intersection we just show that fixed
$\skd \in \sembot{\wfkd{\ctx}{\kd}}[\sctx]$ the set is a CUPER,
by \ih{} each $\sembot{\wfty{
  \ctx \cons \annkd{\var}{\kd}
}{\ty}}[\pair{\sctx}{\skd}]$ is a CUPER.

\item \RuleTyVar{label}
  \[
    \sembot{\RuleTyVar{concl}}[\sctx] =
    \sctx\left(\var\right)
    \in \sembot{\wfkd{\ctx}{\kd}}[\sctx]
  \] by assumption.

\item \RuleTyRef{label}
\[
  \sembot{\RuleTyRef{concl}}[\sctx]
  = \sembot{\wfty{\ctx}{\ty}}[\sctx]
  \in \sembot{\wfkd{\ctx}{\kdtype}}[\sctx]
  = \pper{\dinf}
\]
By \ih{} $\sembot{\wfty{\ctx}{\ty}}[\sctx]$ is a CUPER.

\item \RuleTyApp{label}
\[
  \sembot{\RuleTyApp{concl}}[\sctx]
  = \sembot{\wftm{\ctx}{\tyalt}{\tyfun*[\var]{\ty}{\kd}}}[\sctx]
  \in \sembot{\wfkd{\ctx}{\kd \subst{\tm}{\var}}}[\sctx]
\]
By \ih{} $\sembot{\wftm{\ctx}{\tyalt}{\tyfun*[\var]{\ty}{\kd}}}[\sctx]
\in \sembot{\wfkd{\ctx}{\tyfun*[\var]{\ty}{\kd}}}[\sctx]
= \sembot{\wfkd{\ctx \cons \annty{\var}{\ty}}{\kd}}[\sctx]
= \sembot*{\wfkd{\ctx}{\kd \subst{\tm}{\var}}}[\sctx]
= \pper{\dinf}$.

\item \RuleKdIndPos{label} and \RuleKdIndStrat{label}
\begin{multline*}
  \sembot{\wftm{\ctx}{\decl}{\ioverline*{k}{\ty} \to \kdtype}}[\sctx] \\
  = \mathsf{fix}_X \scalebox{0.8}{$\displaystyle
  \underbrace{\bigcup_i
  \begin{setbuild}{\pair{
    \pair{\ctor_i}{\ioverline*{j}{\sval}}
  }{
    \pair{\ctor_i}{\ioverline*{j}{\svalalt}}
  }}
    \pair{\sval_j}{\svalalt_j} \in \sembot{\wfty{
      \ctx \cons
      \annkd{\tyctor}{\kdfun{\ioverline*{k}{\ty}}{\kdtype}} \cons
      \ioverline*{j - 1}{\annty{\var}{\tyalt}}
    }{\tyalt_j}}[\pair{\sctx}{X}]
  \end{setbuild} \cup \left\{ \pair{\bot}{\bot} \right\}}_\alpha$} \\
  \in \sembot{\wfkd{\ctx}{\ioverline*{k}{\ty} \to \kdtype}}[\sctx]
  = \pper{\dinf}
\end{multline*}
First, we show that if $X \in \pper{\dinf}$ then $\alpha \in
\pper{\dinf}$, since $\alpha$ is the union of disjoint $C_i$ we just
need to show that each $C_i$ is a CUPER.
\begin{enumerate}
\item \textbf{Symmetry}:
Fix $\pair{\pair{\ctor_i}{\ioverline*{j}{\sval}}
  }{
  \pair{\ctor_i}{\ioverline*{j}{\svalalt}}
}$ in the set, then $\pair{\sval_j}{\svalalt_j} \in
\sembot*{\wfty{\ctx \cons
    \annkd{\tyctor}{\kdfun{\ioverline*{k}{\ty}}{\kdtype}} \cons \\ \ioverline*{j - 1}{\annty{\var}{\tyalt}}}{\tyalt_j}}[\pair{\sctx}{X}]$ by definition.
By \ih{}, symmetry of this PER gives $\pair{\svalalt_j}{\sval_j} \in
\sembot*{\wfty{\ctx \cons \annkd{\tyctor}{\kdfun{\ioverline*{k}{\ty}}{\kdtype}} \cons \ioverline*{j - 1}{\annty{\var}{\tyalt}}}{\tyalt_j}}[\pair{\sctx}{X}]$, hence
$\pair{\pair{\ctor_i}{\ioverline*{j}{\svalalt}}}{\pair{\ctor_i}{\ioverline*{j}{\sval}}}$
in the set.

\item \textbf{Transitivity}:
Fix
$\pair{\pair{\ctor_i}{\ioverline*{j}{\sval}}}{\pair{\ctor_i}{\ioverline*{j}{\svalalt}}}
$ in the set and
$\pair{\pair{\ctor_i}{\ioverline*{j}{\svalalt}}}{\pair{\ctor_i}{\ioverline*{j}{\svalaltalt}}}
$ in the set,
then $\pair{\sval_j}{\svalalt_j}$ and
$\pair{\svalalt_j}{\svalaltalt_j}$ are in
$\sembot{\wfty{\ctx \cons \annkd{\tyctor}{\kdfun{\ioverline*{k}{\ty}}{\kdtype}} \cons \ioverline*{j - 1}{\annty{\var}{\tyalt}}}{\tyalt_j}}[\pair{\sctx}{X}]$.
By \ih{}, transitivity gives $\pair{\sval_j}{\svalaltalt_j} \in
\sembot{\wfty{\ctx \cons \annkd{\tyctor}{\kdfun{\ioverline*{k}{\ty}}{\kdtype}} \cons \ioverline*{j - 1}{\annty{\var}{\tyalt}}}{\tyalt_j}}[\pair{\sctx}{X}]$, hence
$\pair{\pair{\ctor_i}{\ioverline*{j}{\sval}}}{\pair{\ctor_i}{\ioverline*{j}{\svalaltalt}}}$
is in the set.

\item \textbf{Pointedness}:
$\pair{\bot}{\bot}$ in the set by definition.

\item \textbf{Completeness}:
Let ${\{\pair{a_k}{b_k}\}_{k \in I}}$ be directed,
then for each $k$, $a_k,b_k$ are built from $\sval^k_j,\svalalt^k_j$
with $\pair{\sval^k_j}{\svalalt^k_j} \in
\sembot{\wfty{\ctx \cons \annkd{\tyctor}{\kdfun{\ioverline*{k}{\ty}}{\kdtype}} \cons \ioverline*{j - 1}{\annty{\var}{\tyalt}}}{\tyalt_j}}[\pair{\sctx}{X}]$ by \ih{}.  Since this PER is
complete and pairing is continuous, we get closure under lubs, hence
$\pair{\Lub_k a_k}{\Lub_k b_k}$ is in the set.

\item \textbf{Uniformity}:
Let $\pair{a}{b}$ in the set,
then $\pair{\sval_j}{\svalalt_j} \in \sembot*{\wfty{\ctx \cons \annkd{\tyctor}{\kdfun{\ioverline*{k}{\ty}}{\kdtype}} \cons \ioverline*{j - 1}{\annty{\var}{\tyalt}}}{\tyalt_j}}[\pair{\sctx}{X}]$
by definition.  By \ih{} and uniformity of this PER,
$\pair{\pi_n(\sval_j)}{\pi_n(\svalalt_j)} \in
\sembot*{\wfty{\ctx \cons \annkd{\tyctor}{\kdfun{\ioverline*{k}{\ty}}{\kdtype}} \cons \ioverline*{j - 1}{\annty{\var}{\tyalt}}}{\tyalt_j}}[\pair{\sctx}{X}]$.
By compatibility of constructors with projection, this gives
$\pair{\pi_n(a)}{\pi_n(b)}$ in the set.
\end{enumerate}

By \Cref{thm:unrefined-semantics-contractive}, each field
interpretation is non-expansive in $X$,
by \Cref{lem:tagged-tuple-projection}, agreement at level $n$
of the fields yields agreement at level $n+1$ of the corresponding
constructor summand.
Hence $\alpha$ is contractive and by \Cref{lem:guarded-fixpoint} the
fixed point exists and is unique.
\end{itemize}
\end{proof}

\begin{proof}[Proof of \Crefsub{thm:per-respecting-types}{refined}]
By induction on the structure of $\wftm{\ctx}{\ty}{\kd}$.
\begin{itemize}
\item \RuleTyNat{label} and \RuleTyBool{label}
  \[
    \sem{\wftm{\ctx}{\tybase}{\kd}}[\sctx] =
    \begin{setbuild}{\pair{\demb{\tybase}{\sval}}{\demb{\tybase}{\sval}}}
      \sval \in \tybase
    \end{setbuild} =
    \sem{\wftm{\ctx}{\tybase}{\kd}}[\sctxalt]
  \]

\item \RuleTyFun{label}
\begin{align*}
  & \sem{\RuleTyFun{concl}}[\sctx] \\
= &
    \begin{setbuild}{\pair{\sfun}{\sfunalt}}[lc@{\;}llc@{\;}lll]
        \forall&\pair{\sval}{\svalalt}
        &\in{}& \sem{\wfty{\ctx}{\ty}}[\sctx].\;
        &\pair{\sfun \app \sval}{\sfunalt \app \svalalt} &\in{}&
        \sem{\wfty{
          \ctx \cons \annty{\var}{\ty}
        }{\tyalt}}[\pair{\sctx}{\sval}]
      \\
        \forall&\pair{\sval\bots}{\svalalt\bots}
        &\in{}& \sembot{\wfty{\ctx}{\ty}}[\projctx{\ctx}{\sctx}].\;
        &\pair{\sfun \app \sval\bots}{\sfunalt \app \svalalt\bots} &\in{}&
        \sembot{\wfty{
          \ctx \cons \annty{\var}{\ty}
        }{\tyalt}}[\projctx{\ctx}{\sctx}]
    \end{setbuild} \\
  & \text{By \ih{} and \Crefsub{thm:per-respecting-types}{unrefined}} \\
= &
    \begin{setbuild}{\pair{\sfun}{\sfunalt}}[lc@{\;}llc@{\;}lll]
        \forall&\pair{\sval}{\svalalt}
        &\in{}& \sem{\wfty{\ctx}{\ty}}[\sctxalt].\;
        &\pair{\sfun \app \sval}{\sfunalt \app \svalalt} &\in{}&
        \sem{\wfty{
          \ctx \cons \annty{\var}{\ty}
        }{\tyalt}}[\pair{\sctxalt}{\sval}]
      \\
        \forall&\pair{\sval\bots}{\svalalt\bots}
        &\in{}& \sembot{\wfty{\ctx}{\ty}}[\projctx{\ctx}{\sctxalt}].\;
        &\pair{\sfun \app \sval\bots}{\sfunalt \app \svalalt\bots} &\in{}&
        \sembot{\wfty{
          \ctx \cons \annty{\var}{\ty}
        }{\tyalt}}[\projctx{\ctx}{\sctxalt}]
    \end{setbuild} \\
= & \sem{\RuleTyFun{concl}}[\sctxalt]
\end{align*}

\item \RuleTyAll{label}
\begin{align*}
  & \sem{\RuleTyAll{concl}}[\sctx] \\
= & \begin{setbuild}{\pair{\sval}{\svalalt}}[c]
      \forall \skd \in \sem{\wfkd{\ctx}{\kd}}[\sctx],\;
      \forall \skd\bots \in
      \sembot{\wfkd{\ctx}{\kd}}[\projctx{\ctx}{\sctx}] \\
      \projty{\ctx}{\sctx}{\kd}{\skd} \subseteq \skd\bots
      \Rightarrow
      \pair{\sval}{\svalalt} \in
      \sem{\wfty{
        \ctx \cons \annkd{\var}{\kd}
      }{\ty}}[\pair{\sctx}{\pair{\skd}{\skd\bots}}]
    \end{setbuild} \\
  & \text{By \ih{}, \Crefsub{thm:semantics-per-respecting-kinds}{refined} and
    \Crefsub{thm:semantics-per-respecting-kinds}{unrefined}} \\
= & \begin{setbuild}{\pair{\sval}{\svalalt}}[c]
      \forall \skd \in \sem{\wfkd{\ctx}{\kd}}[\sctxalt],\;
      \forall \skd\bots \in
      \sembot{\wfkd{\ctx}{\kd}}[\projctx{\ctx}{\sctxalt}] \\
      \projty{\ctx}{\sctxalt}{\kd}{\skd} \subseteq \skd\bots
      \Rightarrow
      \pair{\sval}{\svalalt} \in
      \sem{\wfty{
        \ctx \cons \annkd{\var}{\kd}
      }{\ty}}[\pair{\sctxalt}{\pair{\skd}{\skd\bots}}]
    \end{setbuild} \\
= & \sem{\RuleTyAll{concl}}[\sctxalt]
\end{align*}

\item \RuleTyVar{label}
\[
    \sem{\RuleTyVar{concl}}[\sctx]
  = \sctx\left(\var\right)_1
  = \sctxalt\left(\var\right)_1
  = \sem{\RuleTyVar{concl}}[\sctxalt]
\]

\item \RuleTyRef{label}
\begin{align*}
  & \sem{\RuleTyRef{concl}}[\sctx] \\
= & \begin{setbuild}{
      \pair{\sval}{\svalalt} \in \sem{\wfty{\ctx}{\ty}}[\sctx]
    }
      \semprp{\wfprop{
          \ctx \cons \annty{\var}{\ty}
      }{\prop}}[\pair{\sctx}{\sval}] = \strue
    \end{setbuild} \\
  & \text{By \ih{} and \Cref{thm:refined-per-respecting-propositions}} \\
= & \begin{setbuild}{
      \pair{\sval}{\svalalt} \in \sem{\wfty{\ctx}{\ty}}[\sctxalt]
    }
      \semprp{\wfprop{
          \ctx \cons \annty{\var}{\ty}
      }{\prop}}[\pair{\sctxalt}{\sval}] = \strue
    \end{setbuild} \\
= & \sem{\RuleTyRef{concl}}[\sctxalt]
\end{align*}

\item \RuleTyApp{label}
\begin{align*}
  & \sem{\RuleTyApp{concl}}[\sctx] \\
= & \sem{\wftm{\ctx}{\tyalt}{\kdfun*[{\var}]{\ty}{\kd}}}[\sctx]\left(
    \semtm{\tm}[\projrun{\ctx}{\sctx}]
  \right) \\
  & \text{By \ih{}} \\
= & \sem{\wftm{\ctx}{\tyalt}{\kdfun*[{\var}]{\ty}{\kd}}}[\sctxalt]\left(
    \semtm{\tm}[\projrun{\ctx}{\sctx}]
  \right) \\
  & \text{By \Crefsub{thm:per-respecting-terms}{refined}}
    \pair{
      \semtm{\tm}[\projrun{\ctx}{\sctx}]
    }{
      \semtm{\tm}[\projrun{\ctx}{\sctxalt}]
    } \in \sem{\wfty{\ctx}{\ty}}[\sctx] \\
  & \text{By \Cref{thm:refined-sub-unrefined}}
    \pair{
      \semtm{\tm}[\projrun{\ctx}{\sctx}]
    }{
      \semtm{\tm}[\projrun{\ctx}{\sctxalt}]
    } \in \sembot{\wfty{\ctx}{\ty}}[\projctx{\ctx}{\sctx}] \\
  & \text{By } \sem{\wfkd{\ctx}{\kdfun*[{\var}]{\ty}{\kd}}}[\sctx]
    \text{ respecting} \\
  & \sembot{\wfty{\ctx}{\ty}}[\projctx{\ctx}{\sctx}] \cap \dom{\sem{\wfty{\ctx}{\ty}}[\sctx]}^2 \\
= & \sem{\wftm{\ctx}{\tyalt}{\kdfun*[{\var}]{\ty}{\kd}}}[\sctxalt]\left(
      \semtm{\tm}[\projrun{\ctx}{\sctxalt}]
    \right) \\
= & \sem{\RuleTyApp{concl}}[\sctxalt]
\end{align*}

\item \RuleKdIndPos{label} and \RuleKdIndStrat{label}
\begin{align*}
  & \sem{\wftm{\ctx}{\decl}{\ioverline{k}{\ty} \to \kdtype}}[\sctx] \\
= & \;\mathsf{lfp}_X \left(\scalebox{0.5}{$\displaystyle
      \ioverline{k}{\svalaltalt} \mapsto
      \Bigcup_i \begin{setbuild}{\pair{
        \pair{\ctor_i}{\ioverline{j}{\sval}}
      }{
        \pair{\ctor_i}{\ioverline{j}{\svalalt}}
      }}
        \pair{\sval_j}{\svalalt_j} \in \sem{\wfty{
          \ctx \cons
          \annkd{\tyctor}{\kdfun{\ioverline{k}{\ty}}{\kdtype}} \cons
          \ioverline{j - 1}{\annty{\var}{\tyalt}}
        }{\tyalt_j}}[\pair{\pair{\sctx}{\pair{X}{
            \sembot{\wftm{\ctx}{\decl}{\kdfun{\ioverline{k}{\ty}}{\kdtype}}}[\projctx{\ctx}{\sctx}]
          }}}{\ioverline{j - 1}{\sval}}] \\
      \pair{\svalaltalt_k}{\semtm{\tm_k}[\pair{\projrun{\ctx}{\sctx}}{\ioverline*{j}{\sval}}]}
      \in \sembot{\wfty{
        \ctx \cons \ioverline*{k - 1}{\annty{\var}{\ty}}
      }{\ty_k}}[\projctx{\ctx}{\sctx}]
      \end{setbuild}
    $}\right) \\
  & \text{By \ih{}, \Crefsub{thm:semantics-per-respecting-kinds}{unrefined},
      and \Crefsub{thm:per-respecting-terms}{refined}} \\
= & \;\mathsf{lfp}_X \left(\scalebox{0.5}{$\displaystyle
      \ioverline{k}{\svalaltalt} \mapsto
      \Bigcup_i \begin{setbuild}{\pair{
        \pair{\ctor_i}{\ioverline{j}{\sval}}
      }{
        \pair{\ctor_i}{\ioverline{j}{\svalalt}}
      }}
        \pair{\sval_j}{\svalalt_j} \in \sem{\wfty{
          \ctx \cons
          \annkd{\tyctor}{\kdfun{\ioverline{k}{\ty}}{\kdtype}} \cons
          \ioverline{j - 1}{\annty{\var}{\tyalt}}
        }{\tyalt_j}}[\pair{\pair{\sctxalt}{\pair{X}{
            \sembot{\wftm{\ctx}{\decl}{\kdfun{\ioverline{k}{\ty}}{\kdtype}}}[\projctx{\ctx}{\sctxalt}]
          }}}{\ioverline{j - 1}{\sval}}] \\
      \pair{\svalaltalt_k}{\semtm{\tm_k}[\pair{\projrun{\ctx}{\sctxalt}}{\ioverline*{j}{\sval}}]}
      \in \sembot{\wfty{
        \ctx \cons \ioverline*{k - 1}{\annty{\var}{\ty}}
      }{\ty_k}}[\projctx{\ctx}{\sctxalt}]
      \end{setbuild}
    $}\right) \\
= & \sem{\wftm{\ctx}{\decl}{\ioverline{k}{\ty} \to \kdtype}}[\sctxalt] \\
\end{align*}
\end{itemize}
\end{proof}

\begin{proof}[Proof of \Crefsub{thm:per-respecting-types}{unrefined}]
By induction on the structure of $\wftm{\ctx}{\ty}{\kd}$.
\begin{itemize}
\item \RuleTyNat{label} and \RuleTyBool{label}
\[
    \sembot{\wftm{\ctx}{\tybase}{\kd}}[\sctx]
  = \begin{setbuild}{\pair{\demb{\tybase}{\sval}}{\demb{\tybase}{\sval}}}
      \sval \in \tybase
    \end{setbuild} \cup \left\{\pair{\bot}{\bot}\right\}
  = \sembot{\wftm{\ctx}{\tybase}{\kd}}[\sctxalt]
\]

\item \RuleTyFun{label}
\begin{align*}
  & \sembot{\RuleTyFun{concl}}[\sctx] \\
= & \begin{setbuild}{\pair{\sfun}{\sfunalt}}[l]
      \forall \pair{\sval}{\svalalt} \in \sembot{\wfty{\ctx}{\ty}}[\sctx].\;
      \pair{\sfun \app \sval}{\sfunalt \app \svalalt} \in
      \sembot{\wfty{\ctx \cons \annty{\var}{\ty}}{\tyalt}}[\sctx]
    \end{setbuild} \\
  & \text{By \ih{}} \\
= & \begin{setbuild}{\pair{\sfun}{\sfunalt}}[l]
      \forall \pair{\sval}{\svalalt} \in \sembot{\wfty{\ctx}{\ty}}[\sctxalt].\;
      \pair{\sfun \app \sval}{\sfunalt \app \svalalt} \in
      \sembot{\wfty{\ctx \cons \annty{\var}{\ty}}{\tyalt}}[\sctxalt]
    \end{setbuild} \\
= & \sembot{\RuleTyFun{concl}}[\sctxalt]
\end{align*}

\item \RuleTyAll{label}
\begin{align*}
  & \sembot{\RuleTyAll{concl}}[\sctx] \\
= & \begin{setbuild}{\pair{\sval}{\svalalt}}[l]
      \forall \skd \in \sembot{\wfkd{\ctx}{\kd}}[\sctx].\;
      \pair{\sval}{\svalalt} \in
      \sembot{\wfty{
        \ctx \cons \annkd{\var}{\kd}
      }{\ty}}[\pair{\sctx}{\skd}]
    \end{setbuild} \\
& \text{By \ih{} and \Crefsub{thm:semantics-per-respecting-kinds}{unrefined}} \\
= & \begin{setbuild}{\pair{\sval}{\svalalt}}[l]
      \forall \skd \in \sembot{\wfkd{\ctx}{\kd}}[\sctxalt].\;
      \pair{\sval}{\svalalt} \in
      \sembot{\wfty{
        \ctx \cons \annkd{\var}{\kd}
      }{\ty}}[\pair{\sctxalt}{\skd}]
    \end{setbuild} \\
= & \sembot{\RuleTyAll{concl}}[\sctxalt]
\end{align*}

\item \RuleTyVar{label}
\[
    \sembot{\RuleTyVar{concl}}[\sctx]
  = \sctx\left(\var\right)
  = \sctxalt\left(\var\right)
  = \sembot{\RuleTyVar{concl}}[\sctxalt]
\]

\item \RuleTyRef{label}
\begin{align*}
  & \sembot{\RuleTyRef{concl}}[\sctx] \\
= & \sembot{\wfty{\ctx}{\ty}}[\sctx] \\
  & \text{By \ih{}} \\
= & \sembot{\wfty{\ctx}{\ty}}[\sctxalt] \\
= & \sembot{\RuleTyRef{concl}}[\sctxalt]
\end{align*}

\item \RuleTyApp{label}
\begin{align*}
  & \sembot{\RuleTyApp{concl}}[\sctx] \\
= & \sembot{\wftm{\ctx}{\tyalt}{\kdfun*[{\var}]{\ty}{\kd}}}[\sctx] \\
  & \text{By \Crefsub{thm:semantics-per-respecting-kinds}{unrefined}} \\
= & \sembot{\wftm{\ctx}{\tyalt}{\kdfun*[{\var}]{\ty}{\kd}}}[\sctxalt] \\
= & \sembot{\RuleTyApp{concl}}[\sctxalt]
\end{align*}

\item \RuleKdIndPos{label} and \RuleKdIndStrat{label}
\begin{align*}
  & \sembot{\wftm{\ctx}{\decl}{\ioverline{k}{\ty} \to \kdtype}}[\sctx] \\
= & \;\mathsf{fix}_X \left(\scalebox{0.7}{$\displaystyle
    \bigcup_i \begin{setbuild}{\pair{
      \pair{\ctor_i}{\ioverline{j}{\sval}}
    }{
      \pair{\ctor_i}{\ioverline{j}{\svalalt}}
    }}
      \pair{\sval_j}{\svalalt_j} \in \sembot{\wfty{
        \ctx \cons
        \annkd{\tyctor}{\kdfun{\ioverline{k}{\ty}}{\kdtype}} \cons
        \ioverline{j - 1}{\annty{\var}{\tyalt}}
      }{\tyalt_j}}[\pair{\sctx}{X}]
    \end{setbuild} \cup \left\{ \pair{\bot}{\bot} \right\}$}\right) \\
  & \text{By \ih{}} \\
= & \;\mathsf{fix}_X \left(\scalebox{0.7}{$\displaystyle
    \bigcup_i \begin{setbuild}{\pair{
      \pair{\ctor_i}{\ioverline{j}{\sval}}
    }{
      \pair{\ctor_i}{\ioverline{j}{\svalalt}}
    }}
      \pair{\sval_j}{\svalalt_j} \in \sembot{\wfty{
        \ctx \cons
        \annkd{\tyctor}{\kdfun{\ioverline{k}{\ty}}{\kdtype}} \cons
        \ioverline{j - 1}{\annty{\var}{\tyalt}}
      }{\tyalt_j}}[\pair{\sctxalt}{X}]
    \end{setbuild} \cup \left\{ \pair{\bot}{\bot} \right\}$}\right) \\
= & \sembot{\wftm{\ctx}{\decl}{\ioverline{k}{\ty} \to \kdtype}}[\sctxalt]
\end{align*}

\end{itemize}
\end{proof}

\begin{proof}[Proof of \Cref{thm:refined-sub-unrefined}]
By induction on the structure of $\wftm{\ctx}{\ty}{\kd}$.

\begin{itemize}
\item \RuleTyNat{label} and \RuleTyBool{label}
\begin{align*}
  & \sem{\wftm{\ctx}{\tybase}{\kd}}[\sctx] \\
= & \begin{setbuild}{\pair{\demb{\tybase}{\sval}}{\demb{\tybase}{\sval}}}
  \sval \in \tybase
\end{setbuild} \\
\subseteq & \begin{setbuild}{\pair{\demb{\tybase}{\sval}}{\demb{\tybase}{\sval}}}
  \sval \in \tybase
\end{setbuild} \cup \left\{\pair{\bot}{\bot}\right\} \\
= & \sembot{\wftm{\ctx}{\tybase}{\kd}}[\projctx{\ctx}{\sctx}]
\end{align*}

\item \RuleTyFun{label}
\begin{align*}
  & \sem{\RuleTyFun{concl}}[\sctx] \\
= &
    \begin{setbuild}{\pair{\sfun}{\sfunalt}}[lc@{\;}llc@{\;}lll]
        \forall&\pair{\sval}{\svalalt}
        &\in{}& \sem{\wfty{\ctx}{\ty}}[\sctx].\;
        &\pair{\sfun \app \sval}{\sfunalt \app \svalalt} &\in{}&
        \sem{\wfty{
          \ctx \cons \annty{\var}{\ty}
        }{\tyalt}}[\pair{\sctx}{\sval}]
      \\
        \forall&\pair{\sval\bots}{\svalalt\bots}
        &\in{}& \sembot{\wfty{\ctx}{\ty}}[\projctx{\ctx}{\sctx}].\;
        &\pair{\sfun \app \sval\bots}{\sfunalt \app \svalalt\bots} &\in{}&
        \sembot{\wfty{
          \ctx \cons \annty{\var}{\ty}
        }{\tyalt}}[\projctx{\ctx}{\sctx}]
    \end{setbuild} \\
\subseteq & \begin{setbuild}{\pair{\sfun}{\sfunalt}}[l]
      \forall \pair{\sval\bots}{\svalalt\bots}
      \in \sembot{\wfty{\ctx}{\ty}}[\projctx{\ctx}{\sctx}].\;
      \pair{\sfun \app \sval\bots}{\sfunalt \app \svalalt\bots} \in
      \sembot{\wfty{
        \ctx \cons \annty{\var}{\ty}
      }{\tyalt}}[\projctx{\ctx}{\sctx}]
    \end{setbuild} \\
& \text{Dropping the first conjunct} \\
= & \sembot{\RuleTyFun{concl}}[\projctx{\ctx}{\sctx}]
\end{align*}

\item \RuleTyAll{label}
\begin{align*}
  & \sem{\RuleTyAll{concl}}[\sctx] \\
= & \begin{setbuild}{\pair{\sval}{\svalalt}}[c]
      \forall \skd \in \sem{\wfkd{\ctx}{\kd}}[\sctx],\;
      \forall \skd\bots \in
      \sembot{\wfkd{\ctx}{\kd}}[\projctx{\ctx}{\sctx}] \\
      \projty{\ctx}{\sctx}{\kd}{\skd} \subseteq \skd\bots
      \Rightarrow
      \pair{\sval}{\svalalt} \in
      \sem{\wfty{
        \ctx \cons \annkd{\var}{\kd}
      }{\ty}}[\pair{\sctx}{\pair{\skd}{\skd\bots}}]
    \end{setbuild} \\
& \text{Instantiating at the constant empty-relation element $\skd_\emptyset \in
  \sem{\wfkd{\ctx}{\kd}}[\sctx]$, for which} \\
& \text{$\projty{\ctx}{\sctx}{\kd}{\skd_\emptyset} = \emptyset \subseteq \skd\bots$
  holds for every $\skd\bots$, and applying \ih{}} \\
\subseteq & \begin{setbuild}{\pair{\sval}{\svalalt}}[l]
      \forall \skd\bots \in
      \sembot{\wfkd{\ctx}{\kd}}[\projctx{\ctx}{\sctx}].\;
      \pair{\sval}{\svalalt} \in
      \sembot{\wfty{
        \ctx \cons \annkd{\var}{\kd}
      }{\ty}}[\pair{\projctx{\ctx}{\sctx}}{\skd\bots}]
    \end{setbuild} \\
= & \sembot{\RuleTyAll{concl}}[\projctx{\ctx}{\sctx}]
\end{align*}

\item \RuleTyVar{label}
\begin{align*}
            &\; \projty{\ctx}{\sctx}{\kd}{\sem{\RuleTyVar{concl}}[\sctx]} \\
  =         &\; \projty{\ctx}{\sctx}{\kd}{\sctx\left(\var\right)_1}       \\
            &\; \text{By assumption}                                      \\
  \subseteq &\; \sctx\left(\var\right)_2                                  \\
  =         &\; \projctx{\ctx}{\sctx}\left(\var\right)                    \\
  =         & \sembot{\RuleTyVar{concl}}[\projctx{\ctx}{\sctx}]
\end{align*}

\item \RuleTyRef{label}
\begin{align*}
  & \sem{\RuleTyRef{concl}}[\sctx] \\
= & \begin{setbuild}{
      \pair{\sval}{\svalalt} \in \sem{\wfty{\ctx}{\ty}}[\sctx]
    }
      \semprp{\wfprop{
          \ctx \cons \annty{\var}{\ty}
      }{\prop}}[\pair{\sctx}{\sval}] = \strue
    \end{setbuild} \\
\subseteq & \sem{\wfty{\ctx}{\ty}}[\sctx] \\
& \text{By \ih{}} \\
\subseteq & \sembot{\wfty{\ctx}{\ty}}[\projctx{\ctx}{\sctx}] \\
= & \sembot{\RuleTyRef{concl}}[\projctx{\ctx}{\sctx}]
\end{align*}

\item \RuleTyApp{label}
\begin{align*}
  & \;\projty{\ctx}{\sctx}{\kd \subst{\tm}{\var}}{
      \sem{\RuleTyApp{concl}}[\sctx]
  } \\
= & \;\projty{\ctx}{\sctx}{\kd \subst{\tm}{\var}}{
     \sem{\wftm{\ctx}{\tyalt}{\kdfun*[{\var}]{\ty}{\kd}}}[\sctx]\left(\semtm{\tm}[\projrun{\ctx}{\sctx}] \right)
  } \\
\subseteq & \;\projty{\ctx}{\sctx}{\kdfun*[{\var}]{\ty}{\kd}}{
     \sem{\wftm{\ctx}{\tyalt}{\kdfun*[{\var}]{\ty}{\kd}}}[\sctx]
  } \\
& \text{By \ih{}} \\
\subseteq & \sembot{\wftm{\ctx}{\tyalt}{\kdfun*[{\var}]{\ty}{\kd}}}[\projctx{\ctx}{\sctx}] \\
= & \sembot{\RuleTyApp{concl}}[\projctx{\ctx}{\sctx}]
\end{align*}

\item \RuleKdIndPos{label} and \RuleKdIndStrat{label}
  We note the definitions
  \begin{align*}
    & \sem{\wftm{\ctx}{\decl}{\ioverline{k}{\ty} \to \kdtype}}[\sctx] \\
= & \;\mathsf{lfp}_X \left(\scalebox{0.5}{$\displaystyle
    \underbrace{\ioverline*{k}{\svalaltalt} \mapsto
      % \underbrace{
        \Bigcup_i \begin{setbuild}{\pair{
      \pair{\ctor_i}{\ioverline*{j}{\sval}}
    }{
      \pair{\ctor_i}{\ioverline*{j}{\svalalt}}
    }}
      \pair{\sval_j}{\svalalt_j} \in \sem{\wfty{
        \ctx \cons
        \annkd{\tyctor}{\kdfun{\ioverline*{k}{\ty}}{\kdtype}} \cons
        \ioverline*{j - 1}{\annty{\var}{\tyalt}}
      }{\tyalt_j}}[\pair{\pair{\sctx}{\pair{X}{
          \sembot{\wftm{\ctx}{\decl}{\kdfun{\ioverline*{k}{\ty}}{\kdtype}}}[\projctx{\ctx}{\sctx}]
        }}}{\ioverline*{j - 1}{\sval}}] \\
      \pair{\svalaltalt_k}{\semtm{\tm_k}[\pair{\projrun{\ctx}{\sctx}}{\ioverline*{j}{\sval}}]}
      \in \sembot{\wfty{
        \ctx \cons \ioverline*{k - 1}{\annty{\var}{\ty}}
      }{\ty_k}}[\projctx{\ctx}{\sctx}]
  \end{setbuild}
% }_{\beta}
  }_{\alpha}$}\right) \\
  & \sembot{\wftm{\ctx}{\decl}{\ioverline{k}{\ty} \to
    \kdtype}}[\projctx{\ctx}{\sctx}] \\
  = & \;\mathsf{fix}_X \left(\scalebox{0.7}{$\displaystyle
    \bigcup_i \begin{setbuild}{\pair{
      \pair{\ctor_i}{\ioverline{j}{\sval}}
    }{
      \pair{\ctor_i}{\ioverline{j}{\svalalt}}
    }}
      \pair{\sval_j}{\svalalt_j} \in \sembot{\wfty{
        \ctx \cons
        \annkd{\tyctor}{\kdfun{\ioverline{k}{\ty}}{\kdtype}} \cons
        \ioverline{j - 1}{\annty{\var}{\tyalt}}
      }{\tyalt_j}}[\pair{\sctx}{X}]
    \end{setbuild} \cup \left\{ \pair{\bot}{\bot} \right\}$}\right)
  \end{align*}
  First we observe that if \[ \projty{\ctx}{\sctx}{\ioverline*{k}{\ty} \to
    \kdtype}{X} \subseteq \sembot{\wftm{\ctx}{\decl}{\ioverline*{k}{\ty}
      \to \kdtype}}[\projctx{\ctx}{\sctx}] \] then by \ihs{}.
  \[ \projty{\ctx}{\sctx}{\ioverline*{k}{\ty} \to \kdtype}{\alpha(X)}
  \subseteq \sembot*{\wftm{\ctx}{\decl}{\ioverline*{k}{\ty} \to
      \kdtype}}[\projctx{\ctx}{\sctx}] \]

  Now we proceed by cases on the formation rule:
  \begin{itemize}
  \item \RuleKdIndPos{label}:
  Since $\alpha$ is a monotone operator we can apply the pre-fixed
  point induction principle, by definition:
  \begin{multline*}
    \projty{\ctx}{\sctx}{\ioverline*{k}{\ty} \to \kdtype}{
      \ioverline{k}{\svalaltalt} \mapsto
      \sembot{\wftm{\ctx}{\decl}{\ioverline*{k}{\ty} \to
          \kdtype}}[\projctx{\ctx}{\sctx}]
    } \\
    \subseteq \sembot{\wftm{\ctx}{\decl}{\ioverline*{k}{\ty} \to
        \kdtype}}[\projctx{\ctx}{\sctx}]
  \end{multline*}
  From the previous observation,
  \begin{multline*}
  \alpha(\ioverline{k}{\svalaltalt}
  \mapsto \sembot{\wftm{\ctx}{\decl}{\ioverline{k}{\ty} \to
      \kdtype}}[\projctx{\ctx}{\sctx}])
  \\\subseteq
  \sembot{\wftm{\ctx}{\decl}{\ioverline{k}{\ty} \to
      \kdtype}}[\projctx{\ctx}{\sctx}]
  \end{multline*}
  Since the least fixed point is bounded above by any pre-fixed point,
  it follows that
  \[ \projty{\ctx}{\sctx}{\ioverline*{k}{\ty} \to \kdtype}{
    \mathsf{lfp}_X(\alpha)
  } \subseteq \sembot{\wftm{\ctx}{\decl}{\ioverline*{k}{\ty} \to
      \kdtype}}[\projctx{\ctx}{\sctx}] \]

  \item \RuleKdIndStrat{label}:
    Instead here since we lack monotonicity, we will show it by
    subdividing the construction by the stratification index
    \begin{align*}
      P(n) \defeq \ioverline*{k}{\svalaltalt} \mapsto \begin{cases}
        \sem{
          \wftm{\ctx}{D}{\ioverline*{k}{\tyalt} \to \kdtype}
        }[\sctx] \app \ioverline*{k}{\svalaltalt} & \text{if } n >
                                                    \dret{\snat}{\semtm{m}[\projrun{\ctx}{\sctx}]
                                                    \app
                                                    \ioverline*{k}{\svalaltalt}}
        \\
        \emptyset & \text{otherwise}
      \end{cases}
    \end{align*}
    Clearly $\forall n \in \mathbb{N}. \; P(n) \leq \sem{
      \wftm{\ctx}{D}{\ioverline*{k}{\tyalt} \to \kdtype}
    }[\sctx]$.

    By induction on $n$ that $\forall n \in \mathbb{N}. \;
    \projty{\ctx}{\sctx}{\ioverline*{k}{\ty} \to \kdtype}{P(n)}
    \subseteq \sembot*{\wftm{\ctx}{D}{\ioverline*{k}{\ty} \to
        \kdtype}}[\projctx{\ctx}{\sctx}]$. Since
    each $P(n) = \alpha(P(n - 1))$ for $n > 0$, the inductive case
    follows from the previous observation. The base case is trivial since
    $P(0) = \emptyset$.
  \end{itemize}

\end{itemize}
\end{proof}

\begin{proof}[Proof of \Cref{thm:unrefined-equality-respects-refined-equality}]
From \Cref{thm:refined-sub-unrefined} $\pair{\sval}{\sval'},
\pair{\svalalt}{\svalalt'} \in \sembot{\wftm{\ctx}{\ty}{\kdtype}}[\projctx{\ctx}{\sctx}]$
\begin{enumerate}[label={(\roman*)}]
\item{$(\Rightarrow)$} By transitivity and symmetry
   $\pair{\sval'}{\svalalt'} \in \sembot{\wftm{\ctx}{\ty}{\kdtype}}[\projctx{\ctx}{\sctx}]$.
\item{$(\Leftarrow)$} By transitivity and symmetry
   $\pair{\sval}{\svalalt} \in \sembot{\wftm{\ctx}{\ty}{\kdtype}}[\projctx{\ctx}{\sctx}]$.
\end{enumerate}
\end{proof}

\begin{proof}[Proof of \Crefsub{thm:per-respecting-terms}{refined}]
By induction on the structure of $\wftm{\ctx}{\tm}{\ty}$.

\begin{itemize}
\item \RuleTmNat{label} and \RuleTmBool{label}
\begin{align*}
  & \pair{\semtm{b}[\projrun{\ctx}{\sctx}]}{\semtm{b}[\projrun{\ctx}{\sctxalt}]} \\
= & \; \pair{\demb{\tybase}{b}}{\demb{\tybase}{b}} \\
\in & \begin{setbuild}{\pair{\demb{\tybase}{b}}{\demb{\tybase}{b}}}
    b \in \tybase
  \end{setbuild} \\
= & \sem{\wfty{\ctx}{\tybase}}[\sctx]
\end{align*}

\item \RuleTmVar{label}
\begin{align*}
  & \pair{\semtm{\var}[\projrun{\ctx}{\sctx}]}{\semtm{\var}[\projrun{\ctx}{\sctxalt}]} \\
= & \; \pair{\sctx(\var)}{\sctxalt(\var)} \\
& \text{By assumption} \\
\in & \; \sem{\wfty{\ctx}{\ty}}[\sctx]
\end{align*}

\item \RuleTmLam{label}
\begin{align*}
  & \pair{\semtm{\tmlam{\var}{\tm}}[\projrun{\ctx}{\sctx}]}{\semtm{\tmlam{\var}{\tm}}[\projrun{\ctx}{\sctxalt}]} \\
= & \pair{
    \demb{\to}{\sval \mapsto \semtm{\tm}[\pair{\projrun{\ctx}{\sctx}}{\sval}]}
}{
    \demb{\to}{\sval \mapsto \semtm{\tm}[\pair{\projrun{\ctx}{\sctxalt}}{\sval}]}
} \\
& \text{We need to show}\\
\in & \sem{\wfty{\ctx}{\tyfun*[\var]{\ty}{\tyalt}}}[\sctx] \\
= &
    \begin{setbuild}{\pair{\sfun}{\sfunalt}}[lc@{\;}llc@{\;}llr]
        \forall&\pair{\sval}{\svalalt}
        &\in{}& \sem{\wfty{\ctx}{\ty}}[\sctx].\;
        &\pair{\sfun \app \sval}{\sfunalt \app \svalalt} &\in{}&
        \sem{\wfty{
          \ctx \cons \annty{\var}{\ty}
        }{\tyalt}}[\pair{\sctx}{\sval}] & \text{(i)}
      \\
        \forall&\pair{\sval\bots}{\svalalt\bots}
        &\in{}& \sembot{\wfty{\ctx}{\ty}}[\projctx{\ctx}{\sctx}].\;
        &\pair{\sfun \app \sval\bots}{\sfunalt \app \svalalt\bots} &\in{}&
        \sembot{\wfty{
          \ctx \cons \annty{\var}{\ty}
        }{\tyalt}}[\projctx{\ctx}{\sctx}] & \text{(ii)}
    \end{setbuild}
\end{align*}
It is enough to show that fixed arbitrary $\pair{\sval}{\svalalt} \in
\sem{\wfty{\ctx}{\ty}}[\sctx]$ and $\pair{\sval\bots}{\svalalt\bots}
\in \sembot*{\wfty{\ctx}{\ty}}[\projctx{\ctx}{\sctx}]$ the conditions
(i) and (ii) hold. We show them separately.
\begin{enumerate}[label={(\roman*)}]
\item \begin{align*}
  & \pair{
    \demb{\to}{\sval \mapsto
    \semtm{\tm}[\pair{\projrun{\ctx}{\sctx}}{\sval}]}
    \app \sval
}{
    \demb{\to}{\sval \mapsto
    \semtm{\tm}[\pair{\projrun{\ctx}{\sctxalt}}{\sval}]}
    \app \svalalt
} \\
= & \pair{
    \semtm{\tm}[\pair{\projrun{\ctx}{\sctx}}{\sval}]
}{
    \semtm{\tm}[\pair{\projrun{\ctx}{\sctxalt}}{\svalalt}]
} \\
& \text{By \ih{}} \\
\in & \sem{\wfty{\ctx \cons \annty{\var}{\ty}}{\tyalt}}[\pair{\sctx}{\sval}]
\end{align*}

\item
\begin{align*}
  & \pair{
    \demb{\to}{\sval \mapsto
    \semtm{\tm}[\pair{\projrun{\ctx}{\sctx}}{\sval}]}
    \app \sval\bots
}{
    \demb{\to}{\sval \mapsto
    \semtm{\tm}[\pair{\projrun{\ctx}{\sctxalt}}{\sval}]}
    \app \svalalt\bots
} \\
= & \pair{
    \semtm{\tm}[\pair{\projrun{\ctx}{\sctx}}{\sval\bots}]
}{
    \semtm{\tm}[\pair{\projrun{\ctx}{\sctxalt}}{\svalalt\bots}]
} \\
  & \text{By \Crefsub{thm:per-respecting-terms}{unrefined}
    and \Cref{thm:refined-runtime-projection-is-unrefined-runtime-interpretation}} \\
\in & \sembot{\wfty{\ctx \cons \annty{\var}{\ty}}{\tyalt}}[\projctx{\ctx}{\sctx}]
\end{align*}
\end{enumerate}

\item \RuleTmApp{label}
\begin{align*}
& \pair{\semtm{\tm \app \tmalt}[\projrun{\ctx}{\sctx}]
}{
  \semtm{\tm \app \tmalt}[\projrun{\ctx}{\sctxalt}]
} \\
= & \pair{
  \semtm{\tm}[\projrun{\ctx}{\sctx}] \app \semtm{\tmalt}[\projrun{\ctx}{\sctx}]
}{
  \semtm{\tm}[\projrun{\ctx}{\sctxalt}] \app \semtm{\tmalt}[\projrun{\ctx}{\sctxalt}]
} \\
& \text{We need to show} \\
\in & \sem{\wfty{\ctx}{\tyalt \subst{\tmalt}{\var}}}[\sctx]
\end{align*}
By \ih{} applied to \tmalt and \tm we have, respectively,
\begin{align*}
  \pair{
    \semtm{\tmalt}[\projrun{\ctx}{\sctx}]
  }{
    \semtm{\tmalt}[\projrun{\ctx}{\sctxalt}]
  } & \in \sem{\wfty{\ctx}{\ty}}[\sctx], \text{ and}
\end{align*}
\begin{align*}
&\pair{
  \semtm{\tm}[\projrun{\ctx}{\sctx}]
}{
  \semtm{\tm}[\projrun{\ctx}{\sctxalt}]
} \\
\in & \sem{\wfty{\ctx}{\tyfun*[\var]{\ty}{\tyalt}}}[\sctx] \\
= &
  \begin{setbuild}{\pair{\sfun}{\sfunalt}}[lc@{\;}llc@{\;}llr]
      \forall&\pair{\sval}{\svalalt}
      &\in{}& \sem{\wfty{\ctx}{\ty}}[\sctx].\;
      &\pair{\sfun \app \sval}{\sfunalt \app \svalalt} &\in{}&
      \sem{\wfty{
        \ctx \cons \annty{\var}{\ty}
      }{\tyalt}}[\pair{\sctx}{\sval}] & \text{(i)}
    \\
      \forall&\pair{\sval\bots}{\svalalt\bots}
      &\in{}& \sembot{\wfty{\ctx}{\ty}}[\projctx{\ctx}{\sctx}].\;
      &\pair{\sfun \app \sval\bots}{\sfunalt \app \svalalt\bots} &\in{}&
      \sembot{\wfty{
        \ctx \cons \annty{\var}{\ty}
      }{\tyalt}}[\projctx{\ctx}{\sctx}] & \text{(ii)}
  \end{setbuild}.
\end{align*}
Finally, we can show
\begin{align*}
& \pair{
  \semtm{\tm}[\projrun{\ctx}{\sctx}] \app \semtm{\tmalt}[\projrun{\ctx}{\sctx}]
}{
  \semtm{\tm}[\projrun{\ctx}{\sctxalt}] \app \semtm{\tmalt}[\projrun{\ctx}{\sctxalt}]
} \\
& \text{By (i)} \\
\in & \sem{\wfty{
  \ctx \cons \annty{\var}{\ty}
}{\tyalt}}[\pair{\sctx}{\semtm{\tmalt}[\projrun{\ctx}{\sctx}]}] \\
& \text{By \Cref{thm:substitution-refined-ty-tm}} \\
= & \sem{\wfty{\ctx}{\tyalt \subst{\tmalt}{\var}}}[\sctx]
\end{align*}

\item \RuleTmTyLam{label}
\begin{align*}
& \pair{
  \semtm{\tmLam{\var}{\tm}}[\projrun{\ctx}{\sctx}]
}{
  \semtm{\tmLam{\var}{\tm}}[\projrun{\ctx}{\sctxalt}]
} \\
= & \pair{
  \semtm{\tm}[\projrun{\ctx}{\sctx}]
}{
  \semtm{\tm}[\projrun{\ctx}{\sctxalt}]
} \\
& \text{We need to show} \\
\in & \sem{\wfty{\ctx}{\tyall{\var}{\kd}{\ty}}}[\sctx] \\
= & \begin{setbuild}{\pair{\sval}{\svalalt}}[c]
  \forall \skd \in \sem{\wfkd{\ctx}{\kd}}[\sctx],\;
  \forall \skd\bots \in
  \sembot{\wfkd{\ctx}{\kd}}[\projctx{\ctx}{\sctx}] \\
  \projty{\ctx}{\sctx}{\kd}{\skd} \subseteq \skd\bots
  \Rightarrow
  \pair{\sval}{\svalalt} \in
  \sem{\wfty{
    \ctx \cons \annkd{\var}{\kd}
  }{\ty}}[\pair{\sctx}{\pair{\skd}{\skd\bots}}]
\end{setbuild}
\end{align*}
It is enough to show that for arbitrary fixed $\skd \in
\sem{\wfkd{\ctx}{\kd}}[\sctx]$ and $\skd\bots \in
\sembot*{\wfkd{\ctx}{\kd}}[\projctx{\ctx}{\sctx}]$ with
$\projty{\ctx}{\sctx}{\kd}{\skd} \subseteq \skd\bots$ the pair belongs
to the body interpretation, which is exactly the \ih{} for \tm in the
extended context.
\begin{align*}
& \pair{
  \semtm{\tm}[\projrun{\ctx}{\sctx}]
}{
  \semtm{\tm}[\projrun{\ctx}{\sctxalt}]
} \\
& \text{By \ih{}} \\
\in & \sem{\wfty{
  \ctx \cons \annkd{\var}{\kd}
}{\ty}}[\pair{\sctx}{\pair{\skd}{\skd\bots}}]
\end{align*}

\item \RuleTmTyApp{label}
\begin{align*}
& \pair{
  \semtm{\tm \tmapp \ty}[\projrun{\ctx}{\sctx}]
}{
  \semtm{\tm \tmapp \ty}[\projrun{\ctx}{\sctxalt}]
} \\
= & \pair{
  \semtm{\tm}[\projrun{\ctx}{\sctx}]
}{
  \semtm{\tm}[\projrun{\ctx}{\sctxalt}]
} \\
& \text{We need to show} \\
\in & \sem{\wfty{\ctx}{\tyalt \subst{\ty}{\var}}}[\sctx]
\end{align*}
By \ih{} we have
\begin{align*}
& \pair{
  \semtm{\tm}[\projrun{\ctx}{\sctx}]
}{
  \semtm{\tm}[\projrun{\ctx}{\sctxalt}]
} \\
\in & \sem{\wfty{\ctx}{\tyall{\var}{\kd}{\tyalt}}}[\sctx] \\
= & \begin{setbuild}{\pair{\sval}{\svalalt}}[c]
  \forall \skd \in \sem{\wfkd{\ctx}{\kd}}[\sctx],\;
  \forall \skd\bots \in
  \sembot{\wfkd{\ctx}{\kd}}[\projctx{\ctx}{\sctx}] \\
  \projty{\ctx}{\sctx}{\kd}{\skd} \subseteq \skd\bots
  \Rightarrow
  \pair{\sval}{\svalalt} \in
  \sem{\wfty{
    \ctx \cons \annkd{\var}{\kd}
  }{\tyalt}}[\pair{\sctx}{\pair{\skd}{\skd\bots}}]
\end{setbuild}.
\end{align*}
The interpretations of the type argument \ty form a valid index: by
\Crefsub{thm:type-interpretation-well-formed}{refined}
$\sem{\wftm{\ctx}{\ty}{\kd}}[\sctx] \in \sem{\wfkd{\ctx}{\kd}}[\sctx]$,
by \Crefsub{thm:type-interpretation-well-formed}{unrefined}
$\sembot*{\wftm{\ctx}{\ty}{\kd}}[\projctx{\ctx}{\sctx}] \in
\sembot*{\wfkd{\ctx}{\kd}}[\projctx{\ctx}{\sctx}]$, and by
\Cref{thm:refined-sub-unrefined}
$\projty{\ctx}{\sctx}{\kd}{\sem{\wftm{\ctx}{\ty}{\kd}}[\sctx]} \subseteq
\sembot{\wftm{\ctx}{\ty}{\kd}}[\projctx{\ctx}{\sctx}]$. Instantiating
the universal quantifier at $\sem{\wftm{\ctx}{\ty}{\kd}}[\sctx]$ and
$\sembot{\wftm{\ctx}{\ty}{\kd}}[\projctx{\ctx}{\sctx}]$, whose
implication premise holds, gives
\begin{align*}
& \pair{
  \semtm{\tm}[\projrun{\ctx}{\sctx}]
}{
  \semtm{\tm}[\projrun{\ctx}{\sctxalt}]
} \\
\in & \sem{\wfty{
  \ctx \cons \annkd{\var}{\kd}
}{\tyalt}}[\pair{\sctx}{
  \pair{
    \sem{\wftm{\ctx}{\ty}{\kd}}[\sctx]
  }{
    \sembot{\wftm{\ctx}{\ty}{\kd}}[\projctx{\ctx}{\sctx}]
  }
}] \\
& \text{By \Cref{thm:substitution-refined-ty-ty}} \\
= & \sem{\wfty{\ctx}{\tyalt \subst{\ty}{\var}}}[\sctx]
\end{align*}

\item \RuleTmRefl{label}
By \ih{} the pair lies in $\sem{\wfty{\ctx}{\ty}}[\sctx]$. Write
$\sval_\tm \defeq \semtm{\tm}[\projrun{\ctx}{\sctx}]$ for its first
component. Partial reflexivity gives $\pair{\sval_\tm}{\sval_\tm} \in
\sem{\wfty{\ctx}{\ty}}[\sctx]$, and by \Cref{thm:refined-sub-unrefined}
$\pair{\sval_\tm}{\sval_\tm} \in \sembot{\wfty{\ctx}{\ty}}[\projctx{\ctx}{\sctx}]$;
since $\tm$ does not mention $\var$, this is exactly
$\semprp*{\wfprop{\ctx \cons \annty{\var}{\ty}}{\var \propeq_\ty
\tm}}[\pair*{\sctx}{\\\sval_\tm}] = \strue$. Hence the pair, whose first
component is $\sval_\tm$, satisfies the refinement predicate, so
\begin{align*}
& \pair{
  \semtm{\tm}[\projrun{\ctx}{\sctx}]
}{
  \semtm{\tm}[\projrun{\ctx}{\sctxalt}]
} \\
\in & \begin{setbuild}{\pair{\sval}{\svalalt} \in \sem{\wfty{\ctx}{\ty}}[\sctx]}
  \semprp{\wfprop{
      \ctx \cons \annty{\var}{\ty}
  }{\var \propeq_\ty \tm}}[\pair{\sctx}{\sval}] = \strue
\end{setbuild} \\
= & \sem{\wfty{\ctx}{\tyref{\var}{\ty}{\var \propeq_\ty \tm}}}[\sctx]
\end{align*}

\item \RuleTmSub{label}
\begin{align*}
& \pair{
  \semtm{\tm}[\projrun{\ctx}{\sctx}]
}{
  \semtm{\tm}[\projrun{\ctx}{\sctxalt}]
} \\
& \text{By \ih{}} \\
\in & \sem{\wfty{\ctx}{\ty}}[\sctx] \\
& \text{By \Crefsub{thm:subtyping-semantics}{refined}} \\
\subseteq & \sem{\wfty{\ctx}{\tyalt}}[\sctx]
\end{align*}

\item \RuleTmIntro{label}
Let $D \defeq \decl$. The constructor is interpreted by the constant
function
\[
  \semtm{\ctor(i)_D}[\projrun{\ctx}{\sctx}]
  = \demb{\to}{\ioverline*{j}{\sval} \mapsto \pair{\ctor_i}{\ioverline*{j}{\sval}}},
\]
so the two components of the required pair coincide. Since the type is
a function type, it suffices to fix related arguments
\[
  \pair{\sval_j}{\svalalt_j} \in \sem{\wfty{
    \ctx \cons \ioverline*{j - 1}{\annty{\var}{\tyalt \subst{D}{\tyctor}}}
  }{\tyalt_j \subst{D}{\tyctor}}}[\pair{\sctx}{\ioverline*{j - 1}{\sval}}]
\]
and show that the application lands in the codomain,
\[
  \pair{
    \pair{\ctor_i}{\ioverline*{j}{\sval}}
  }{
    \pair{\ctor_i}{\ioverline*{j}{\svalalt}}
  } \in \sem{\wfty{
    \ctx \cons \ioverline*{j}{\annty{\var}{\tyalt\subst{D}{\tyctor}}}
  }{D \tmApp \ioverline*{k}{\tm}}}[\pair{\sctx}{\ioverline*{j}{\sval}}].
\]
The interpretation of the function type also imposes the analogous
condition on unrefined-related arguments; this is precisely
\Crefsub{thm:per-respecting-terms}{unrefined} applied to $\ctor(i)_D$, so
we are left with the refined condition above.
Let $\mathcal{F}$ be the operator defining the refined interpretation
of $D$. As that interpretation is a fixed point of $\mathcal{F}$, the
codomain above equals
\[
  \mathcal{F}\!\left(
    \sem{\wftm{\ctx}{D}{\kdfun{\ioverline*{k}{\ty}}{\kdtype}}}[\sctx]
  \right) \app
  \ioverline*{k}{\semtm{\tm}[\pair{\projrun{\ctx}{\sctx}}{\ioverline*{j}{\sval}}]},
\]
whose $i$-th disjunct is
\[
  \scalebox{0.65}{$\displaystyle
    \begin{setbuild}{\pair{
      \pair{\ctor_i}{\ioverline{j}{\sval}}
    }{
      \pair{\ctor_i}{\ioverline{j}{\svalalt}}
    }}[c@{\quad}r]
      \pair{\sval_j}{\svalalt_j} \in \sem{\wfty{
        \ctx \cons \annkd{\tyctor}{\kdfun{\ioverline{k}{\ty}}{\kdtype}} \cons
        \ioverline{j - 1}{\annty{\var}{\tyalt}}
      }{\tyalt_j}}[\pair{\pair{\sctx}{\pair{
          \sem{\wftm{\ctx}{D}{\kdfun{\ioverline*{k}{\ty}}{\kdtype}}}[\sctx]
        }{
          \sembot{\wftm{\ctx}{D}{\kdfun{\ioverline*{k}{\ty}}{\kdtype}}}[\projctx{\ctx}{\sctx}]
        }}}{\ioverline{j - 1}{\sval}}] & \text{(i)} \\
      \pair{\svalaltalt_k}{\semtm{\tm_k}[\pair{\projrun{\ctx}{\sctx}}{\ioverline*{j}{\sval}}]}
      \in \sembot{\wfty{
        \ctx \cons \ioverline*{k - 1}{\annty{\var}{\ty}}
      }{\ty_k}}[\projctx{\ctx}{\sctx}] & \text{(ii)}
    \end{setbuild}
  $}
\]
The required pair lies in this disjunct. Condition~(ii) holds: its two
sides coincide, and the common index denotation lies in the
interpretation of the corresponding index type, so partial reflexivity
relates it to itself. For condition~(i), each argument pair satisfies,
by assumption,
\begin{align*}
  & \pair{\sval_j}{\svalalt_j} \\
& \text{by assumption} \\
\in {} & \scalebox{1}{$\displaystyle \sem{\wfty{
      \ctx \cons \ioverline*{j - 1}{\annty{\var}{\tyalt \subst{D}{\tyctor}}}
    }{\tyalt_j \subst{D}{\tyctor}}}[\pair{\sctx}{\ioverline*{j - 1}{\sval}}]$} \\
& \text{By \Cref{thm:substitution-refined-ty-ty}} \\
= {} & \scalebox{0.8}{$\displaystyle \sem{\wfty{
      \ctx \cons \annkd{\tyctor}{\kdfun{\ioverline*{k}{\ty}}{\kdtype}} \cons
      \ioverline*{j - 1}{\annty{\var}{\tyalt}}
    }{\tyalt_j}}[\pair{\pair{\sctx}{\pair{
        \sem{\wftm{\ctx}{D}{\kdfun{\ioverline*{k}{\ty}}{\kdtype}}}[\sctx]
      }{
        \sembot{\wftm{\ctx}{D}{\kdfun{\ioverline*{k}{\ty}}{\kdtype}}}[\projctx{\ctx}{\sctx}]
      }}}{\ioverline*{j - 1}{\sval}}]$}
\end{align*}
which is condition~(i). Hence both conditions hold, the pair lies in
the disjunct, and therefore in the codomain.

\item \RuleTyElim{label}
We use the following abbreviations:
\begin{align*}
D &\defeq \decl \\
e(\sval, r) &\defeq
\begin{cases}
  \semtm{c_i}[\projrun{\ctx}{\sctx}] \app
  \ioverline*{j}{\svalaltalt} \app
  \ioverline*{j}{
    \tmel{\tyctor}{\tyalt}{\pair{\projrun{\ctx}{\sctx}}{\ioverline{j-1}{\svalaltalt}}}{\svalaltalt_j}{r}
  }
  & \text{if } \sval = \pair{
      \ctor_i
    }{
      \ioverline*{j}{\svalaltalt}
    } \\
  \bot & \text{otherwise}
\end{cases}
\end{align*}

\begin{align*}
  & \pair{
    \semtm{\tmelim{D}{\ioverline*{i}{c}}}[\projrun{\ctx}{\sctx}]
  }{
    \semtm{\tmelim{D}{\ioverline*{i}{c}}}[\projrun{\ctx}{\sctxalt}]
  } \\
= & \pair{
\demb{\to}{\mathsf{lfp}_r
\ioverline{k}{\sval} \cons   % indexes
\svalalt             \mapsto % value
e(\svalalt, r)
}}{
\demb{\to}{\mathsf{lfp}_{r'}
\ioverline{k}{\sval} \cons   % indexes
\svalalt             \mapsto % value
e(\svalalt, r')
}} \\
\defeq & \pair{r}{r'}
\end{align*}
This pair exists and is an element of $\dinf$ as it is the least fixed
point of continuous operators in $\dinf$. We name the two components
$r$ and $r'$.

What we need to show is that this pair lies in
\begin{align*}
\sem{\wfty{\ctx}{
  \tyfun{\left(\ioverline*{k}{\annty{\var}{\ty}}\right)}{
    \tyfun*[{\varalt}]{D \tmApp {\ioverline*{k}{\var}}}{\tyaltalt}
  }
}}[\sctx]
\end{align*}

We distinguish two cases based on whether the type is stratified or
positive:
\begin{itemize}

\item \textbf{Positive case:}

We define the collection of all those elements of the domain for which
the property holds.
\begin{align*}
  P \defeq \ioverline*{k}{\svalaltalt} \mapsto
  \begin{setbuild}{\pair{\sval}{\svalalt}}
    \pair{\sval}{\svalalt} \in \sem{
      \wftm{\ctx}{D}{\ioverline*{k}{\tyalt} \to \kdtype}
    }[\sctx] \app \ioverline*{k}{\svalaltalt} \\
  \pair{e(\sval, r)}{e(\svalalt, r')}
  \in \sem{\wfty{
    \ctx \cons \ioverline*{k}{\annty{\var}{\tyalt}}
    \cons \annty{\varaltalt}{D \tmapp \ioverline*{k}{\var}}
  }{
    \tyaltalt
  }}[\pair{\pair{\sctx}{\ioverline*{k}{\svalaltalt}}}{\sval}]
  \end{setbuild}
\end{align*}

The goal is to show that $P =
\sem{\wftm{\ctx}{D}{\ioverline*{k}{\tyalt} \to \kdtype}}[\sctx]$.

By definition we already have $P \leq
\sem{\wftm{\ctx}{D}{\ioverline*{k}{\tyalt} \to \kdtype}}[\sctx]$, we
need to show the other inclusion; since
$\sem{\wftm{\ctx}{D}{\ioverline*{k}{\tyalt} \to \kdtype}}[\sctx]$ is
defined as the least fixed point of $\mathcal{F}$, it is enough to show
that $P \geq \mathcal{F}(P)$, i.e. that $P$ is a pre-fixed point of
$\mathcal{F}$.

\begin{align*}
  & \mathcal{F}(P) \\
={} & \scalebox{0.55}{$\displaystyle
  \ioverline{k}{\svalaltalt} \mapsto
  \Bigcup_i \begin{setbuild}{\pair{
    \pair{\ctor_i}{\ioverline{j}{\sval}}
  }{
    \pair{\ctor_i}{\ioverline{j}{\svalalt}}
  }}
    \pair{\sval_j}{\svalalt_j} \in \sem{\wfty{
      \ctx \cons
      \annkd{\tyctor}{\kdfun{\ioverline{k}{\ty}}{\kdtype}} \cons
      \ioverline{j - 1}{\annty{\var}{\tyalt}}
    }{\tyalt_j}}[\pair{\pair{\sctx}{\pair{P}{
        \sembot{\wftm{\ctx}{\decl}{\kdfun{\ioverline{k}{\ty}}{\kdtype}}}[\projctx{\ctx}{\sctx}]
      }}}{\ioverline{j - 1}{\sval}}] \\
      \pair{\svalaltalt_k}{\semtm{\tm_k}[\pair{\projrun{\ctx}{\sctx}}{\ioverline*{j}{\sval}}]}
      \in \sembot{\wfty{
        \ctx \cons \ioverline*{k - 1}{\annty{\var}{\ty}}
      }{\ty_k}}[\projctx{\ctx}{\sctx}]
  \end{setbuild}
$}
\end{align*}

Before continuing, note that by \Cref{def:type-operator-lattice} the
order on type operators compares them pointwise, but only at
indices in the refined domain $\dom*{\sem*{\wfty{\ctx \cons
\ioverline*{k - 1}{\annty{\var}{\ty}}}{\ty_k}}[\pair{\sctx}{\ioverline*{k - 1}{\svalaltalt}}]}$.
Fixed arbitrary $\svalaltalt_k
\in \sem{\wfty{\ctx \cons \ioverline*{k - 1}{\annty{\var}{\ty}}
}{\ty_k}}[\pair{\sctx}{\ioverline*{k - 1}{\svalaltalt}}]
$, we need to show that
if $\pair*{
\pair*{C_i}{\ioverline{j}{\sval}}
}{ \\
\pair*{C_i}{\ioverline{j}{\svalalt}}
} \in \mathcal{F}(P) \app
\ioverline{k}{\svalaltalt}$, then
$\pair{
\pair{C_i}{\ioverline{j}{\sval}}
}{
\pair{C_i}{\ioverline{j}{\svalalt}}
} \in P \app \ioverline{k}{\svalaltalt}$.

To do this it is enough to show that
\begin{align*}
  & \pair{
    e(\pair{C_i}{\ioverline*{j}{\sval}}, r)
  }{
    e(\pair{C_i}{\ioverline*{j}{\svalalt}}, r')
  } \\
= & \vpair{
    \semtm{c_i}[\projrun{\ctx}{\sctx}] \app
    \ioverline{j}{\svalaltalt} \app
    \ioverline{j}{
      \tmel{\tyctor}{\tyalt}{\pair{\projrun{\ctx}{\sctx}}{\ioverline{j-1}{\svalaltalt}}}{\svalaltalt_j}{r}
    }
  }{
    \semtm{c_i}[\projrun{\ctx}{\sctxalt}] \app
    \ioverline{j}{\svalaltalt} \app
    \ioverline{j}{
      \tmel{\tyctor}{\tyalt}{\pair{\projrun{\ctx}{\sctxalt}}{\ioverline{j-1}{\svalaltalt}}}{\svalaltalt_j}{r'}
    }
  } \\
\in & \sem{\wfty{
    \ctx \cons \ioverline*{k}{\annty{\var}{\tyalt}}
    \cons \annty{\varaltalt}{D \tmapp \ioverline*{k}{\var}}
  }{
    \tyaltalt
  }}[\pair{\pair{\sctx}{\ioverline*{k}{\svalaltalt}}}{\pair{C_i}{\ioverline*{j}{\sval}}}]
\end{align*}
We apply \Cref{lem:per-respecting-eliminator} to each mention of
$\mathsf{tmEl}$ with the
approximant $P$, which is below
$\sem{\wftm{\ctx}{D}{\kdfun*{({\ioverline*{k}{\ann{\var}{\ty}}})}{\kdtype}}}[\sctx]$,
by $\mathcal{F}$ being monotonic it follows that $\mathcal{F}(P)
\leq
\mathcal{F}(\sem{\wftm{\ctx}{D}{\kdfun*{({\ioverline*{k}{\ann{\var}{\ty}}})}{\kdtype}}}[\sctx])
= \sem{\wftm{\ctx}{D}{\kdfun*{({\ioverline*{k}{\ann{\var}{\ty}}})}{\kdtype}}}[\sctx]$,
thus by \ih{} on $c_i$ we have that the pair is in the interpretation of the type, as required.

Since $P$ is therefore a pre-fixed
point of $\mathcal{F}$, the least fixed point
$\sem*{\wftm{\ctx}{D}{\kdfun*{({\ioverline*{k}{\ann{\var}{\ty}}})}{\kdtype}}}[\sctx]$
is below $P$, giving the equality.

\item \textbf{Stratified case:}
  We prove it by induction on the metric. We define the $n$-th
  approximation of the type as:
  \begin{align*}
    P(n) \defeq \ioverline*{k}{\svalaltalt} \mapsto \begin{cases}
      \sem{
        \wftm{\ctx}{D}{\ioverline*{k}{\tyalt} \to \kdtype}
      }[\sctx] \app \ioverline*{k}{\svalaltalt} & \text{if } n >
                                                  \dret{\snat}{\semtm{m}[\projrun{\ctx}{\sctx}]
                                                  \app
                                                  \ioverline*{k}{\svalaltalt}}
      \\
      \emptyset & \text{otherwise}
    \end{cases}
  \end{align*}
  Clearly $\forall n \in \mathbb{N}. \; P(n) \leq \sem{
    \wftm{\ctx}{D}{\ioverline*{k}{\tyalt} \to \kdtype}
  }[\sctx]$.
  As in the positive case, by \Cref{def:type-operator-lattice} it
  suffices to range over refined-domain indices, where the motive
  interpretation is defined. We show that $\forall n \in \mathbb{N}$
  \begin{align*}
    \forall \svalaltalt_k &\in \sem{\wfty{\ctx \cons \ioverline*{k - 1}{\annty{\var}{\ty}}
    }{\ty_k}}[\pair{\sctx}{\ioverline*{k - 1}{\svalaltalt}}] \\
    \forall \pair{\sval}{\svalalt} &\in P(n) \app
    \ioverline*{k}{\svalaltalt}. \\
    \pair{e(\sval, r)}{e(\svalalt, r')} &\in \sem{\wfty{
      \ctx \cons \ioverline*{k}{\annty{\var}{\tyalt}}
      \cons \annty{\varaltalt}{D \tmapp \ioverline*{k}{\var}}
    }{
      \tyaltalt
    }}[\pair{\pair{\sctx}{\ioverline*{k}{\svalaltalt}}}{\sval}]
  \end{align*}
  By induction over $n$:
  \begin{itemize}
    \item $n = 0$ is trivial, since $P(0)$ is empty.
    \item $n = m + 1$:
    Let $\pair{\sval}{\svalalt} = \pair{
      \pair{C_i}{\ioverline*{j}{\sval}}
    }{
      \pair{C_i}{\ioverline*{j}{\svalalt}}
    }$
    We can show

    \begin{align*}
      & \pair{
        e(\pair{C_i}{\ioverline*{j}{\sval}}, r)
      }{
        e(\pair{C_i}{\ioverline*{j}{\svalalt}}, r')
      } \\
    = & \vpair{
        \semtm{c_i}[\projrun{\ctx}{\sctx}] \app
        \ioverline{j}{\svalaltalt} \app
        \ioverline{j}{
          \tmel{\tyctor}{\tyalt}{\pair{\projrun{\ctx}{\sctx}}{\ioverline{j-1}{\svalaltalt}}}{\svalaltalt_j}{r}
        }
      }{
        \semtm{c_i}[\projrun{\ctx}{\sctxalt}] \app
        \ioverline{j}{\svalaltalt} \app
        \ioverline{j}{
          \tmel{\tyctor}{\tyalt}{\pair{\projrun{\ctx}{\sctxalt}}{\ioverline{j-1}{\svalaltalt}}}{\svalaltalt_j}{r'}
        }
      } \\
    \in & \sem{\wfty{
        \ctx \cons \ioverline*{k}{\annty{\var}{\tyalt}}
        \cons \annty{\varaltalt}{D \tmapp \ioverline*{k}{\var}}
      }{
        \tyaltalt
      }}[\pair{\pair{\sctx}{\ioverline*{k}{\svalaltalt}}}{\pair{C_i}{\ioverline*{j}{\sval}}}]
    \end{align*}
    We apply \Cref{lem:per-respecting-eliminator} to each mention of
    $\mathsf{tmEl}$ with the
    approximant $P(m)$, which is below
    $\sem{\wftm{\ctx}{D}{\kdfun*{({\ioverline*{k}{\ann{\var}{\ty}}})}{\kdtype}}}[\sctx]$
    and by the \ih{} they satisfy the required property, thus by \ih{}
    on $c_i$ we have that the pair is in the interpretation of the type,
    as required.
  \end{itemize}

\end{itemize}

\end{itemize}
\end{proof}

\begin{proof}[Proof of \Cref{thm:refined-runtime-projection-is-unrefined-runtime-interpretation}]
By induction on the structure of $\wfctx{\ctx}$.
\begin{itemize}
\item \RuleCtxEmpty{label}
\begin{align*}
      & \pair{\projrun{\ctxemp}{\sunit}}{\projrun{\ctxemp}{\sunit}} \\
={}   & \pair{\sunit}{\sunit} \\
\in{} & \unrefass{\wfctx{\ctxemp}}{\projctx{\ctxemp}{\sunit}}
\end{align*}

\item \RuleCtxTy{label}
\begin{align*}
& \pair{\projrun{\ctx \cons \annty{\var}{\ty}}{\sctx \cons
  \sval}}{\projrun{\ctx \cons \annty{\var}{\ty}}{\sctxalt \cons
  \svalalt}} \\
={} & \pair{\projrun{\ctx}{\sctx} \cons \sval}{\projrun{\ctx}{\sctxalt} \cons
      \svalalt} \\
& \text{By \ih{} and \Cref{thm:refined-sub-unrefined}} \\
\in{} & \unrefass{\wfctx{\ctx}}{\projctx{\ctx}{\sctx}} \times
        \sembot{\wfty{\ctx}{\ty}}[\projctx{\ctx}{\sctx}] \\
= & \unrefass{\wfctx{\ctx \cons \annty{\var}{\ty}}}{\projctx{\ctx \cons
  \annty{\var}{\ty}}{\sctx \cons \sval}}
\end{align*}

\item \RuleCtxKd{label}
  \begin{align*}
& \pair{\projrun{\ctx \cons \annkd{\var}{\kd}}{\sctx \cons \skd}}{\projrun{\ctx \cons \annkd{\var}{\kd}}{\sctxalt \cons
  \skdalt}} \\
    ={} & \pair{\projrun{\ctx}{\sctx}}{\projrun{\ctx}{\sctxalt}} \\
& \text{By \ih{}} \\
    \in{} & \unrefass{\wfctx{\ctx}}{\projctx{\ctx}{\sctx}} \\
= & \unrefass{\wfctx{\ctx \cons \annkd{\var}{\kd}}}{\projctx{\ctx
    \cons
    \annkd{\var}{\kd}}{\sctx \cons \skd}}
  \end{align*}

\end{itemize}
\end{proof}

\begin{proof}[Proof of \Crefsub{thm:per-respecting-terms}{unrefined}]

By induction on the structure of $\wftm{\ctx}{\tm}{\ty}$.
\begin{itemize}
\item \RuleTmNat{label} and \RuleTmBool{label}
\begin{align*}
      & \pair{\semtm{b}[\sctxalt]}{\semtm{b}[\sctxaltalt]} \\
  ={} & \pair{\demb{\tybase}{b}}{\demb{\tybase}{b}} \\
\in{} & \begin{setbuild}{
           \pair{\demb{\tybase}{b}}{\demb{\tybase}{b}}
         }
           b \in \tybase
         \end{setbuild} \cup \{\pair{\bot}{\bot}\} \\
={} & \sembot{\wfty{\ctx}{\tybase}}[\sctx]
\end{align*}

\item \RuleTmVar{label}
\begin{align*}
      & \pair{\semtm{\var}[\sctxalt]}{\semtm{\var}[\sctxaltalt]} \\
  ={} & \pair{\sctxalt(\var)}{\sctxaltalt(\var)} \\
      & \text{By assumption} \\
\in{} & \sembot{\wfty{\ctx}{\ty}}[\sctx]
\end{align*}

\item \RuleTmLam{label}
\begin{align*}
  & \pair{\semtm{\tmlam{\var}{\tm}}[\sctxalt]}{\semtm{\tmlam{\var}{\tm}}[\sctxaltalt]} \\
= & \pair{
    \demb{\to}{\sval \mapsto
      \semtm{\tm}[\pair{\sctxalt}{\sval}]}
  }{
    \demb{\to}{\sval \mapsto
      \semtm{\tm}[\pair{\sctxaltalt}{\sval}]}
  } \\
& \text{We need to show} \\
\in & \sembot{\wfty{\ctx}{\tyfun*[\var]{\ty}{\tyalt}}}[\sctx] \\
= & \begin{setbuild}{\pair{\sfun}{\sfunalt}}[l]
  \forall \pair{\sval}{\svalalt} \in
  \sembot{\wfty{\ctx}{\ty}}[\sctx].\;
  \pair{\sfun \app \sval}{\sfunalt \app \svalalt} \in
  \sembot{\wfty{\ctx \cons \annty{\var}{\ty}}{\tyalt}}[\sctx]
\end{setbuild}
\end{align*}
Fix arbitrary
$\pair{\sval}{\svalalt} \in
\sembot{\wfty{\ctx}{\ty}}[\sctx]$. Hence
\begin{align*}
  & \pair{
    \demb{\to}{\sval \mapsto
      \semtm{\tm}[\pair{\sctxalt}{\sval}]}
    \app \sval
  }{
    \demb{\to}{\sval \mapsto
      \semtm{\tm}[\pair{\sctxaltalt}{\sval}]}
    \app \svalalt
  } \\
= & \pair{
    \semtm{\tm}[\pair{\sctxalt}{\sval}]
}{
    \semtm{\tm}[\pair{\sctxaltalt}{\svalalt}]
  } \\
& \text{By \ih{}} \\
\in & \sembot{\wfty{\ctx \cons \annty{\var}{\ty}}{\tyalt}}[\sctx]
\end{align*}

\item \RuleTmApp{label}
\begin{align*}
& \pair{\semtm{\tm \app \tmalt}[\sctxalt]
}{
  \semtm{\tm \app \tmalt}[\sctxaltalt]
} \\
= & \pair{
  \semtm{\tm}[\sctxalt] \app
  \semtm{\tmalt}[\sctxalt]
}{
  \semtm{\tm}[\sctxaltalt] \app
  \semtm{\tmalt}[\sctxaltalt]
} \\
& \text{We need to show} \\
\in & \sembot{\wfty{\ctx}{\tyalt \subst{\tmalt}{\var}}}[\sctx]
\end{align*}
By the \ih{} applied to $\tmalt$ and $\tm$ we have, respectively,
\begin{align*}
  \pair{
    \semtm{\tmalt}[\sctxalt]
  }{
    \semtm{\tmalt}[\sctxaltalt]
  } & \in \sembot{\wfty{\ctx}{\ty}}[\sctx]
\end{align*}
and
\begin{align*}
&\pair{
  \semtm{\tm}[\sctxalt]
}{
  \semtm{\tm}[\sctxaltalt]
} \\
\in & \sembot{\wfty{\ctx}{\tyfun*[\var]{\ty}{\tyalt}}}[\sctx] \\
= & \begin{setbuild}{\pair{\sfun}{\sfunalt}}[l]
  \forall \pair{\sval}{\svalalt} \in
  \sembot{\wfty{\ctx}{\ty}}[\sctx].\;
  \pair{\sfun \app \sval}{\sfunalt \app \svalalt} \in
  \sembot{\wfty{\ctx \cons \annty{\var}{\ty}}{\tyalt}}[\sctx]
\end{setbuild}.
\end{align*}
Instantiating the universal quantifier with the argument pair gives
\begin{align*}
& \pair{
  \semtm{\tm}[\sctxalt] \app
  \semtm{\tmalt}[\sctxalt]
}{
  \semtm{\tm}[\sctxaltalt] \app
  \semtm{\tmalt}[\sctxaltalt]
} \\
\in & \sembot{\wfty{
  \ctx \cons \annty{\var}{\ty}
      }{\tyalt}}[\sctx] \\
& \text{By noticing that the unrefined semantics of types is independent of
    terms} \\
= & \sembot{\wfty{\ctx}{\tyalt \subst{\tmalt}{\var}}}[\sctx]
\end{align*}

\item \RuleTmTyLam{label}
\begin{align*}
& \pair{
  \semtm{\tmLam{\var}{\tm}}[\sctxalt]
}{
  \semtm{\tmLam{\var}{\tm}}[\sctxaltalt]
} \\
= & \pair{
  \semtm{\tm}[\sctxalt]
}{
  \semtm{\tm}[\sctxaltalt]
} \\
& \text{We need to show} \\
\in & \sembot{\wfty{\ctx}{\tyall{\var}{\kd}{\ty}}}[\sctx] \\
= & \begin{setbuild}{\pair{\sval}{\svalalt}}[l]
  \forall \skd \in \sembot{\wfkd{\ctx}{\kd}}[\sctx].\;
  \pair{\sval}{\svalalt} \in
  \sembot{\wfty{
    \ctx \cons \annkd{\var}{\kd}
  }{\ty}}[\pair{\sctx}{\skd}]
\end{setbuild}
\end{align*}
For arbitrary fixed
$\skd \in \sembot{\wfkd{\ctx}{\kd}}[\sctx]$.
\begin{align*}
& \pair{
  \semtm{\tm}[\sctxalt]
}{
  \semtm{\tm}[\sctxaltalt]
} \\
& \text{By \ih{}} \\
\in & \sembot{\wfty{
  \ctx \cons \annkd{\var}{\kd}
}{\ty}}[\pair{\sctx}{\skd}]
\end{align*}

\item \RuleTmTyApp{label}
\begin{align*}
& \pair{
  \semtm{\tm \tmapp \ty}[\sctxalt]
}{
  \semtm{\tm \tmapp \ty}[\sctxaltalt]
} \\
= & \pair{
  \semtm{\tm}[\sctxalt]
}{
  \semtm{\tm}[\sctxaltalt]
} \\
& \text{We need to show} \\
\in & \sembot{\wfty{\ctx}{\tyalt \subst{\ty}{\var}}}[\sctx]
\end{align*}
By the \ih{} we have
\begin{align*}
& \pair{
  \semtm{\tm}[\sctxalt]
}{
  \semtm{\tm}[\sctxaltalt]
} \\
\in & \sembot{\wfty{\ctx}{\tyall{\var}{\kd}{\tyalt}}}[\sctx] \\
= & \begin{setbuild}{\pair{\sval}{\svalalt}}[l]
  \forall \skd \in \sembot{\wfkd{\ctx}{\kd}}[\sctx].\;
  \pair{\sval}{\svalalt} \in
  \sembot{\wfty{
    \ctx \cons \annkd{\var}{\kd}
  }{\tyalt}}[\pair{\sctx}{\skd}].
\end{setbuild}
\end{align*}
By \Crefsub{thm:type-interpretation-well-formed}{unrefined}, the
unrefined interpretation of the type argument is a valid index.
Instantiating the universal quantifier at
$\sembot{\wftm{\ctx}{\ty}{\kd}}[\sctx]$ gives
\begin{align*}
& \pair{
  \semtm{\tm}[\sctxalt]
}{
  \semtm{\tm}[\sctxaltalt]
} \\
\in & \sembot{\wfty{
  \ctx \cons \annkd{\var}{\kd}
}{\tyalt}}[\pair{\sctx}{
  \sembot{\wftm{\ctx}{\ty}{\kd}}[\sctx]
      }] \\
& \text{By \Cref{thm:substitution-unrefined-ty-ty}} \\
= & \sembot{\wfty{\ctx}{\tyalt \subst{\ty}{\var}}}[\sctx]
\end{align*}

\item \RuleTmRefl{label}
\begin{align*}
& \pair{
  \semtm{\tm}[\sctxalt]
}{
  \semtm{\tm}[\sctxaltalt]
} \\
& \text{By the \ih{}} \\
\in & \sembot{\wfty{\ctx}{\ty}}[\sctx] \\
= & \sembot{\wfty{\ctx}{\tyref{\var}{\ty}{\var \propeq_\ty \tm}}}[\sctx]
\end{align*}

\item \RuleTmSub{label}
\begin{align*}
& \pair{
  \semtm{\tm}[\sctxalt]
}{
  \semtm{\tm}[\sctxaltalt]
} \\
& \text{By the \ih{}} \\
\in & \sembot{\wfty{\ctx}{\ty}}[\sctx] \\
& \text{By \Crefsub{thm:subtyping-semantics}{unrefined}} \\
= & \sembot{\wfty{\ctx}{\tyalt}}[\sctx]
\end{align*}

\item \RuleTmIntro{label}
Let $D \defeq \decl$. The constructor is interpreted by the constant
function
\[
  \semtm{\ctor(i)_D}[\sctxalt]
  = \demb{\to}{\ioverline*{j}{\sval} \mapsto \pair{\ctor_i}{\ioverline*{j}{\sval}}},
\]
so the two components of the required pair coincide. Since the type is
a function type, it suffices to fix related arguments
\[
  \pair{\sval_j}{\svalalt_j} \in \sembot{\wfty{
    \ctx \cons \ioverline*{j - 1}{\annty{\var}{\tyalt \subst{D}{\tyctor}}}
  }{\tyalt_j \subst{D}{\tyctor}}}[\sctx]
\]
and show that the application lands in the codomain,
\[
  \pair{
    \pair{\ctor_i}{\ioverline*{j}{\sval}}
  }{
    \pair{\ctor_i}{\ioverline*{j}{\svalalt}}
  } \in \sembot{\wfty{
    \ctx \cons \ioverline*{j}{\annty{\var}{\tyalt\subst{D}{\tyctor}}}
  }{D \tmApp \ioverline*{k}{\tm}}}[\sctx].
\]

Let $\mathcal{F}$ be the operator defining the unrefined interpretation
of $D$. As that interpretation is a fixed point of $\mathcal{F}$, the
codomain above equals
\[
  \mathcal{F}\!\left(
    \sembot{\wftm{\ctx}{D}{\kdfun{\ioverline*{k}{\ty}}{\kdtype}}}[\sctx]
  \right)
\]
whose $i$-th disjunct is
\[
  \scalebox{0.95}{$\displaystyle
    \begin{setbuild}{\pair{
      \pair{\ctor_i}{\ioverline{j}{\sval}}
    }{
      \pair{\ctor_i}{\ioverline{j}{\svalalt}}
    }}[c@{\quad}r]
      \pair{\sval_j}{\svalalt_j} \in \sembot{\wfty{
        \ctx \cons \annkd{\tyctor}{\kdfun{\ioverline*{k}{\ty}}{\kdtype}} \cons
        \ioverline{j - 1}{\annty{\var}{\tyalt}}
      }{\tyalt_j}}[\pair{\sctx}{\sembot{\wftm{\ctx}{D}{\kdfun{\ioverline*{k}{\ty}}{\kdtype}}}[\sctx]}]
      \negspace{6}
    \end{setbuild}$}
\]
Each argument pair satisfies, by assumption,
\begin{align*}
  & \pair{\sval_j}{\svalalt_j} \\
& \text{by assumption} \\
\in {} & \scalebox{1}{$\displaystyle \sembot{\wfty{
      \ctx \cons \ioverline*{j - 1}{\annty{\var}{\tyalt \subst{D}{\tyctor}}}
    }{\tyalt_j \subst{D}{\tyctor}}}[\sctx]$} \\
& \text{By \Cref{thm:substitution-unrefined-ty-ty}} \\
= {} & \scalebox{1}{$\displaystyle \sembot{\wfty{
      \ctx \cons \annkd{\tyctor}{\kdfun{\ioverline*{k}{\ty}}{\kdtype}} \cons
      \ioverline*{j - 1}{\annty{\var}{\tyalt}}
       }{\tyalt_j}}[
       \pair{\sctx}{\sembot{\wftm{\ctx}{D}{\kdfun{\ioverline*{k}{\ty}}{\kdtype}}}[\sctx]}
       ]$}
\end{align*}
Hence the pair lies in the disjunct, and therefore in the codomain.

\item \RuleTyElim{label}
We use the following abbreviations:
\begin{align*}
D &\defeq \decl \\
e(\sval, r) &\defeq
\begin{cases}
  \semtm{c_i}[\sctxalt] \app
  \ioverline*{j}{\svalaltalt} \app
  \ioverline*{j}{
    \tmel{\tyctor}{\tyalt}{\pair{\sctxalt}{\ioverline{j-1}{\svalaltalt}}}{\svalaltalt_j}{r}
  }
  & \text{if } \sval = \pair{
      \ctor_i
    }{
      \ioverline*{j}{\svalaltalt}
    } \\
  \bot & \text{otherwise}
\end{cases}
\end{align*}

\begin{align*}
  & \pair{
    \semtm{\tmelim{D}{\ioverline*{i}{c}}}[\sctxalt]
  }{
    \semtm{\tmelim{D}{\ioverline*{i}{c}}}[\sctxaltalt]
  } \\
= & \pair{
\demb{\to}{\mathsf{lfp}_r
\ioverline{k}{\sval} \cons   % indexes
\svalalt             \mapsto % value
e(\svalalt, r)
}}{
\demb{\to}{\mathsf{lfp}_{r'}
\ioverline{k}{\sval} \cons   % indexes
\svalalt             \mapsto % value
e(\svalalt, r')
}} \\
\defeq & \pair{r}{r'}
\end{align*}
This pair exists and is an element of $\dinf$ as it is the least fixed
point of continuous operators in $\dinf$. We name the two components
$r$ and $r'$.

What we need to show is that this pair lies in
\begin{align*}
\sembot{\wfty{\ctx}{
  \tyfun{\left(\ioverline*{k}{\annty{\var}{\ty}}\right)}{
    \tyfun*[{\varalt}]{D \tmApp {\ioverline*{k}{\var}}}{\tyaltalt}
  }
}}[\sctx]
\end{align*}

By \Cref{lem:unref-per-respecting-eliminator}, and the fact that all
unrefined types are complete the least fixed point of the operator
defining the eliminator semantics is contained in the unrefined
interpretation of the type of the eliminator.
\end{itemize}
\end{proof}

\begin{proof}[Proof of \Cref{thm:consistency}]
Now suppose by contradiction that there exists a closed term $\tm$ such
that $\wftm{\ctxemp}{\tm}{\tyall{\var}{\kdtype}{\var}}$, then:
\begin{align*}
    & \pair{\semtm{\tm}[\sunit]}{\semtm{\tm}[\sunit]} \\
    & \text{By \Crefsub{thm:per-respecting-terms}{refined}} \\
\in & \sem{\wfty{\ctxemp}{\tyall{\var}{\kdtype}{\var}}}[\sunit] \\
={} & \begin{setbuild}{\pair{\sval}{\svalalt}}[c]
  \forall \skd \in \sem{\wfkd{\ctxemp}{\kdtype}}[\sunit],\;
  \forall \skd\bots \in \sembot{\wfkd{\ctxemp}{\kdtype}}[\sunit] \\
  \projty{\ctxemp}{\sunit}{\kdtype}{\skd} \subseteq \skd\bots
  \Rightarrow
  \pair{\sval}{\svalalt} \in
  \sem{\wfty{
    \ctxemp \cons \annkd{\var}{\kdtype}
  }{\var}}[\pair{\sunit}{\pair{\skd}{\skd\bots}}]
\end{setbuild} \\
={} & \begin{setbuild}{\pair{\sval}{\svalalt}}[c]
  \forall \skd \in \per{\dinf},\;
  \forall \skd\bots \in \pper{\dinf} \\
  \skd \subseteq \skd\bots
  \Rightarrow
  \pair{\sval}{\svalalt} \in \skd
\end{setbuild}
\end{align*}
Instantiating at $\skd = \emptyset \in \per{\dinf}$ and
$\skd\bots = \{\pair{\bot}{\bot}\} \in \pper{\dinf}$, for which
$\emptyset \subseteq \{\pair{\bot}{\bot}\}$, any
$\pair{\sval}{\svalalt}$ in the set above would have to satisfy
$\pair{\sval}{\svalalt} \in \emptyset$, which is impossible. Hence
$\sem*{\wfty{\ctxemp}{\tyall{\var}{\kdtype}{\var}}}[\sunit] = \emptyset$,
contradicting $\pair{\semtm{\tm}[\sunit]}{\semtm{\tm}[\sunit]}$ being an
element of it.
\end{proof}

\begin{proof}[Proof of \Cref{thm:refined-per-respecting-propositions}]
We proceed by induction on the structure of $\wfprop{\ctx}{\prop}$.
\begin{itemize}
\item \RulePropEq{label}
\begin{align*}
& \semprp{\wfprop{\ctx}{\tm \propeq_\ty \tmalt}}[\sctx] \\
={} & \begin{cases}
    \strue & \text{if } \pair{
      \semtm{\tm}[\projrun{\ctx}{\sctx}]
    }{
      \semtm{\tmalt}[\projrun{\ctx}{\sctx}]
    } \in
    \sembot{\wfty{\ctx}{\ty}}[\projctx{\ctx}{\sctx}] \\
    \sfalse & \text{otherwise}
\end{cases} \\
& \text{By \Crefsub{thm:per-respecting-terms}{refined} and
  \Cref{thm:unrefined-equality-respects-refined-equality}} \\
={} & \begin{cases}
    \strue & \text{if } \pair{
      \semtm{\tm}[\projrun{\ctx}{\sctxalt}]
    }{
      \semtm{\tmalt}[\projrun{\ctx}{\sctxalt}]
    } \in
    \sembot{\wfty{\ctx}{\ty}}[\projctx{\ctx}{\sctxalt}] \\
    \sfalse & \text{otherwise}
  \end{cases} \\
={} & \semprp{\wfprop{\ctx}{\tm \propeq_\ty \tmalt}}[\sctxalt]
\end{align*}

\item \RulePropAnd{label}
\begin{align*}
& \semprp{\wfprop{\ctx}{\prop \propand \propalt}}[\sctx] \\
={}& \begin{cases}
  \semprp{\wfprop{\ctx}{\propalt}}[\sctx]
  & \text{if } \semprp{\wfprop{\ctx}{\prop}}[\sctx] = \strue \\
  \sfalse & \text{otherwise}
\end{cases} \\
& \text{By \ih{}} \\
={} & \begin{cases}
  \semprp{\wfprop{\ctx}{\propalt}}[\sctxalt]
  & \text{if } \semprp{\wfprop{\ctx}{\prop}}[\sctxalt] = \strue \\
  \sfalse & \text{otherwise}
\end{cases} \\
={} & \semprp{\wfprop{\ctx}{\prop \propand \propalt}}[\sctxalt]
\end{align*}

\item \RulePropNot{label}
\begin{align*}
& \semprp{\wfprop{\ctx}{\propnot \prop}}[\sctx] \\
={} & \begin{cases}
  \sfalse & \text{if } \semprp{\wfprop{\ctx}{\prop}}[\sctx] = \strue \\
  \strue & \text{otherwise}
\end{cases} \\
& \text{By \ih{}} \\
={} & \begin{cases}
  \sfalse & \text{if } \semprp{\wfprop{\ctx}{\prop}}[\sctxalt] = \strue \\
  \strue & \text{otherwise}
\end{cases} \\
={} & \semprp{\wfprop{\ctx}{\propnot \prop}}[\sctxalt]
\end{align*}

\end{itemize}
\end{proof}

\begin{proof}[Proof of \Crefsub{thm:subtyping-semantics}{refined}]
By structural induction on the derivation of
$\wfsub{\ctx}{\ty}{\tyalt}$:
\begin{itemize}
\item \RuleSubRefl{label}
\begin{align*}
& \sem{\wfty{\ctx}{\ty}}[\sctx] \\
= & \sem{\wfty{\ctx}{\ty}}[\sctx]
\end{align*}

\item \RuleSubTop{label}
\begin{align*}
& \sem{\wfty{\ctx}{\ty}}[\sctx] \\
= & \begin{setbuild}{
\pair{\sval}{\svalalt} \in \sem{\wfty{\ctx}{\ty}}[\sctx]
}
\semprp{\wfprop{\ctx \cons \annty{\var}{\ty}}{\tmtrue \propeq_\tybool \tmtrue}}[\pair{\sctx}{\sval}]
\end{setbuild} \\
= & \sem{\wfty{\ctx}{\tyref{\var}{\ty}{\tmtrue \propeq_\tybool
\tmtrue}}}[\sctx]
\end{align*}

\item \RuleSubImpl{label}
\begin{align*}
& \sem{\wfty{\ctx}{\tyref{\var}{\ty}{\prop}}}[\sctx] \\
= & \begin{setbuild}{
\pair{\sval}{\svalalt} \in \sem{\wfty{\ctx}{\ty}}[\sctx]
}
\semprp{\wfprop{\ctx \cons \annty{\var}{\ty}}{\prop}}[\pair{\sctx}{\sval}]
\end{setbuild} \\
& \text{by \Cref{ass:smt-soundness}} \\
\subseteq & \begin{setbuild}{
\pair{\sval}{\svalalt} \in \sem{\wfty{\ctx}{\ty}}[\sctx]
}
\semprp{\wfprop{\ctx \cons \annty{\var}{\ty}}{\propalt}}[\pair{\sctx}{\sval}]
\end{setbuild} \\
= & \sem{\wfty{\ctx}{\tyref{\var}{\ty}{\propalt}}}[\sctx]
\end{align*}

\item \RuleSubFun{label}
\begin{align*}
&\sem{\wfty{\ctx}{\tyfun*[\var]{\ty}{\tyaltalt}}}[\sctx] \\
=&
\begin{setbuild}{\pair{\sfun}{\sfunalt}}[lc@{\;}llc@{\;}llr]
    \forall&\pair{\sval}{\svalalt}
    &\in{}& \sem{\wfty{\ctx}{\ty}}[\sctx].\;
    &\pair{\sfun \app \sval}{\sfunalt \app \svalalt} &\in{}&
    \sem{\wfty{
      \ctx \cons \annty{\var}{\ty}
    }{\tyaltalt}}[\pair{\sctx}{\sval}]
  \\
    \forall&\pair{\sval\bots}{\svalalt\bots}
    &\in{}& \sembot{\wfty{\ctx}{\ty}}[\projctx{\ctx}{\sctx}].\;
    &\pair{\sfun \app \sval\bots}{\sfunalt \app \svalalt\bots} &\in{}&
    \sembot{\wfty{
      \ctx \cons \annty{\var}{\ty}
    }{\tyaltalt}}[\projctx{\ctx}{\sctx}]
\end{setbuild} \\
& \text{By \ih{} and \Crefsub{thm:subtyping-semantics}{unrefined}} \\
\subseteq & \begin{setbuild}{\pair{\sfun}{\sfunalt}}[lc@{\;}llc@{\;}llr]
    \forall&\pair{\sval}{\svalalt}
    &\in{}& \sem{\wfty{\ctx}{\tyalt}}[\sctx].\;
    &\pair{\sfun \app \sval}{\sfunalt \app \svalalt} &\in{}&
    \sem{\wfty{
      \ctx \cons \annty{\var}{\ty}
    }{\tyaltalt}}[\pair{\sctx}{\sval}]
  \\
    \forall&\pair{\sval\bots}{\svalalt\bots}
    &\in{}& \sembot{\wfty{\ctx}{\tyalt}}[\projctx{\ctx}{\sctx}].\;
    &\pair{\sfun \app \sval\bots}{\sfunalt \app \svalalt\bots} &\in{}&
    \sembot{\wfty{
      \ctx \cons \annty{\var}{\ty}
    }{\tyaltalt}}[\projctx{\ctx}{\sctx}]
\end{setbuild} \\
& \text{By \ih{} and \Cref{thm:semantic-narrowing}} \\
\subseteq & \begin{setbuild}{\pair{\sfun}{\sfunalt}}[lc@{\;}llc@{\;}llr]
    \forall&\pair{\sval}{\svalalt}
    &\in{}& \sem{\wfty{\ctx}{\tyalt}}[\sctx].\;
    &\pair{\sfun \app \sval}{\sfunalt \app \svalalt} &\in{}&
    \sem{\wfty{
      \ctx \cons \annty{\var}{\tyalt}
    }{\tyaltaltalt}}[\pair{\sctx}{\sval}]
  \\
    \forall&\pair{\sval\bots}{\svalalt\bots}
    &\in{}& \sembot{\wfty{\ctx}{\tyalt}}[\projctx{\ctx}{\sctx}].\;
    &\pair{\sfun \app \sval\bots}{\sfunalt \app \svalalt\bots} &\in{}&
    \sembot{\wfty{
      \ctx \cons \annty{\var}{\tyalt}
    }{\tyaltaltalt}}[\projctx{\ctx}{\sctx}]
\end{setbuild} \\
= & \sem{\wfty{\ctx}{\tyfun*[\var]{\tyalt}{\tyaltaltalt}}}[\sctx]
\end{align*}

\item \RuleSubFlatten{label}

\begin{align*}
& \sem{\wfty{\ctx}{\tyref{\var}{\tyref{\varalt}{\ty}{\propalt}}{\prop}}}[\sctx] \\
= & \begin{setbuild}{
  \pair{\sval}{\svalalt} \in \sem{\wfty{\ctx}{\tyref{\varalt}{\ty}{\propalt}}}[\sctx]
}
  \semprp{\wfprop{\ctx \cons \annty{\var}{\tyref{\varalt}{\ty}{\propalt}}}{\prop}}[\pair{\sctx}{\sval}]
\end{setbuild} \\
= & \begin{setbuild}{
  \pair{\sval}{\svalalt} \in \sem{\wfty{\ctx}{\ty}}[\sctx]
}[c@{\;}r]
  \semprp{\wfprop{\ctx \cons
      \annty{\var}{\tyref{\varalt}{\ty}{\propalt}}}{\prop}}[\pair{\sctx}{\sval}]
  &\text{(i)}\\
  \semprp{\wfprop{\ctx \cons \annty{\varalt}{\ty}}{\propalt}}[\pair{\sctx}{\sval}]
  &\text{(ii)}\\
\end{setbuild} \\
& \text{By \Cref{thm:substitution-refined-prop-tm}, the well-formedness hypothesis} \\
& \text{on the right-hand type, and the definition of } \propand \\
& \text{if (ii) holds then the value is in } \dom{\sem{\wfty{\ctx}{\tyref{\varalt}{\ty}{\propalt}}}[\sctx]}
  \\
= & \begin{setbuild}{
  \pair{\sval}{\svalalt} \in \sem{\wfty{\ctx}{\ty}}[\sctx]
}
\semprp{\wfprop{\ctx \cons \annty{\varalt}{\ty}}{
    \propalt \propand \prop \subst{\varalt}{\var}
}}[\pair{\sctx}{\sval}]
\end{setbuild} \\
  = & \sem{\wfty{\ctx}{\tyref{\varalt}{\ty}{\propalt \propand \prop \subst{\varalt}{\var}}}}[\sctx]
\end{align*}

\item \RuleSubIx{label}:
\begin{align*}
  & \sem{\wfty{\ctx}{\ty \app \ioverline*{k}{\tm}}}[\sctx] \\
= & \sem{\wftm{\ctx}{\ty}{\kdfun{\ioverline*{k}{\tyalt}}{\kdtype}}}[\sctx]
    \app
    \ioverline*{k}{\semtm{\tm}[\projrun{\ctx}{\sctx}]} \\
& \text{By \Crefsub{thm:per-respecting-terms}{refined}}
    \semtm{\tm_k}[\projrun{\ctx}{\sctx}], \semtm{\tmalt_k}[\projrun{\ctx}{\sctx}]
    \in \dom{\sem{\wfty{\ctx}{\ty_k}}[\sctx]} \\
& \text{By \Cref{ass:smt-soundness}}
    \pair{\semtm{\tm_k}[\projrun{\ctx}{\sctx}]}{\semtm{\tmalt_k}[\projrun{\ctx}{\sctx}]}
  \in \sembot{\wfty{\ctx}{\ty_k}}[\projctx{\ctx}{\sctx}], \\
& \text{\Cref{thm:refined-sub-unrefined}, and kind being invariant
  under related arguments} \\
= & \sem{\wftm{\ctx}{\ty}{\kdfun{\ioverline*{k}{\tyalt}}{\kdtype}}}[\sctx]
    \app
    \ioverline*{k}{\semtm{\tmalt}[\projrun{\ctx}{\sctx}]} \\
= & \sem{\wfty{\ctx}{\ty \app \ioverline*{k}{\tmalt}}}[\sctx]
\end{align*}

\item \RuleSubRefine{label}
\begin{align*}
  & \sem{\wfty{\ctx}{\tyref{\var}{\ty}{\prop}}}[\sctx] \\
= & \begin{setbuild}{
  \pair{\sval}{\svalalt} \in \sem{\wfty{\ctx}{\ty}}[\sctx]
}
  \semprp{\wfprop{\ctx \cons \annty{\var}{\ty}}{\prop}}[\pair{\sctx}{\sval}]
\end{setbuild} \\
& \text{By \ih{} and \Crefsub{thm:semantic-narrowing}{prop}} \\
\subseteq & \begin{setbuild}{
  \pair{\sval}{\svalalt} \in \sem{\wfty{\ctx}{\tyalt}}[\sctx]
}
  \semprp{\wfprop{\ctx \cons \annty{\var}{\tyalt}}{\prop}}[\pair{\sctx}{\sval}]
\end{setbuild} \\
= & \sem{\wfty{\ctx}{\tyref{\var}{\tyalt}{\prop}}}[\sctx]
\end{align*}

\item \RuleSubForget{label}
\begin{align*}
& \sem{\wfty{\ctx}{\tyref{\var}{\ty}{\prop}}}[\sctx] \\
= & \begin{setbuild}{
  \pair{\sval}{\svalalt} \in \sem{\wfty{\ctx}{\ty}}[\sctx]
}
  \semprp{\wfprop{\ctx \cons \annty{\var}{\ty}}{\prop}}[\pair{\sctx}{\sval}]
\end{setbuild} \\
\subseteq & \sem{\wfty{\ctx}{\ty}}[\sctx]
\end{align*}

\item \RuleSubTrans{label}
\begin{align*}
& \sem{\wfty{\ctx}{\ty}}[\sctx] \\
& \text{By \ih{} on the first premise, which is well-formed by the second} \\
\subseteq & \sem{\wfty{\ctx}{\tyalt}}[\sctx] \\
& \text{By \ih{} on the third premise} \\
\subseteq & \sem{\wfty{\ctx}{\tyaltalt}}[\sctx]
\end{align*}

\item \RuleSubAll{label}
\begin{align*}
& \sem{\wfty{\ctx}{\tyall{\var}{\kd}{\ty}}}[\sctx] \\
= & \begin{setbuild}{\pair{\sval}{\svalalt}}[c]
    \forall \skd \in \sem{\wfkd{\ctx}{\kd}}[\sctx],\;
    \forall \skd\bots \in
    \sembot{\wfkd{\ctx}{\kd}}[\projctx{\ctx}{\sctx}] \\
    \projty{\ctx}{\sctx}{\kd}{\skd} \subseteq \skd\bots
    \Rightarrow
    \pair{\sval}{\svalalt} \in
    \sem{\wfty{\ctx \cons \annkd{\var}{\kd}}{\ty}}[\pair{\sctx}{\pair{\skd}{\skd\bots}}]
\end{setbuild} \\
& \text{The conditions on $\skd$ and $\skd\bots$ are those of
  $\pair{\sctx}{\pair{\skd}{\skd\bots}} \in \dom{\sem{\wfctx{\ctx
  \cons \annkd{\var}{\kd}}}}$, so by \ih} \\
\subseteq & \begin{setbuild}{\pair{\sval}{\svalalt}}[c]
    \forall \skd \in \sem{\wfkd{\ctx}{\kd}}[\sctx],\;
    \forall \skd\bots \in
    \sembot{\wfkd{\ctx}{\kd}}[\projctx{\ctx}{\sctx}] \\
    \projty{\ctx}{\sctx}{\kd}{\skd} \subseteq \skd\bots
    \Rightarrow
    \pair{\sval}{\svalalt} \in
    \sem{\wfty{\ctx \cons \annkd{\var}{\kd}}{\tyalt}}[\pair{\sctx}{\pair{\skd}{\skd\bots}}]
\end{setbuild} \\
= & \sem{\wfty{\ctx}{\tyall{\var}{\kd}{\tyalt}}}[\sctx]
\end{align*}

\end{itemize}
\end{proof}

\begin{proof}[Proof of \Crefsub{thm:subtyping-semantics}{unrefined}]
By structural induction on the derivation of
$\wfsub{\ctx}{\ty}{\tyalt}$:
\begin{itemize}
\item \RuleSubRefl{label}
\begin{align*}
  & \sembot{\wfty{\ctx}{\ty}}[\sctx] \\
  = & \sembot{\wfty{\ctx}{\ty}}[\sctx]
\end{align*}

\item \RuleSubTop{label}
\begin{align*}
  & \sembot{\wfty{\ctx}{\ty}}[\sctx] \\
  = & \sembot{\wfty{\ctx}{\tyref{\var}{\ty}{\tmtrue \propeq_\tybool
      \tmtrue}}}[\sctx]
\end{align*}
\item \RuleSubImpl{label}
\begin{align*}
  & \sembot{\wfty{\ctx}{\tyref{\var}{\ty}{\prop}}}[\sctx] \\
  = & \sembot{\wfty{\ctx}{\ty}}[\sctx] \\
  = & \sembot{\wfty{\ctx}{\tyref{\var}{\ty}{\propalt}}}[\sctx]
\end{align*}

\item \RuleSubFun{label}
\begin{align*}
  & \sembot{\wfty{\ctx}{\tyfun*[\var]{\ty}{\tyaltalt}}}[\sctx] \\
= & \begin{setbuild}{\pair{\sfun}{\sfunalt}}[l]
    \forall \pair{\sval}{\svalalt} \in \sembot{\wfty{\ctx}{\ty}}[\sctx].\;
    \pair{\sfun \app \sval}{\sfunalt \app \svalalt} \in
    \sembot{\wfty{
      \ctx \cons \annty{\var}{\ty}
    }{\tyaltalt}}[\sctx]
  \end{setbuild} \\
  & \text{By \ih{} on the first premise and \Crefsub{thm:semantic-narrowing}{unrefined}} \\
  & \sembot{\wfty{\ctx \cons \annty{\var}{\ty}}{\tyaltalt}}[\sctx] =
    \sembot{\wfty{\ctx \cons \annty{\var}{\tyalt}}{\tyaltalt}}[\sctx] \\
= & \begin{setbuild}{\pair{\sfun}{\sfunalt}}[l]
    \forall \pair{\sval}{\svalalt} \in \sembot{\wfty{\ctx}{\tyalt}}[\sctx].\;
    \pair{\sfun \app \sval}{\sfunalt \app \svalalt} \in
    \sembot{\wfty{
      \ctx \cons \annty{\var}{\tyalt}
    }{\tyaltalt}}[\sctx]
  \end{setbuild} \\
  & \text{By \ih{} on the second premise} \\
= & \begin{setbuild}{\pair{\sfun}{\sfunalt}}[l]
    \forall \pair{\sval}{\svalalt} \in \sembot{\wfty{\ctx}{\tyalt}}[\sctx].\;
    \pair{\sfun \app \sval}{\sfunalt \app \svalalt} \in
    \sembot{\wfty{
      \ctx \cons \annty{\var}{\tyalt}
    }{\tyaltaltalt}}[\sctx]
  \end{setbuild} \\
= & \sembot{\wfty{\ctx}{\tyfun*[\var]{\tyalt}{\tyaltaltalt}}}[\sctx]
\end{align*}

\item \RuleSubFlatten{label}
\begin{align*}
  & \sembot{\wfty{\ctx}{\tyref{\var}{\tyref{\varalt}{\ty}{\propalt}}{\prop}}}[\sctx] \\
  = & \sembot{\wfty{\ctx}{\tyref{\varalt}{\ty}{\propalt}}}[\sctx] \\
  = & \sembot{\wfty{\ctx}{\ty}}[\sctx] \\
  = & \sembot{\wfty{\ctx}{\tyref{\varalt}{\ty}{\propalt \propand \prop \subst{\varalt}{\var}}}}[\sctx]
\end{align*}

\item \RuleSubIx{label}:
\begin{align*}
  & \sembot{\wfty{\ctx}{\ty \app \ioverline*{k}{\tm}}}[\sctx] \\
= & \sembot{\wftm{\ctx}{\ty}{\kdfun{\ioverline*{k}{\tyalt}}{\kdtype}}}[\sctx] \\
= & \sembot{\wfty{\ctx}{\ty \app \ioverline*{k}{\tmalt}}}[\sctx] \\
\end{align*}

\item \RuleSubRefine{label}:
\begin{align*}
  & \sembot{\wfty{\ctx}{\tyref{\var}{\ty}{\prop}}}[\sctx] \\
= & \sembot{\wfty{\ctx}{\ty}}[\sctx] \\
  & \text{By \ih{}} \\
= & \sembot{\wfty{\ctx}{\tyalt}}[\sctx] \\
= & \sembot{\wfty{\ctx}{\tyref{\var}{\tyalt}{\prop}}}[\sctx]
\end{align*}

\item \RuleSubForget{label}
\begin{align*}
  & \sembot{\wfty{\ctx}{\tyref{\var}{\ty}{\prop}}}[\sctx] \\
= & \sembot{\wfty{\ctx}{\ty}}[\sctx]
\end{align*}

\item \RuleSubTrans{label}
\begin{align*}
  & \sembot{\wfty{\ctx}{\ty}}[\sctx] \\
  & \text{By \ih{} on the first premise, which is well-formed by the second} \\
= & \sembot{\wfty{\ctx}{\tyalt}}[\sctx] \\
  & \text{By \ih{} on the third premise} \\
= & \sembot{\wfty{\ctx}{\tyaltalt}}[\sctx]
\end{align*}

\item \RuleSubAll{label}
\begin{align*}
  & \sembot{\wfty{\ctx}{\tyall{\var}{\kd}{\ty}}}[\sctx] \\
= & \begin{setbuild}{\pair{\sval}{\svalalt}}[l]
    \forall \skd \in \sembot{\wfkd{\ctx}{\kd}}[\sctx].\;
    \pair{\sval}{\svalalt} \in
    \sembot{\wfty{\ctx \cons \annkd{\var}{\kd}}{\ty}}[\pair{\sctx}{\skd}]
  \end{setbuild} \\
  & \text{By \ih{} at each $\pair{\sctx}{\skd} \in \dom{\sembot{\wfctx{\ctx \cons \annkd{\var}{\kd}}}}$} \\
= & \begin{setbuild}{\pair{\sval}{\svalalt}}[l]
    \forall \skd \in \sembot{\wfkd{\ctx}{\kd}}[\sctx].\;
    \pair{\sval}{\svalalt} \in
    \sembot{\wfty{\ctx \cons \annkd{\var}{\kd}}{\tyalt}}[\pair{\sctx}{\skd}]
  \end{setbuild} \\
= & \sembot{\wfty{\ctx}{\tyall{\var}{\kd}{\tyalt}}}[\sctx]
\end{align*}

\end{itemize}
\end{proof}

\begin{proof}[Proof of \Cref{thm:unrefined-semantics-contractive}]
We abbreviate $\ctxaltalt \defeq \ctx \cons \annkd{\var}{\kd} \cons
\ctxalt$ and $\sctxaltalt(X) \defeq \pair{\pair{\sctx}{X}}{\sctxalt}$.
By induction on the derivation of $\wftm{\ctxaltalt}{\ty}{\kdalt}$.

\begin{itemize}

\item \RuleTyNat{label} and \RuleTyBool{label}:
\[
\sembot{\wfty{\ctxaltalt}{\tybase}}[\sctxaltalt(A)]
= \begin{setbuild}{\pair{\demb{\tybase}{b}}{\demb{\tybase}{b}}}
  b \in \tybase
\end{setbuild} \cup \{ \pair{\bot}{\bot} \}
= \sembot{\wfty{\ctxaltalt}{\tybase}}[\sctxaltalt(B)]
\]
The operator is contractive as it is constant.

\item \RuleTyVar{label}:
  $\sembot{\wfty{\ctxaltalt}{\varalt}}[\sctxaltalt(A)] = \sctxaltalt(A)(\varalt)$
  and $\sembot{\wfty{\ctxaltalt}{\varalt}}[\sctxaltalt(B)] = \sctxaltalt(B)(\varalt)$,
  if $\var = \varalt$ then the operator is the identity, otherwise it
  is constant and hence contractive.

\item \RuleTyFun{label}:
\begin{multline*}
  \sembot{\wfty{\ctxaltalt}{\tyfun*[\varalt]{\ty}{\tyalt}}}[\sctxaltalt(C)] \\
= \begin{setbuild}{\pair{\sfun}{\sfunalt}}[l]
    \forall \pair{\sval}{\svalalt} \in
    \sembot{\wfty{\ctxaltalt}{\ty}}[\sctxaltalt(C)].\;
    \pair{\sfun \app \sval}{\sfunalt \app \svalalt} \in
    \sembot{\wfty{
      \ctxaltalt \cons \annty{\varalt}{\ty}
    }{\tyalt}}[\sctxaltalt(C)]
\end{setbuild}
\end{multline*}

Let $\pair{\sfun}{\sfunalt} \in
\sembot{\wfty{\ctxaltalt}{\tyfun*[\varalt]{\ty}{\tyalt}}}[\sctxaltalt(A)]$;
we show that $\pair*{\pi_{n + 1}(\sfun)}{\pi_{n + 1}(\sfunalt)} \in
\sembot*{\wfty{\ctxaltalt}{\tyfun*[\varalt]{\ty}{\tyalt}}}[\sctxaltalt(B)]$.
\begin{align*}
         & \pair{\sval}{\svalalt}
           \in \sembot{\wfty{\ctxaltalt}{\ty}}[\sctxaltalt(B)] \\
         & \text{By the \ih{} they agree at least up to level $n$} \\
\implies & \pair{\pi_n(\sval)}{\pi_n(\svalalt)}
           \in \sembot{\wfty{\ctxaltalt}{\ty}}[\sctxaltalt(A)] \\
\implies & \pair{\sfun \app \pi_n(\sval)}{\sfunalt \app \pi_n(\svalalt)}
           \in \sembot{\wfty{\ctxaltalt \cons \annty{\varalt}{\ty}}{\tyalt}}[\sctxaltalt(A)] \\
         & \text{By the \ih{} they agree at least up to level $n$} \\
\implies & \pair*{\pi_n(\sfun \app \pi_n(\sval))}{\pi_n(\sfunalt \app \pi_n(\svalalt))}
           \in \sembot*{\wfty{\ctxaltalt \cons \annty{\varalt}{\ty}}{\tyalt}}[\sctxaltalt(B)] \\
         & \text{By the fact that } \pi_{n+1}(\sfun)\app\sval = \pi_n(\sfun\app\pi_n(\sval)) \text{ in the bilimit} \\
\implies & \pair*{\pi_n(\sfun \app \pi_n(\sval))}{\pi_n(\sfunalt \app \pi_n(\svalalt))}
           \in \sembot*{\wfty{\ctxaltalt \cons \annty{\varalt}{\ty}}{\tyalt}}[\sctxaltalt(B)] \\
\implies & \pair{\pi_{n + 1}(\sfun) \app \sval}{\pi_{n + 1}(\sfunalt) \app \svalalt}
           \in \sembot*{\wfty{\ctxaltalt \cons \annty{\varalt}{\ty}}{\tyalt}}[\sctxaltalt(B)] \\
\implies & \pair{\pi_{n + 1}(\sfun)}{\pi_{n + 1}(\sfunalt)}
           \in \sembot{\wfty{\ctxaltalt}{\tyfun*[\varalt]{\ty}{\tyalt}}}[\sctxaltalt(B)]
\end{align*}
The other direction is symmetric. Hence the operator is contractive.

\item \RuleTyAll{label}:
\begin{multline*}
\sembot{\wfty{\ctxaltalt}{\tyall{\varalt}{\kdaltalt}{\ty}}}[\sctxaltalt(C)] \\
= \begin{setbuild}{\pair{\sval}{\svalalt}}[l]
  \forall \skd
    \in
  \sembot{\wfkd{\ctxaltalt}{\kdaltalt}}[\sctxaltalt].\;
  \pair{\sval}{\svalalt}
    \in
  \sembot{
    \wfty{
      \ctxaltalt \cons \annkd{\varalt}{\kdaltalt}
    }{\ty}
  }[\pair{\sctxaltalt(C)}{\skd}]
\end{setbuild}
\end{multline*}

We can apply the \ih{} on each instance of the body operator; the
alternative that holds depends only on the derivation of $\ty$, so it
is the same for every $\skd$.

If every instance is the identity, then the whole operator is the
identity. Otherwise every instance is contractive. Hence for every
$\skd$ the corresponding body interpretations agree at level $n+1$.

\item \RuleTyRef{label}:
\[\sembot{\wfty{\ctxaltalt}{\tyref{\varalt}{\ty}{\prop}}}[\sctxaltalt(C)] \\
  = \sembot{\wfty{\ctxaltalt}{\ty}}[\sctxaltalt(C)]\]
By the \ih{} the operator is either the identity or contractive.

\item \RuleTyApp{label}:
\[\sembot{\wftm{\ctxaltalt}{\ty \app \tm}{\kd \subst{\tm}{\varalt}}}[\sctxaltalt(C)]
  = \sembot{\wftm{\ctxaltalt}{\tyalt}{\tyfun*[\var]{\ty}{\kd}}}[\sctxaltalt(C)] \]
By the \ih{} the operator is either the identity or contractive.

\item \RuleKdIndPos{label} and \RuleKdIndStrat{label}:
\begin{multline*}
  \sembot{\wftm{\ctxaltalt}{\decl}{\ioverline*{k}{\ty} \to \kdtype}}[\sctxaltalt(C)] \\
  = \mathsf{fix}_X \scalebox{0.9}{$\displaystyle
  {\bigcup_i
  \begin{setbuild}{\pair{
    \pair{\ctor_i}{\ioverline*{j}{\sval}}
  }{
    \pair{\ctor_i}{\ioverline*{j}{\svalalt}}
  }}
    \pair{\sval_j}{\svalalt_j} \in \sembot{\wfty{
      \ctxaltalt \cons
      \annkd{\tyctor}{\kdfun{\ioverline*{k}{\ty}}{\kdtype}} \cons
      \ioverline*{j - 1}{\annty{\var}{\tyalt}}
    }{\tyalt_j}}[\pair{\sctxaltalt(C)}{X}]
  \end{setbuild} \cup \left\{ \pair{\bot}{\bot} \right\}}$}
\end{multline*}

Let $\Phi_C(X)$ denote the operator under the fixed point computed in
$\sctxaltalt(C)$. We first show that if $A_n = B_n$ and $X_n = Y_n$
then $\Phi_A(X)_{n + 1} = \Phi_B(Y)_{n + 1}$: let $
\pair{\pair{\ctor_i}{\ioverline*{j}{\sval_j}}}{\pair{\ctor_i}{\ioverline*{j}{\svalalt_j}}}
\in \Phi_A(X)$, we show that its $\pair{\pi_{n +
1}(\pair{\ctor_i}{\ioverline*{j}{\sval_j}})}{\pi_{n +
1}(\pair{\ctor_i}{\ioverline*{j}{\svalalt_j}})}$ is in $\Phi_B(Y)$.

\begin{align*}
  & \pair{\sval_j}{\svalalt_j}
    \in \sembot*{\wfty{
      \ctxaltalt \cons
      \annkd{\tyctor}{\kdfun{\ioverline*{k}{\ty}}{\kdtype}} \cons
      \ioverline*{j - 1}{\annty{\var}{\tyalt}}
    }{\tyalt_j}}[\pair*{\sctxaltalt(A)}{X}] \\
  & \text{By the \ih{}, applied once for $\var$ and once for
    $\tyctor$, they agree at least up to level $n$} \\
\implies & \pair*{\pi_n(\sval_j)}{\pi_n(\svalalt_j)}
    \in \sembot*{\wfty{
      \ctxaltalt \cons
      \annkd{\tyctor}{\kdfun{\ioverline*{k}{\ty}}{\kdtype}} \cons
      \ioverline*{j - 1}{\annty{\var}{\tyalt}}
    }{\tyalt_j}}[\pair*{\sctxaltalt(B)}{Y}] \\
  & \text{By \Cref{lem:tagged-tuple-projection} and the fact that
    $\pi_n(a) = \pi_n(b)$ implies $\pi_m(a) = \pi_m(b)$ for $m \le n$}
  \\
\implies & \pair{
  \pi_{n+1}
  \!\left(
    \pair{\ctor_i}{\ioverline*{j}{\sval_j}}
  \right)
}{
  \pi_{n+1}
  \!\left(
    \pair{\ctor_i}{\ioverline*{j}{\svalalt_j}}
  \right)
} \in \Phi_B(Y)
\end{align*}

The other direction is symmetric. Hence if $A_n = B_n$ and $X_n =
Y_n$ then $\Phi_A(X)_{n + 1} = \Phi_B(Y)_{n + 1}$; in particular,
taking $A = B$, each $\Phi_C$ is contractive.

We show by induction on $m$ that $A_n = B_n$ implies
$\Phi_A^m(\{\pair{\bot}{\bot}\})_{n + 1} =
\Phi_B^m(\{\pair{\bot}{\bot}\})_{n + 1}$: the base case is trivial,
and the inductive step follows from the previous argument, as
agreement at level $n + 1$ implies agreement at level $n$.

Since all CUPERs agree at level $0$, contractivity of $\Phi_C$ gives
$(\Phi_C^m(S))_m = (\Phi_C^m(\{\pair{\bot}{\bot}\}))_m$ for every $S$
by induction on $m$; instantiating $S$ with the fixed point,
$(\mathsf{fix}_X\,\Phi_C(X))_m = (\Phi_C^m(\{\pair{\bot}{\bot}\}))_m$.
Taking $m = n + 1$, the two fixed points agree at level $n + 1$.
Hence the operator is contractive.
\end{itemize}
\end{proof}

\begin{proof}[Proof of \Cref{thm:monotonicity-types}]
We use the following abbreviations
\begin{itemize}
\item $\ctxalt \defeq \ctx \cons
  \annkd{\tyctor}{\kd}
  \cons \ioverline{i}{\ann{\var}{\tyalt}}$
\item $\sctxalt(x) = \pair{
  \pair{\sctx}{\pair{x}{\skd\bots}}
}{
  \ioverline{i}{\sval}
}$
\end{itemize}

by induction on the structure of $\wftm{\ctxalt}{\tyaltalt}{\kdalt}$,
\begin{itemize}

\item \RuleTyNat{label} and \RuleTyBool{label} when $p \in \{ \pospos, \posneg \}$
  \[
      \sem{\wfty{\ctxalt}{\tybase}}[\sctxalt(\skd)]
    = \sem{\wfty{\ctxalt}{\tybase}}[\sctxalt(\skdalt)]
  \]
\item \RuleTyFun{label} when $p = \pospos$
\begin{align*}
  & \sem{\wfty{\ctxalt}{\tyfun*[\varalt]{\tyalt}{\tyaltalt}}}[\sctxalt(\skd)] \\
= & \begin{setbuild}{\pair{\sfun}{\sfunalt}}[l]
      \forall \pair{\sval}{\svalalt} \in \sem{\wfty{\ctxalt}{\tyalt}}[\sctxalt(\skd)].\;
      \pair{\sfun \app \sval}{\sfunalt \app \svalalt} \in
      \sem{\wfty{\ctxalt \cons \annty{\varalt}{\tyalt}}{\tyaltalt}}[\pair{\sctxalt(\skd)}{\sval}] \\
      \forall \pair{\sval\bots}{\svalalt\bots} \in
      \sembot{\wfty{\ctxalt}{\tyalt}}[\projctx{\ctxalt}{\sctxalt(\skd)}].\;
      \pair{\sfun \app \sval\bots}{\sfunalt \app \svalalt\bots} \in
      \sembot{\wfty{
        \ctxalt \cons \annty{\varalt}{\tyalt}
      }{\tyaltalt}}[\projctx{\ctxalt}{\sctxalt(\skd)}]
  \end{setbuild} \\
  & \text{By \ih{}} \sem{\wfty{\ctxalt}{\tyalt}}[\sctxalt(\skdalt)] \subseteq
    \sem{\wfty{\ctxalt}{\tyalt}}[\sctxalt(\skd)] \\
\subseteq & \begin{setbuild}{\pair{\sfun}{\sfunalt}}[l]
      \forall \pair{\sval}{\svalalt} \in \sem{\wfty{\ctxalt}{\tyalt}}[\sctxalt(\skdalt)].\;
      \pair{\sfun \app \sval}{\sfunalt \app \svalalt} \in
      \sem{\wfty{\ctxalt \cons \annty{\varalt}{\tyalt}}{\tyaltalt}}[\pair{\sctxalt(\skd)}{\sval}] \\
      \forall \pair{\sval\bots}{\svalalt\bots} \in
      \sembot{\wfty{\ctxalt}{\tyalt}}[\projctx{\ctxalt}{\sctxalt(\skd)}].\;
      \pair{\sfun \app \sval\bots}{\sfunalt \app \svalalt\bots} \in
      \sembot{\wfty{
        \ctxalt \cons \annty{\varalt}{\tyalt}
      }{\tyaltalt}}[\projctx{\ctxalt}{\sctxalt(\skd)}]
  \end{setbuild} \\
& \text{Notice } \projctx{\ctxalt}{\sctxalt(\skd)} = \projctx{\ctxalt}{\sctxalt(\skdalt)} \\
= & \begin{setbuild}{\pair{\sfun}{\sfunalt}}[l]
      \forall \pair{\sval}{\svalalt} \in \sem{\wfty{\ctxalt}{\tyalt}}[\sctxalt(\skdalt)].\;
      \pair{\sfun \app \sval}{\sfunalt \app \svalalt} \in
      \sem{\wfty{\ctxalt \cons \annty{\varalt}{\tyalt}}{\tyaltalt}}[\pair{\sctxalt(\skd)}{\sval}] \\
      \forall \pair{\sval\bots}{\svalalt\bots} \in
      \sembot{\wfty{\ctxalt}{\tyalt}}[\projctx{\ctxalt}{\sctxalt(\skdalt)}].\;
      \pair{\sfun \app \sval\bots}{\sfunalt \app \svalalt\bots} \in
      \sembot{\wfty{
        \ctxalt \cons \annty{\varalt}{\tyalt}
      }{\tyaltalt}}[\projctx{\ctxalt}{\sctxalt(\skdalt)}]
  \end{setbuild} \\
& \text{By \ih{}} \sem{\wfty{\ctxalt \cons \annty{\varalt}{\tyalt}}{\tyaltalt}}[\pair{\sctxalt(\skdalt)}{\sval}] \supseteq
    \sem{\wfty{\ctxalt \cons \annty{\varalt}{\tyalt}}{\tyaltalt}}[\pair{\sctxalt(\skd)}{\sval}] \\
\subseteq & \begin{setbuild}{\pair{\sfun}{\sfunalt}}[l]
      \forall \pair{\sval}{\svalalt} \in \sem{\wfty{\ctxalt}{\tyalt}}[\sctxalt(\skdalt)].\;
      \pair{\sfun \app \sval}{\sfunalt \app \svalalt} \in
      \sem{\wfty{\ctxalt \cons \annty{\varalt}{\tyalt}}{\tyaltalt}}[\pair{\sctxalt(\skdalt)}{\sval}] \\
      \forall \pair{\sval\bots}{\svalalt\bots} \in
      \sembot{\wfty{\ctxalt}{\tyalt}}[\projctx{\ctxalt}{\sctxalt(\skdalt)}].\;
      \pair{\sfun \app \sval\bots}{\sfunalt \app \svalalt\bots} \in
      \sembot{\wfty{
        \ctxalt \cons \annty{\varalt}{\tyalt}
      }{\tyaltalt}}[\projctx{\ctxalt}{\sctxalt(\skdalt)}]
  \end{setbuild} \\
= & \sem{\wfty{\ctxalt}{\tyfun*[\varalt]{\tyalt}{\tyaltalt}}}[\sctxalt(\skdalt)]
\end{align*}

\item \RuleTyFun{label} when $p = \posneg$
\begin{align*}
  & \sem{\wfty{\ctxalt}{\tyfun*[\varalt]{\tyalt}{\tyaltalt}}}[\sctxalt(\skdalt)] \\
= & \begin{setbuild}{\pair{\sfun}{\sfunalt}}[l]
      \forall \pair{\sval}{\svalalt} \in \sem{\wfty{\ctxalt}{\tyalt}}[\sctxalt(\skdalt)].\;
      \pair{\sfun \app \sval}{\sfunalt \app \svalalt} \in
      \sem{\wfty{\ctxalt \cons \annty{\varalt}{\tyalt}}{\tyaltalt}}[\pair{\sctxalt(\skdalt)}{\sval}] \\
      \forall \pair{\sval\bots}{\svalalt\bots} \in
      \sembot{\wfty{\ctxalt}{\tyalt}}[\projctx{\ctxalt}{\sctxalt(\skdalt)}].\;
      \pair{\sfun \app \sval\bots}{\sfunalt \app \svalalt\bots} \in
      \sembot{\wfty{
        \ctxalt \cons \annty{\varalt}{\tyalt}
      }{\tyaltalt}}[\projctx{\ctxalt}{\sctxalt(\skdalt)}]
  \end{setbuild} \\
& \text{By \ih{}} \sem{\wfty{\ctxalt}{\tyalt}}[\sctxalt(\skd)] \subseteq
    \sem{\wfty{\ctxalt}{\tyalt}}[\sctxalt(\skdalt)] \\
\subseteq & \begin{setbuild}{\pair{\sfun}{\sfunalt}}[l]
      \forall \pair{\sval}{\svalalt} \in \sem{\wfty{\ctxalt}{\tyalt}}[\sctxalt(\skd)].\;
      \pair{\sfun \app \sval}{\sfunalt \app \svalalt} \in
      \sem{\wfty{\ctxalt \cons \annty{\varalt}{\tyalt}}{\tyaltalt}}[\pair{\sctxalt(\skdalt)}{\sval}] \\
      \forall \pair{\sval\bots}{\svalalt\bots} \in
      \sembot{\wfty{\ctxalt}{\tyalt}}[\projctx{\ctxalt}{\sctxalt(\skdalt)}].\;
      \pair{\sfun \app \sval\bots}{\sfunalt \app \svalalt\bots} \in
      \sembot{\wfty{
        \ctxalt \cons \annty{\varalt}{\tyalt}
      }{\tyaltalt}}[\projctx{\ctxalt}{\sctxalt(\skdalt)}]
  \end{setbuild} \\
& \text{Notice } \projctx{\ctxalt}{\sctxalt(\skd)} =
  \projctx{\ctxalt}{\sctxalt(\skdalt)} \\
= & \begin{setbuild}{\pair{\sfun}{\sfunalt}}[l]
      \forall \pair{\sval}{\svalalt} \in \sem{\wfty{\ctxalt}{\tyalt}}[\sctxalt(\skd)].\;
      \pair{\sfun \app \sval}{\sfunalt \app \svalalt} \in
      \sem{\wfty{\ctxalt \cons \annty{\varalt}{\tyalt}}{\tyaltalt}}[\pair{\sctxalt(\skdalt)}{\sval}] \\
      \forall \pair{\sval\bots}{\svalalt\bots} \in
      \sembot{\wfty{\ctxalt}{\tyalt}}[\projctx{\ctxalt}{\sctxalt(\skd)}].\;
      \pair{\sfun \app \sval\bots}{\sfunalt \app \svalalt\bots} \in
      \sembot{\wfty{
        \ctxalt \cons \annty{\varalt}{\tyalt}
      }{\tyaltalt}}[\projctx{\ctxalt}{\sctxalt(\skd)}]
  \end{setbuild} \\
& \text{By \ih{}} \sem{\wfty{\ctxalt \cons \annty{\varalt}{\tyalt}}{\tyaltalt}}[\pair{\sctxalt(\skd)}{\sval}] \supseteq
    \sem{\wfty{\ctxalt \cons \annty{\varalt}{\tyalt}}{\tyaltalt}}[\pair{\sctxalt(\skdalt)}{\sval}] \\
\subseteq & \begin{setbuild}{\pair{\sfun}{\sfunalt}}[l]
      \forall \pair{\sval}{\svalalt} \in \sem{\wfty{\ctxalt}{\tyalt}}[\sctxalt(\skd)].\;
      \pair{\sfun \app \sval}{\sfunalt \app \svalalt} \in
      \sem{\wfty{\ctxalt \cons \annty{\varalt}{\tyalt}}{\tyaltalt}}[\pair{\sctxalt(\skd)}{\sval}] \\
      \forall \pair{\sval\bots}{\svalalt\bots} \in
      \sembot{\wfty{\ctxalt}{\tyalt}}[\projctx{\ctxalt}{\sctxalt(\skd)}].\;
      \pair{\sfun \app \sval\bots}{\sfunalt \app \svalalt\bots} \in
      \sembot{\wfty{
        \ctxalt \cons \annty{\varalt}{\tyalt}
      }{\tyaltalt}}[\projctx{\ctxalt}{\sctxalt(\skd)}]
  \end{setbuild} \\
= & \sem{\wfty{\ctxalt}{\tyfun*[\varalt]{\tyalt}{\tyaltalt}}}[\sctxalt(\skd)]
\end{align*}

\item \RuleTyAll{label} when $p = \pospos$

\begin{align*}
  & \sem{\wfty{\ctxalt}{\tyall{\var}{\kd}{\tyalt}}}[\sctxalt(\skd)] \\
= & \begin{setbuild}{\pair{\sval}{\svalalt}}[c]
  \forall \skdaltalt \in \sem{\wfkd{\ctxalt}{\kd}}[\sctxalt(\skd)],\;
  \forall \skdaltalt\bots \in
  \sembot{\wfkd{\ctxalt}{\kd}}[\projctx{\ctxalt}{\sctxalt(\skd)}] \\
  \projty{\ctxalt}{\sctxalt(\skd)}{\kd}{\skdaltalt} \subseteq \skdaltalt\bots
  \Rightarrow
  \pair{\sval}{\svalalt} \in
  \sem{\wfty{
    \ctxalt \cons \annkd{\var}{\kd}
  }{\tyalt}}[\pair{\sctxalt(\skd)}{\pair{\skdaltalt}{\skdaltalt\bots}}]
\end{setbuild} \\
& \text{By \Cref{thm:monotonicity-kinds} and }
\projctx{\ctxalt}{\sctxalt(\skd)} =
\projctx{\ctxalt}{\sctxalt(\skdalt)} \\
\subseteq & \begin{setbuild}{\pair{\sval}{\svalalt}}[c]
  \forall \skdaltalt \in \sem{\wfkd{\ctxalt}{\kd}}[\sctxalt(\skdalt)],\;
  \forall \skdaltalt\bots \in
  \sembot{\wfkd{\ctxalt}{\kd}}[\projctx{\ctxalt}{\sctxalt(\skdalt)}] \\
  \projty{\ctxalt}{\sctxalt(\skdalt)}{\kd}{\skdaltalt} \subseteq \skdaltalt\bots
  \Rightarrow
  \pair{\sval}{\svalalt} \in
  \sem{\wfty{
    \ctxalt \cons \annkd{\var}{\kd}
  }{\tyalt}}[\pair{\sctxalt(\skd)}{\pair{\skdaltalt}{\skdaltalt\bots}}]
\end{setbuild} \\
& \text{By \ih{}} \\
\subseteq & \begin{setbuild}{\pair{\sval}{\svalalt}}[c]
  \forall \skdaltalt \in \sem{\wfkd{\ctxalt}{\kd}}[\sctxalt(\skdalt)],\;
  \forall \skdaltalt\bots \in
  \sembot{\wfkd{\ctxalt}{\kd}}[\projctx{\ctxalt}{\sctxalt(\skdalt)}] \\
  \projty{\ctxalt}{\sctxalt(\skdalt)}{\kd}{\skdaltalt} \subseteq \skdaltalt\bots
  \Rightarrow
  \pair{\sval}{\svalalt} \in
  \sem{\wfty{
    \ctxalt \cons \annkd{\var}{\kd}
  }{\tyalt}}[\pair{\sctxalt(\skdalt)}{\pair{\skdaltalt}{\skdaltalt\bots}}]
\end{setbuild} \\
= & \sem{\wfty{\ctxalt}{\tyall{\var}{\kd}{\tyalt}}}[\sctxalt(\skdalt)]
\end{align*}

\item \RuleTyAll{label} when $p = \posneg$

\begin{align*}
  & \sem{\wfty{\ctxalt}{\tyall{\var}{\kd}{\tyalt}}}[\sctxalt(\skdalt)] \\
= & \begin{setbuild}{\pair{\sval}{\svalalt}}[c]
  \forall \skdaltalt \in \sem{\wfkd{\ctxalt}{\kd}}[\sctxalt(\skdalt)],\;
  \forall \skdaltalt\bots \in
  \sembot{\wfkd{\ctxalt}{\kd}}[\projctx{\ctxalt}{\sctxalt(\skdalt)}] \\
  \projty{\ctxalt}{\sctxalt(\skdalt)}{\kd}{\skdaltalt} \subseteq \skdaltalt\bots
  \Rightarrow
  \pair{\sval}{\svalalt} \in
  \sem{\wfty{
    \ctxalt \cons \annkd{\var}{\kd}
  }{\tyalt}}[\pair{\sctxalt(\skdalt)}{\pair{\skdaltalt}{\skdaltalt\bots}}]
\end{setbuild} \\
& \text{By \Cref{thm:monotonicity-kinds} and }
\projctx{\ctxalt}{\sctxalt(\skdalt)} =
\projctx{\ctxalt}{\sctxalt(\skd)} \\
\subseteq & \begin{setbuild}{\pair{\sval}{\svalalt}}[c]
  \forall \skdaltalt \in \sem{\wfkd{\ctxalt}{\kd}}[\sctxalt(\skd)],\;
  \forall \skdaltalt\bots \in
  \sembot{\wfkd{\ctxalt}{\kd}}[\projctx{\ctxalt}{\sctxalt(\skd)}] \\
  \projty{\ctxalt}{\sctxalt(\skd)}{\kd}{\skdaltalt} \subseteq \skdaltalt\bots
  \Rightarrow
  \pair{\sval}{\svalalt} \in
  \sem{\wfty{
    \ctxalt \cons \annkd{\var}{\kd}
  }{\tyalt}}[\pair{\sctxalt(\skdalt)}{\pair{\skdaltalt}{\skdaltalt\bots}}]
\end{setbuild} \\
& \text{By \ih{}} \\
\subseteq & \begin{setbuild}{\pair{\sval}{\svalalt}}[c]
  \forall \skdaltalt \in \sem{\wfkd{\ctxalt}{\kd}}[\sctxalt(\skd)],\;
  \forall \skdaltalt\bots \in
  \sembot{\wfkd{\ctxalt}{\kd}}[\projctx{\ctxalt}{\sctxalt(\skd)}] \\
  \projty{\ctxalt}{\sctxalt(\skd)}{\kd}{\skdaltalt} \subseteq \skdaltalt\bots
  \Rightarrow
  \pair{\sval}{\svalalt} \in
  \sem{\wfty{
    \ctxalt \cons \annkd{\var}{\kd}
  }{\tyalt}}[\pair{\sctxalt(\skd)}{\pair{\skdaltalt}{\skdaltalt\bots}}]
\end{setbuild} \\
= & \sem{\wfty{\ctxalt}{\tyall{\var}{\kd}{\tyalt}}}[\sctxalt(\skd)]
\end{align*}

\item \RuleTyVar{label} when $p \in \{ \pospos, \posneg \}$
\begin{align*}
  & \sem{\wfty{\ctxalt}{\var}}[\sctxalt(\skd)] \\
= &\; \sctxalt(\skd)(\var) \\
  & \text{By assumption} \\
\subseteq / \supseteq &\; \sctxalt(\skdalt)(\var) \\
= & \sem{\wfty{\ctxalt}{\var}}[\sctxalt(\skdalt)]
\end{align*}

\item \RuleTyRef{label} when $p \in \{ \pospos, \posneg \}$
\begin{align*}
  & \sem{\wfty{\ctxalt}{\tyalt}}[\sctxalt(\skd)] \\
= & \begin{setbuild}{
  \pair{\sval}{\svalalt} \in \sem{\wfty{\ctxalt}{\ty}}[\sctxalt(\skd)]
}
  \semprp{\wfprop{
    \ctxalt \cons \annty{\var}{\ty}
  }{\prop}}[\pair{\sctxalt(\skd)}{\sval}]
\end{setbuild} \\
  & \text{Notice that $\semprp{-}$ is invariant under kind assignments} \\
= & \begin{setbuild}{
  \pair{\sval}{\svalalt} \in \sem{\wfty{\ctxalt}{\ty}}[\sctxalt(\skd)]
}
  \semprp{\wfprop{
    \ctxalt \cons \annty{\var}{\ty}
  }{\prop}}[\pair{\sctxalt(\skdalt)}{\sval}]
\end{setbuild} \\
& \text{By \ih{}} \\
\subseteq / \supseteq
& \begin{setbuild}{
  \pair{\sval}{\svalalt} \in \sem{\wfty{\ctxalt}{\ty}}[\sctxalt(\skdalt)]
}
  \semprp{\wfprop{
    \ctxalt \cons \annty{\var}{\ty}
  }{\prop}}[\pair{\sctxalt(\skdalt)}{\sval}]
\end{setbuild}
= \sem{\wfty{\ctxalt}{\tyalt}}[\sctxalt(\skdalt)]
\end{align*}

\item \RuleTyApp{label} when $p \in \{ \pospos, \posneg \}$
\begin{align*}
  & \sem{\wftm{\ctxalt}{{\tyalt} \app {\tm}}{\kd}}[\sctxalt(\skd)] \\
= & \sem{\wftm{\ctxalt}{\tyalt}{\kdfun*[\var]{\ty}{\kd}}}[\sctxalt(\skd)] \app \semtm{\tm}[\projrun{\ctxalt}{\sctxalt(\skd)}] \\
  & \text{By noticing that } \projrun{\ctxalt}{\sctxalt(\skd)} = \projrun{\ctxalt}{\sctxalt(\skdalt)} \\
= & \sem{\wftm{\ctxalt}{\tyalt}{\kdfun*[\var]{\ty}{\kd}}}[\sctxalt(\skd)] \app \semtm{\tm}[\projrun{\ctxalt}{\sctxalt(\skdalt)}] \\
  & \text{By \ih{}} \\
\leq / \geq & \sem{\wftm{\ctxalt}{\tyalt}{\kdfun*[\var]{\ty}{\kd}}}[\sctxalt(\skdalt)] \app \semtm{\tm}[\projrun{\ctxalt}{\sctxalt(\skdalt)}] \\
= & \sem{\wfty{\ctxalt}{{\tyalt} \app {\tm}}}[\sctxalt(\skdalt)]
\end{align*}

\item \RuleKdIndPos{label} when $p \in \{\pospos, \posneg \}$
Since
\begin{align*}
  & \sem{\wftm{\ctxalt}{\decl}{\ioverline*{k}{\ty} \to \kdtype}}[\sctxalt(\skd)] \\
= & \;\mathsf{lfp}_X \left(\scalebox{0.5}{$\displaystyle
      \ioverline{k}{\svalaltalt} \mapsto
      \Bigcup_i \begin{setbuild}{\pair{
        \pair{\ctor_i}{\ioverline{j}{\sval}}
      }{
        \pair{\ctor_i}{\ioverline{j}{\svalalt}}
      }}
        \pair{\sval_j}{\svalalt_j} \in \sem{\wfty{
          \ctxalt \cons
          \annkd{\tyctor}{\kdfun{\ioverline{k}{\ty}}{\kdtype}} \cons
          \ioverline{j - 1}{\annty{\var}{\tyalt}}
        }{\tyalt_j}}[\pair{\pair{\sctxalt(\skd)}{\pair{X}{
            \sembot{\wftm{\ctxalt}{\decl}{\kdfun{\ioverline{k}{\ty}}{\kdtype}}}[\projctx{\ctxalt}{\sctxalt(\skd)}]
          }}}{\ioverline{j - 1}{\sval}}] \\
      \pair{\svalaltalt_k}{\semtm{\tm_k}[\pair{\projrun{\ctxalt}{\sctxalt(\skd)}}{\ioverline*{j}{\sval}}]}
      \in \sembot{\wfty{
        \ctxalt \cons \ioverline*{k - 1}{\annty{\var}{\ty}}
      }{\ty_k}}[\projctx{\ctxalt}{\sctxalt(\skd)}]
      \end{setbuild}
    $}\right)
\end{align*}
\begin{align*}
  & \sem{\wftm{\ctxalt}{\decl}{\ioverline*{k}{\ty} \to \kdtype}}[\sctxalt(\skdalt)] \\
= & \;\mathsf{lfp}_X \left(\scalebox{0.5}{$\displaystyle
      \ioverline{k}{\svalaltalt} \mapsto
      \Bigcup_i \begin{setbuild}{\pair{
        \pair{\ctor_i}{\ioverline{j}{\sval}}
      }{
        \pair{\ctor_i}{\ioverline{j}{\svalalt}}
      }}
        \pair{\sval_j}{\svalalt_j} \in \sem{\wfty{
          \ctxalt \cons
          \annkd{\tyctor}{\kdfun{\ioverline{k}{\ty}}{\kdtype}} \cons
          \ioverline{j - 1}{\annty{\var}{\tyalt}}
        }{\tyalt_j}}[\pair{\pair{\sctxalt(\skdalt)}{\pair{X}{
            \sembot{\wftm{\ctxalt}{\decl}{\kdfun{\ioverline{k}{\ty}}{\kdtype}}}[\projctx{\ctxalt}{\sctxalt(\skdalt)}]
          }}}{\ioverline{j - 1}{\sval}}] \\
      \pair{\svalaltalt_k}{\semtm{\tm_k}[\pair{\projrun{\ctxalt}{\sctxalt(\skdalt)}}{\ioverline*{j}{\sval}}]}
      \in \sembot{\wfty{
        \ctxalt \cons \ioverline*{k - 1}{\annty{\var}{\ty}}
      }{\ty_k}}[\projctx{\ctxalt}{\sctxalt(\skdalt)}]
      \end{setbuild}
    $}\right)
\end{align*}

Since they are both defined as the least fixed point of a monotone
functional, it suffices to show that for all $X$ the functionals
respect $\leq / \geq$ pointwise; for this it is enough to show
that each component of the functional respects the order $\leq / \geq$:
\[ \scalebox{0.6}{$\begin{array}{c}
\sem{\wfty{
  \ctxalt \cons
  \annkd{\tyctor}{\kdfun{\ioverline*{k}{\ty}}{\kdtype}} \cons
  \ioverline*{j - 1}{\annty{\var}{\tyalt}}
}{\tyalt_j}}[\pair{\pair{\sctxalt(\skd)}{\pair{X}{
    \sembot{\wftm{\ctxalt}{\decl}{\kdfun{\ioverline*{k}{\ty}}{\kdtype}}}[\projctx{\ctxalt}{\sctxalt(\skd)}]
  }}}{\ioverline*{j - 1}{\sval}}] \\ \subseteq / \supseteq \\
\sem{\wfty{
  \ctxalt \cons
  \annkd{\tyctor}{\kdfun{\ioverline*{k}{\ty}}{\kdtype}} \cons
  \ioverline*{j - 1}{\annty{\var}{\tyalt}}
}{\tyalt_j}}[\pair{\pair{\sctxalt(\skdalt)}{\pair{X}{
    \sembot{\wftm{\ctxalt}{\decl}{\kdfun{\ioverline*{k}{\ty}}{\kdtype}}}[\projctx{\ctxalt}{\sctxalt(\skdalt)}]
}}}{\ioverline*{j - 1}{\sval}}]
\end{array}$} \] which follows from the \ih{} and the fact that
$\projrun{\ctxalt}{\sctxalt(\skd)} = \projrun{\ctxalt}{\sctxalt(\skdalt)}$ and
$\projctx{\ctxalt}{\sctxalt(\skd)} = \projctx{\ctxalt}{\sctxalt(\skdalt)}$

\item \RuleKdIndStrat{label} when $p \in \{\pospos, \posneg \}$ the
only possible case is that $\positive{\tyctor}{-} = \posnone$ and
hence
\begin{multline*}
\sem{\wftm{\ctxalt}{\declstrat}{\ioverline*{k}{\ty} \to \kdtype}}[\sctxalt(\skd)]
\\ =
\sem{\wftm{\ctxalt}{\declstrat}{\ioverline*{k}{\ty} \to \kdtype}}[\sctxalt(\skdalt)]
\end{multline*}
as the semantics of the kind is independent of the refined kind
assignment of $\tyctor$.

\end{itemize}
\end{proof}

\begin{proof}[Proof of \Cref{thm:monotonicity-kinds}]
We use the following abbreviations
\begin{itemize}
\item $\ctxalt \defeq \ctx \cons
  \annkd{\tyctor}{\kdalt}
  \cons \ioverline{i}{\ann{\var}{\tyalt}}$
\item $\sctxalt(x) = \pair{
  \pair{\sctx}{\pair{x}{\skd\bots}}
}{
  \ioverline{i}{\sval}
}$
\end{itemize}

We prove the case for $p = \posneg$, the case for $p = \pospos$ is
identical but with the inequalities reversed.

First, we notice that all kinds have the shape
$\kdfun{\ioverline*{k}{\ty}}{\kdtype}$, thus if $\positive{\tyctor}{\kd}
\leq \posneg$ then $\forall k\; \positive{\tyctor}{\ty_k} \leq \pospos$.
Now, we can proceed by induction on $k$:
\begin{itemize}
\item $k = 0$, then $\kd = \kdtype$, and
  $\sem{\wfkd{\ctxalt}{\kdtype}}[\sctxalt(\skd)] = \per{\dinf} =
  \sem{\wfkd{\ctxalt}{\kdtype}}[\sctxalt(\skdalt)]$.
\item $k > 0$, then $\kd = \kdfun{\ty}{\kd}$, and
\begin{align*}
  & \sem{\wfkd{\ctxalt}{\kdfun{\ty}{\kd}}}[\sctxalt(\skd)] \\
= & \begin{setbuild}{\sfun}
  \sfun \in \dinf \to \left(\dinf \to \right)^{\arity{\kd}} \per{\dinf} \\
  \forall \sval \in \dom{\sem{\wfty{\ctxalt}{\ty}}[\sctxalt(\skd)]} \;
  \sfun(\sval) \in \sem{\wfkd{\ctxalt \cons \annty{\var}{\ty}}{\kd}}[\pair{\sctxalt(\skd)}{\sval}] \\
  \forall \pair{\sval}{\svalalt} \in \left(\sembot{\wfty{\ctxalt}{\ty}}[\projctx{\ctxalt}{\sctxalt(\skd)}] \cap \dom{\sem{\wfty{\ctxalt}{\ty}}[\sctxalt(\skd)]}^2\right) \;
  \sfun(\sval) = \sfun(\svalalt)
\end{setbuild} \\
  & \text{By \Cref{thm:monotonicity-types} and }
  \projctx{\ctxalt}{\sctxalt(\skd)} = \projctx{\ctxalt}{\sctxalt(\skdalt)} \\
\supseteq & \begin{setbuild}{\sfun}
  \sfun \in \dinf \to \left(\dinf \to \right)^{\arity{\kd}} \per{\dinf} \\
  \forall \sval \in \dom{\sem{\wfty{\ctxalt}{\ty}}[\sctxalt(\skdalt)]} \;
  \sfun(\sval) \in \sem{\wfkd{\ctxalt \cons \annty{\var}{\ty}}{\kd}}[\pair{\sctxalt(\skd)}{\sval}] \\
  \forall \pair{\sval}{\svalalt} \in \left(\sembot{\wfty{\ctxalt}{\ty}}[\projctx{\ctxalt}{\sctxalt(\skdalt)}] \cap \dom{\sem{\wfty{\ctxalt}{\ty}}[\sctxalt(\skdalt)]}^2\right) \;
  \sfun(\sval) = \sfun(\svalalt)
\end{setbuild} \\
& \text{By \ih{}} \\
\supseteq & \begin{setbuild}{\sfun}
  \sfun \in \dinf \to \left(\dinf \to \right)^{\arity{\kd}} \per{\dinf} \\
  \forall \sval \in \dom{\sem{\wfty{\ctxalt}{\ty}}[\sctxalt(\skdalt)]} \;
  \sfun(\sval) \in \sem{\wfkd{\ctxalt \cons \annty{\var}{\ty}}{\kd}}[\pair{\sctxalt(\skdalt)}{\sval}] \\
  \forall \pair{\sval}{\svalalt} \in \left(\sembot{\wfty{\ctxalt}{\ty}}[\projctx{\ctxalt}{\sctxalt(\skdalt)}] \cap \dom{\sem{\wfty{\ctxalt}{\ty}}[\sctxalt(\skdalt)]}^2\right) \;
  \sfun(\sval) = \sfun(\svalalt)
\end{setbuild} \\
= & \sem{\wfkd{\ctxalt}{\kdfun{\ty}{\kd}}}[\sctxalt(\skdalt)]
\end{align*}

\end{itemize}

\end{proof}

\begin{proof}[Proof of \Cref{lem:per-respecting-eliminator}]
By induction on the structure of $\ty$:
\begin{itemize}
\item If $\ty = \tyctor \app \ioverline*{k}{\tm}$,
\begin{align*}
& \vpair{
  \tmel{\tyctor}{\tyctor \app \ioverline*{k}{\tm}}{
    \sctx_\rho
  }{
    \semtm{\tmaltaltalt}[\sctx_\rho]
  }{r}
}{
  \tmel{\tyctor}{\tyctor \app \ioverline*{k}{\tm}}{
    \sctxalt_\rho
  }{
    \semtm{\tmaltaltalt}[\sctxalt_\rho]
  }{r'}
} \\
= & \vpair{
  r \app \ioverline{k}{\semtm{\tm}[\sctx_\rho]} \app
  \semtm{\tmaltaltalt}[\sctx_\rho]
}{
  r' \app \ioverline{k}{\semtm{\tm}[\sctxalt_\rho]} \app
  \semtm{\tmaltaltalt}[\sctxalt_\rho]
} \\
& \text{As $\tmaltaltalt : \tyctor \app \ioverline*{k}{\tm}$ gives
  $\semtm{\tmaltaltalt} \in \skd \app \ioverline{k}{\semtm{\tm}}$, the realizer hypothesis yields} \\
\in & \sem{\wfty{
  \ctx \cons
  \ioverline*{k}{\annty{\var}{\ty}} \cons
  \annty{\varalt}{D \tmapp \ioverline*{k}{\var}}
}{
  \tyaltalt
}}[\pair{\pair{\sctx}{
  \ioverline*{k}{\semtm{\tm}[\sctx_\rho]}
}}{
  \semtm{\tmaltaltalt}[\sctx_\rho]
}] \\
& \text{By \Cref{thm:substitution-refined-ty-tm}} \\
= & \sem{\wfty{
  \ctx \cons \ctxalt(j - 1)
}{
  \tyel{\tyctor}{\tyaltalt}{\tmaltaltalt}{\tyctor \app \ioverline*{k}{\tm}}
}}[\pair{\pair{\sctx}{\pair{\skd}{\skd\bots(\sctx)}}}{\ioverline{j - 1}{\sval}}]
\end{align*}
\item If $\ty \in \{\tybase, \ty \app \ioverline*{n}{\tm}\}$

\begin{align*}
& \vpair{
  \tmel{\tyctor}{\ty}{
    \sctx_\rho
  }{
    \semtm{\tmaltaltalt}[\sctx_\rho]
  }{r}
}{
  \tmel{\tyctor}{\ty}{
    \sctxalt_\rho
  }{
    \semtm{\tmaltaltalt}[\sctxalt_\rho]
  }{r'}
} \\
= & \pair{\sunit}{\sunit} \\
\in & \sem{\wfty{
  \ctx \cons \ctxalt(j - 1)
}{
  \tyunit
}}[\pair{\pair{\sctx}{\pair{\skd}{\skd\bots(\sctx)}}}{\ioverline{j - 1}{\sval}}] \\
= & \sem{\wfty{
  \ctx \cons \ctxalt(j - 1)
}{
  \tyel{\tyctor}{\tyaltalt}{\tmaltaltalt}{\ty}
}}[\pair{\pair{\sctx}{\pair{\skd}{\skd\bots(\sctx)}}}{\ioverline{j - 1}{\sval}}]
\end{align*}

\item If $\ty = \tyref{\var}{\ty}{\prop}$
\begin{align*}
& \vpair{
  \tmel{\tyctor}{\tyref{\var}{\ty}{\prop}}{
    \sctx_\rho
  }{
    \semtm{\tmaltaltalt}[\sctx_\rho]
  }{r}
}{
  \tmel{\tyctor}{\tyref{\var}{\ty}{\prop}}{
    \sctxalt_\rho
  }{
    \semtm{\tmaltaltalt}[\sctxalt_\rho]
  }{r'}
} \\
= & \vpair{
  \tmel{\tyctor}{\ty}{
    \sctx_\rho
  }{
    \semtm{\tmaltaltalt}[\sctx_\rho]
  }{r}
}{
  \tmel{\tyctor}{\ty}{
    \sctxalt_\rho
  }{
    \semtm{\tmaltaltalt}[\sctxalt_\rho]
  }{r'}
} \\
& \text{By induction hypothesis} \\
\in & \sem{\wfty{
  \ctx \cons \ctxalt(j - 1)
}{
  \tyel{\tyctor}{\tyaltalt}{\tmaltaltalt}{\ty}
}}[\pair{\pair{\sctx}{\pair{\skd}{\skd\bots(\sctx)}}}{\ioverline{j - 1}{\sval}}] \\
= & \sem{\wfty{
  \ctx \cons \ctxalt(j - 1)
}{
  \tyel{\tyctor}{\tyaltalt}{\tmaltaltalt}{\tyref{\var}{\ty}{\prop}}
}}[\pair{\pair{\sctx}{\pair{\skd}{\skd\bots(\sctx)}}}{\ioverline{j - 1}{\sval}}]
\end{align*}

\item If $\ty = \tyfun[{\var}]{\ty}{\tyalt}$
Fixed $\pair{\svalaltalt}{\svalaltaltalt} \in \sem{\wfty{
  \ctx \cons \ctxalt(j - 1)
}{
  \ty
}}[\pair{\pair{\sctx}{\pair{\skd}{\skd\bots(\sctx)}}}{\ioverline{j - 1}{\sval}}]$ by \ih{}
\begin{align*}
\text{(i)} & \vpair{\tmel{\tyctor}{\tyalt}{
    \pair{\sctx_\rho}{\svalaltalt}
  }{
    \semtm{\tmaltaltalt \app \var}[\pair{\sctx_\rho}{\svalaltalt}]
  }{r}
}{\tmel{\tyctor}{\tyalt}{
    \pair{\sctxalt_\rho}{\svalaltaltalt}
  }{
    \semtm{\tmaltaltalt \app \var}[\pair{\sctxalt_\rho}{\svalaltaltalt}]
  }{r'}
} \\
\in & \sem{\wfty{
  \ctx \cons \ctxalt(j - 1) \cons
  \annty{\var}{\ty}
}{
  \tyel{\tyctor}{\tyaltalt}{\tmaltaltalt \app \var}{\tyalt}
}}[\pair{\pair{\pair{\sctx}{\pair{\skd}{\skd\bots(\sctx)}}}{\ioverline{j - 1}{\sval}}}{\svalaltalt}]
\end{align*}
And
fixed $\pair{\svalaltalt\bots}{\svalaltaltalt\bots} \in \sembot{\wfty{
  \ctx \cons \ctxalt(j - 1)
}{
  \ty
}}[\projctx{
  \ctx \cons \ctxalt(j - 1)
}{\pair{\pair{\sctx}{\pair{\skd}{\skd\bots(\sctx)}}}{\ioverline{j - 1}{\sval}}}]$
by \Cref{lem:unref-per-respecting-eliminator}
\begin{align*}
\text{(ii)} & \vpair{\tmel{\tyctor}{\tyalt}{
    \pair{\sctx_\rho}{\svalaltalt\bots}
  }{
    \semtm{\tmaltaltalt \app \var}[\pair{\sctx_\rho}{\svalaltalt\bots}]
  }{r}
}{\tmel{\tyctor}{\tyalt}{
    \pair{\sctxalt_\rho}{\svalaltaltalt\bots}
  }{
    \semtm{\tmaltaltalt \app \var}[\pair{\sctxalt_\rho}{\svalaltaltalt\bots}]
  }{r'}
} \\
\in & \sembot{\wfty{
  \ctx \cons \ctxalt(j - 1) \cons
  \annty{\var}{\ty}
}{
  \tyel{\tyctor}{\tyaltalt}{\tmaltaltalt \app \var}{\tyalt}
}}[\pair{\projctx{
  \ctx \cons \ctxalt(j - 1)
}{\pair{\pair{\sctx}{\pair{\skd}{\skd\bots(\sctx)}}}{\ioverline{j - 1}{\sval}}}}{\svalaltalt\bots}]
\end{align*}
Hence,
\begin{align*}
& \vpair{
  \tmel{\tyctor}{\tyfun[{\var}]{\ty}{\tyalt}}{
    \sctx_\rho
  }{
    \semtm{\tmaltaltalt}[\sctx_\rho]
  }{r}
}{
  \tmel{\tyctor}{\tyfun[{\var}]{\ty}{\tyalt}}{
    \sctxalt_\rho
  }{
    \semtm{\tmaltaltalt}[\sctxalt_\rho]
  }{r'}
} \\
= & \vpair{\demb{\to}{
  \svalaltalt \mapsto \tmel{\tyctor}{\tyalt}{
    \pair{\sctx_\rho}{\svalaltalt}
  }{
    \semtm{\tmaltaltalt}[\sctx_\rho] \app \svalaltalt
  }{r}
}}{\demb{\to}{
  \svalaltalt \mapsto \tmel{\tyctor}{\tyalt}{
    \pair{\sctxalt_\rho}{\svalaltalt}
  }{
    \semtm{\tmaltaltalt}[\sctxalt_\rho] \app \svalaltalt
  }{r'}
}} \\
& \text{We notice that $\semtm{\tm \app \var}[\pair{\sctx}{\sval}] =
  \semtm{\tm}[\sctx] \app \sval$ if $x$ is the last
  bound variable} \\
= & \vpair{\demb{\to}{
  \svalaltalt \mapsto \tmel{\tyctor}{\tyalt}{
    \pair{\sctx_\rho}{\svalaltalt}
  }{
    \semtm{\tmaltaltalt \app \var}[\pair{\sctx_\rho}{\svalaltalt}]
  }{r}
}}{\demb{\to}{
  \svalaltalt \mapsto \tmel{\tyctor}{\tyalt}{
    \pair{\sctxalt_\rho}{\svalaltalt}
  }{
    \semtm{\tmaltaltalt \app \var}[\pair{\sctxalt_\rho}{\svalaltalt}]
  }{r'}
}} \\
& \text{By (i) and (ii)} \\
\in & \sem{\wfty{
  \ctx \cons \ctxalt(j - 1)
}{
  \tyel{\tyctor}{\tyaltalt}{\tmaltaltalt}{\tyfun[{\var}]{\ty}{\tyalt}}
}}[\pair{\pair{\sctx}{\pair{\skd}{\skd\bots(\sctx)}}}{\ioverline{j - 1}{\sval}}]
\end{align*}

\item $\ty = \tyall{\var}{\kd}{\ty}$

First, we notice that for any
$\skdalt \in \sem{\wfkd{\ctx \cons \ctxalt(j - 1)}{\kd}}[\pair{\pair{\sctx}{\pair{\skd}{\skd\bots(\sctx)}}}{\ioverline{j - 1}{\sval}}]$, \\
$\skdalt\bots \in \sembot{\wfkd{\ctx \cons \ctxalt(j - 1)}{\kd}}[\pair{\pair{\sctx}{\pair{\skd}{\skd\bots(\sctx)}}}{\ioverline{j - 1}{\sval}}]$,
by IH
\begin{align*}
& \vpair{\tmel{\tyctor}{\ty}{
    \projrun{
      \ctx \cons \ctxalt(j - 1) \cons
      \annkd{\var}{\kd}
    }{\pair{\pair{\pair{\sctx}{\pair{\skd}{\skd\bots(\sctx)}}}{\ioverline{j - 1}{\sval}}}{\pair{\skdalt}{\skdalt\bots}}}
  }{
    \semtm{\tmaltaltalt \app \var}[\sctx_\rho]
  }{r}
}{
  \tmel{\tyctor}{\ty}{
    \projrun{
      \ctx \cons \ctxalt(j - 1) \cons
      \annkd{\var}{\kd}
    }{\pair{\pair{\pair{\sctxalt}{\pair{\skd}{\skd\bots(\sctxalt)}}}{\ioverline{j - 1}{\svalalt}}}{\pair{\skdalt}{\skdalt\bots}}}
  }{
    \semtm{\tmaltaltalt \app \var}[\sctxalt_\rho]
  }{r'}
} \\
\in & \sem{\wfty{
  \ctx \cons \ctxalt(j - 1) \cons
  \annkd{\var}{\kd}
}{
  \tyel{\tyctor}{\tyaltalt}{\tmaltaltalt \app \var}{\ty}
}}[\pair{\pair{\pair{\sctx}{\pair{\skd}{\skd\bots(\sctx)}}}{\ioverline{j - 1}{\sval}}}{\pair{\skdalt}{\skdalt\bots}}] \\
\end{align*}

\begin{align*}
& \vpair{
  \tmel{\tyctor}{\tyall{\var}{\kd}{\ty}}{
    \sctx_\rho
  }{
    \semtm{\tmaltaltalt}[\sctx_\rho]
  }{r}
}{
  \tmel{\tyctor}{\tyall{\var}{\kd}{\ty}}{
    \sctxalt_\rho
  }{
    \semtm{\tmaltaltalt}[\sctxalt_\rho]
  }{r'}
} \\
= & \vpair{\tmel{\tyctor}{\ty}{
    \sctx_\rho
  }{
    \semtm{\tmaltaltalt}[\sctx_\rho]
  }{r}
}{
  \tmel{\tyctor}{\ty}{
    \sctxalt_\rho
  }{
    \semtm{\tmaltaltalt}[\sctxalt_\rho]
  }{r'}
} \\
& \text{$\projrun{\ctx \cons \annkd{\var}{\kd}}{\pair{\sctx}{x}} = \projrun{\ctx}{\sctx}$} \\
= & \vpair{\tmel{\tyctor}{\ty}{
    \sctx_\rho
  }{
    \semtm{\tmaltaltalt \app \var}[\sctx_\rho]
  }{r}
}{
  \tmel{\tyctor}{\ty}{
    \sctxalt_\rho
  }{
    \semtm{\tmaltaltalt \app \var}[\sctxalt_\rho]
  }{r'}
} \\
& \text{By the first observation} \\
\in & \sem{\wfty{
  \ctx \cons \ctxalt(j - 1)
}{
  \tyall{\var}{\kd}{
  \tyel{\tyctor}{\tyaltalt}{\tmaltaltalt \app \var}{\ty}}
}}[{\pair{\pair{\sctx}{\pair{\skd}{\skd\bots(\sctx)}}}{\ioverline{j - 1}{\sval}}}] \\
= & \sem{\wfty{
  \ctx \cons \ctxalt(j - 1)
}{
  \tyel{\tyctor}{\tyaltalt}{\tmaltaltalt}{\tyall{\var}{\kd}{\ty}}
}}[\pair{\pair{\sctx}{\pair{\skd}{\skd\bots(\sctx)}}}{\ioverline{j - 1}{\sval}}]
\end{align*}

\end{itemize}
\end{proof}

\begin{proof}[Proof of \Cref{lem:unref-per-respecting-eliminator}]
By induction on the structure of $\ty$:
\begin{itemize}
\item If $\ty = \tyctor \app \ioverline*{k}{\tm}$,
\begin{align*}
& \vpair{
  \tmel{\tyctor}{\tyctor \app \ioverline*{k}{\tm}}{\sctx_\rho}{\semtm{\tmaltaltalt}[\sctx_\rho]}{r}
}{
  \tmel{\tyctor}{\tyctor \app \ioverline*{k}{\tm}}{\sctxalt_\rho}{\semtm{\tmaltaltalt}[\sctxalt_\rho]}{r'}
} \\
= & \vpair{
  r \app \ioverline{k}{\semtm{\tm}[\sctx_\rho]} \app \semtm{\tmaltaltalt}[\sctx_\rho]
}{
  r' \app \ioverline{k}{\semtm{\tm}[\sctxalt_\rho]} \app \semtm{\tmaltaltalt}[\sctxalt_\rho]
} \\
& \text{By \Crefsub{thm:per-respecting-terms}{unrefined}, since $\sctx_\rho, \sctxalt_\rho$ are related,
  the index terms give a related pair} \\
& \text{$\pair{\ioverline{k}{\semtm{\tm}[\sctx_\rho]}}{\ioverline{k}{\semtm{\tm}[\sctxalt_\rho]}}$, and, as
  $\tmaltaltalt : \tyctor \app \ioverline*{k}{\tm}$,} \\
& \text{$\pair{\semtm{\tmaltaltalt}[\sctx_\rho]}{\semtm{\tmaltaltalt}[\sctxalt_\rho]} \in
  \skd\bots(\sctx) \app \ioverline*{k}{\semtm{\tm}[\sctx_\rho]}$; the realizer hypothesis yields} \\
\in & \sembot{\wfty{
  \ctx \cons
  \ioverline*{k}{\annty{\var}{\ty}} \cons
  \annty{\varalt}{\tyctor \app \ioverline*{k}{\var}}
}{
  \tyaltalt
}}[\pair{\sctx}{\skd\bots(\sctx)}] \\
& \text{Since $\sembot{-}$ does not depend on the term assignment} \\
= & \sembot{\wfty{
  \ctx \cons \ctxalt(j - 1)
}{
  \tyel{\tyctor}{\tyaltalt}{\tmaltaltalt}{\tyctor \app \ioverline*{k}{\tm}}
}}[\pair{\pair{\sctx}{\skd\bots(\sctx)}}{\sctxaltalt}]
\end{align*}

\item If $\ty \in \{\tybase, \ty \app \ioverline*{n}{\tm}\}$,
\begin{align*}
& \vpair{
  \tmel{\tyctor}{\ty}{\sctx_\rho}{\semtm{\tmaltaltalt}[\sctx_\rho]}{r}
}{
  \tmel{\tyctor}{\ty}{\sctxalt_\rho}{\semtm{\tmaltaltalt}[\sctxalt_\rho]}{r'}
} \\
= & \pair{\sunit}{\sunit} \\
\in & \sembot{\wfty{
  \ctx \cons \ctxalt(j - 1)
}{\tyunit}}[\pair{\pair{\sctx}{\skd\bots(\sctx)}}{\sctxaltalt}] \\
= & \sembot{\wfty{
  \ctx \cons \ctxalt(j - 1)
}{\tyel{\tyctor}{\tyaltalt}{\tmaltaltalt}{\ty}}}[\pair{\pair{\sctx}{\skd\bots(\sctx)}}{\sctxaltalt}]
\end{align*}

\item If $\ty = \tyref{\var}{\ty}{\prop}$,
\begin{align*}
& \vpair{
  \tmel{\tyctor}{\tyref{\var}{\ty}{\prop}}{\sctx_\rho}{\semtm{\tmaltaltalt}[\sctx_\rho]}{r}
}{
  \tmel{\tyctor}{\tyref{\var}{\ty}{\prop}}{\sctxalt_\rho}{\semtm{\tmaltaltalt}[\sctxalt_\rho]}{r'}
} \\
= & \vpair{
  \tmel{\tyctor}{\ty}{\sctx_\rho}{\semtm{\tmaltaltalt}[\sctx_\rho]}{r}
}{
  \tmel{\tyctor}{\ty}{\sctxalt_\rho}{\semtm{\tmaltaltalt}[\sctxalt_\rho]}{r'}
} \\
& \text{By \ih{}} \\
\in & \sembot{\wfty{
  \ctx \cons \ctxalt(j - 1)
}{
  \tyel{\tyctor}{\tyaltalt}{\tmaltaltalt}{\ty}
}}[\pair{\pair{\sctx}{\skd\bots(\sctx)}}{\sctxaltalt}] \\
= & \sembot{\wfty{
  \ctx \cons \ctxalt(j - 1)
}{
  \tyel{\tyctor}{\tyaltalt}{\tmaltaltalt}{\tyref{\var}{\ty}{\prop}}
}}[\pair{\pair{\sctx}{\skd\bots(\sctx)}}{\sctxaltalt}]
\end{align*}

\item If $\ty = \tyfun[{\var}]{\ty}{\tyalt}$,
fixed $\pair{\svalaltalt\bots}{\svalaltaltalt\bots} \in \sembot{\wfty{
  \ctx \cons \ctxalt(j - 1)
}{\ty}}[\pair{\pair{\sctx}{\skd\bots(\sctx)}}{\sctxaltalt}]$
by \ih{}
\begin{align*}
& \vpair{\tmel{\tyctor}{\tyalt}{
    \pair{\sctx_\rho}{\svalaltalt\bots}
  }{
    \semtm{\tmaltaltalt \app \var}[\pair{\sctx_\rho}{\svalaltalt\bots}]
  }{r}
}{\tmel{\tyctor}{\tyalt}{
    \pair{\sctxalt_\rho}{\svalaltaltalt\bots}
  }{
    \semtm{\tmaltaltalt \app \var}[\pair{\sctxalt_\rho}{\svalaltaltalt\bots}]
  }{r'}
} \\
\in & \sembot{\wfty{
  \ctx \cons \ctxalt(j - 1) \cons
  \annty{\var}{\ty}
}{
  \tyel{\tyctor}{\tyaltalt}{\tmaltaltalt \app \var}{\tyalt}
}}[\pair{\pair{\sctx}{\skd\bots(\sctx)}}{\sctxaltalt}]
\end{align*}
Hence,
\begin{align*}
& \vpair{
  \tmel{\tyctor}{\tyfun[{\var}]{\ty}{\tyalt}}{\sctx_\rho}{\semtm{\tmaltaltalt}[\sctx_\rho]}{r}
}{
  \tmel{\tyctor}{\tyfun[{\var}]{\ty}{\tyalt}}{\sctxalt_\rho}{\semtm{\tmaltaltalt}[\sctxalt_\rho]}{r'}
} \\
= & \vpair{\demb{\to}{
  \svalaltalt \mapsto \tmel{\tyctor}{\tyalt}{
    \pair{\sctx_\rho}{\svalaltalt}
  }{
    \semtm{\tmaltaltalt}[\sctx_\rho] \app \svalaltalt
  }{r}
}}{\demb{\to}{
  \svalaltalt \mapsto \tmel{\tyctor}{\tyalt}{
    \pair{\sctxalt_\rho}{\svalaltalt}
  }{
    \semtm{\tmaltaltalt}[\sctxalt_\rho] \app \svalaltalt
  }{r'}
}} \\
& \text{We notice that $\semtm{\tm \app \var}[\pair{\sctx}{\sval}] =
  \semtm{\tm}[\sctx] \app \sval$ if $x$ is the last bound variable} \\
= & \vpair{\demb{\to}{
  \svalaltalt \mapsto \tmel{\tyctor}{\tyalt}{
    \pair{\sctx_\rho}{\svalaltalt}
  }{
    \semtm{\tmaltaltalt \app \var}[\pair{\sctx_\rho}{\svalaltalt}]
  }{r}
}}{\demb{\to}{
  \svalaltalt \mapsto \tmel{\tyctor}{\tyalt}{
    \pair{\sctxalt_\rho}{\svalaltalt}
  }{
    \semtm{\tmaltaltalt \app \var}[\pair{\sctxalt_\rho}{\svalaltalt}]
  }{r'}
}} \\
& \text{By the above} \\
\in & \sembot{\wfty{
  \ctx \cons \ctxalt(j - 1)
}{
  \tyel{\tyctor}{\tyaltalt}{\tmaltaltalt}{\tyfun[{\var}]{\ty}{\tyalt}}
}}[\pair{\pair{\sctx}{\skd\bots(\sctx)}}{\sctxaltalt}]
\end{align*}

\item $\ty = \tyall{\var}{\kd}{\ty}$

First, we notice that for any
$\skdalt \in \sembot{\wfkd{\ctx \cons \ctxalt(j - 1)}{\kd}}[\pair{\pair{\sctx}{\skd\bots(\sctx)}}{\sctxaltalt}]$,
since the runtime projection does not depend on the
newly bound kind variable, by IH
\begin{align*}
& \vpair{\tmel{\tyctor}{\ty}{
    \sctx_\rho
  }{
    \semtm{\tmaltaltalt \app \var}[\sctx_\rho]
  }{r}
}{
  \tmel{\tyctor}{\ty}{
    \sctxalt_\rho
  }{
    \semtm{\tmaltaltalt \app \var}[\sctxalt_\rho]
  }{r'}
} \\
\in & \sembot{\wfty{
  \ctx \cons \ctxalt(j - 1) \cons \annkd{\var}{\kd}
}{
  \tyel{\tyctor}{\tyaltalt}{\tmaltaltalt \app \var}{\ty}
}}[\pair{\pair{\pair{\sctx}{\skd\bots(\sctx)}}{\sctxaltalt}}{\skdalt}]
\end{align*}

\begin{align*}
& \vpair{
  \tmel{\tyctor}{\tyall{\var}{\kd}{\ty}}{\sctx_\rho}{\semtm{\tmaltaltalt}[\sctx_\rho]}{r}
}{
  \tmel{\tyctor}{\tyall{\var}{\kd}{\ty}}{\sctxalt_\rho}{\semtm{\tmaltaltalt}[\sctxalt_\rho]}{r'}
} \\
= & \vpair{
  \tmel{\tyctor}{\ty}{\sctx_\rho}{\semtm{\tmaltaltalt}[\sctx_\rho]}{r}
}{
  \tmel{\tyctor}{\ty}{\sctxalt_\rho}{\semtm{\tmaltaltalt}[\sctxalt_\rho]}{r'}
} \\
& \text{$\projrun{\ctx \cons \annkd{\var}{\kd}}{\pair{\sctx}{x}} = \projrun{\ctx}{\sctx}$} \\
= & \vpair{
  \tmel{\tyctor}{\ty}{\sctx_\rho}{\semtm{\tmaltaltalt \app \var}[\sctx_\rho]}{r}
}{
  \tmel{\tyctor}{\ty}{\sctxalt_\rho}{\semtm{\tmaltaltalt \app \var}[\sctxalt_\rho]}{r'}
} \\
& \text{By the first observation} \\
\in & \sembot{\wfty{
  \ctx \cons \ctxalt(j - 1)
}{
  \tyall{\var}{\kd}{\tyel{\tyctor}{\tyaltalt}{\tmaltaltalt \app \var}{\ty}}
}}[\pair{\pair{\sctx}{\skd\bots(\sctx)}}{\sctxaltalt}] \\
= & \sembot{\wfty{
  \ctx \cons \ctxalt(j - 1)
}{
  \tyel{\tyctor}{\tyaltalt}{\tmaltaltalt}{\tyall{\var}{\kd}{\ty}}
}}[\pair{\pair{\sctx}{\skd\bots(\sctx)}}{\sctxaltalt}]
\end{align*}
\end{itemize}
\end{proof}

%%% Local Variables:
%%% mode: LaTeX
%%% TeX-master: "main"
%%% TeX-engine: default
%%% End:

\end{document}

\begin{comment}
  Move annotation layer
\end{comment}

%%% Local Variables:
%%% mode: LaTeX
%%% TeX-master: t
%%% TeX-engine: default
%%% End: